\documentclass[
12pt, 
english, 
onehalfspacing, 
liststotoc, 
parskip, 
headsepline, 
a4paper]{MastersDoctoralThesis} 

\usepackage{amsmath}
\usepackage{amssymb}
\usepackage{amsfonts}
\usepackage{amsbsy}
\usepackage{mathrsfs}
\usepackage{bm}
\usepackage{relsize}
\usepackage{enumitem}
\newcounter{numln}
\usepackage{multirow}
\usepackage{adjustbox}
\usepackage{url}
\usepackage[T1]{fontenc}
\usepackage{pbsi}
\usepackage{float}

\usepackage{booktabs} \usepackage{multirow}

\usepackage{multicol}

\usepackage{cite}
\usepackage{ragged2e}

\usepackage{afterpage}
\newlist{numitemise}{itemize}{2}
\setlist[numitemise]{wide}
\setlist[numitemise, 1]{labelindent=0pt,labelwidth=2em, label=\stepcounter{numln}\makebox[2em]{\thenumln.\hfill}, leftmargin=\dimexpr\labelwidth+\labelsep\relax}
\setlist[numitemise, 2]{labelindent=\dimexpr -2em-\labelsep\relax, labelwidth=\dimexpr 2em+\labelsep\relax, label=\stepcounter{numln}\makebox[\dimexpr\labelwidth + \labelsep\relax]{\thenumln.\hfill\textbullet}, leftmargin=\dimexpr\leftmargin+2\labelsep\relax}

\usepackage{mathpazo}

\definecolor{darkblue}{rgb}{0.0, 0.0, 0.62}
\definecolor{deepmagenta}{rgb}{0.8, 0.0, 0.7}
\definecolor{darkred}{rgb}{0.55, 0.0, 0.0}
\definecolor{violetryb}{rgb}{0.53, 0.0, 0.69}
\definecolor{royalpurple}{rgb}{0.47, 0.32, 0.66}
\definecolor{regalia}{rgb}{0.32, 0.18, 0.5}
\definecolor{purpleheart}{rgb}{0.41, 0.21, 0.61}
\definecolor{plum}{rgb}{0.56, 0.27, 0.52}
\definecolor{deeppurple}{rgb}{0.41, 0.16, 0.38}

\RequirePackage[numbers,sort&compress]{natbib}
\AtBeginDocument{\hypersetup{breaklinks=true,citecolor=darkred,linkcolor=darkblue,linktoc=page}}

\usepackage{setspace}
\usepackage{graphicx}
\usepackage{wasysym}
\usepackage{comment}
\usepackage{textgreek}
\usepackage{mathtools}
\usepackage{commath}
\usepackage{color}

\usepackage{wrapfig}
\usepackage{makecell}
\usepackage{datetime}

\newdateformat{monthyeardate}{\monthname[\THEMONTH], \THEYEAR}

\thesistitle{{Non-parametric Reconstruction Of\\
		Some Cosmological Parameters}} 

\supervisor{Prof. Narayan Banerjee} 

\examiner{} 

\degree{Doctor of Philosophy} 

\author{Purba Mukherjee} 
\addresses{} 

\subject{Physical Sciences} \keywords{} \university{\href{https://www.iiserkol.ac.in/}{IISER Kolkata}} \department{Department of Physical Sciences}

\AtBeginDocument{
\hypersetup{pdftitle=\ttitle} \hypersetup{pdfauthor=\authorname} \hypersetup{pdfkeywords=\keywordnames} }

\begin{document}
\frontmatter 

\pagestyle{plain} 

\begin{titlepage}
\begin{center}

\textsc{\Large Doctoral Thesis}\\[0.75cm] \HRule \\[0.4cm] {\huge \textbf{\ttitle}\par}\vspace{0.4cm} \HRule \\[1.5cm] 

\begin{center}
	\large{Purba Mukherjee \\
	(~14IP011~)}\\ \vspace{0.5cm}
	\large{{\emph{under the supervision of}\\ Prof. Narayan Banerjee}}\\ \vspace{0.5cm}
    \large{Department of Physical Sciences\\
    Indian Institute of Science Education and Research\\
     Kolkata\vspace{0.0cm}}
\end{center}

\vspace{0.5cm}

{\centering
	\includegraphics[width=0.3\textwidth]{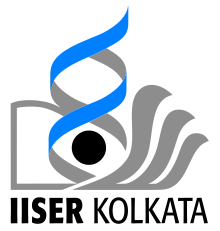}}
	
\vspace{1.cm}

\textit{ A thesis submitted in fulfilment of the requirements for 
the degree of\\ \degreename \ in the \deptname \ at the\\ \ Indian Institute of 
Science Education and Research Kolkata }

\vspace{0.6cm}

{\large March, 2022}\\[4cm] 

\vfill
\end{center}
\end{titlepage}

\sloppy
\begin{declaration}
\noindent \par\vspace{10pt}
\vspace{0.5cm}
\begin{FlushLeft}
	\justify
	\par I, Ms. \textbf{Purba Mukherjee} Registration No. \textbf{14IP011} dated \textbf{24th July 2014}, a student of the Department of 
	Physical Sciences of the Integrated PhD Programme of IISER Kolkata, hereby declare that this thesis is my 
	own work and, to the best of my knowledge, it neither contains materials previously published or written 
	by any other person, nor has it been submitted for any degree/diploma or any other academic award anywhere 
	before. I have used the originality checking service to prevent inappropriate copying. \vskip 0.3cm
	
	\par I also declare that all copyrighted material incorporated into this thesis is in compliance with the 
	Indian Copyright Act, 1957 (amended in 2012) and that I have received written permission from the copyright 
	owners for my use of their work. \vskip 0.3cm
	
	\par I hereby grant permission to IISER Kolkata to store the thesis in a database which can be accessed by others.\\
	
\end{FlushLeft}
\vspace{0.1cm}
\begin{flushleft}
	Date: {March 22, 2022} 
\end{flushleft}

\vspace{0.5cm}

\begin{flushright}
\begin{figure}[h!]  
	\begin{flushright}

	\end{flushright}
\end{figure} 
\textbf{Purba Mukherjee} \\
	
	Department of Physical Sciences \\
	
	Indian Institute of Science Education and Research Kolkata\\
	
	Mohanpur 741246, West Bengal, India.
\end{flushright}

\vspace{3.0cm}

\end{declaration}

\cleardoublepage

\clearpage{}\thispagestyle{plain}
\null\vfil
{\noindent \huge \textbf{Certificate from the Supervisor} \par}\vspace{10pt}
\noindent
\vspace{0.5cm}
\begin{FlushLeft}
	\justify 	
	This is to certify that the thesis entitled ``\textbf{Non-parametric Reconstruction Of Some Cosmological Parameters}'' submitted by Ms. \textbf{Purba Mukherjee}, Registration No. 
	\textbf{14IP011} dated 
	\textbf{24th July 2014}, a student of the Department of Physical Sciences of the Integrated PhD Programme/ PhD Programme of IISER Kolkata, is based upon his/her own research 
	work under my supervision. This is also to certify that neither the thesis nor any part of it has been submitted for any degree/diploma or any other academic award 
	anywhere before. In my opinion, the thesis fulfils the requirement for the award of the degree of Doctor of Philosophy. \\
	
\end{FlushLeft}

~~
\vspace{0.1cm}
\begin{flushleft}
	Date: {March 22, 2022} 
\end{flushleft}

\begin{flushright}
	
\begin{figure}[h!]  
\begin{flushright}

\end{flushright}
\end{figure}

\textbf{Narayan Banerjee} \\
	
	Professor \\
	
	Department of Physical Sciences \\
	
	Indian Institute of Science Education and Research Kolkata\\
	
	Mohanpur 741246, West Bengal, India.
	
\end{flushright}
\clearpage{}
\cleardoublepage

\clearpage{}\thispagestyle{plain}
\null\vfil
{\noindent \huge \textbf{Acknowledgements} \par}\vspace{10pt}
\sloppy
This journey would not have been possible without the constant help and support of many people. This is to express my gratitude towards them.

First and foremost, I would like to express my sincere gratitude to my supervisor, Prof. Narayan Banerjee, for his guidance, continuous support, and encouragement throughout my doctorate studies at IISER Kolkata. I am indebted to him for his infinite patience and belief in me, which has made this thesis possible, helping me develop a better understanding of the subject and mentoring me to build an aptitude for scientific research. It was a lifetime opportunity to work under his supervision. I would like to thank him for his incredible teachings, helpful discussions
and invaluable advice that has kept on motivating me in the way of life.

I am grateful to Dr. Ananda Dasgupta for his constant guidance, support, and cooperation. It was my privilege to learn the basics of programming from him. I want to thank him for providing me with every possible solution and new ideas to deal with the problems I faced in python.

I would also like to thank my research progress committee members, Dr. Golam Mortuza Hossain and Prof. Rajesh Kumble Nayak, for their insightful comments and suggestions. I acknowledge financial support from IISER Kolkata and express my heartfelt gratitude to all the academic and non-academic staff members of the IISER Kolkata family. My sincere appreciation for the office staff of the Department of Physical Sciences, IISER Kolkata, and Assistant Librarian, IISER-K Library, for their cooperation and assistance. I would also like to thank ``Dirac Supercomputing Facility", IISER Kolkata, for providing the necessary computational support during my research.

I convey special gratitude towards my seniors and collaborators, Ankan Mukherjee and Soumya Chakrabarti. At times they have been like an elder brother to me. I am sincerely thankful to my fellow friends and co-researchers Basab, Shibendu, Shreya, Sourav, Subhadip and Tanima for making my stay at IISER Kolkata a wonderful and memorable experience. I will cherish the memories I made here for the rest of my life.

I thank my seniors Anushree Datta, Avijit Chowdhury, Chiranjeeb Singha, Gopal Sardar, Nandan Roy, Nivedita Bhadra, Sachin Pandey, Santanu Tripathi, Sayak Ray, Soumyajit Seth, Srijita Sinha, Subhajit Barman for their valuable bits of advice. I am thankful to all my friends Aabir, Alakananda, Ananya, Arnab, Arpan, Brotoraj, Debanjana, Darshana, Diganta, Fareeha, Madhura, Maitraya, Pampa, Prashanti, Ramita, Roshan, Sambo, Saptarshi, Saswata, Satyaki, Samit, Saumya, Siddhartha, Soumi, Souvick, Srijit, Subarna, Subhankar, Sujoy, Swarup and Tista.

Finally, I express my sincere gratitude to my parents for their unconditional love, undeterred support, encouragement, inspiration, and sacrifices. Without them, it would not have been possible to pursue this dream. I wish that someday I would be a better daughter, give them a lot more peace, happiness and make them proud. I cannot end without thanking my grandparents for their love, care, and comfort. I express my deepest affection for my partners in crime, Paromita and Banibrata, who have always been by my side. I sincerely thank them for their wholehearted support and humour over the years. I wish them a lot of happiness and success in life.

\clearpage{}
\cleardoublepage

\clearpage{}\thispagestyle{plain}
\null\vfil
{\noindent \huge \textbf{Abstract} \par}\vspace{10pt}

The present thesis is devoted to the non-parametric reconstruction of some cosmological parameters using diverse observational data sets adopting the Gaussian Process regression. The Universe is assumed to be spatially homogeneous and isotropic, thus described by the FLRW metric. An assessment of the cosmic distance-duality relation, reconstruction of the kinematical quantities and exploring the possibility of a non-gravitational interaction in the cosmic dark sector, between dark energy and dark matter, have been carried out. In case of chapters \ref{ch2:chap2}, \ref{ch4:chap4} and \ref{ch5:chap5} a spatially flat Universe has been assumed at the outset. We have relaxed this flatness assumption in chapter \ref{ch7:chap7} and investigated the possible effect of a non-zero spatial curvature from the Planck 2018 survey.

The first chapter provides a brief introduction to cosmology. In the second chapter, a reconstruction of the cosmic distance duality relation (CDDR) has been discussed. As cosmography is strongly dependent on the validity of CDDR, the reliability of CDDR is evaluated with the increasing quality and quantity of present observational data. Indication towards a non-violation of CDDR in the late-time Universe is obtained.

In the third chapter, a non-parametric reconstruction of the cosmic deceleration and jerk parameters have been discussed. Reconstruction for the effective equation of state parameter has also been carried out for the model-independent datasets. A fitting function for the reconstructed $j$ as a polynomial in $z$, in the low redshift range $0<z<1$, has been obtained. 

The fourth chapter is devoted to exploring the possibility of a non-gravitational interaction between dark matter and the dark energy sector. A non-parametric reconstruction of interaction function $Q$ in the cosmic dark sector has been attempted. Three cases for the dark energy EoS have been considered. These are the decaying vacuum energy $\Lambda$ with $w = -1$, the $w$CDM model and the Chevallier-Polarski-Linder (CPL) parametrization of dark energy. An analytic expression for the reconstructed $Q$ as a polynomial in $z$ has been provided. The evolution of the dark matter and dark energy density parameters ${\Omega}_m$ and ${\Omega}_D$ have also been checked in the presence of this interaction term. The thermodynamics considerations are also studied in the presence of this interacting scenario.

In the fifth chapter, attempts have been made to revisit a non-parametric reconstruction of the cosmic deceleration parameter $q$ using various combinations of recently updated background datasets,  following an improved analysis. The growth rate measurements from the redshift-space distortions have been utilized to investigate the effect of matter perturbations. 

The reconstructed quantities mimic the $\Lambda$CDM behaviour in the very low redshift range. Results indicate that the $\Lambda$CDM model is well consistent and included at the 2$\sigma$ level in the domain of all the reconstructions.  

Finally, chapter six contains the concluding remarks and relevant discussion regarding the overall work presented in the thesis.\clearpage{}
\cleardoublepage

\clearpage{}\thispagestyle{plain}
\null\vfil
{\noindent \huge \textbf{Preface} \par}\vspace{10pt}
\noindent

	\begin{FlushLeft}
\justify  
		
		The research work contained in this thesis was carried out at the Department of Physical Sciences,  Indian Institute of Science Education and Research Kolkata, India.
		\vskip 0.5 cm
		
		Chapter 1 contains an introduction to cosmology and focuses on the reconstruction methods from observational data. The succeeding chapters are based on the following papers:

		\begin{itemize}

			\item Chapter 2
			
			\textbf{Purba Mukherjee} and Ankan Mukherjee, {\it ``Assessment of the cosmic distance duality relation using Gaussian process"}, \href{https://doi.org/10.1093/mnras/stab1054}{Mon. Not. R. Astron. Soc. \textbf{504}, 3938 (2021)}.

			\item Chapter 3
			
			\textbf{Purba Mukherjee} and Narayan Banerjee, {\it ``Non-parametric reconstruction of the cosmological jerk parameter"},  \href{https://doi.org/10.1140/epjc/s10052-021-08830-5}{Eur. Phys. J. C \textbf{81}, 36 (2021)}.

			\item Chapter 4
			
			\textbf{Purba Mukherjee} and Narayan Banerjee, {\it  ``Non-parametric reconstruction of interaction in the cosmic dark sector''}, \href{https://doi.org/10.1103/PhysRevD.103.123530}{Phys. Rev. D \textbf{103}, 123530 (2021)}.

			\item Chapter 5  

			\textbf{Purba Mukherjee} and Narayan Banerjee, {\it ``Revisiting a non-parametric reconstruction of the deceleration parameter from combined background and the growth rate data"},  \href{https://doi.org/10.1016/j.dark.2022.100998}{Phys. Dark Univ. \textbf{36}, 100998 (2022)}.

		\end{itemize}    
		
		Lastly, chapter 6 contains the concluding remarks and relevant discussion regarding  the overall work presented in the thesis.
		
	\end{FlushLeft}
\clearpage{}
\cleardoublepage

\tableofcontents

\begin{abbreviations}{l l } \textbf{2dFGRS}~~~~ &	Two-degree-Field Galaxy Redshift Survey \\[0.2cm]
	
	\textbf{6dFGS}~~~~ &	Six-degree-Field Galaxy Survey \\[0.2cm]
	
	\textbf{AP}~~~~ & Alcock-Paczynski \\[0.2cm]
	
	\textbf{BAO}~~~~ & Baryon Acoustic Oscillations \\[0.2cm]
	
	\textbf{BBC}~~~~ & BEAMS with Bias Corrections \\[0.2cm]
	
	\textbf{BD}~~~~ & Brans-Dicke \\[0.2cm]
	
	\textbf{BOSS}~~~~ & Baryon Oscillation Spectroscopic Survey \\[0.2cm]
	
	\textbf{CC}~~~~ & Cosmic Chronometer 	\\[0.2cm]
	
	\textbf{CDDR}~~~~ & Cosmic Distance-Duality Relation 	\\[0.2cm]
	
	\textbf{CDM}~~~~ & Cold Dark Matter 	\\[0.2cm]
	
	\textbf{CL}~~~~ & Confidence Level 	\\[0.2cm]
	
	\textbf{CMB}~~~~ & Cosmic Microwave Background		\\[0.2cm]
	
	\textbf{CMBR}~~~~ & Cosmic Microwave Background	Radiation	\\[0.2cm]
	
	\textbf{COBE}~~~~ & Cosmic Background Explorer		\\[0.2cm]
	
	\textbf{CPL}~~~~ & Chevallier-Polarski-Linder		\\[0.2cm]
	
	\textbf{DE}~~~~ & Dark Energy			\\[0.2cm]
	
	\textbf{DDE}~~~~ & Dynamical Dark Energy		\\[0.2cm]
	
	\textbf{DES}~~~~ & Dark Energy Survey				\\[0.2cm]
	
	\textbf{DGP}~~~~ & Dvali-Gabadadze-Porrati				\\[0.2cm]
	
	\textbf{DM}~~~~ & Dark Matter						\\[0.2cm]
	
	\textbf{DMR}~~~~ & Differential Microwave Radiometer		\\[0.2cm]
	
	\textbf{eBOSS}~~~~ & extended Baryon Oscillation Spectroscopic Survey	\\[0.2cm]
	
	\textbf{EoS}~~~~ & Equation of State				\\[0.2cm]
	
	\textbf{FLRW}~~~~ & Friedmann-Lema\^{i}tre-Robertson-Walker	\\[0.2cm]
	
	\textbf{GA}~~~~ & Genetic Algorithm	\\[0.2cm] 

	\textbf{GP}~~~~ & Gaussian Process	\\[0.2cm] 
	
	\textbf{GR}~~~~ & General Relativity	\\[0.2cm]
	
	\textbf{GRB}~~~~ & Gamma Ray Burst	\\[0.2cm]
	
	\textbf{HDE}~~~~ & Holographic Dark Energy			\\[0.2cm]
	
	\textbf{HST}~~~~ & Hubble Space Telescope			\\[0.2cm]
	
	\textbf{IMF}~~~~ & Initial Mass Function		\\[0.2cm]
	
	\textbf{IR}~~~~ & Infra-Red						\\[0.2cm]
	
	\textbf{JBP}~~~~ & Jassal-Bagla-Padmanabhan			\\[0.2cm]
	
	\textbf{JLA}~~~~ & Joint Light-curve Analysis	\\[0.2cm]
	
	\textbf{LISA}~~~~ & Laser Interferometer Space Antenna	\\[0.2cm]
	
	\textbf{LRG}~~~~ & Luminous Red Galaxies	\\[0.2cm]
	
	\textbf{LRS}~~~~ & Local Regression Smoothing		\\[0.2cm]
	
	\textbf{Ly$\alpha$}~~~~ & Lyman-$\alpha$		\\[0.2cm]
	
	\textbf{MCMC}~~~~ & Markov Chain Monte Carlo			\\[0.2cm]
	
	\textbf{MCT}~~~~ & Multi-Cycle Treasury		\\[0.2cm]
	
	\textbf{MGS}~~~~ & Main Galaxy Sample		\\[0.2cm]
	
	\textbf{MH}~~~~ & Metropolis-Hastings		\\[0.2cm]
	
	\textbf{PCA}~~~~ & Principal Component Analysis	\\[0.2cm] 
	
	\textbf{QSO}~~~~ & Quasi-Stellar Objects	\\[0.2cm] 
	
	\textbf{RSD}~~~~ & Redshift-Space Distortions		\\[0.2cm]
	
	\textbf{SDSS}~~~~ & Sloan Digital Sky Survey		\\[0.2cm]
	
	\textbf{SGL}~~~~ & Strong Gravitational Lensing		\\[0.2cm]
	
	\textbf{SNIa}~~~~ & Type Ia Supernovae				\\[0.2cm]
	
	\textbf{SNLS}~~~~ & Supernova Legacy Survey				\\[0.2cm]
	
	\textbf{SPS}~~~~ & Stellar Population Synthesis		\\[0.2cm]
	
	\textbf{SZ}~~~~ & Sunyaev-Zeldovich		\\[0.2cm]
	
	\textbf{UV}~~~~ & Ultra-Violet						\\[0.2cm]
	
	\textbf{WMAP}~~~~ & Wilkinson Microwave Anisotropy Probe	
\end{abbreviations}

\listoffigures 

\listoftables

\clearpage{}\addchaptertocentry{List of Publications}
\thispagestyle{plain}
\null\vfil
{\noindent \huge \textbf{List of Publications} \par}\vspace{8pt}
\begin{FlushLeft}
\justify
\underline{\emph{This thesis is based on the following papers}}

1.   \textbf{Purba Mukherjee} and Narayan Banerjee, {\it ``Non-parametric reconstruction of the cosmological jerk parameter"}, \href{https://doi.org/10.1140/epjc/s10052-021-08830-5}{Eur. Phys. J. C \textbf{81}, 36 (2021)}; \href{https://arxiv.org/abs/2007.10124}{arXiv:2007.10124}~[astro-ph.CO].\\

2.  \textbf{Purba Mukherjee} and Ankan Mukherjee, {\it  ``Assessment of the cosmic distance  duality relation using Gaussian process''},  \href{https://doi.org/10.1093/mnras/stab1054}{Mon. Not. R. Astron. Soc. \textbf{504}, 3938 (2021)}; \href{https://arxiv.org/abs/2104.06066}{arXiv:2104.06066}~[astro-ph.CO].\\

3.  \textbf{Purba Mukherjee} and Narayan Banerjee, {\it ``Nonparametric reconstruction of interaction in the cosmic dark sector"}, \href{https://doi.org/10.1103/PhysRevD.103.123530}{Phys. Rev. D \textbf{103}, 123530 (2021)}; \href{https://arxiv.org/abs/2105.09995}{arXiv:2105.09995}~[astro-ph.CO].\\

4.  \textbf{Purba Mukherjee} and Narayan Banerjee, {\it ``Revisiting a non-parametric reconstruction of the deceleration parameter from combined background and the growth rate data''}, \href{https://doi.org/10.1016/j.dark.2022.100998}{Phys. Dark Univ. \textbf{36}, 100998 (2022)}; \href{https://arxiv.org/abs/2007.15941}{arXiv:2007.15941}~[astro-ph.CO].\\

\underline{\emph{Other publications}}

1.  \textbf{Purba Mukherjee}, Ankan Mukherjee, Harvinder Kaur Jassal, Ananda Dasgupta and Narayan Banerjee, {\it ``Holographic dark energy: constraints on the interaction from diverse observational data sets''} , \href{https://doi.org/10.1140/epjp/i2019-12504-7}{Euro. Phys. J. Plus  \textbf{134}, 147 (2019)}; \href{https://arxiv.org/abs/1710.02417}{arXiv:1710.02417}~[astro-ph.CO].\\

2.  \textbf{Purba Mukherjee} and Soumya Chakrabarti, {\it ``Exact solutions and accelerating universe in modified Brans-Dicke theories''}, \href{https://doi.org/10.1140/epjc/s10052-019-7201-x}{Eur. Phys. J. C \textbf{79}, 681 (2019)}; \href{https://arxiv.org/abs/1908.01564}{arXiv:1908.01564}~[gr-qc].\\

3.  \textbf{Purba Mukherjee} and Narayan Banerjee, {\it ``Constraining the curvature density parameter in cosmology"}, \href{https://doi.org/10.1103/PhysRevD.105.063516}{Phys. Rev. D \textbf{105}, 063516 (2022)}; \href{https://arxiv.org/abs/2202.07886}{arXiv:2202.07886}~[astro-ph.CO].

\end{FlushLeft} 
\clearpage{}
\cleardoublepage

\dedicatory{\bsifamily{\vspace{4cm} To  Mani \& Baba \ldots}}

\mainmatter 

\pagestyle{thesis}

\makeatletter
\def\blfootnote{\gdef\@thefnmark{}\@footnotetext}
\makeatother
\sloppy
\clearpage{}\chapter{Introduction} \label{ch1}

{\it Cosmology} deals with the scientific study of the evolution of the Universe, from the origin to its ultimate fate. The {\it standard model of cosmology} is based on the {\it cosmological principle}, which states that the Universe is spatially homogeneous and isotropic on sufficiently large scales. The presence of large-scale structure, like galaxies and galaxy clusters, indicate that the Universe appears inhomogeneous when observed at length scales smaller than around 350 Mpc\footnote{1 pc = $3.09 \times 10^{16}$ m = 3.26 light-years (ly); 1 Mpc = $10^6$ pc} \cite{inhom1,inhom2}, and the level of anisotropies in the Universe is smaller than roughly one part in $10^5$ as observed from the Cosmic Microwave Background Radiation \cite{hudodcmb}. 

The standard cosmological model assumes that {\it Einstein's theory of General Relativity} is the correct description of gravity. The first exact cosmological solution to the Einstein's field equations was initially provided by Einstein\cite{einstein_lambda} in 1917, assuming the Universe to be static in time. The beginning of the twentieth century was marked with technological advances reaching a new level to provide evidence for the existence of galaxies besides our own that are moving away from us \cite{expan}. Friedmann\cite{fried, fried2}, Lemat\^{i}re\cite{lem}, Robertson\cite{robert} and Walker\cite{walker} presented new solutions to the Einstein field equations of an expanding Universe that resulted in the formulation of the {\it Friedmann-Lema\^{i}tre-Robertson-Walker} (FLRW) metric as a key for the standard cosmological model. 

If $\lambda_{\text{em}}$ and $\lambda_{\text{obs}}$ are the wavelength of light at the points of emission and observation in the Universe, the cosmological redshift is defined as 
\begin{equation}
z  = \frac{\lambda_{\text{obs}} - \lambda_{\text{em}}}{\lambda_{\text{em}}}.
\end{equation}
This redshift $z$ is directly proportional to the velocity $v$ of these receding galaxies. 

In 1929, Edwin Hubble\cite{hubble_ref}, noticed a systematic pattern of how incoming light from different galaxies is redshifted, and proposed that the recessional velocity $v$ is related to the distance $r$ of a galaxy from us, via an empirical formula~\cite{hubble_ref2} 
\begin{equation} \label{ch1:Hubble's law}
v = H ~r.
\end{equation}
This is known as {\it Hubble's Law}, where $H$ is a constant of proportionality known as the Hubble parameter, quoted in units of km Mpc$^{-1}$ s$^{-1}$. Its value at the present epoch $t_0$ is around $H_0 \approx 70$ km Mpc$^{-1}$ s$^{-1}$. The reciprocal of $H$ is known as the {\it Hubble time}. $H$ measures the relative rate of expansion which dictates the time evolution of the Universe. 

One of the most important tools that have contributed significantly to our understanding regarding the composition and evolution of the Universe is the {\it Cosmic Microwave Background Radiation} (CMBR) \cite{cmb_relic}. This relic radiation was first detected by Penzias and Wilson\cite{cmb_wilson} in 1965. Further investigations revealed that the CMBR has a thermal blackbody spectrum at a uniform temperature of $T_{0} \approx 2.725$ K at the present epoch, corresponding to the microwave part of the electromagnetic spectrum, which affects the formation of galaxies \cite{peebles_cmb}. The presence of anisotropies in the relic radiation, first predicted in 1967 by Sachs and Wolfe\cite{sachs_wolfe}, showed that cosmic structures have originated from these primordial inhomogeneities. The subsequent years focused on understanding how these anisotropies have led to the growth of large scale structures in the Universe \cite{lss2, lss3, lss4, lss5}. 

The following years witnessed remarkable developments in theoretical and observational sectors. In the late 1970s and early 80s, the theory of cosmic {\it inflation} by Starobinsky\cite{staro2_infla, staro3_infla}, Guth\cite{guth} and Linde\cite{linde_infla} provided an explanation for the origin of large scale structure in the Universe. Inflation is a theory of an exponential expansion in the early Universe. As a solution to the {\it horizon problem}, {\it flatness problem} and {\it monopole problem}, inflationary models was first employed by Guth\cite{guth} in 1981. Inflation predicts that the large scale structure in the Universe has originated via a gravitational collapse of perturbations that were formed from quantum fluctuations in the inflationary epoch.

During the late 1990s, two independent projects, the Supernova Cosmology Project \cite{perl1999} (led by Perlmutter) and the High-Z Supernova Search Team \cite{riess1998, schmidt, riess1999} (led by Schmidt and Riess), measured the luminosity distances of the type Ia supernovae, standard candles having fixed intrinsic brightness, and found that the galaxies and galaxy clusters are moving apart from one other in an accelerated rate. It was expected that the recessional velocity would be decelerating due to the gravitational attraction of the matter distribution in the Universe. But to their surprise, the observed supernovae appeared dimmer than expected. For the last two decades, this accelerated expansion of the Universe has been the most bewildering observation in cosmology. A variety of theoretical models have been proposed, either in the form of an additional field called {\it dark energy} in the matter sector or in the form of modifying the theory of gravity itself, as an explanation to the late-time acceleration \cite{varun, quint_wd, de_rev, brax}, and this list is ever increasing in the absence of a universally accepted one. 

In this chapter, the background dynamics of the standard cosmological model has been discussed, followed by a brief description on the late-time accelerated expansion of the Universe, along with an introduction to the statistical analysis methods utilized for reconstruction in cosmology. For more insight into the foundations of modern cosmology, one can refer to some standard literature \cite{book1, book2, book3, book4, book5, book6, akr}.

\section{Friedmann Cosmology}

As the Universe is homogeneous, two particles separated by $\mathbf{r}$ distance in the physical frame of reference can be 
transformed to some coordinate system, known as comoving coordinates, such that
\begin{equation}
\mathbf{r} = a(t) \mathbf{x},
\end{equation} 
where $\mathbf{x}$ is the comoving distance, i.e., the distance of separation between the same two points in the comoving frame. The concept of homogeneity 
ensures that $a$ is essentially a function of time alone, known as the {\it scale factor} of the Universe. It is a measure of the length scale of the Universe that determines how the physical separations are growing with time as the coordinate distances $\mathbf{r}$ are by definition fixed.

The infinitesimal distance element in a spatially homogeneous and isotropic Universe is given by the {\it Friedmann-Lema\^{i}tre-Robertson-Walker} 
(FLRW) metric
\begin{equation} 
\dif s^2 = - \dif t^2 + a^2(t) \left(\frac{\dif r^2}{1-kr^2} + r^2 \dif \theta^2 + r^2 \sin^2\theta \dif \phi^2 \right), \label{ch1:frw}
\end{equation}
where $t$ is cosmic time, $a(t)$ is the scale factor and $(r$, $\theta$, $\phi )$ are the ``comoving" spatial coordinates in spherical polar system. 
The isotropy and homogeneity of the space section demand the spatial curvature to be a constant, which can be scaled to pick up values from $-1, +1, 0$ 
corresponding to an open, closed, or flat Universe, respectively and is denoted by the curvature index $k$. 

The metric evolves according to the {\it Einstein field equation} which relates the geometry of spacetime to the distribution of matter within it. Varying the 
Einstein-Hilbert action  
\begin{equation}
	\mathcal{S}_\text{EH} = \int \sqrt{-g} \left(\frac{R}{16 \pi G} + \mathcal{L}_m \right) \mathrm{d}^4 x ,
\end{equation}
with respect to the metric $g_{\mu \nu}$, yields the Einstein field equations 
\begin{equation}
G_{\mu \nu} \equiv R_{\mu \nu} - \frac{1}{2} g_{\mu \nu} R = 8 \pi G T_{\mu \nu} . \label{ch1:einstein}
\end{equation} 
Here $G$ is the Newtonian gravitational constant, $g \equiv \text{det}(g_{\mu \nu})$ is the determinant of $g_{\mu \nu}$, $R = g^{\mu \nu} R_{\mu \nu}$ 
is the Ricci scalar obtained by contracting the Ricci tensor $R_{\mu \nu}$ which contributes to the Lagrangian density of the gravitational sector, 
$\mathcal{L}_m$ is the Lagrangian density of matter sector, $G_{\mu \nu}$ is the Einstein tensor and $T_{\mu \nu}$ is the energy-momentum tensor of 
the matter distribution in the Universe, given by
\begin{equation}
	T^{\mu}_{\nu} = -2 \frac{\partial \mathcal{L}_m}{\partial g_{\mu \nu}} + \delta^{\mu}_{\nu} \mathcal{L}_m .
\end{equation}

This distribution of matter is in the form of a {\it perfect fluid}, described by the stress-energy tensor
\begin{equation}
T^\mu_\nu = (\rho + p) u^{\mu} u_{\nu} + p ~\delta^{\mu}_{\nu} 
= \left(\begin{matrix}
-\rho & 0 & 0 & 0 \\
0 & p & 0 & 0 \\
0 & 0 & p & 0 \\
0 & 0 & 0 & p \\
\end{matrix}\right), 
\end{equation}
where $\rho$ and $p$ are the energy density and pressure of the perfect fluid. Einstein's equation tells how the presence of matter curves spacetime. 

\subsubsection*{The Friedmann equations}

Applying the FLRW metric \eqref{ch1:frw} to the Einstein field equation \eqref{ch1:einstein}, we arrive at the two basic equations of cosmology, 
the {\it Friedmann equations}, that govern the time evolution of the scale factor $a(t)$ are
\begin{equation}
3\left(\frac{\dot{a}}{a}\right)^2 + 3\frac{k}{a^2} = 8 \pi G \rho,	\label{ch1:fried1}
\end{equation} 
and
\begin{equation}
2\frac{\ddot{a}}{a} + \frac{\dot{a}^2 + k}{a^2} = 8 \pi G p .\label{ch1:fried2}
\end{equation}

Here, an overhead `dot' denotes derivatives w.r.t. the cosmic time $t$.

From Hubble's law, one can identify the Hubble parameter $H$ as,
\begin{equation}
H(t) = \frac{\dot{a}}{a}. \label{ch1:Hubble}
\end{equation}

The Friedmann equations can be rewritten in terms of $H$ and its derivatives as,
\begin{equation}
H^2 + \frac{k}{a^2} = \frac{8 \pi G}{3} \rho , \label{ch1:fried1new}
\end{equation}
and
\begin{equation}
2 \dot{H} + 3H^2 + \frac{k}{a^2} = - 8 \pi G p. \label{ch1:fried2new}
\end{equation}

\subsubsection*{The continuity equation}

The contracted Bianchi identity, $G^{\mu \nu}_{~~~; \mu} = 0$, yields the {\it continuity equation} that gives the evolution of energy 
density $\rho(t)$ in the Universe.
\begin{equation}
\dot{\rho} + 3 H ( \rho + p) = 0. \label{ch1:continuity}
\end{equation}
This is not an independent equation as it can be derived from Eq. \eqref{ch1:fried1} and \eqref{ch1:fried2}.

\subsubsection*{The equation of state}

The total energy content of the Universe is assumed to behave like a perfect fluid with the energy density $\rho$ and the pressure $p$ related through an {\it equation of state} (EoS) as
\begin{equation}
p = w ~ \rho ,
\end{equation}
where $w$ is the {\it equation of state parameter}.

The total energy density $\rho$ in equations \eqref{ch1:fried1} and \eqref{ch1:fried1new} is composed of different components $\rho_i$'s, each having 
their respective EoS parameter $w_i$. If $w_i$ for any component is known, the corresponding $\rho_i$ can be determined from an integration of equation 
\eqref{ch1:continuity} as
\begin{equation}
	\rho_i \propto e^{-3\int \left[1+w_i(a)\right] \dif a} .
\end{equation}

In the case of pressureless non-relativistic matter (also termed as `dust') we have $w =0$ as $p=0$ and for radiation (relativistic particles) $w=\frac{1}{3}$ 
as $p = \frac{1}{3}\rho$. We can calculate the energy density for matter or dust as $\rho_m \propto \frac{1}{a^3}$, and for radiation to be $\rho_r \propto 
\frac{1}{a^4}$ respectively. 

\subsubsection*{Critical density, density parameter and spatial curvature}

For a given $H$, there is a particular value of $\rho$, which makes the geometry of the Universe flat. This is known as the {\it critical 
density} $\rho_c$, defined as
\begin{equation} 
\rho_c(t) = \frac{3H^2}{8 \pi G}. \label{ch1:critical}
\end{equation}

Since, $G = 6.67 \times 10^{-11}$ m$^3$ kg$^{-1}$ s$^{-2}$ and $H_0$ can be scaled to a dimensionless form $h = \frac{H_0}{100 \text{ km} \text{ Mpc}^{-1} 
\text{ s}^{-1}}$, we can compute the present value of the critical density, $\rho_{c0} = \rho_c(t_0)$, as
\begin{equation}
\rho_{c0} = 1.88 ~h^2 \times 10^{-26} \text{ kg m$^{-3}$} .
\end{equation}

The {\it density parameter} $\Omega$  is defined as the ratio of the actual (or observed) density $\rho$  to the critical density 
$\rho _c$ of the Friedmann Universe,
\begin{equation}
\Omega \equiv \frac{\rho}{\rho_c}.
\end{equation}

Substituting equation \eqref{ch1:critical} in \eqref{ch1:fried1new} gives,
\begin{equation}
\Omega -  1 = \frac{k}{a^2 H^2}.
\end{equation}

This $\Omega$ determines the spatial geometry of the Universe. We have the following possible choices, 
\begin{itemize}
\item If $\rho > \rho_c$~ or ~ $\Omega>1$ , ~ $k = +1$ $\implies$ spatial geometry is closed,

\item If $\rho < \rho_c$~ or ~ $\Omega<1$ , ~ $k = -1$ $\implies$ spatial geometry is open,

\item If $\rho = \rho_c$~ or ~ $\Omega=1$ , ~ $k = 0$ ~~~$\implies$ spatial geometry is flat.
\end{itemize}

The individual energy density components in the Universe are given by, $\Omega_m = \frac{\rho_m}{\rho_c}$ for matter and $\Omega_r = \frac{\rho_r}
{\rho_c}$ for radiation. We also denote the density parameter associated with the curvature term as $\Omega_k = -\frac{k}{a^2 H^2}$.

The initial value of $\Omega_k$ has to be tantalizingly close to zero for correctly describing the present state of the evolution, which indicates that the Universe essentially starts with a zero spatial curvature. This is known as the flatness or fine-tuning problem of the standard cosmological model, which is believed to be taken care of by an early accelerated expansion called {\it inflation}\cite{guth}. The monograph by Liddle and Lyth\cite{book2} provides a brief but systematic description in this context. However, if $\Omega_k$ is negligible, but $k$ itself is non-zero, it may reappear in the course of evolution and makes its presence felt as the Universe evolves. Recent cosmological observations like Wilkinson Microwave Anisotropy Probe (WMAP) and {\it Planck} satellite indicate that the Universe tends to be spatially flat \cite{ref1, ref2, ref3, planck_cmb, planck}.

\subsubsection*{Distance measures in cosmology \label{ch1:distance_measure}}

Distance measure turns out as one of the most crucial tasks involved in cosmography, helping us establish a standard relation between the observational data with 
theoretical models. Distance measures are often used to relate some observable quantity (such as luminosity of a distant star, or angular 
size of acoustic peaks in the CMB power spectrum) to some other quantity that is not directly observable but more 
convenient for calculations. 

The redshift of spectral lines which justifies the notion of an expanding Universe, can be related to the scale factor. If we receive light from 
a distant object with a redshift of $z$ at the present epoch $t_0$ (i.e., $z=0$), then the scale factor at the time $t$ when the object originally 
emitted that light is given by
\begin{equation}
a(t) = \frac{a_0}{1+z} , \label{ch1:redshift}
\end{equation} 
where $a_0$ is scale factor at the present epoch $z=0$. It is convenient to use the redshift $z$ for studying the dynamics and evolution of the Universe, instead of cosmic time $t$, as $z$ is a dimensionless observational quantity. 

The {\it Hubble distance} is defined as,
\begin{equation}
d_{H}={\frac {c}{H_{0}}} \approx 3000 ~h^{-1}{\text{ Mpc }} \approx 9.26 \times 10^{25}~h^{-1}{\text{ m }}.
\end{equation} 
Here $c$ is the speed of light, $H_{0}$ is the Hubble parameter at present epoch, which can be scaled to a dimensionless form as $h = \frac{H_0}
{100 \text{ km} \text{ Mpc}^{-1} \text{ s}^{-1}}$.

The Hubble parameter $H(z)$ can again be represented in a dimensionless way, called the {\it reduced Hubble parameter}, given by 
\begin{equation}
E(z) = \frac{H(z)}{H_0}.
\end{equation}

Cosmologists use different measures for distance from the observer to an object at redshift $z$ along the line of sight. 
These are the comoving distance, the transverse comoving distance, the luminosity distance and the angular diameter distance. 
The {\it comoving distance} $d_{C}$ between two observers, both moving with the Hubble flow that accounts for the expansion of the Universe and does not change with time, is defined as 
\begin{equation} 
d_{C}(z)=d_{H}\int _{0}^{z}{\frac {\dif z'}{E(z')}}. \label{ch1:comoving_d}
\end{equation}

Two comoving objects at redshift $z$, separated by an angle $\delta \theta$, are said to cover a distance $d_P \delta\theta $. The {\it transverse} or {\it angular comoving distance}, also known as the physical distance, $d_P$, is defined as
\begin{equation} 
d_{P}(z)={\begin{cases}{\frac {d_{H}}{\sqrt {\Omega _{k}}}}\sinh \left({\frac {{\sqrt {\Omega _{k}}}d_{C}(z)}{d_{H}}}\right),&\Omega _{k}>0\\
d_{C}(z), &\Omega _{k}=0\\{\frac {d_{H}}{\sqrt {|\Omega _{k}|}}}\sin \left({\frac {{\sqrt {|\Omega _{k}|}}d_{C}(z)}{d_{H}}}\right).&
\Omega _{k}<0\end{cases}} \label{ch1:physical_d}
\end{equation}

An object of size $l$ at redshift $z$ that appears to have an angular size $\delta\theta$, has the {\it angular diameter distance} of 
$d_A \approx \frac{l}{\delta\theta}$, under the assumption of Euclidean geometry. This $d_A$ is related to $d_P$ by
\begin{equation}
d_A(z)  = \frac{d_P(z)}{1+z} . \label{ch1:angular_diamter_d}
\end{equation}  

It is interesting to note that for low redshift ($z<<1$) observations $d_A \approx d_P$. For a distant object, we have $d_A < d_P$. Therefore a 
distant object appears to have a larger at angular extent.

If the intrinsic luminosity $L$ of a distant object is known, we can calculate its {\it luminosity distance} $d_{L}(z)= \sqrt{ {\frac{L}{4\pi S}}}$ 
by measuring the flux $S$. The luminosity distance $d_L$ is related to $d_P$ by
\begin{equation}
d_L(z) = d_P(1+z) .  \label{ch1:luminosity_d}
\end{equation}

In a static Universe, $d_L = d_P$. For a nearby object (i.e. $z << 1$), $d_L \approx d_P$. However, for an object at a long distance $d_L > d_P$, it appears 
to be farther away than it really is.

All the distance measures in cosmology that are discussed above, can be written in their respective dimensionless or normalized forms as given below.
\begin{align}
D_C \equiv \frac{d_C}{d_H} , ~~~~D~~ \equiv \frac{d_P}{d_H} , ~~~~D_A \equiv \frac{d_A}{d_H} , ~~~~D_L \equiv \frac{d_L}{d_H}. \label{ch1:normalized_d}
\end{align}

The luminosity distance $d_L$ and the angular diameter distance $d_A$ are connected through the {\it cosmic distance-duality relation} (CDDR) given as,
\begin{equation}
d_L = d_A (1 + z)^2 . \label{ch1:cddr}
\end{equation} 

The CDDR was first given by Etherington\cite{etherin1993} in the context of a FLRW metric, and is often recognized as Etherington's reciprocity theorem. 
Cosmography is strongly dependent on the validity of CDDR.

\section{The accelerated expansion}

Equations \eqref{ch1:fried1new} and \eqref{ch1:fried2new} can be combined to derive a third equation which is independent of the curvature index 
$k$ and describes the acceleration of the scale factor as
\begin{equation}
\frac{\ddot{a}}{a} =  -\frac{4 \pi G}{3}\left(\rho+3p\right). \label{ch1:acceleration}
\end{equation}

Considering a Taylor expansion of the scale factor $a$ about the present epoch $t_0$, we get
\begin{equation} 
a(t) = a(t_0 ) ~+~ \dot{a}(t_0 )~ \left[t - t_0 \right] ~+~ \frac{1}{2}~ \ddot{a}(t_0 )~ \left[t - t_0 \right]^2 ~+~ \cdots . \label{ch1:taylor_a}
\end{equation}

On dividing equation \eqref{ch1:taylor_a} throughout by $a_0$, we can write
\begin{equation}
\frac{a(t)}{a(t_0)} = 1 ~+~ H_0~ \left[t - t_0 \right] ~+~ \frac{q_0}{2}~H_0^2~ \left[t - t_0 \right]^2 ~+~ \cdots ,
\end{equation} 
where
\begin{equation}
q_0 = - \frac{ \ddot{a}(t_0)}{a_0 H_0^2} , \label{ch1:q_0}
\end{equation} 
is called the deceleration parameter at the present epoch.

The larger the value of $q_0$, the more rapid is the deceleration. Further, equation \eqref{ch1:q_0} can be generalised for all $t$ and the deceleration 
parameter $q(t)$, is defined as 
\begin{equation}
q(t) = - \frac{ \ddot{a}}{a H^2}. \label{ch1:q_t}
\end{equation} 
If $q<0$, then the Universe is accelerating whereas if $q>0$, we get a decelerating expansion. 

The knowledge of standard cosmology suggests that all the known components in the energy budget of the Universe (dark matter, baryonic matter, 
relativistic particles like photons, and neutrinos) respect the strong energy condition, $\rho + 3p > 0$. In this case equation \eqref{ch1:acceleration} 
results in $\ddot{a} < 0$, i.e. the expansion should be decelerated. But surprisingly, the Universe has experienced two distinct periods of accelerated 
expansion, an early exponential {\it inflation}, and the late-time cosmic acceleration. Between these two phases of accelerated expansion, there 
prevailed a phase of decelerated expansion.

Hubble's observation of an expanding Universe indicated that the Universe has originated from an initial singularity called the {\it Big Bang}. 
It is based on two main assumptions, the cosmological principle and the universality of Einstein's gravity. The Big Bang model provides the 
best description to how the Universe expanded from an initial state of zero-volume, infinite density and high temperature, as well as offers a 
comprehensive explanation for a broad range of observed phenomena, like the abundance of light elements, existence of the CMBR and large-scale 
structure. Despite its enormous success, the Big Bang model could not explain the reasons behind an isotropic CMB temperature sky even for causally 
disconnected regions, an almost flat space section, and the absence of magnetic monopoles. Guth\cite{guth} in 1981 proposed inflationary models as 
a theoretical solution to the {\it horizon}, {\it flatness} and {\it monopole} problem. The early Universe has experienced an accelerated exponential 
expansion in the inflationary period, $10^{-36}$ s after the Big Bang that ended around $10^{-33}$ to $10^{-32}$ s, followed by a decelerated expansion.

The late-time accelerated expansion of the Universe was first discovered in 1998 at redshifts $z<1$ by two supernova observing groups, the Supernova 
Cosmology Project \cite{perl1999} and the High-Z Supernova Search Team \cite{riess1998, schmidt, riess1999}, individually. The type Ia supernovae are 
fairly reliable standard candles with known intrinsic brightness and can be distinguished in a wide range of distance. As the Universe expands, the 
distance between the object and the observer increases. So, the radiated photons get redshifted. The observed brightness of these objects and the redshift 
of the observed photons give a measurement for the expansion of the Universe. 

The brightness of a supernova can be expressed in terms of its absolute 
magnitude and can thus be used in the cosmic luminosity distance $d_L$ determination. The apparent $(L_{\text{ap}})$ and intrinsic $(L)$ 
luminosities are related to the luminosity distance $d_L$ as
\begin{equation}
L_\text{ap} = \frac{L}{4 \pi d_L^2} .
\end{equation}

Supernovae luminosity distance $d_L$ measurements are tabulated at different redshift $z$ in the form of distance modulus $\mu_B$, defined as the difference 
between the apparent magnitude $m_B$ and the absolute magnitude $M_B$ of the B-band (wavelength band of blue line) of the observed spectrum, given by 
\begin{equation}
\mu_B = m_B - M_B = 5 \log_{10} \frac{d_L}{\text{1 Mpc}} + 25 .
\end{equation}

This $d_L (z)$ can be expressed in terms of the present values of the Hubble parameter $H_0$ and the deceleration parameter $q_0$ as 
\begin{equation}
d_L (z) = \frac{c z}{H_0} \left\lbrace 1+ \frac{z}{2} \left[1-q_0\right] +	\mathcal{O}(z^2) \right\rbrace.
\end{equation}
Both the supernova groups measured the luminosity distances and observed the dimming of supernovae. The measured $d_L$'s are higher than their expected 
values, indicating a negative $q_0$. This confirmed that the light sources are receding away from each other at an accelerated rate.

Apart from type-Ia supernova \cite{ref4, suzuki, betoule}, observational data from the Baryon Acoustic Oscillations (BAO) \cite{sdssbao, sdssbao2}, 
WMAP \cite{ref5}, {\it Planck} \cite{ref6, ref7, planck_cmb, planck} satellite, Dark Energy Survey (DES) \cite{des} also confirm the occurrence 
of a smooth transition, from a past decelerated to the present accelerated expansion of the Universe, at some intermediate redshift $z \approx 0.5$ 
\cite{zt_ref0, zt_ref1, zt_ref2, zt_ref3, zt_ref4, zt_ref5, zt_ref6, zt_ref7}. For comprehensive reviews on different aspects of the accelerated 
expansion of the Universe we refer to \cite{book_acc, haridasu2017a&a, rubin2016apj}.

\subsection{Need for an exotic component}

The late-time cosmic acceleration puzzle is one of the most compelling problems in physics. For the Universe to undergo an accelerated expansion, 
gravity has to be repulsive. This is an extremely astonishing behaviour, as known observed matter forms satisfy the primary feature that 
\textit{gravity is attracting}. But the accelerated expansion invokes the possibility of repulsive gravity at cosmological scales. Attempts to 
find a reasonable explanation to this puzzle led to two distinct possibilities. The implication is either gravity behaves far differently than 
what we think, or that the Universe comprises of some mysterious component with exotic gravitational properties giving rise to an effective
negative pressure. 

If we consider the Universe to be solely dominated by matter in the present epoch (i.e. $p = 0$), we can calculate the value of $q_0$ from Eq. 
\eqref{ch1:critical} and \eqref{ch1:acceleration} to be
\begin{equation}
q_0 = \frac{\Omega_0}{2}.  \label{ch1:q_0 with Omega_0}
\end{equation}

For the Universe to accelerate, $q_0<0$ implies $\Omega_0<0$, which is impossible in the case of ordinary matter. Therefore, cosmologists postulate the 
existence of a new exotic component, called \textit{dark energy}, which satisfies the inequality $\rho + 3 p < 0$ (see equation \eqref{ch1:acceleration}), 
based on the assumption that GR is the appropriate theory of gravity. As an alternative approach, one can look for suitable modifications to the theory of 
gravity, where the late-time cosmic acceleration can be realized without introducing this dark energy sector to the energy budget of the Universe. 

According to the Planck mission, visible baryonic matter that makes up stars and galaxies contributes to only around 5\% of the total energy 
budget of the Universe. About 26\% of this energy budget resides in the form of dark matter, that is responsible for the formation of large scale 
structures in the Universe and for explaining the motion of galaxies and clusters. The remaining 69\% is in the form of dark energy, which accounts 
for the late-time accelerated expansion. Thus, a significant fraction of research for understanding the physics of this current accelerating Universe and 
unveiling its mysteries have become a prime focus.

\section{Modelling the late-time cosmic acceleration} \label{ch1:models}

To accommodate the exotic component in the energy budget of the Universe within the realm of GR, we rewrite the Friedmann equations 
\eqref{ch1:fried1new} and \eqref{ch1:fried2new} as,
\begin{align}
H^2 + \frac{k}{a^2} &= \frac{8 \pi G}{3} \left(\rho_m + \rho_r + \rho_d \right), \label{ch1:fried1new_withDE} \\
2 \dot{H} + 3H^2 + \frac{k}{a^2} &= - 8 \pi G \left(p_r + p_d \right). \label{ch1:fried2new_withDE}
\end{align}

The total energy density $\rho$ in RHS has been split into individual components $\rho_i$'s, where $i=m$ signifies the contribution from 
non-relativistic baryons and dark matter, $i=r$ is radiation, i.e., the contribution from relativistic particles like photons, and $i=d$ 
stands for contribution from dark energy. The non-relativistic baryonic and cold dark matter has negligible kinetic energy compared to 
the rest mass energy. With this approximation we can consider the matter sector to be pressureless, i.e., $p_m = 0$. The pressure arising 
from radiation is denoted as $p_r$, and $p_d$ is the pressure contribution from the dark energy sector. Therefore, the total pressure 
is given by, $p = p_r+p_d$.

It is convenient to introduce the density of individual components in a dimensionless way by scaling them with the critical density 
$\rho_c$, defined in equation \eqref{ch1:critical}. We define
\begin{equation}
\Omega_i = \frac{\rho_i}{\rho_c}.
\end{equation} 

We can write down the constraint equation for this model from Eq. \eqref{ch1:fried1new_withDE} as,
\begin{equation}
\Omega_m + \Omega_k+ \Omega_r + \Omega_d = 1. \label{ch1:DEconstriant}
\end{equation} 
Here $\Omega_k = -\frac{k}{a^2 H^2}$ is the contribution from the spatial curvature.

For the known components like matter and radiation, their respective EoS are $w_m = 0$ and $w_r = \frac{1}{3}$ respectively. 
The equation of state for dark energy, $w_d$, is 
\begin{equation}
w_d = \frac{p_d}{\rho_d}.
\end{equation} 

We represent the density parameters at the current epoch $z=0$ as $\Omega_{m0},~\Omega_{r0},~\Omega_{k0}$ and $\Omega_{d0}$. For pressureless matter 
$\Omega_m = \Omega_{m0} (1+z)^3$, for radiation $\Omega_r = \Omega_{r0} (1+z)^4$, and in case of the spatial curvature $\Omega_{k} = \Omega_{k0}(1+z)^2$, 
utilizing the conventional normalization of the scale factor $a_0=1$. Therefore, the reduced Hubble parameter for this model is
\begin{equation}
\small E^2(z) = \Omega_{m0}(1+z)^3 + \Omega_{k0}(1+z)^2 + \Omega_{r0}(1+z)^4 + \Omega_{d0} \exp{\left[ 3 \mathlarger{\int}_{0}^{z} \frac{1+w_d(x)}{1+x} \dif x \right]} . \label{ch1:DEmodeling}
\end{equation}
Here $\Omega_{d} = \frac{\Omega_{d0}}{E^2(z)} e^{{  \mathlarger{\int}_{0}^{z} \frac{3\left[1+w_d(x)\right]}{1+x} \dif x }}$~ represents the contribution from dark energy.

The effective EoS for this composite model is given by,
\begin{equation}
w_{\text{eff}} = \frac{p}{\rho} = \frac{p_r + p_d}{\rho_m+\rho_r+\rho_d}. \label{ch1:weff}
\end{equation} 
The effective EoS parameter $w_{\text{eff}}$ needs to be less than $-\frac{1}{3}$ in order to have an accelerated Universe, by recalling equation 
\eqref{ch1:acceleration}.

The deceleration parameter $q$ is given by
\begin{equation}
q = \frac{1}{2}\sum_{\substack{i}}\Omega_i \left(1+ 3 ~w_i\right) .
\end{equation}

As the late-time Universe has negligible contribution from the radiation compared to the other components, we can estimate $q$ for a 
spatially flat Universe composed of pressureless matter and dark energy as,
\begin{equation}
q \approx \frac{1}{2}\left(1+3~w_{d} ~\Omega_{d}\right). \label{ch1:q_and_omega_d}
\end{equation} 

Therefore, $w_{d} < -\frac{1}{3}\Omega_{d}^{-1}$ is the limiting condition for the occurrence of late-time cosmic acceleration. The recent 
cosmological observations from Planck 2018 data release suggest that $\Omega_{d0} = \Omega_d(z = 0) \approx 0.7$. Therefore, the value of DE 
EoS should be, $w_{d0} = w_{d}(z = 0) \lesssim - 0.5$ at the present epoch. 

There are different theoretical prescriptions for dark energy. However, none of them have been universally accepted, each having its own flaws and 
limitations. The simplest model of dark energy is the {\it cosmological constant} $\Lambda$, having an equation of state $w=-1$.

\section{Cosmological Constant}

The cosmological constant, $\Lambda$ was introduced into the field equation by Einstein in 1917 for obtaining a static cosmological solution with $a = 
\frac{1}{\sqrt{\Lambda}}$, known as Einstein's static Universe. Following Hubble's discovery of an expanding Universe in 1929, Einstein regretted his 
idea and called $\Lambda$ his greatest mistake \cite{einstein_lambda}. $\Lambda$ was reintroduced in 1981, as a possible candidate to explain the early 
exponential expansion  with $ a(t) \propto \exp \left( \sqrt{{\Lambda}/{3}} ~t \right)$, in the context of inflation. Described by Padmanabhan\cite{padma} 
as the ``\textit{weight of the vacuum}", $\Lambda$ has an inconsistent record being often accepted or rejected. After the discovery of the late-time cosmic 
acceleration in 1998, $\Lambda$ has been reconsidered as the most popular and simplest possible candidate for dark energy \cite{planck, planck_cmb, sahni, 
carroll, padma, peebles, frieman2008araa, amendola2010prl,mehrabi2018prd}. 

A Universe composed of pressureless cold dark matter (CDM) and the cosmological constant ($\Lambda$) as dark energy is called the $\Lambda$CDM model. 
It is frequently referred to as the {\it standard model of cosmology}. The constant energy density associated with $\Lambda$ is, 
\begin{equation}
\rho_\Lambda = \frac{\Lambda}{8 \pi G} . \label{ch1:rho_CC}
\end{equation}

By considering the fluid equation for $\rho_\Lambda$, we see that for $\rho_\Lambda$ to be a constant by definition, we must have 
\begin{equation}
p_\Lambda = - \rho_\Lambda. \label{ch1:p_CC}
\end{equation} 
The cosmological constant has an effective negative pressure, with the equation of state parameter $w_\Lambda = -1$.

On adding the contributions of energy density and pressure due to $\Lambda$, in the acceleration equation \eqref{ch1:acceleration}, it can be seen 
that a sufficiently large positive value of cosmological constant with $\Lambda > 4 \pi G(\rho_m+\rho_r+3p_r)$, can successfully drive the accelerated 
expansion of the Universe. 

Although $\Lambda$ leads to an accelerated expansion, it has its own share of problems \cite{carroll, sahni, peebles, padma, paddy, varuncqg, veltenepjc, 
wein}. Observationally, $\Lambda$ is of the order of $H_0^2$ in magnitude. This roughly corresponds to a critical density 
${\rho_\Lambda}^{\text{obs}} \approx 10^{-47}$ (GeV)$^4$. Theoretically, $\Lambda$ can be estimated from the concept of quantum field theory. Defining 
an empty space as a collection of quantum fields and assuming that the zero-point fluctuations of such vacuum fields contribute to $\Lambda$, the value 
of the vacuum energy density computed is approximately ${\rho_\Lambda}^{\text{th}} \approx 10^{74}$ (GeV)$^4$. This theoretically predicted value of 
$\Lambda$ overwhelmingly mismatches the observationally required one, with $\frac{{\rho_\Lambda}^{\text{th}}}{{\rho_\Lambda}^{\text{obs}}} \approx 10^{121}$, 
often referred to as the \textit{cosmological constant problem} \cite{peebles}.

Recent observations confirm that the present values of the dark matter and dark energy densities are comparable, having almost the same order of magnitude. 
This seems to indicate that we are presently living in a very special period of cosmic history. This coincidental unit order density ratio at the present 
epoch, which requires a set of finely-tuned initial conditions in the early Universe, is known as the \textit{coincidence problem} of standard cosmology 
\cite{veltenepjc}. 

For a detailed account on the various inconsistencies with $\Lambda$, we refer to the famous work of Weinberg\cite{wein}.

\section{Possible Alternatives to $\Lambda$}

As the easiest choice of $\Lambda$ as the dark energy runs into trouble, therefore, the quest for dark energy has been alive along all possible ways. 
This section is devoted to the study of some of these possible alternatives for $\Lambda$, as driver of the accelerated expansion of the Universe.

\subsection{Models with constant DE EoS}

For a phenomenological study of dark energy, cosmologists consider a constant value for the dark energy EoS, which is not necessarily restrictive to $-1$. Dark 
energy models defined with constant DE EoS are known as {\it Quiescence} \cite{varun}. As an example, one can consider a Universe composed of CDM and DE, where 
the latter is described by a constant EoS parameter $w$. This is known as the $w$CDM model. It is utilized to investigate the observational evidence for any 
possible deviation from the standard $\Lambda$CDM model. In case of the $w$CDM model, $\rho_d$ no longer remains constant, thus allowing an evolution of the 
dark energy density with redshift.

\subsection{Models with variable DE EoS}

In order to resolve the cosmic coincidence problem, DE models with an evolution required attention. It is assumed that the dark energy EoS $w_d$ has varied in time 
during the evolutionary history of the Universe. This led to the introduction of various dynamical dark energy (DDE) models with a time-dependent EoS parameter.  
The Chevallier-Polarski-Linder (CPL) model, given by the functional form $w(z) = w_0 + w_1 \frac{z}{1+z}$ \cite{cpl_main,linder} is the most popular and widely used 
DDE model, represented as a function of redshift $z$.

\begin{table*}[t!]
	\caption{{\small Table showing different dark energy models with EoS $w(z)$ evolving a function of redshift, that have been studied in the literature.}} 
	\begin{center}
		\resizebox{0.95\textwidth}{!}{\renewcommand{\arraystretch}{1.4} \setlength{\tabcolsep}{16pt}\centering 
			\begin{tabular}{l c c } 
				\hline
				\hline
				\textbf{Model} & $w(z)$ & \textbf{References}\\
				\hline				
				Chevallier-Polarski-Linder (CPL) & $w_0 + w_1 \frac{z}{1+z}$ & \cite{cpl_main, linder}				\\
				Jassal-Bagla-Padmanabhan (JBP) & $w_0 + w_1 \frac{z}{(1+z)^2}$ & \cite{jbp, jbp2}			\\
				Barboza-Alcaniz parametrization & $w_0+ w_1 \frac{z(1+z)}{1+z^2}$  &  \cite{barboza}      \\
				Wetterich parametrization & $\frac{w_0}{1+w_2 \ln(1+z)}$ & \cite{wetterich}      \\
				Ma-Zhang parametrization & $w_0 +w_1\left(\frac{\ln(2+z)}{1+z} - \ln 2\right)$ &  \cite{mazhang}  \\  
				\hline
				\hline
			\end{tabular}
		}
	\end{center}
	\label{tab-wz-param}
\end{table*}

Several other alternatives models \cite{jbp, jbp2, barboza, wetterich, mazhang} have been proposed with an evolving DE EoS $w(z)$, shown in Table \ref{tab-wz-param}.

\subsection{Scalar field DE models}

A scalar field rolling down a potential can give rise to an acceleration, which serves as a possible candidate for dark energy. Introducing a scalar 
field $\phi$ associated with a potential $V(\phi)$ makes the vacuum energy dynamical that helps in alleviating the cosmic coincidence problem. 
The scalar field models of dark energy includes \textit{Quintessence}, \textit{Phantom fields}, \textit{Tachyon fields} and \textit{K-essence} models. 

\subsubsection*{Quintessence}

The quintessence scalar field model is the most popular description for dynamical dark energy. A spatially homogeneous time-dependent scalar field $\phi$ 
associated with a potential $V(\phi)$, minimally coupled to the matter field is considered. This $\phi$ has negative pressure and it slowly rolls down the 
potential $V(\phi)$. 

For a scalar field $\phi$ with Lagrangian density $\mathcal{L}_\phi = \frac{1}{2} \partial^\mu \phi \partial_\mu \phi - V(\phi )$, the relevant action 
is given by
\begin{equation}
S = \int  \mathrm{d}^4 x ~\sqrt{-g} ~[-X -V (\phi)],
\end{equation} 
where $X = \frac{1}{2} \partial^\mu \phi \partial_\mu \phi$ is the kinetic term.

On varying the action with respect to the metric $g_{\mu \nu}$, the stress-energy tensor for $\phi$ takes the form of a fluid, with
\begin{align}
\rho_\phi &= \frac{\dot{\phi}^2}{2} + V(\phi) , \\
p_\phi &= \frac{\dot{\phi}^2}{2} - V(\phi) .
\end{align} 
Here $\frac{\dot{\phi}^2}{2}$ is the kinetic part, and $V(\phi)$ is the potential term, respectively.

On varying the action with respect to $\phi$, we find equation of motion for the scalar field $\phi$ as
\begin{equation}
\ddot{\phi} + 3 H \phi + \frac{dV}{d\phi} = 0.
\end{equation}

The DE EoS parameter $w_d$ for the quintessence scalar field $\phi$ is given by 
\begin{equation}
w_d = \frac{p_\phi}{\rho_\phi} = \frac{\dot{\phi}^2 - 2~V(\phi)}{\dot{\phi}^2 + 2~V(\phi)} ,
\end{equation} 

such that $w_d$ has an evolution that ranges between $-1 \leq w_d \leq 1$ for a real scalar field $\phi$ and a positive definite $V(\phi)$. Depending  on the nature 
of potential $V(\phi)$, quintessence models are classified into three different classes.
\begin{itemize}
\item When $V(\phi) << \dot{\phi}^2, ~w_d \approx 1$, then $\rho_d \propto a^{-6}$,  which is equivalent to the stiff matter and  does not contribute to dark energy.

\item When $V(\phi) >> \dot{\phi}^2, ~w_d \approx -1$, then $\rho_d \approx$ constant,  which is equivalent to the cosmological constant.

\item For intermediate cases $-1 < w_d < 1$, then $\rho_d \propto a^{-m}$ and the accelerated expansion can be realized in the limit $0 \leq m < 2$ \cite{quint_wd}. 
\end{itemize}

\begin{table*}[htb!]
\caption{{\small Different quintessence scalar field potentials $V(\phi)$ that have been studied in the existing literature.}}
\begin{center}
	\resizebox{0.95\textwidth}{!}{\renewcommand{\arraystretch}{1.4} \setlength{\tabcolsep}{65pt} \centering 
	\begin{tabular}{c c}
	\hline
	\hline
	$V(\phi)$ & \textbf{References}\\
	\hline
	$V_{0} ~e^{-\lambda \phi}$	& \cite{quint_ratra, quint_wett}	\\
	$V_{0} \phi^{-\alpha} , ~\alpha >0 $ & \cite{quint_ratra}	\\
	$V_{0} \sinh^{-\alpha}\left(\lambda \phi \right) $ & \cite{sahni, quint6} \\ 
	$V_{0} \left( e^{\alpha \phi} + e^{\beta \phi} \right) $  &   \cite{quint1, quint_barreio} \\
	$M^{4 + \alpha}	\phi^{-\alpha}$ & \cite{quint2}  \\
	$V_{0} \left( e^{M_{p}/\phi}-1 \right)$  & \cite{quint2, quint4} \\ 
	$V_{0} \left( \cosh \phi -1 \right)^\alpha $ & \cite{quint_wang} 	\\ 
	$V_{0} \left[ 1 + \cos \left( \phi/f \right) \right]$ &  \cite{quint_kim}  \\
	$V_{0}~ \phi^{4}, m^{2} \phi^{2}$  & \cite{quint_frieman}   \\
	$V_{0} e^{\lambda \phi^{2}}/\phi^{\alpha}, ~\alpha>0$  & \cite{quint_brax, quint_brax2} \\
	$V_{0} e^{- \lambda \phi} \left(1 + A \sin \nu \phi\right)$ & \cite{quint_dodel} \\ 
	$V_{0} e^{\lambda \phi} \left[(\phi - B )^\alpha + A \right] $	& \cite{quint_skordis} \\
	\hline 
	\hline
	\end{tabular}
	}
\end{center}
\label{tab-quint-pot}
\end{table*}

The idea of a quintessence scalar field was first introduced by Ratra and Peebles\cite{quint_ratra} and Wetterich\cite{quint_wett} in the context of cosmic 
inflation. To gain insight into the diverse amount of work in the context of late-time cosmic acceleration with different types of quintessence potentials, we 
refer all readers to the following references \cite{quint1, quint2, quint3, quint4, quint5, quint6, quint8, quint9, quint10, quint11, quint12, quint13, 
quint14, quint15, quint16, quint17}. A comprehensive list of the different scalar field potentials studied in the context of quintessence models are given in 
Table \ref{tab-quint-pot}.

\subsubsection*{Phantom field}

To explain the late-time cosmic acceleration, Caldwell\cite{phantom_cald} introduced the phantom field model of DE. The kinetic term $X$ has negative 
signature in this particular scenario. For a phantom field $\phi$ with Lagrangian density $\mathcal{L}_\phi = -X-V(\phi)$, the relevant action is given by
\begin{equation}
S = \int  \mathrm{d}^4 x ~\sqrt{-g} ~[-X -V (\phi)],
\end{equation} 
with $X = -\frac{1}{2} \partial^\mu \phi \partial_\mu \phi$ as the kinetic term.  

The energy density $\rho_\phi$ and the pressure $p_\phi$ of the phantom field, obtained from the stress-energy tensor for $\phi$ are
\begin{align}
\rho_\phi &= -\frac{\dot{\phi}^2}{2} + V(\phi) , \\
p_\phi &= -\frac{\dot{\phi}^2}{2} - V(\phi) ,
\end{align} 
where $-\frac{\dot{\phi}^2}{2}$ is the kinetic term and $V(\phi)$ is the potential term.

The EoS parameter for dark energy described by the phantom field $\phi$, is 
\begin{equation}
w_d = \frac{p_\phi}{\rho_\phi} = \frac{\dot{\phi}^2 + 2~V(\phi)}{\dot{\phi}^2 - 2~V(\phi)} ,
\end{equation}
and for $V(\phi) >> \dot{\phi}^2$, $w_d < -1$. A phantom field rolls up the potential due to its negative kinetic energy, which leads to a very rapid expansion 
of the Universe up to an infinite extent within a finite time. This scenario is known as \textit{Big Rip}, where both the proper volume and the rate of expansion 
become infinite at a finite future. 

For $\dot{\phi}= 0$, the equation of state parameter $w_d = -1$, and the cosmological constant scenario is restored in case of the phantom DE model. 

The scalar field models in which the evolution of EoS parameter mimics the phantom field are called \textit{quintom} models \cite{quintom1,quintom2,
quintom3,quintom4}.

\subsubsection*{Tachyon field}

The theoretical foundation for the concept of a tachyon scalar field model stems from string theory. Tachyons are theoretically postulated particles, having 
negative squared masses that travel with speeds greater than the speed of light. During the decay time of D-branes, a pressureless gas having finite energy 
density is formed, which resembles classical dust \cite{tach1, tach2, tach3, tach4, tach5, tach6, tach7, tach8}. It is interesting to note that tachyons have 
an EoS parameter smoothly varying in the range $-1 < w_d < 0$. This leads cosmologists to consider tachyons as a viable candidate for dark energy \cite{tach_gibbons}. 
The late-time cosmic acceleration can be generated by considering the tachyon field dark energy models \cite{tach9, tach10, tach11, tach12, tach13, tach14}. 
The negative squared tachyon mass rests on the maxima of its associated scalar field potential and is subjected to very small perturbations. This leads to a 
condensation of the tachyon state, characterized by rolling down from the maxima and achieving a real mass.

The relevant action for a tachyon field $\phi$ is given as,
\begin{equation}
S = - \int \mathrm{d}^4 x ~ V (\phi) \sqrt{-\text{det}\left(g_{\alpha \beta} + \partial_\alpha \phi \partial_\beta \phi \right)},
\end{equation} 
where $V (\phi)$ is the tachyon field potential. The wave equation takes the form

\begin{equation}
\frac{\ddot{\phi}}{1-\dot{\phi}^2} + 3 H \dot{\phi} + \frac{1}{V} \frac{dV}{d\phi} = 0 .
\end{equation}

The energy density $\rho_\phi$ and pressure $p_\phi$ of the tachyon field are
\begin{align}
\rho_\phi &= \frac{V(\phi)}{\sqrt{1-\dot{\phi}^2}} , \\
p_\phi &= - V(\phi) \sqrt{1-\dot{\phi}^2} .
\end{align} 
The DE EoS parameter for a tachyon field is given by,
\begin{equation}
w_d = \dot{\phi}^2 - 1.
\end{equation}

The allowed range for $\dot{\phi}^2$ is $0 < \dot{\phi}^2 < 1$. So the DE EoS for tachyon field varies in the range $-1 < w_d < 0$. The condition for an 
accelerated expansion of the Universe is $\dot{\phi}^2 < \frac{2}{3}$. 

\subsubsection*{K-essence}

The K-essence scalar field model, named the K-inflation \cite{K1, K2}, was formulated to describe an inflationary model of the early Universe. 
In contrast to the quintessence models where the potential energy term gives rise to an accelerated expansion, in the K-essence scalar field models, 
the kinetic part has a dominating contribution to the energy density, which drives the late-time cosmic acceleration. Chiba \textit{et al}\cite{K_chiba} 
first introduced the idea of a K-essence scalar field to model the accelerated expansion of the Universe. It was later generalized by Armendariz-Picon 
\textit{et al}\cite{K_picon1, K_picon2} and known as the K-essence models of dark energy.

The K-essence scalar field action has a general form
\begin{equation}
S = \int \mathrm{d}^4 x ~\sqrt{-g} ~\mathcal{L}\left(\phi, X \right), 
\end{equation}
where the Lagrangian density $\mathcal{L}$ is an arbitrary function of the K-essence scalar field $\phi$ and its kinetic term $X = \frac{1}{2} \partial^\mu \phi \partial_\mu \phi$.

On varying this $\mathcal{L}$ with respect to the metric, we obtain the energy momentum tensor in the form
\begin{equation}
T_{\mu \nu} = 2 X \mathcal{L}_{,X} u_\mu u_\nu  + g_{\mu\nu} \mathcal{L} ,
\end{equation} 
with $\mathcal{L}_{,X} = \frac{\partial \mathcal{L}}{\partial X}$ and the 4-momentum vector $u_\mu = \frac{\partial_\mu \phi }{\sqrt{2 X}}$.

The pressure $p$ is given by the Lagrangian density, $p = \mathcal{L}(X, \phi)$, and the energy density is given by 
\begin{equation}
	\rho = 2 X \mathcal{L}_{,X} - \mathcal{L}.
\end{equation}
So, the EoS parameter in case of a K-essence DE model takes the form
\begin{equation}
w_d = \frac{\mathcal{L}}{2 X \mathcal{L}_{,X} - \mathcal{L}}.
\end{equation} 
The K-essence model can reproduce the cosmological constant ($w_d = -1$) for the particular condition $2 X \mathcal{L}_{,X}= 0$ \cite{K3}. For general 
discussions on the K-essence scalar field model we refer readers to the references \cite{K4, K5, K6, K7}.

\subsection{Holographic Dark Energy}

The idea of holographic dark energy (HDE) stems from thermodynamics, namely the \textit{holographic principle} in quantum gravity theory. 't Hooft\cite{holo1} 
and Susskind\cite{holo2} conjectured that ``\textit{any phenomena within a volume can be explained by the set of degrees of freedom residing on its 
boundary, and the degrees of freedom are determined by the area of the boundary rather than the volume}''. This idea is based on the black hole entropy 
bound, suggested by Bekenstein\cite{holo3, holo4}. The formation of a black hole leads to a connection between the short distance ultraviolet (UV) 
cut-off, to a long distance infrared (IR) cut-off by the constraint such that the total quantum zero-point energy of the system should not exceed the mass 
of black holes of the same size \cite{holo5}. This can be expressed by the inequality
\begin{equation}
	L^3 \rho_{\Lambda} \leq L M_p^2 ,
\end{equation}
where $M_p = (8\pi G)^{-2}$ is the reduced Planck mass, $\rho_{\Lambda}$ is the quantum zero-point energy density determined by the UV cut-off and $L$ 
is the length scale of the system size. The length for which this inequality saturates is the IR cut-off. 

In the context of dark energy, the holographic principle was first introduced by Li\cite{holo6} with the holographic energy density, $\rho_H$, given by 
\begin{equation}
	\rho_{H} =  3 C^2 M_p^2 / L^2 ,
\end{equation}
where $C^2$ is a dimensionless coupling parameter. For holographic dark energy, the system size is the observable Universe and thus the IR cut-off is 
the cosmological horizon. Detailed studies on different HDE models can be found in the references \cite{holo6, holo7, holo8, holo9, holo10, holo11, holo12}.

\subsection{Modified gravity models}

As an alternative explanation to the phenomenon of late-time cosmic acceleration, there exists another class of models that modifies Einstein's GR. 
A vast range of modified gravity theories now exist in the literature \cite{mod_grav, mod_grav2, mod_grav3, mod_grav4, mod_grav5, mod_grav6, 
mod_grav7} which can give rise to an accelerated expansion of the Universe without recourse to a dark energy component. This list includes $f (R)$ 
gravity \cite{fr1, fr2, fr3, fr4, fr5, fr6, fr7, fr8, fr9, fr10}, scalar-tensor theories \cite{coupled_quint, bd_nb, st1, st2, st3, st4, st5, st6, 
bd_nb_pavon, bd_nb_pavon2, anjan_bd, st9, sudipta_bd, st10, purba_bd}, 
braneworld models like the Dvali-Gabadadze-Porrati (DGP) \cite{brane, brane2} model, $f(T)$ gravity \cite{ft} and $f(T,T_G)$ gravity \cite{ftg}, Galilean 
gravity \cite{galilean}, Gauss-Bonnet gravity \cite{gaussbonnet1, gaussbonnet2, gaussbonnet3}, and some extended theories \cite{extendedbd, quintinflation}. 
However, these models are mostly unsuitable for accurately explaining the local astronomical observations like the bending of light rays, perihelion 
precession of Mercury, and time dilation by the gravitational field of the Sun. A few commonly used modified gravity theories are discussed below.

\subsubsection*{f(R) gravity}

The action in $f(R)$ gravity is modified by replacing the Ricci scalar $R$ with an analytic function of $R$, as
\begin{equation}
S = \int \mathrm{d}^4 x ~ \sqrt{-g}~ \left[ f(R) + \mathcal{L}_m \right] ,
\end{equation} 
where $f (R)$ is an arbitrary function of $R$ and $\mathcal{L}_m$ is matter Lagrangian density. The modified field equations for the $f(R)$ gravity models are,
\begin{equation}
\frac{\partial f}{\partial R} R_{\mu \nu} -\frac{f}{2} g_{\mu \nu} - \left( \nabla_\mu \nabla_\nu - g_{\mu \nu} \square \right) \frac{\partial f}{\partial R} = 8 \pi G T_{\mu \nu} .
\end{equation} 
$T_{\mu \nu}$ is the stress-energy tensor of the matter distribution. 

The $f(R)$ gravity theories are capable of modelling early inflation or late-time acceleration depending on the chosen functional form. Models with $f(R) = R^2$ 
are successful in producing inflationary scenarios, whereas models with $f (R) = \frac{1}{R^n}$ for $n > 0$ are proposed to drive the late-time accelerated 
expansion. For more details on cosmological dynamics in $f (R)$ gravity theories we refer to the references \cite{fR1, fR2, fR3, fR4, fR5, fr3, fR7, fr9, 
fR10, fR11, fR12, fR13, fR14, fR15, fR16, fR17}.

\subsubsection*{Scalar-tensor theories}

Scalar-tensor theories in gravitation and cosmology are based on the idea of a non-minimal coupling between the scalar field and the geometry. Brans-Dicke 
theory \cite{bransdicke} is the simplest option among all possible existing scalar-tensor theories, where a scalar field $\phi$ is coupled 
to the Ricci scalar $R$. The Lagrangian density for $\phi$ is given by
\begin{equation}
\mathcal{L}_{\phi} = \frac{\phi R}{2}-\frac{\omega_{BD}}{2\phi}(\nabla \phi)^2 .
\end{equation}
This $\omega_{BD}$ is called the Brans-Dicke parameter.

Banerjee and Pavon\cite{bd_nb_pavon} have shown that the cosmic acceleration can directly be generated from the BD theory without introducing any exotic 
component in the matter sector, for suitable lower negative values of $\omega_{BD} \sim \mathcal{O}(1)$. This necessary criterion for a low $\omega_{BD}$ 
in the cosmological scenario contradicts the local astronomy, which demands a high value of $\omega_{BD}$. The BD theory can produce a non-decelerated expansion 
in the presence of an additional minimally coupled scalar field \cite{bd_nb_pavon2}.  The possibility for a late-time acceleration in BD theory for some 
specific choice of an additional potential has been explored by Sen and Sen\cite{anjan_bd}. Moreover, a non-minimal coupling between matter and the BD scalar 
field can account for a smooth transition to an accelerated phase of expansion from a decelerated one for very high values of $\omega_{BD}$ \cite{sudipta_bd}.

\section{Reconstruction methods in Cosmology} \label{ch1:recon_methods}

The absence of a consensus model for cosmic acceleration presents a challenge in connecting theory with observations. So, there have 
been attempts towards building dark energy models right from the observations. This leads to a reverse way of looking at the evolution. Rather 
than trying to find the evolution from a given matter sector using Einstein field equations, one uses the evolutionary history that directly 
fits with observations to find out the possible distribution of matter. {\it Reconstruction} is a kind of {\it reverse engineering} technique 
with broad applications in modern cosmology.

Normally physical quantities like the dark energy equation of state parameter $w_d$ \cite{saini, varun}, the quintessence potential $V(\phi)$ 
\cite{staro, huterer1, huterer2} occupies the central stage of interest in this game of reconstruction. A recent trend of reconstruction ignores 
any dynamical equation and attempts to find out the kinematical quantities like the Hubble parameter $H$ and its higher derivatives like the 
deceleration parameter $q$, jerk parameter $j$, directly from observations. 

The {\it kinematic} or {\it cosmographic quantities} are usually defined as the time derivatives of the scale factor $a$. For convenience, we 
redefine these kinematic quantities as a function of redshift $z$ (defined in equation (\ref{ch1:redshift})) instead of cosmic time $t$, as $z$ is 
a dimensionless observational quantity \cite{quint_wd}. The kinematic quantities related to the expansion of the Universe are,  

\begin{itemize}
	\item[(i)]  \textit{Hubble parameter}, 
	\begin{equation} \label{ch1:Hdef}
	H = \frac{\dot{a}}{a} = \frac{1}{1+z} \frac{\dif z}{\dif t}.
	\end{equation}	
	\item[(ii)]  \textit{Deceleration parameter}, 
	\begin{equation} \label{ch1:qdef}
	q = -\frac{ \ddot{a}}{a H^2} = -1 +(1+z)\frac{H'}{H}.
	\end{equation}	
	\item[(iii)]  \textit{Jerk parameter}, 
	\begin{equation} \label{ch1:jdef}
	j = \frac{ \dddot{a}}{a H^3} =  1 - 2(1+z)\frac {H'}{H} + (1+z)^2 \frac{\left[ H'^2 + H~H''\right]}{H^2}.
	\end{equation}	
\end{itemize}
Throughout this thesis, a `dot' denotes derivative with respect to the cosmic time $t$ whereas a `prime' stands for derivative with respect 
to the redshift $z$.

The dark energy equation of state parameter $w_d$, the effective equation of state $w_{ \text{eff}}$, the density parameters $\Omega_i$'s 
in the Friedmann equations constitute the {\it dynamical parameters} of the Universe. For a Universe having interaction in the dark sector, 
namely dark matter and dark energy, the interaction function $Q$, which describes the rate of transfer of energy between dark 
matter and dark energy, also serves as a dynamical variable.  

With dark energy having an equation of state $w(z)$ (ignoring the contribution from radiation), we can write $E(z)$ by integrating the 
Friedmann equation \eqref{ch1:fried1new_withDE} as,
\begin{equation} \label{ch1:Ez}
\begin{split}
E^2(z) = \Omega_{m0}(1+z)^3 + \Omega_{k0}(1+z)^2  + ~~~~~~~~~~~~~~~~~~~~~~~~~~~~~~~~~~~~~~~~~~~~~~~~~~~\\ 
~~~~~~~~~~~~~~~~~~~~~~~~~+(1-\Omega_{m0}-\Omega_{k0}) \exp{\left[ 3 \int_{0}^{z} \frac{1+w(x)}{1+x} \dif x \right]}. 
\end{split}
\end{equation} 
$\Omega_{m0}$ and $\Omega_{k0}$ are the normalized density parameters for the matter sector and the spatial curvature at the present epoch. On differentiating equation \eqref{ch1:Ez}, we get
\begin{equation}
w(z) = \frac{2(1 + z)E~E' - 3~E^2 + \Omega_{k0}(1 + z)^2}{3\left[ E^2 - \Omega_{m0}(1 + z)^3 - \Omega_{k0}(1 + z)^2 \right]}.
\end{equation}

Assuming that the dark energy is due to a scalar field $\phi$, the associated scalar potential $V(\phi)$ can also be written as a function of 
$z$, as
\begin{equation}
V\left[\phi(z)\right] = \frac{1}{8 \pi G} \left[ 3H^2 - (1+z)H H'\right] - \frac{3}{16 \pi G} \Omega_{m0} H_0^2 (1+z)^3.
\end{equation}

The prime advantage of reconstruction through the kinematical quantities is that, neither it assumes any theory of gravity 
(like GR, $f(R)$ gravity, scalar-tensor theory, etc.) nor does it assume a given matter distribution like a quintessence field or some exotic fluid 
via the equation of state. The basic \textit{a priori} assumption is that the Universe is spatially homogeneous and isotropic, and thus described by 
the FLRW metric. Thus, reconstruction through kinematical quantities might lead to some novel understanding of the distribution of matter and the 
possible interaction amongst themselves.

Any reconstruction in cosmology can be done in two possible ways. One is called the {\it parametric reconstruction} where the quantity to be 
reconstructed, e.g. $f$, is parametrized as a function of the redshift $z$ as $f \equiv f(z)$. A suitable ansatz like $f(z) = \sum_{\substack{i}} 
f_i ~ z^i$, is chosen and the values of the parameters $f_i$'s are estimated with the help of observational data \cite{param1, param2, quint10, 
quint9, linder, cpl_main}. However, reconstruction with this parametric approach normally is a bit biased as the quantities depend on $z$ in a 
given way, according to the functional form chosen. A more robust form is a {\it non-parametric reconstruction} which attempts to build up the actual 
functional form of $f$ w.r.t. $z$ directly from the observational data \cite{quint8, nonparam1, nonparam2, gp1, gp2, gp3, pca1, pca2, pca3, lrs1, lrs2, 
ga2, ga1, arjona, gp_seikel}.

Minimizing the $\chi^2$ function through optimization, or maximizing the likelihood function via marginalization, are the two statistical methods 
involved in cosmological reconstruction for obtaining the best-fit parameter values. Estimating the error associated with these best-fit values 
requires calculating the parameter covariance matrix from the $\chi^2$ function. Alternatively, a Markov Chain Monte Carlo (MCMC) analysis 
\cite{mcmc_book} can be done to obtain constraints on the parameter values via a marginalization of the posterior probability distribution 
over the parameter space. These are fundamental methods common to both the parametric and non-parametric approaches. 

There are several methods for implementing a non-parametric reconstruction in cosmology. These include the Principal Component Analysis (PCA) 
\cite{pca1, pca2, pca3}, Local Regression Smoothing (LRS) \cite{lrs1, lrs2}, Genetic Algorithms (GA) \cite{ga2, ga1, arjona} and Gaussian 
Process (GP) \cite{gp1, gp2, gp3, gp_seikel}. This thesis is devoted to studying the non-parametric reconstruction of some cosmological parameters 
by adopting the GP formalism. 

\subsection{$\chi^2$ minimization}

To estimate the cosmological parameter values from observational data, the $\chi^2$-statistics is adopted. If we have a set of observational 
data $(x_i, f_i)$; $i=\left\lbrace1,\cdots,n\right\rbrace$ with additional noise $\sigma_{i}$ associated to the measurement at each $x_i$, we 
can assume a theoretical function as $f(x_i,\left\lbrace \theta \right\rbrace) ~\forall ~i$, to describe this dataset with $\left\lbrace \theta 
\right\rbrace$ set of parameters. The $\chi^2$ function is generally defined as
\begin{equation}
\chi^2 = \mathlarger{\mathlarger{\sum}}_{\substack{i}} \frac{\left[f_i - f(x_i,\left\lbrace \theta \right\rbrace)\right]^2}{\sigma_{i}^2}.
\end{equation}

In cosmological data analysis, $x_i$ are represented by the redshift $z$. The $f_i$'s are generally given by the 
Hubble parameter measurements $H$, or different distance measures like the comoving distances $d_C$, the distance modulus compilation $\mu$, or 
some composite function of the various cosmological distance measures the at different $z$.

In case we have a complicated description of data-set with an observational noise given by some general covariance matrix $\bm{\Sigma}$ instead 
of a set of diagonalized variances $\sigma_{i}$ at respective $x_i$, we can rewrite the $\chi^2$ function as
\begin{equation}
\chi^2 = \mathlarger{\mathlarger{\sum}}_{\substack{i,j}} \left[f_i - f(x_i,\left\lbrace \theta \right\rbrace)\right]^{\text{\small T}} \Sigma_{ij}^{-1} \left[f_j - 
f(x_j,\left\lbrace \theta \right\rbrace)\right].
\end{equation} 
The superscript `$\text{\small T}$' denotes the transpose of any matrix.

For statistical analysis with a combination of $m$ data sets, the $\chi^2$ associated with individual data sets are added up to define the 
combined $\chi_{_{\text{ tot}}}^2$ as
\begin{equation}
\chi_{_{\text{ tot}}}^2 = \mathlarger{\sum}_{\substack{m}} ~\chi^2_{_m} ,
\end{equation}
where $m$ denotes the datasets taken into account for that particular combination.

To obtain the best-fit parameter $\left\lbrace \theta \right\rbrace$ values, we minimize the $\chi^2$ function. 

We define another quantity, the reduced $\chi_\nu^2$, defined as
\begin{equation}
\chi^2_{\nu} = \frac{\chi^2}{\nu},
\end{equation} 
where $\nu$ signifies the degrees of freedom. 

While fitting a function to some data, the necessary condition that needs to be checked for preventing any over-fitting is 
\begin{equation}
\chi^2_\nu < 1.
\end{equation} 
For computing the minimized $\chi^2$ we use \texttt{scipy}\footnote{\url{https://www.scipy.org}}, the numerical optimization package in python. For handling matrices and arrays in python, we use the fundamental package for scientific computing, \texttt{numpy}\footnote{ \url{https://numpy.org}}.

\subsection{Error propagation rule}

To estimate the error associated with the best-fit parameter values obtained from the $\chi^2$ minimization, we need to calculate the parameter covariance matrix $\mathbf{C}$, given by
\begin{equation}
\mathbf{C} = \left(\frac{\partial^2\chi^2}{\partial\theta_i\partial\theta_j}\right)^{-1}\Bigg|_{\{\hat{\theta}\}}.
\end{equation}

The variance $\sigma^2_{\theta_i}$ is given by the diagonal terms of the matrix for parameter $\theta_i$, and the off-diagonal terms 
are the covariance $\text{cov}(\theta_i,\theta_j)$ between two sets of parameters $\left\lbrace \theta_i, \theta_j \right\rbrace$ 
associated to that corresponding term.

Obtaining the best-fit reconstructed function defined by the parameters $\left\lbrace \theta \right\rbrace$, such that $f \equiv f\left(\left\lbrace 
\theta \right\rbrace\right)$, along with the error uncertainties $\sigma_f$, requires knowledge about the uncertainties associated with the parameter 
values $\left\lbrace \sigma_\theta \right\rbrace$. The Taylor series expansion of $f(\left\lbrace \theta \right\rbrace)$ around the best-fitting 
parameter values, $\left\lbrace \hat{\theta} \right\rbrace$ is given by
\begin{equation}
f\left(\left\lbrace \theta \right\rbrace\right) = \left. f(\left\lbrace \theta \right\rbrace\right)\Big\vert_{\left\lbrace\hat{\theta}\right\rbrace} 
+ \mathlarger{\mathlarger{\sum}}_{\substack{i}} \left(\frac{\partial f}{\partial \theta_i}\right)\Bigg\vert_{{\left\lbrace\hat{\theta}\right\rbrace}} \Delta\theta_i + 
\mathcal{O}\left(\Delta\theta_i^2\right)+ \cdots ,
\end{equation}
where $\Delta\theta_i \equiv (\theta_i - \hat{\theta_i})$, being negligibly small, are the only significant terms that contributes to the probability 
density. The higher-order terms $\mathcal{O}\left(\Delta\theta_i^2\right)$ can be ignored as the probability rapidly falls for higher derivatives of 
the function $f$ from the best-fit.

Thus, we can estimate the variance of $f(\left\lbrace \theta \right\rbrace)$ as
\begin{equation}
\sigma_f^2=\mathlarger{\mathlarger{\sum}}_{\substack{i}} \sigma_{\theta_i}^2\left(\frac{\partial f}{\partial\theta_i}\right)^2\Bigg|_{\{\hat{\theta}\}}+\mathlarger{\mathlarger{\sum}}_{\substack{ij,i\neq j}}\text{cov}(\theta_i,\theta_j)\left(\frac{\partial f}{\partial\theta_i}\frac{\partial f}{\partial\theta_j}\right)\Bigg|_{\{\hat{\theta}\}}.
\end{equation}

\subsection{Maximum likelihood analysis}

The likelihood, like $\chi^2$, is also a function of the parameters $\left\lbrace \theta \right\rbrace \equiv \theta_i,$ 
for $i =\left\lbrace 1 \cdots n  \right\rbrace$. In Bayes' theorem, the {\it posterior probability distribution} of parameters 
$\left\lbrace \theta \right\rbrace$ is expressed as
\begin{equation}
p( \left\lbrace \theta \right\rbrace \vert D, I) = \frac{p( \left\lbrace \theta \right\rbrace \vert I)~ p(D \vert \left\lbrace \theta \right\rbrace 
, I)}{p(D \vert I)}	,
\end{equation} 
where $I$ is the proposition representing the prior information, and $D$ represents the observational data, $p(D \vert \left\lbrace \theta \right\rbrace, 
I)$ is the probability of obtaining $D$ if $\left\lbrace \theta \right\rbrace$ is given according to $I$, $p(D \vert \left\lbrace \theta \right\rbrace, 
I)$ is the {\it likelihood} $\mathcal{L}$, $p( \left\lbrace \theta \right\rbrace \vert I)$ is the prior probability and $p(D \vert I)$ is called the global 
likelihood which serves as a normalization factor.
\begin{equation}
p(D|I) = \int_{\substack{\theta_1}}\cdots\int_{\substack{\theta_n}} p( \left\lbrace \theta \right\rbrace \vert I)~p(D \vert \left\lbrace \theta \right\rbrace, 
I)~d \theta_1 \cdots ~d \theta_n,
\end{equation}
such that
\begin{equation}
\int_{\substack{\theta_1}}\cdots\int_{\substack{\theta_n}} p( \left\lbrace \theta \right\rbrace \vert D, I)~d \theta_1 \cdots ~d \theta_n = 1.
\end{equation}

The likelihood $\mathcal{L}$ is related to the $\chi^2$ function as,
\begin{equation}
\mathcal{L}(\left\lbrace \theta \right\rbrace) = \exp \left(-\frac{\chi^2}{2}\right). \label{ch1:likelihood}
\end{equation}

From equation \eqref{ch1:likelihood} we can infer that the minimized value of $\chi^2$ corresponds to the maximized likelihood function $\mathcal{L}$. 
Thus, the $\chi^2$ minimization is equivalent to the maximization of likelihood. 

In this thesis, we obtain the best-fit parameter values and the associated error uncertainties, mostly employing a Markov Chain Monte Carlo (MCMC) analysis, 
for sampling the posterior probability distribution over the parameter space. This approach provides a more efficient way of exploring parameter space via random 
walks from one set of parameter values to the next, adopting the Metropolis-Hastings (MH) algorithm. The MH rule compares the likelihood of new vs old set of 
parameter values and determines whether these random walks are to be accepted or rejected. The algorithm generally drifts towards the highest likelihood regions, 
where a fit to the data is best. It then meanders around that region of the parameter space, exploring the shape of $\mathcal{L}$ in the vicinity of the maximum. 
This exploration maps out the posterior probability of all parameter values via maximization of the marginal likelihood function, or the integrated likelihood, 
which gives the marginalized constraints in the parameter space. Thus, the MCMC analysis is similar to the $\chi^2$ minimization, where one marginalizes over 
the parameters instead of optimizing them. 

We adopt a python implementation of the ensemble sampler for MCMC, the \texttt{emcee}\footnote{ \url{https://github.com/dfm/emcee}}, introduced by Foreman-Mackey 
\textit{et al}\cite{emcee}. We plot the two dimensional confidence contours showing the uncertainties along with the one dimensional marginalized posterior 
probability distributions, using the \texttt{GetDist}\footnote{ \url{https://github.com/cmbant/getdist}} module of python, developed by Lewis\cite{getdist}.

\section{Gaussian process}

A Gaussian Process (GP) involves an indexed collection of random variables having a Multivariate Normal distribution. GPs can be used to infer 
a distribution over functions directly. The distribution of a GP is the joint distribution of all random variables, which is a distribution over 
functions within a continuous domain. For a given set of Gaussian-distributed observational data, we use GP to reconstruct the most probable 
underlying continuous function describing that data and also obtain the associated confidence levels, without limiting to any particular 
parametrization ansatz. 

Let us consider a function $f$ formed from a GP. The value of $f$, when evaluated at some point $x$, is a Gaussian random variable with mean $\mu(x)$ 
and variance $\text{var}(x)$. The function value at $x$ is dependent on the function value at some other point $\tilde{x}$ (especially when $x$ and $\tilde{x}$ are close to each other) and is related by a covariance 
function, $\text{cov}(f(x), f(\tilde{x})) = \kappa(x, \tilde{x})$, which correlates values of the function at $x$ and $\tilde{x}$ separated by $\vert x-\tilde{x} 
\vert$ distance units.

Therefore, the distribution of functions can be described by the following quantities
\begin{eqnarray}
\mu(x) &=& \mathcal{E}[f(x)], \\
\kappa(x,\tilde{x}) &=&  \mathcal{E}[(f(x)-\mu(x))(f(\tilde{x})-\mu(\tilde{x}))],\\
\text{var}(x) &=& \kappa(x,x) ,
\end{eqnarray}
where $\mathcal{E}$ denotes the expectation.

The Gaussian process is written as 
\begin{equation}
f(x) \sim \mathcal{GP} (\mu(x), \kappa(x, \tilde{x})), \label{ch1:fx_gp}
\end{equation}  
where $\mathcal{GP}$ represents a Gaussian Process.

The covariance function $\kappa(x, \tilde{x})$ depends on a set of free parameters, called the \textit{hyperparameters}, namely the  characteristic length 
scale $l$ and the signal variance $\sigma_f$. The hyperparameter $l$ roughly corresponds to the distance one needs to move in input space before the function 
value changes significantly, while $\sigma_f$ describes typical changes in the function value. Different choices for the covariance function may have different 
effects on the reconstruction. A wide range of possible covariance functions is available in the literature \cite{rw, mackay, william}. As a standard choice 
one may consider the squared exponential covariance, 
\begin{equation} \label{ch1:sqexp}
\kappa(x, \tilde{x}) = \sigma_f^2 \exp \left[ - \frac{(x-\tilde{x})^2}{2l^2}\right].
\end{equation}

Another possible choice is the Mat\'{e}rn $\nu$ covariance, with $\nu \equiv \left(p+\frac{1}{2}\right)$, given by
\begin{equation}\label{ch1:matern}
\begin{split}
\kappa_{\nu = p+\frac{1}{2}}(x,\tilde{x}) = \sigma_f^2 \exp \left( \frac{-\sqrt{2p+1}}{l} \vert x - \tilde{x} \vert \right) \frac{p!}{(2p)!} \times ~~~~~~~~~~~~~~~~~~~~~~~~~~~~\\~~~~~~~~~~~~~~~~~~~~~~~~~~~~~~~~~~~~~~~~~~ \times \mathlarger{\mathlarger{\sum}}_{i=0}^{p}~ \frac{(p+i)!}{i!(p-i)!} \left(\frac{2\sqrt{2p+1}}{l} 
\vert x - \tilde{x} \vert \right)^{p-i}.
\end{split}
\end{equation} 
Here $p$ denotes the order of the Mat\'{e}rn covariance function.

The squared exponential covariance is extensively used in cosmology. For a reconstruction involving an $n$th order derivative, the Mat\'{e}rn $\nu$ 
covariance works well if $\nu > n$. 

Given a data set $\mathcal{D}$ of $n$ observations, $\mathcal{D} = \left\lbrace(x_i, y_i)\vert_{i = 1, . . . , n}\right\rbrace$, we attempt to 
reconstruct a function $f$ that describes this data. We consider a set of input points $\mathbf{X} = \lbrace x_i \rbrace$, and the covariance matrix 
$\mathbf{K} =\kappa (\mathbf{X}, \mathbf{X}) $ is given by $[\kappa(\mathbf{X}, \mathbf{X})]_{ij} = \kappa(x_i, x_j)$. For GPs, any $x_i$ is assigned 
a random variable $f(x_i)$, and the joint distribution of a finite number of these variables $\lbrace f(x_1),\dots,f(x_n)\rbrace$ is itself Gaussian.
\begin{equation}
\mathbf{f} \sim \mathcal{GP}(\boldsymbol{\mu}, \mathbf{K}) \label{ch1:function} ,
\end{equation}
with $\boldsymbol{\mu} = (\mu({x}_1),\cdots,\mu({x}_n))$ and $\mathbf{f} = (f({x}_1),\cdots,f({x}_n))$ respectively. 

Excluding observational data, one can use the covariance matrix $\mathbf{K}$ to generate a Gaussian vector $\mathbf{f}^{*}$ of function values at 
$\mathbf{X}^{*}$ with $f^{*}_i = f(x^{*}_i)$, such that
\begin{equation}\label{ch1:prior}
\mathbf{f}^{*} \sim \mathcal{GP} \left( \boldsymbol{\mu}^{*} , \mathbf{K}^{**} \right).
\end{equation} 
Here $\boldsymbol{\mu}^{*}$ is the a priori assumed mean of $\mathbf{f}^{*}$, and $\mathbf{K}^{**} = \kappa(\mathbf{X}^{*}, \mathbf{X}^{*})$.

Observational data $(x_i, y_i)$ can also be described by GPs, assuming Gaussian error distribution. The actual observations are considered 
to be scattered around the underlying function, i.e. $y_i = f(x_i)+\epsilon_i$, where Gaussian noise $\epsilon_i$ with variance $\sigma_i^2$ is 
assumed. This variance needs to be added to the covariance matrix
\begin{equation}\label{ch1:gpdata}
\mathbf{y} \sim \mathcal{GP} \left( \boldsymbol{\mu}, \mathbf{K} + \mathbf{C} \right),
\end{equation} 
where $\mathbf{C}$ is the covariance matrix of the data. For uncorrelated data, we use $\mathbf{C} = \sigma_i^2 \mathbf{I}$. The above two GPs, given by equation 
\eqref{ch1:prior} for $\mathbf{f}^*$ and Eq. \eqref{ch1:gpdata} for $\mathbf{y}$ can be combined in the joint distribution,
\begin{equation}
\begin{bmatrix}
\mathbf{y} \\
\mathbf{f}^*
\end{bmatrix} \sim \mathcal{GP} \left( \begin{bmatrix}
\boldsymbol{\mu} \\
\boldsymbol{\mu}^*
\end{bmatrix} \begin{bmatrix}
\mathbf{K} + \mathbf{C} & \mathbf{K}^* \\
{\mathbf{K}^{*}}^{\text{\small T}} & \mathbf{K}^{**} 	\end{bmatrix}\right),
\end{equation} 
where $\mathbf{K}^* = \kappa(\mathbf{X},\mathbf{X}^*)$ and ${\mathbf{K}^{*}}^{\text{\small T}} = \kappa(\mathbf{X}^*,\mathbf{X})$ respectively. 

Using the standard rules for conditioning Gaussian functions, their predictive distribution is given by
\begin{equation}\label{ch1:posterior}
\mathbf{f}^* \vert \mathbf{X}^*, \mathbf{X}, \mathbf{y} \sim \mathcal{GP}	\left({\overline{\boldsymbol{\mu}}}^*, \boldsymbol{\Sigma}^* \right),
\end{equation} 
where
\begin{equation}
{\overline{\boldsymbol{\mu}}}^* = \boldsymbol{\mu}^* + {\mathbf{K}^{*}}^{\text{\small T}} [\mathbf{K} + \mathbf{C}]^{-1} (\mathbf{y} - 
\boldsymbol{\mu}),
\end{equation} 
and
\begin{equation}
\boldsymbol{\Sigma}^* =  \mathbf{K}^{**} - {\mathbf{K}^{*}}^{\text{\small T}} [\mathbf{K}+\mathbf{C}]^{-1} \mathbf{K}^*	,
\end{equation} 
are the mean and covariance of $\mathbf{f}^*$ respectively. Equation \eqref{ch1:posterior} is the posterior distribution of the function 
given the data \eqref{ch1:gpdata} and the prior \eqref{ch1:prior}.

Although equation \eqref{ch1:posterior} covers noise in training the data, it is still a distribution over noise-free predictions $\mathbf{f}^*$. 
Contribution from the noise can be included into our final predictions $\mathbf{y}^*$, by adding $\mathbf{C}$ to $\boldsymbol{\Sigma}^*$ in equation 
\eqref{ch1:posterior}. Therefore, 
\begin{equation}\label{ch1:posterior_noise}
\mathbf{y}^* \vert \mathbf{X}^*, \mathbf{X}, \mathbf{y} \sim \mathcal{GP}	\left({\overline{\boldsymbol{\mu}}}^*, \boldsymbol{\Sigma}^*+\mathbf{C} \right).
\end{equation} 

To apply the above equations for reconstructing a function, we need to estimate the hyperparameters $\sigma_f$ and $l$. They can be trained by maximizing 
the marginal likelihood. The marginal likelihood depends only on the locations $\mathbf{X}$ of the observations, but not on $\mathbf{X}^*$, where we want 
to reconstruct the function. For a Gaussian prior $\mathbf{f}\vert \mathbf{X}, \sigma_f , l \sim \mathcal{GP} (\boldsymbol{\mu}, \mathbf{K})$ 
and with $\mathbf{y} \vert \mathbf{f} \sim \mathcal{GP} (\mathbf{f}, \mathbf{C})$, the log marginal likelihood is given by
\begin{equation}\label{ch1:lnlike}
\ln \mathcal{L} = -	\frac{1}{2}(\mathbf{y}-\boldsymbol{\mu})^{\text{\small T}} [\mathbf{K} + \mathbf{C}]^{-1} ~(\mathbf{y}-\boldsymbol{\mu}) -\frac{1}{2}\ln 
\vert \mathbf{K} + \mathbf{C} \vert - \frac{n}{2}\ln 2\pi.
\end{equation}
The hyperparameters $\sigma_f$ and $l$ are obtained by maximizing this log marginal likelihood. 

This approach also provides a vigorous way for estimating derivatives of the function from a covariance between the function and its derivative, and another 
between the derivatives for the reconstruction. These covariances can be obtained by differentiating the original covariance function $\kappa(x,\tilde{x})$. 
Given the Gaussian Process for $f(x)$ (from equation \eqref{ch1:fx_gp}), the Gaussian Processes for the first and second derivatives for $f(x)$ are consequently 
given by
\begin{align}
f~'(x) &\sim \mathcal{GP} \left(\mu'(x), \frac{\partial^2 \kappa(x, \tilde{x})}{\partial x \partial \tilde{x}}\right), \label{ch1:f1x_gp}\\
f~''(x) &\sim \mathcal{GP} \left(\mu''(x), \frac{\partial^4 \kappa(x, \tilde{x})}{\partial x^2 \partial \tilde{x}^2}\right). \label{ch1:f2x_gp}
\end{align}

Gaussian processes have widely been applied in cosmological data analysis. A non-parametric reconstruction using GP has been utilized in 
\cite{gp1, gp2, gp3, gp_seikel, gp4, gp5, gp6, purba_cddr, purba_q, purba_j, purba_int, purba_ok, yangguocai, gp9, gp10, gp11, gp12, gp13, gp14, gp15, bilicki, lin, 
xia, nunes, carlos, jesus_nonpara, levisaid2021, zhang_li, li_rsd}. We refer to the publicly available GP website\footnote{ \url{http://www.gaussianprocess.org}} 
for more details of the method. In a pedagogical introduction to GP, Seikel, Clarkson and Smith\cite{gp_seikel} developed the publicly available \texttt{GaPP}\footnote{\url{https://github.com/carlosandrepaes/GaPP}} code which has been utilized in the present context.

\section{Observational datasets utilized in reconstruction} \label{ch1:obsdata}

In this thesis, we have used different combinations of background datasets like the Cosmic Chronometer (CC) measurements of the Hubble parameter, recent 
compilations of the Type Ia Supernova distance modulus data (SN), the Pantheon Supernova compilation of CANDELS and CLASH Multi-Cycle Treasury (MCT) programs 
obtained by the Hubble Space Telescope (HST), the radial and volume-averaged Baryon Acoustic Oscillation (BAO) data, the Cosmic Microwave Background (CMB) Shift parameter data and 
the growth rate $f\sigma_8$ measurements from Redshift-Space Distortions (RSD), caused by the peculiar motions of galaxies \cite{kaiser} for the non-parametric 
reconstruction in cosmology. A summary of the datasets is given below.

\subsection{Cosmic Chronometers}

The Hubble parameter $H(z)$ can directly be estimated by calculating the differential ages of galaxies \cite{cc_101,cc_102,cc_103,cc_104,cc_106,cc_105}, 
usually called cosmic chronometer (CC) as
\begin{equation} \label{ch1:ccH}
H(z) = -\frac{1}{1+z} \frac{\dif z}{\dif t}.
\end{equation}

\begin{table*}[t!]
	\caption{{\small The Cosmic Chronometer Hubble parameter $H$ measurements (in units of km s$^{-1}$ Mpc$^{-1}$) and their errors $\sigma_H$ 
			at redshift $z$ obtained from the differential age method (CC).}} 
	\begin{center}
		\resizebox{0.99\textwidth}{!}{\renewcommand{\arraystretch}{1.35} \setlength{\tabcolsep}{13pt}\centering 
			\begin{tabular}{l c c c c c} 
				\hline
				\hline
				\textbf{Index} & $z$ & $H \pm \sigma_H$ \textbf{(CCB)} & $H \pm \sigma_H$ \textbf{(CCM)} & $H \pm \sigma_H$ \textbf{(CCH)} & \textbf{References}\\
				\hline				
				1 & 0.07 & 69 $\pm$ 19.6 & $\cdots$ & $\cdots$ & \cite{cc_101}\\
				2 & 0.09 & $\cdots$ & $\cdots$ & 69 $\pm$ 12 & \cite{cc_102}\\
				3 & 0.12 & 68.6 $\pm$ 26.2 & $\cdots$ & $\cdots$ & \cite{cc_101}\\
				4 & 0.17 & $\cdots$ & $\cdots$ & 83 $\pm$ 8 & \cite{cc_102}\\
				5 & 0.1797 & 75 $\pm$ 4 & 81 $\pm$ 5 & $\cdots$ & \cite{cc_103}\\
				6 & 0.1993 & 75 $\pm$ 5 & 81 $\pm$ 6 & $\cdots$ & \cite{cc_103}\\
				7 & 0.2 & 72.9 $\pm$ 29.6 & $\cdots$ & $\cdots$ & \cite{cc_101}\\
				8 & 0.27 & $\cdots$ & $\cdots$  & 77 $\pm$ 14 &\cite{cc_102}\\
				9 & 0.28 & 88.8 $\pm$ 36.6 & $\cdots$ & $\cdots$ & \cite{cc_101}\\ 
				10 & 0.3519 & 83 $\pm$ 14 & 88 $\pm$ 16 & $\cdots$ & \cite{cc_103}\\
				11 & 0.3802 & 83 $\pm$ 13.5 & 89.2 $\pm$ 14.1 & $\cdots$ & \cite{cc_104}\\
				12 & 0.4 & $\cdots$ & $\cdots$ & 95 $\pm$ 17 & \cite{cc_102}\\
				13 & 0.4004 & 77.0 $\pm$ 10.2 & 82.8 $\pm$ 10.6 & $\cdots$ & \cite{cc_104}\\
				14 & 0.4247 & 87.1 $\pm$ 11.2 & 93.7 $\pm$ 11.7 & $\cdots$ & \cite{cc_104}\\
				15 & 0.4497 & 92.8 $\pm$ 12.9 & 99.7 $\pm$ 13.4 & $\cdots$ & \cite{cc_104}\\
				16 & 0.47 & 89 $\pm$ 49.6 & $\cdots$ & $\cdots$ & \cite{cc_106}\\
				17 & 0.4783 & 80.9 $\pm$ 9.0 &  86.6 $\pm$ 8.7  & $\cdots$ & \cite{cc_104}\\
				18 & 0.48 & $\cdots$ & $\cdots$ & 97 $\pm$ 62 & \cite{cc_102}\\
				19 & 0.5929 & 104 $\pm$ 13 & 110 $\pm$ 15 & $\cdots$ & \cite{cc_103}\\
				20 & 0.6797 & 92 $\pm$ 8 & 98 $\pm$ 10 & $\cdots$ & \cite{cc_103}\\
				21 & 0.7812 & 105 $\pm$ 12 & 88 $\pm$ 11 & $\cdots$ & \cite{cc_103}\\
				22 & 0.8754 & 125 $\pm$ 17 & 124 $\pm$ 17 & $\cdots$ & \cite{cc_103}\\
				23 & 0.88 & $\cdots$ & $\cdots$ & 90 $\pm$ 40 & \cite{cc_102}\\
				24 & 0.9 & $\cdots$ & $\cdots$ & 117 $\pm$ 23 & \cite{cc_102}\\
				25 & 1.037 & 154 $\pm$ 20 & 113 $\pm$ 15 & $\cdots$ & \cite{cc_103}\\
				26 & 1.3 & $\cdots$ & $\cdots$ & 168 $\pm$ 17 & \cite{cc_102}\\
				27 & 1.363 & 160 $\pm$ 33.6 & 160 $\pm$ 33.6 & $\cdots$ & \cite{cc_105}\\
				28 & 1.43 & $\cdots$ & $\cdots$ & 177 $\pm$ 18 & \cite{cc_102}\\
				29 & 1.53 & $\cdots$ & $\cdots$ & 140 $\pm$ 14 & \cite{cc_102}\\
				30 & 1.75 & $\cdots$ & $\cdots$ & 202 $\pm$ 40 & \cite{cc_102}\\
				31 & 1.965 & 186.5 $\pm$ 50.4 & 186.5 $\pm$ 50.4 & $\cdots$ & \cite{cc_105}\\
				\hline
				\hline
			\end{tabular}
		}
	\end{center}
	\label{ch1:tabcc}
\end{table*}

These CC measurements are independent of the Cepheid distance scale and do not rely on any particular cosmological model \cite{jimenez2002}, yet  
are subject to other sources of systematic uncertainties, like the ones associated with the modelling of stellar ages that are carried out through 
the so-called stellar population synthesis (SPS) techniques. Given a pair of ensembles of passively evolving galaxies at two different redshifts, one can possibly infer $\frac{\dif z}{\dif t}$ from observations, on assuming a concrete SPS model \cite{cc_103, cc_104, sps2}. 
Therefore, one can obtain direct information about the Hubble parameter at different $z$. 

The authors in \cite{cc_103,cc_104} provide 13 $H(z)$ values obtained by considering the BC03 \cite{bc03} and MaStro \cite{mastro} SPS models, which 
we shall refer to as the CCB and CCM compilation, respectively. In \cite{cc_101,cc_106} the authors provide only 5 $H(z)$ values obtained with the 
BC03 model, and have been added to the CCB compilation.  In case of \cite{cc_105}, the combined MaStro/BC03 values are available for 2 $H(z)$ measurements. 
An alternative SPS different from the MaStro and BC03 models, is assumed in \cite{cc_102}, hereafter referred to as the CCH compilation, consisting of 
11 $H(z)$ values. Table \ref{ch1:tabcc} includes almost all the CC data reported in various surveys so far.

\subsection{Type Ia Supernovae}

The type Ia supernovae are widely accepted as {\it standard candles} to measure cosmological distances for their consistent absolute luminosity \cite{riess1998, perl1999}. 
Standard candles are a class of distinguishable objects with defined intrinsic brightness. They can be distinctly identified for a wide range of distance. 

For the supernova data, we consider the recent Pantheon compilation by Scolnic {\it et al}\cite{pan}. Usually, the samples are presented as 
distance modulus $\mu$ with errors tabulated at different redshift $z$. The numerical dataset of the full Pantheon SNIa catalogue is publicly available\footnote{\url{http://dx.doi.org/10.17909/T95Q4X}}$^,$\footnote{\url{https://archive.stsci.edu/prepds/ps1cosmo/index.html}}. 
The Pantheon compilation is presently the largest spectroscopically confirmed SNIa data, which consists of 1048 supernovae from different 
surveys, comprising the Sloan Digital Sky Survey (SDSS) \cite{kessler2009}, SN Legacy Survey (SNLS) \cite{conley2011}, various low-z samples 
viz. the Pan-STARRS1 Medium Deep Survey \cite{rest2014}, the Harvard Smithsonian Center for Astrophysics SN surveys \cite{hicken2009}, the 
Carnegie SN Project \cite{strit2011} and some high-z data from the HST cluster SN survey \cite{suzuki}, GOODS \cite{zt_ref1} 
and CANDELS/CLASH survey \cite{rodney2014, graur2014}. The total number of SNIa in the Pantheon dataset is 1048, which is about twice that of the 
Union 2.1 \cite{suzuki} compilation, and is about $40\%$ more than the Joint Light-curve Analysis (JLA) \cite{betoule} compilation. 

The distance modulus of SNIa can be derived from the observation of light curves through the empirical relation given by Tripp\cite{tripp}
\begin{equation}\label{ch1:trip}
\mu = m^*_B + \alpha X_1 - \beta C - M_B + \Delta_M + \Delta_B,
\end{equation}
where $X_1$ and $C$ are the stretch and colour correction parameters, $m^*_B$ is the observed apparent magnitude and $M_B$ is the absolute magnitude 
in the B-band for SNIa while $\alpha$ and $\beta$ are two nuisance parameters characterizing the luminosity-stretch, and luminosity-colour relations 
respectively. $\Delta_M$ is a distance correction based on the host-galaxy mass of the SNIa and $\Delta_B$ is a distance correction based on predicted 
biases from simulations. Usually, the nuisance parameters $\alpha$ and $\beta$ are optimized simultaneously with the cosmological model parameters or 
are marginalized over. 

\subsubsection*{Pantheon SNIa distance modulus compilation}

In the Pantheon compilation by Scolnic \textit{et al}\cite{pan}, based on the new approach called BEAMS with Bias Corrections (BBC) \cite{beams} the nuisance 
parameters in Eq. \eqref{ch1:trip} are retrieved, and the corrected apparent magnitude $m_B = m^{*}_B + \alpha X_1 - \beta C$ along with $\Delta_M$ and 
$\Delta_B$ corrections are reported. So, the observed distance modulus is reduced to the difference between the corrected apparent magnitude $m_B$ and 
the absolute magnitude $M_B$, i.e. \begin{equation}
\mu = m_B - M_B.
\end{equation} 
Constraints on $M_B$ have been obtained by considering it a free parameter in our analysis. We have marginalized over the Pantheon data for $M_B$ in combination 
with the Hubble data. 

The distance modulus of each supernova can be estimated as
\begin{equation}\label{ch1:mu}
\mu(z) = 5 \log_{10} \frac{d_L(z)}{\text{Mpc}} + 25,
\end{equation}
where $d_L$ is the luminosity distance defined in Eq. \eqref{ch1:luminosity_d}.

The theoretical apparent magnitude is given by
\begin{equation} \label{ch1:m}
m = \mu + M_B = 5 \log_{10} D_L + \mathcal{M} ,
\end{equation} 
where, $\mathcal{M} = M_B - 5\log_{10} \frac{H_0}{c} + 25$ is a nuisance parameter that captures the degeneracy between the absolute magnitude $M_B$ of SNIa 
and the Hubble parameter at present epoch $H_0$.

One can rewrite the luminosity distance $d_L$ in a normalized dimensionless way as
\begin{equation} \label{ch1:DL}
D_L(z)  \equiv 10^{\frac{m - \mathcal{M}}{5}} = \frac{H_0}{c} 10^{\frac{\mu - 25}{5}}.
\end{equation}

This normalized comoving distance $D_L$ is related to the dimensionless transverse comoving distance $D$ as,
\begin{equation} \label{ch1:D_from_DL}
D(z) = \frac{D_L}{1+z}.	
\end{equation}

The statistical uncertainty $\mathbf{C}_{\text{\tiny stat}}$ and systematic uncertainty $\mathbf{C}_{\text{\tiny sys}}$ are also included 
in our calculation. The total uncertainty matrix of the apparent magnitude $m$ is given by, 
\begin{equation} \label{ch1:mu_error}
\bm{\Sigma}_\mu = \mathbf{C}_{\text{\tiny stat}} + \mathbf{C}_{\text{\tiny sys}} .
\end{equation}

The uncertainty in $D(z)$ is propagated from the uncertainties of $m$ and $\mathcal{M}$ using the standard error propagation rule. 
The formula for obtaining the uncertainty in $D(z)$ i.e., $\bm{\Sigma}_D$ from the uncertainties in $\mu$ and $H_0$ is 
\begin{equation}  \label{ch1:sne_sigD}
\bm{\Sigma}_D = \mathbf{D}_1 \bm{\Sigma}_\mu {\mathbf{D}_1}^{\text{\small T}} + \sigma_{H_{0}}^2 \mathbf{D}_2 \mathbf{D}_2^{\text{\small T}},
\end{equation}
where $\sigma_{H_0}$ is the uncertainty in $H_0$, $\bm{\Sigma}_\mu$ is the covariance matrix for $\mu$ and the superscript 
`${\text{\small T}}$' denotes the transpose of any matrix. $\mathbf{D}_1$ and $\mathbf{D}_2$ are the Jacobian matrices, given by
\begin{eqnarray}
\mathbf{D}_1 &=& \text{diag}\left(\frac{\ln 10}{5} \mathbf{D}\right), \\
\mathbf{D}_2 &=& \text{diag}\left(\frac{1}{H_0} \mathbf{D}\right),
\end{eqnarray}
where $\mathbf{D}$ is a vector whose components are the normalized comoving distances of all the SNIa.

\subsubsection*{Pantheon SNIa, CANDELS and CLASH Multi-Cycle Treasury compilation}

Riess \textit{et al}\cite{mct} provides the most recent binned expansion rate $E(z_i)$ dataset that compress information very effectively regarding the 1048 
supernovae at $z < 1.5$ of the Pantheon compilation (which includes 740 SNIa of the JLA compilation), and 15 SNIa at $z > 1$ of the CANDELS and CLASH 
MCT programs obtained by the HST, 9 of which are in the range $1.5 < z < 2.3$.

\begin{table}[t!]
	\caption{{\small $E(z)$ obtained from the Pantheon+MCT compilation via an inversion of ${E^{-1}(z)}$ data reported in Table 6 
			of Ref \cite{mct}. Note the difference in the estimate of $E(z=1.5)$, from the actual quoted value. We pick up the mean 
			value obtained from inversion of quoted ${E^{-1}(z)}$. The covariance matrix has been included in our analysis.}}
	\begin{center}
		\resizebox{0.7\textwidth}{!}{\renewcommand{\arraystretch}{1.4} \setlength{\tabcolsep}{38pt}\centering 	
			\begin{tabular}{l c c} 
				\hline
				\hline
				\textbf{Index} & $z$ & $E(z)$ \\ 
				\hline
				1 & 0.07 & 0.997 $\pm$ 0.023 \\
				2 & 0.20 & 1.111 $\pm$ 0.021 \\
				3 & 0.35 & 1.127 $\pm$ 0.037 \\ 
				4 & 0.55 & 1.366 $\pm$ 0.062 \\
				5 & 0.9 & 1.524 $\pm$ 0.121 \\
				6 & 1.5 & 2.924 $\pm$ 0.675 \\
				\hline
				\hline
			\end{tabular} \label{ch1:tabmct}
	} \end{center}
\end{table}

The raw SNIa measurements have been converted into data on $E(z)$ by parametrizing $E^{-1} (z)$, for six different redshifts in the range $z \in [0.07, 1.5]$, 
assuming a spatially flat Universe. The corresponding values of $E^{-1} (z_i )$ are Gaussian in a very good approximation as shown in the work of Riess 
\textit{et al}\cite{mct} which also contains the corresponding correlation matrix. In this thesis, we have adopted the mean value obtained from an inversion 
of the quoted ${E^{-1}(z)}$ and the inverse covariance matrix.

\subsection{Baryon Acoustic Oscillations}

Baryon acoustic oscillations are regular, periodic fluctuations in the matter power spectrum caused by acoustic density waves in the primordial plasma 
of the early Universe. BAO are widely used as ``\textit{standard rulers}" to measure the distances in cosmology. The length of this standard ruler is 
given by the maximum distance the acoustic waves could travel in the primordial plasma, prior to recombination. The BAO provides a measurement of the 
angular diameter distance $d_A$ as a function of redshift, and is also used to directly measure the expansion rate $H(z)$ in the line of sight.

The early Universe was composed of photons, baryons and dark matter, in which the photons and baryons were tightly coupled. The early Universe was 
homogeneous except for the presence of tiny fluctuations. As the Universe expands it becomes cooler and less dense. So, these fluctuations grew due 
to gravity. Acoustic waves are generated as the photon-baryon fluid is attracted and fall onto the over-densities via compressions and rarefactions. 
These acoustic waves have propagated until the Universe became cool enough for the electrons and protons to recombine. From then on, the baryons and 
photons started to decouple. The time when these decoupled baryons are released from the drag of photons is known as the drag epoch, $z_d$. The 
decoupled photons expanded freely while the acoustic waves restricted the baryons in a scale given by the size of the horizon at $z_d$. Progressively, 
the baryons fell into dark matter potential wells. The two components, baryons and dark matter, attracted one another and their over-densities. 
As all galaxies in the Universe have formed from these matter over-densities. Thus, BAO provide a characteristic scale, which can be measured in 
either the galaxy correlation function or the galaxy power spectrum as functions of redshift.

Before the recombination epoch, the tightly coupled photon-baryon mixture formed via Thomson scattering was in a hot plasma state. The radiation pressure 
and the gravitational attraction acted as two competing forces, thus setting up the oscillations in plasma. A single spherical over-density in the tightly 
coupled photon-baryon plasma would propagate with a speed $c_s$, given by
\begin{equation}
c_s = \frac{c}{\sqrt{3\left(1+\frac{3~\rho_b}{4 ~\rho_\gamma} \right)}}, \label{ch1:cs}
\end{equation}
where $\rho_b = \rho_{b0}(1+z)^3$ is the baryon density and $\rho_\gamma = \rho_{\gamma 0}(1+z)^4$ is the photon density \cite{cs_ref}. 
The ratio $\frac{\rho_b}{\rho_\gamma}$ in equation \eqref{ch1:cs} can be replaced by the ratio of their respective density parameters as, 
$\frac{\rho_b}{\rho_\gamma} \equiv \frac{1}{(1+z)}~\frac{\Omega_{b0}}{\Omega_{\gamma 0}}$, where $\Omega_{b0}$ is the present value of the 
baryon density parameter and $\Omega_{\gamma0}$ is the present value of the photon density parameter.

The photons decoupled from the baryons during recombination and propagated freely, forming the CMBR. The baryons 
became neutral at recombination, and the spherical shells formed by BAO remained imprinted as the surface of \textit{last scattering} on the distribution 
of baryonic matter in the Universe. The distance travelled by the acoustic wave in the time interval starting from the beginning of matter formation to a 
given redshift $z$ is called the comoving sound horizon $r_s$ at redshift $z$.
\begin{equation}
r_s(z) = \int_{z}^{\infty} \frac{c_s(z)}{H(z)} ~\dif z \label{ch1:rs}.
\end{equation}

For the BAO data, this comoving sound horizon ($r_s$) is measured at two definite redshift values, the photon-electron decoupling epoch $(z_* )$ and the photon 
drag epoch $(z_d)$. We rewrite $r_s(z_*) \equiv r_*$ and $r_s(z_d) \equiv r_d$ as the comoving sound horizon at the photon decoupling and the photon drag epoch 
\cite{rdrag_def} respectively, as a matter of notation. Recent observations from the Planck probe \cite{planck} reveals $z_* \approx 1090$ and $z_d \approx 
1060$. Considering the background cosmology described by a $\Lambda$CDM model, the Planck data measures $r_* \approx 145$ and $r_d \approx 147.5$, both in 
units of Mpc.

From equations \eqref{ch1:cs} and \eqref{ch1:rs}, the comoving sound horizon at drag epoch $r_d$ is given by
\begin{equation} 
r_d = \frac{c}{\sqrt{3}} \int_{z_d}^{\infty} \frac{1}{H(z) \left[1+\frac{3\Omega_{b0}}{4 \Omega_{\gamma0}} (1+z)^{-1} \right]^{\frac{1}{2}}} \dif z . \label{ch1:rdrag_def}
\end{equation} 

The acoustic scale at photon decoupling is defined as
\begin{equation}
l_A(z_*) = \pi \frac{d_P(z_*)}{r_s(z_*)} ,
\end{equation}
where $d_P(z_*)$ is the transverse comoving distance at decoupling.

We define another important quantity, namely the dilation scale $D_V$, given by
\begin{equation}
D_V (z)= \left[d_P^2(z) \frac{c z}{H(z)}\right]^\frac{1}{3}. \label{ch1:dilation_d}
\end{equation}
This $D_V$ is also known as the volume-averaged BAO distance, which is a geometric mean of the BAO distance measurements along two transverse 
directions and one radial direction.

We have utilized the volume-averaged $D_V$ BAO measurements and the radial BAO $H(z)$ compilation in the various analysis discussed in the subsequent chapters.

\begin{table}[t!]
	\caption{{\small The Hubble parameter measurements $H(z)$ (in units of km s$^{-1}$ Mpc$^{-1}$) and their errors $\sigma_H$ at redshift $z$ 
			obtained from the radial BAO method ($r$BAO).}}
	\begin{center}
		\resizebox{0.75\textwidth}{!}{\renewcommand{\arraystretch}{1.35} \setlength{\tabcolsep}{22pt}\centering 
			\begin{tabular}{l c c c} 
				\hline
				\hline
				\textbf{Index} & \textbf{$z$} & $H \pm \sigma_H$ & \textbf{References}\\ 
				\hline
				1 & 0.24 & 79.69 $\pm$ 2.99 & \cite{bao_61}\\
				2 & 0.3 & 81.7 $\pm$ 6.22 & \cite{bao_107}\\
				3 & 0.31 & 78.17 $\pm$ 4.74 & \cite{bao_108}\\ 
				4 & 0.34 & 83.8 $\pm$ 3.66 & \cite{bao_61}\\
				5 & 0.35 & 82.7 $\pm$ 8.4 & \cite{bao_77}\\
				6 & 0.36 & 79.93 $\pm$ 3.39 & \cite{bao_108}\\
				7 & 0.38 & 81.5 $\pm$ 1.9 & \cite{alam2017}\\
				8 & 0.40 & 82.04 $\pm$ 2.03 & \cite{bao_108}\\
				9 & 0.43 & 86.45 $\pm$ 3.68 & \cite{bao_61}\\
				10 & 0.44 & 82.6 $\pm$ 7.8 & \cite{blake2012}\\
				11 & 0.44 & 84.81 $\pm$ 1.83 & \cite{bao_108}\\
				12 & 0.48 & 87.79 $\pm$ 2.03 & \cite{bao_108}\\
				13 & 0.51 & 90.4 $\pm$ 1.9 & \cite{alam2017}\\
				14 & 0.52 & 94.35 $\pm$ 2.65 & \cite{bao_108}\\
				15 & 0.56 & 93.33 $\pm$ 2.32 & \cite{bao_108}\\
				16 & 0.57 & 87.6 $\pm$ 7.8 & \cite{bao_2}\\
				17 & 0.57 & 96.8 $\pm$ 3.4 & \cite{anderson2014}\\
				18 & 0.59 & 98.48 $\pm$ 3.19 & \cite{bao_108}\\
				19 & 0.6 & 87.9 $\pm$ 6.1 & \cite{blake2012}\\
				20 & 0.61 & 97.3 $\pm$ 2.1 & \cite{alam2017}\\
				21 & 0.64 & 98.82 $\pm$ 2.99 & \cite{bao_108}\\
				22 & 0.73 & 97.3 $\pm$ 7 & \cite{blake2012}\\
				23 & 0.978 & 113.72 $\pm$ 14.63 & \cite{bao_4}\\
				24 & 1.23 & 131.44 $\pm$ 12.42 & \cite{bao_4}\\
				25 & 1.526 & 148.11 $\pm$ 12.71 & \cite{bao_4}\\
				26 & 1.944 & 172.63 $\pm$ 14.79 & \cite{blake2012}\\
				27 & 2.3 & 224 $\pm$ 8	& \cite{bao_1}\\
				28 & 2.33 & 224 $\pm$ 8 & \cite{bautista2017}\\
				29 & 2.34 & 222 $\pm$ 7 & \cite{bao_9}\\
				30 & 2.36 & 226 $\pm$ 8 & \cite{bao_110}\\
				31 & 2.4 & 227.8 $\pm$ 5.61 & \cite{bao_3}\\
				\hline
				\hline
			\end{tabular} \label{ch1:tabbao}
		}
	\end{center}
\end{table}

The volume-averaged BAO $\frac{D_V}{r_d}$ compilation consists of data from the Six-degree-Field Galaxy Survey (6dFGS) at $z=0.106$ \cite{beutler2011}, 
WiggleZ Dark Energy Survey at $z = 0.44$, $0.6$ and $0.73$  \cite{blake2012, kazin2014}, SDSS Main Galaxy Sample (MGS) at $z = 0.15$ \cite{ross2015}, 
LOWZ and CMASS samples of the Baryon Oscillation Spectroscopic Survey (BOSS) at $z= 0.32, 0.57$ \cite{gilmarin2015}, DR14 galaxy samples of the extended Baryon 
Oscillation Spectroscopic Survey (eBOSS) for Luminous Red Galaxies (LRG) \cite{bautista2018} and quasars \cite{ata2018} at $z = 0.72, 1.52$, 
correlations of Ly$\alpha$ absorption in eBOSS DR14 galaxy sample at $z=2.34$ \cite{agathe} and cross-correlation of Ly$\alpha$ absorption and quasars 
in eBOSS DR14 galaxy sample at $z=2.35$ \cite{blomqvist}. For all the datasets mentioned, we  appropriately utilize the covariance matrices provided in the respective references. When working with the volume-averaged BAO data, we obtained the marginalized constraints on $r_d$ (in units 
of Mpc), considering it a free parameter in the analysis.

An alternative compilation of the Hubble data can be obtained from  the radial BAO peaks in the galaxy power spectrum or from the BAO peaks using the 
Ly-$\alpha$ forest of quasi-stellar objects (QSO), which are based on the clustering of galaxies or quasars (hereafter referred to as $r$BAO). Table 
\ref{ch1:tabbao} includes almost all the $r$BAO data reported in various galaxy surveys like WiggleZ, SDSS MGS, BOSS DR12, eBOSS DR14Q and LRG, Ly-$\alpha$ 
forests, so far \cite{bao_61,bao_1,alam2017,bao_107,bao_108,bautista2017,bao_2,bao_4,bao_3,blake2012,bao_77,anderson2014,bao_9,bao_110}. One may find that 
some of the BAO $H(z)$ data from clustering measurements are correlated as they either belong to the same analysis or there is an overlap between the 
galaxy samples. In our work, we mostly consider the central value and standard deviation of the data unless otherwise mentioned.

\subsection{Cosmic Microwave Background Radiation}

At present, the most powerful source of observational data in cosmology is the Cosmic Microwave Background Radiation (CMBR), first detected by 
Penzias and Wilson\cite{cmb_wilson} in 1965. The CMBR is composed of the relic photons \cite{cmb_relic}, those that decoupled from baryons during recombination 
at some temperature $T \approx 3000K$, with $z_* \approx 1090$ as the redshift of recombination \cite{planck_cmb}. The CMBR resembles a thermal 
black-body spectrum whose temperature evolves with redshift as $T(z) = T_0(1 + z)$ in an expanding Universe, where $T_{0} \approx 2.72548 \pm 0.00057$ 
K \cite{fixsencmb} is the observed CMB temperature at the present epoch. Being the oldest electromagnetic radiation in the Universe that dates back to 
the epoch of recombination, the CMBR provides a lot of information on the early Universe. 

The fundamental observable of the CMBR is its intensity as a function of frequency and direction $\hat{\mathbf{n}}$ on the CMB sky. Despite being isotropic, the 
CMB contains fractional temperature anisotropies at the $10^{-5}$ level \cite{cmbtempf} and fractional polarization at the $10^{-6}$ (or lower) level 
\cite{cmbpolzf}, over a wide range of angular scales. The anisotropies in CMBR, observed by Cosmic Background Explorer (COBE) \cite{cobe}, Differential 
Microwave Radiometer (DMR) \cite{cobedmr}, WMAP \cite{ref5} and Planck satellite \cite{ref6, ref7} are fluctuations 
in the photon temperature and coming at the present epoch from different directions $\hat{\mathbf{n}}$.  

The fluctuations or anisotropies in the CMBR is represented by the function 
\begin{equation}
\Theta(\hat{\mathbf{n}}) = \frac{\Delta T(\hat{\mathbf{n}})}{T} ,
\end{equation} 
where $T$ is the constant average temperature across the CMB sky. These CMB anisotropies can be expanded in terms of the spherical harmonics $Y_{lm}$ 
with coefficients $a_{lm}$ as
\begin{equation}
\Theta(\hat{\mathbf{n}}) = \sum_{l=0}^{\infty} \sum_{m = l}^{-l} a_{lm} Y_{lm}(\hat{\mathbf{n}}) ,
\end{equation}
where, $l$ is the multipole index and is related to the angular separation $\theta$ of the sky as $l \sim \frac{2 \pi}{\theta}$ while $m$ is the phase. 
Thus, a large $l$ corresponds to a small $\theta$ and vice versa.  Assuming these CMB anisotropies to be Gaussian in nature, the multipole moments of the 
temperature field are given by 
\begin{equation}
a_{lm}=\int d\hat{\mathbf{n}}~Y^*_{lm}(\hat{\mathbf{n}})~\Theta(\hat{\mathbf{n}}), 
\end{equation}
with \begin{eqnarray}
\left\langle a^*_{lm}~a_{l'm'} \right\rangle &=& \delta_{ll'}~\delta_{mm'}~C_l ~, \\
\left\langle a_{lm} \right\rangle &=& 0 .
\end{eqnarray}
Here, $Y^*_{lm}$ is the complex conjugate of the spherical harmonics $Y_{lm}$ and $C_l$ is the \textit{angular power spectrum} of the temperature field, defined as
\begin{equation}
C_l = \left\langle a_{lm} a_{lm}^* \right\rangle = \frac{1}{2l+1} \sum_{m} \vert a_{lm} \vert^2 . 
\end{equation}

Finally, the CMB power spectrum is given by
\begin{equation}
\Delta_T^2=\frac{l(l+1)}{2\pi}~C_l~T^2.
\end{equation}

The angular power spectrum is directly related to the statistical two point angular correlation function ${\mathcal{C}}(\theta)$ in two directions, $\hat{\mathbf{n}}$ and 
$\hat{\textbf{n}'}$, averaged over the entire CMB sky as
\begin{equation}
{\mathcal{C}}(\theta) = \left\langle\Theta(\hat{\mathbf{n}})\Theta^{*}(\hat{\textbf{n}'})\right\rangle = \mathlarger{\mathlarger{\sum}}_{l}  \frac{(2l+1)}{4\pi} ~C_{l} 
~ P_{l}(\cos\theta),
\end{equation}
where $\theta = \hat{\mathbf{n}} \cdot \hat{\textbf{n}'}$ is the angular separation between the two directions and $P_{l}$ are Legendre polynomials. 

Hu and Dodelson\cite{hudodcmb} provides an exhaustive review on the CMB temperature and polarization spectrum.

The angular scale of the sound horizon at the last scattering surface is connected to the location of the first acoustic peak of the CMB temperature power spectrum 
anisotropies. It acts as an accurate geometrical probe of dark energy, described by the so-called CMB shift parameter $\mathcal{R}$, which is 
defined as
\begin{equation}
\mathcal{R} = {l^{(1)}_{\text{ref}}}/{l^{(1)}},
\end{equation} 
where $l^{(1)}$ is the  CMB temperature spectrum multipole of the first acoustic peak, and the subscript `ref' stands for the reference flat CDM model with $\Omega_{m0} = 1$. 

The shift parameter $\mathcal{R}$ can not directly be measured from the CMB, but derived from data assuming a fiducial $\Lambda$CDM model as the background 
cosmology. For a spatially flat Universe, the expression generally used for the shift parameter is
\begin{equation}\label{ch1:cmbeqn}
\mathcal{R} = \sqrt{\Omega_{m0}} ~\int_{0}^{z^*} \frac{\dif z}{E(z)} ,
\end{equation}
where $z_*$ = $1089$ is the redshift at recombination.

\begin{table}
	\caption{A compilation of $f\sigma_8$ RSD data reported in different surveys.}
	\label{ch1:tabrsd}
\resizebox{\textwidth}{!}{\renewcommand{\arraystretch}{1.15} \centering \setlength{\tabcolsep}{10pt} 
		\begin{tabular}{l c c c c c} 
			\hline	\hline 
			\textbf{Index} & \textbf{Dataset} & $z$ & $f\sigma_8(z)$ & \textbf{Refs.} & \textbf{Fiducial Cosmology} \\ 
			\hline
			1 & SDSS-LRG & $0.35$ & $0.440\pm 0.050$ & \cite{rsd1}  &$(\Omega_{m0},\Omega_{k0},\sigma_{80})=(0.25,0,0.756)$ \cite{rsd2}\\
			2 & VVDS & $0.77$ & $0.490\pm 0.18$ & \cite{rsd1}  &  $(\Omega_{m0},\Omega_{k0},\sigma_{80})=(0.25,0,0.78)$ \\
			3 & 2dFGRS & $0.17$ & $0.510\pm 0.060$ & \cite{rsd1}   & $(\Omega_{m0},\Omega_{k0})=(0.3,0,0.9)$ \\
			4 & 2MRS &0.02& $0.314 \pm 0.048$ &  \cite{rsd3, rsd4}   & $(\Omega_{m0},\Omega_{k0},\sigma_{80})=(0.266,0,0.65)$ \\
			5 & SNIa+IRAS &0.02& $0.398 \pm 0.065$ & \cite{rsd4, rsd5}   & $(\Omega_{m0},\Omega_{k0},\sigma_{80})=(0.3,0,0.814)$\\
			6 & SDSS-LRG-200 & $0.25$ & $0.3512\pm 0.0583$ & \cite{rsd6}   & $(\Omega_{m0},\Omega_{k0},\sigma_{80})=(0.276,0,0.8)$  \\
			7 & SDSS-LRG-200 & $0.37$ & $0.4602\pm 0.0378$ & \cite{rsd6}   & \\
			8 & SDSS-LRG-60 & $0.25$ & $0.3665\pm0.0601$ & \cite{rsd6}   & $(\Omega_{m0},\Omega_{k0},\sigma_{80})=(0.276,0,0.8)$ \\
			9 & SDSS-LRG-60 & $0.37$ & $0.4031\pm0.0586$ & \cite{rsd6}   & \\
			10 & WiggleZ & $0.44$ & $0.413\pm 0.080$ & \cite{blake2012}    & $(\Omega_{m0},h,\sigma_{80})=(0.27,0.71,0.8)$ \\
			11 & WiggleZ & $0.60$ & $0.390\pm 0.063$ & \cite{blake2012}   &  $C_{ij} \rightarrow$ \cite{blake2012} \\
			12 & WiggleZ & $0.73$ & $0.437\pm 0.072$ & \cite{blake2012}   & \\
			13 & 6dFGS& $0.067$ & $0.423\pm 0.055$ & \cite{rsd8} &   $(\Omega_{m0},\Omega_{k0},\sigma_{80})=(0.27,0,0.76)$ \\
			14 & SDSS-BOSS& $0.30$ & $0.407\pm 0.055$ & \cite{rsd9}  & $(\Omega_{m0},\Omega_{k0},\sigma_{80})=(0.25,0,0.804)$ \\
			15 & SDSS-BOSS& $0.40$ & $0.419\pm 0.041$ & \cite{rsd9} &  \\
			16 & SDSS-BOSS& $0.50$ & $0.427\pm 0.043$ & \cite{rsd9} &   \\
			17 & SDSS-BOSS& $0.60$ & $0.433\pm 0.067$ & \cite{rsd9} &   \\
			18 & Vipers& $0.80$ & $0.470\pm 0.080$ & \cite{rsd10} &   $(\Omega_{m0},\Omega_{k0},\sigma_{80})=(0.25,0,0.82)$  \\
			19 & SDSS-DR7-LRG & $0.35$ & $0.429\pm 0.089$ & \cite{bao_77}  &  $(\Omega_{m0},\Omega_{k0},\sigma_{80})=(0.25,0,0.809)$ \cite{ref1}\\
			20 & GAMA & $0.18$ & $0.360\pm 0.090$ & \cite{rsd13} &   $(\Omega_{m0},\Omega_{k0},\sigma_{80})=(0.27,0,0.8)$ \\
			21& GAMA & $0.38$ & $0.440\pm 0.060$ & \cite{rsd13}  &   \\
			22 & BOSS-LOWZ& $0.32$ & $0.384\pm 0.095$ & \cite{rsd14}   & $(\Omega_{m0},\Omega_{k0},\sigma_{80})=(0.274,0,0.8)$ \\
			23 & SDSS DR10 \& DR11 & $0.32$ & $0.48 \pm 0.10$ & \cite{rsd14}   &  $(\Omega_{m0},\Omega_{k0},\sigma_{80}$)$=(0.274,0,0.8)$ \cite{anderson2014} \\
			24 & SDSS DR10 \& DR11 & $0.57$ & $0.417 \pm 0.045$ & \cite{rsd14} &  \\
			25 & SDSS-MGS & $0.15$ & $0.490\pm0.145$ & \cite{rsd16} & $(\Omega_{m0},h,\sigma_{80})=(0.31,0.67,0.83)$ \\
			26 & SDSS-velocity & $0.10$ & $0.370\pm 0.130$ & \cite{rsd17}   & $(\Omega_{m0},\Omega_{k0},\sigma_{80})=(0.3,0,0.89)$ \cite{rsd18}\\
			27 & FastSound& $1.40$ & $0.482\pm 0.116$ & \cite{rsd19}  &   $(\Omega_{m0},\Omega_{k0},\sigma_{80})=(0.27,0,0.82)$ \cite{rsd20}\\
			28 & SDSS-CMASS & $0.59$ & $0.488\pm 0.060$ & \cite{rsd21} &   $\ \ (\Omega_{m0},h,\sigma_{80})=(0.307115,0.6777,0.8288)$ \\
			29 & BOSS DR12 & $0.38$ & $0.497\pm 0.045$ & \cite{alam2017} &   $(\Omega_{m0},\Omega_{k0},\sigma_{80})=(0.31,0,0.8)$ \\
			30 & BOSS DR12 & $0.51$ & $0.458\pm 0.038$ & \cite{alam2017} &   $C_ij \rightarrow$ \cite{alam2017} \\
			31 & BOSS DR12 & $0.61$ & $0.436\pm 0.034$ & \cite{alam2017} &   \\
			32 & BOSS DR12 & $0.38$ & $0.477 \pm 0.051$ & \cite{rsd23} &   $(\Omega_{m0},h,\sigma_{80})=(0.31,0.676,0.8)$ \\
			33 & BOSS DR12 & $0.51$ & $0.453 \pm 0.050$ & \cite{rsd23} &   $C_ij \rightarrow$ \cite{rsd23} \\
			34 & BOSS DR12 & $0.61$ & $0.410 \pm 0.044$ & \cite{rsd23} &    \\
			35 & Vipers v7& $0.76$ & $0.440\pm 0.040$ & \cite{rsd24} &   $(\Omega_{m0},\sigma_{80})=(0.308,0.8149)$ \cite{planck_cmb} \\
			36 & Vipers v7 & $1.05$ & $0.280\pm 0.080$ & \cite{rsd24} &  \\
			37 & BOSS LOWZ & $0.32$ & $0.427\pm 0.056$ & \cite{rsd25} &   $(\Omega_{m0},\Omega_{k0},\sigma_{80})=(0.31,0,0.8475)$\\
			38 & BOSS CMASS & $0.57$ & $0.426\pm 0.029$ & \cite{rsd25} &   \\
			39 & Vipers  & $0.727$ & $0.296 \pm 0.0765$ & \cite{rsd26} &    $(\Omega_{m0},\Omega_{k0},\sigma_{80})=(0.31,0,0.7)$\\
			40 & 6dFGS+SNIa & $0.02$ & $0.428 \pm 0.0465$ & \cite{rsd27} &   $(\Omega_{m0},h,\sigma_{80})=(0.3,0.683,0.8)$ \\
			41 & Vipers  & $0.6$ & $0.48 \pm 0.12$ & \cite{rsd28} &   $(\Omega_{m0},\Omega_{b0},n_s,\sigma_{80})= (0.3, 0.045, 0.96,0.831)$ \\
			42 & Vipers  & $0.86$ & $0.48 \pm 0.10$ & \cite{rsd28} &   \\
			43 & Vipers PDR-2& $0.60$ & $0.550 \pm 0.120$ & \cite{rsd30}   & $(\Omega_{m0},\Omega_{b0},\sigma_{80})=(0.3,0.045,0.823)$ \\
			44 & Vipers PDR-2& $0.86$ & $0.400 \pm 0.110$ & \cite{rsd30} &  \\
			45 & SDSS DR13  & $0.1$ & $0.48 \pm 0.16$ & \cite{rsd31} &   $(\Omega_{m0},\sigma_{80})=(0.25,0.89)$ \cite{rsd18} \\
			46 & 2MTF & 0.001 & $0.505 \pm 0.085$ &  \cite{rsd32} &   $(\Omega_{m0},\sigma_{80})=(0.3121,0.815)$\\
			47 & Vipers PDR-2 & $0.85$ & $0.45 \pm 0.11$ & \cite{rsd33}    &  $(\Omega_{b0},\Omega_{m0},h)=(0.045,0.30,0.8)$ \\
			48 & BOSS DR12 & $0.31$ & $0.469 \pm 0.098$ & \cite{rsd34}  &   $(\Omega_{m0},h,\sigma_{80})=(0.307,0.6777,0.8288)$\\
			49 & BOSS DR12 & $0.36$ & $0.474 \pm 0.097$ & \cite{rsd34}  &    \\
			50 & BOSS DR12 & $0.40$ & $0.473 \pm 0.086$ & \cite{rsd34}  &   \\
			51 & BOSS DR12 & $0.44$ & $0.481 \pm 0.076$ & \cite{rsd34}  &   \\
			52 & BOSS DR12 & $0.48$ & $0.482 \pm 0.067$ & \cite{rsd34}  &   \\
			53 & BOSS DR12 & $0.52$ & $0.488 \pm 0.065$ & \cite{rsd34}  &   \\
			54 & BOSS DR12 & $0.56$ & $0.482 \pm 0.067$ & \cite{rsd34}  &   \\
			55 & BOSS DR12 & $0.59$ & $0.481 \pm 0.066$ & \cite{rsd34}  &   \\
			56 & BOSS DR12 & $0.64$ & $0.486 \pm 0.070$ & \cite{rsd34}  &   \\
			57 & SDSS DR7 & $0.1$ & $0.376 \pm 0.038$ & \cite{rsd35} &   $(\Omega_{m0},\Omega_{b0},\sigma_{80})=(0.282,0.046,0.817)$ \\
			58 & SDSS-IV & $1.52$ & $0.420 \pm 0.076$ &  \cite{rsd36} &   $(\Omega_{m0},\Omega_{b0} h^2,\sigma_{80})=(0.26479, 0.02258,0.8)$ \\ 
			59 & SDSS-IV & $1.52$ & $0.396 \pm 0.079$ & \cite{rsd37} &  $(\Omega_{m0},\Omega_{b0} h^2,\sigma_{80})=(0.31,0.022,0.8225)$ \\ 
			60 & SDSS-IV & $0.978$ & $0.379 \pm 0.176$ & \cite{bao_4}  & $(\Omega_{m0},\sigma_{80})=(0.31,0.8)$\\
			61 & SDSS-IV & $1.23$  & $0.385 \pm 0.099$ & \cite{bao_4}  &  $C_{ij} \rightarrow $\cite{bao_4} \\
			62 & SDSS-IV & $1.526$ & $0.342 \pm 0.070$ & \cite{bao_4}  &  \\
			63 & SDSS-IV & $1.944$ & $0.364 \pm 0.106$ & \cite{bao_4}  &  \\
			\hline
			\hline
		\end{tabular}
		}
\end{table}

The value of CMB Shift parameter $\mathcal{R} = 1.7488 \pm 0.0074$ is taken from Planck's 2016 release \cite{planck_cmb}. 
Marginalized constraints on the matter density parameter $\Omega_{m0}$ are obtained assuming a fiducial $\Lambda$CDM model for the analysis.

\subsection{Redshift-Space Distortions}

Redshift-space distortions (RSD) are an effect in observational cosmology where the spatial distribution of galaxies appears distorted when their positions 
are looked at as a function of their redshift, rather than as functions of their distances. This effect occurs due to the peculiar velocities of the galaxies, 
causing a Doppler shift in addition to the cosmological redshift. The growth of large structure can not only probe the background evolution of the Universe 
but also distinguish between different cosmological models \cite{ref39, ref41} which may have a similar background evolution but can stand in striking contrast 
to the growth of large scale structure in the Universe.  

A recent compilation of the 63 RSD $f \sigma_8$ measurements, collected by Kazantzidis and Perivolaropoulos\cite{rsdcomp} is used for our analysis. This 
$f \sigma_8$ is called the growth rate of structure. The covariance matrix of 
the 63 $f\sigma_8$ data is assumed to be diagonal except for the WiggleZ, BOSS DR12 and SDSS IV subsets. The individual covariance matrices from the WriggleZ, 
BOSS DR12 and SDSS IV surveys are added to the $f{\sigma_8}$ error uncertainties for obtaining the full covariance matrix of the RSD dataset. 

Table \ref{ch1:tabrsd} includes the 63 $f \sigma_8$ RSD data reported in various surveys \cite{rsd1,rsd2,rsd3,rsd4,rsd5,rsd6,blake2012,rsd8,rsd9,rsd10,bao_77,
ref1,rsd13,rsd14,anderson2014,rsd16,rsd17,rsd18,rsd19,rsd20,rsd21,alam2017,rsd23,rsd24,rsd25,rsd26,rsd27,rsd28,rsd30,rsd31,rsd32,rsd33,rsd34,rsd35,rsd36,rsd37,
bao_4} so far. 

It is important to note that the $f\sigma_{8,\text{obs}}(z)$ data listed in Table \ref{ch1:tabrsd} are obtained assuming a reference $\Lambda$CDM cosmology. 
So, the Alcock-Paczynski (AP) effect \cite{APeffect} should be considered. A rough approximation of this AP effect is given in 
\cite{rsdcomp, macaulay2013, li_rsd} as
\begin{equation}
f\sigma_{8,\text{AP}}(z) \approx \frac{H(z) d_A(z)}{H^{\text{fid}}(z,\Omega_{m0}) d_A^{\text{~fid}}(z,\Omega_{m0}) } f\sigma_{8,\text{obs}}(z), 
\label{ch1:AP}
\end{equation}
where $d_A(z)$ is the angular diameter distance.

\section{Outline of the thesis}

The present work is devoted to the non-parametric reconstruction of some cosmological parameters using diverse observational data sets. The reconstruction is 
done adopting the Gaussian Process. Throughout this work, the GPs are characterized by assuming a zero mean function \textit{a priori}. We consider 
the squared exponential and the Mat\'{e}rn $\nu$  covariance function for employing the GP regression. The reconstruction has been carried out directly 
from the observational data without assuming any particular functional form to start with. So the results are more unbiased, being independent of any 
specific parametrization ansatz. An assessment of the cosmic distance-duality relation, the reconstruction of cosmographical quantities viz. the deceleration 
$q$ and the jerk $j$  parameters have been carried out. Attempts have also been made to investigate the possibility of a non-gravitational interaction 
between dark matter and the dark energy sector. In case of chapters \ref{ch2:chap2}, \ref{ch4:chap4} and \ref{ch5:chap5} a spatially flat Universe has been 
assumed at the outset. We have relaxed this flatness assumption in chapter \ref{ch7:chap7} and investigated the possible effect of a non-zero spatial 
curvature from the Planck 2018 survey.

In the second chapter, the viability of the cosmic distance duality relation (CDDR) has been investigated in a non-parametric approach. The CDDR, given 
in equation \eqref{ch1:cddr}, is the theoretical relation between the luminosity distance $d_L$ and angular diameter distance $d_A$. To test the validity 
of CDDR, we have analysed the following redshift dependence of CDDR, $\eta$, given by
\begin{equation} \label{ch1:eta}
\eta(z) = \frac{d_L}{d_A(1+z)^2}.
\end{equation}

This $\eta$ is called the CDDR ratio. For $\eta = 1$ we recover equation \eqref{ch1:cddr}. Any deviation of $\eta$ from unity indicates a violation of CDDR. 
The distance modulus measurements from the Pantheon SNIa compilation are utilized in reconstructing $d_L$ in the redshift range $0<z<2$. We undertake a 
reconstruction of the Hubble parameter $H(z)$ from the Cosmic Chronometer measurements of the Hubble parameter, followed by another reconstruction of $D_V$ 
from the volume-averaged BAO compilation. The reconstructed $H$ and $D_V$ from the CC and BAO data are combined to obtain $d_A$ in the same redshift range 
$0<z<2$. Finally, the reconstructed functions, viz. $d_L(z)$ and $d_A(z)$, have been combined to obtain the CDDR ratio $\eta(z)$ according to Eq. \eqref{ch1:eta}. 
The reconstruction shows that the theoretical cosmic distance duality relation is in good agreement with the present analysis at the 2$\sigma$ confidence level.

In the third chapter, the cosmological jerk parameter $j$ has been reconstructed in a non-parametric way from observational data independent of a fiducial 
cosmological model. As the deceleration parameter $q$ can now be estimated and is found to be evolving, the next higher order derivative, the jerk 
parameter $j$ should be the focus of attention. There are no assumptions made regarding the theory of gravity or the distribution of matter in the Universe 
to start with. The only basic assumption is that the Universe is spatially flat, homogeneous and isotropic, thus described by the FLRW metric. The jerk 
parameter, defined as $j = \frac{\dddot{a}}{aH^3}$, can be rewritten as a function of the reduced Hubble parameter $E$ and redshift $z$, as 
\begin{eqnarray} \label{jerk}
j(z) &=&   1 - 2(1+z)\frac {E'}{E} + (1+z)^2 \frac{\left(E'^2 + E E''\right)}{E^2} .
\end{eqnarray} 

Model-independent datasets like the CC data and the Pantheon compilation of SNIa distance data are used for this purpose. Finally, $j$ is reconstructed 
utilizing equation \eqref{jerk}. The reconstructed values are consistent with the standard $\Lambda$CDM model within the 2$\sigma$ confidence level.  
Model-dependent data sets like the $r$BAO and the CMB Shift parameter have been included thereafter to examine their possible effect on the reconstruction 
of $j$. The deceleration parameter $q$ can also be reconstructed from the same data sets. This has been utilized to find the effective equation of state 
parameter $w_{\text{ eff}}$ using the model-independent datasets. Further, an approximate fitting function of the reconstructed $j$ has been obtained 
for the model-independent data combinations. 

The fourth chapter investigates the possibility of a non-gravitational interaction between dark matter and dark energy. The CC Hubble data, the 
Pantheon SNIa $E(z)$ compilation of CANDELS and CLASH Multi-Cycle Treasury programs and the $r$BAO Hubble data have been utilized for reconstructing the cosmic 
interaction as a function of redshift. The conservation equation, given in \eqref{ch1:continuity}, for an interacting scenario can be separated into two parts
\begin{eqnarray} 
\dot{\rho_m}+ 3H\rho_m &=& -Q ,\\
\dot{\rho_D} + 3H (1+w)\rho_D &=& Q,
\end{eqnarray} 

where, $w = \frac{p_D}{\rho_D}$ is the equation of state of DE and $Q$ is the interaction term which describes the rate of energy transfer between cosmic dark 
sectors. $Q$ can be represented in a dimensionless form as, $\tilde{Q} = \frac{Q}{ 3 H_0^3} $, which characterizes the interaction. No functional form for 
$\tilde{Q}$ has been assumed. By combining the Friedmann equation and the continuity equations, $\tilde{Q}$ can be represented as
\begin{equation} \label{Q_h}
\begin{split}
\tilde{Q} = \left( \frac{E^2(1+w)}{w} + \frac{(1+z)E^2 w'}{3 w^2} \right) \left[ 2(1+z)E' - 3E \right] + ~~~~~~~~~~~\\ 
~~~~~~~~~~~~~~~~~~~~~~~~- \frac{(1+z)E}{3 w}\left[2(1+z)(E'^2 + E E'') - 4 EE' \right].
\end{split}
\end{equation} 

We reconstruct the function $E(z)$, and its derivatives $E'(z)$ and $E''(z)$ from different combinations of datasets like CC, $r$BAO and Pantheon+MCT compilation. 
Using the reconstructed values of $E(z)$, $E'(z)$ and  $E''(z)$, $\tilde{Q}$ can be obtained once the equation of state parameter $w(z)$ for dark energy is supplied. 
We have investigated $\tilde{Q}$ for three different versions of dark energy (i) an interacting vacuum with $w=-1$, (ii) a $w$CDM model where $w$ is close to $-1$ 
but not exactly equal to that and (iii) the CPL parametrization where $w(z) = w_0 + w_a \frac{z}{1+z}$ \cite{cpl_main}. The possibility of a \textit{no interaction} 
scenario is indicated from the results. Also, the interaction, if any, is not really significant at the present epoch. The direction of energy flow is found to be 
from the dark energy to the dark matter, consistent with the thermodynamic requirement. An analytic expression for the energy transfer rate $Q$, in the 
form of a polynomial in $z$, has been obtained. The evolution of the density parameters ${\Omega}_m$ and ${\Omega}_D$ have also been checked in the presence of 
$\tilde{Q}$. Lastly, attempts have been made for testing the reconstructed interacting model against the laws of thermodynamics.

In the fifth chapter, a non-parametric reconstruction of the cosmic deceleration parameter $q$ has been revisited using various combinations of background 
datasets and the growth rate data. The Pantheon SN distance modulus sample, the CC Hubble parameter compilation including the full systematics, and the BAO 
measurements have been considered as background datasets for the reconstruction of $q$. To investigate the effect of matter perturbations, the growth rate 
measurements from RSD are utilized in reconstructing $q$. The RSD dataset can not only probe the background evolution of the Universe, but also distinguish 
between GR and different modified gravity theories. The deceleration parameter, defined as $q = -\frac{\ddot{a}}{aH^2}$, can be expressed as a function of 
the reduced Hubble parameter $E$ and redshift $z$, as
\begin{eqnarray} 
q(z) &=& -1 + \frac{E'}{E} (1+z) .  \label{q_nonpara}
\end{eqnarray}

Plots of the reconstructed cosmic deceleration parameter $q(z)$ is obtained utilizing equation \eqref{q_nonpara}. The redshift $z_t$, where the transition from 
a past decelerated to a late-time accelerated phase of evolution occurs, is estimated from the reconstructed $q$. A non-parametric reconstruction of $q$ is not new, 
it has been in practice ever since the present acceleration of the Universe became apparent. The purpose of this chapter is to refresh the reconstruction with recent 
datasets, following an improved analysis. The inclusion of the RSD data is a new feature that has been undertaken in this particular work.

Finally, chapter six contains the concluding remarks and relevant discussion regarding the work presented in the thesis.

\cleardoublepage{}
\clearpage{}
\chapter{Assessment of the cosmic distance duality relation}\blfootnote{\begin{flushleft} The work presented in this chapter is based on ``Assessment of the cosmic distance duality relation using Gaussian process", \textbf{Purba Mukherjee} and Ankan Mukherjee, Mon.\ Not.\ R.\ Astron.\ Soc. \textbf{504}, 3938 (2021).\end{flushleft}}  \label{ch2:chap2}
\chaptermark{Assesment of the cosmic distance duality relation}
\section{Introduction}

Among the different definitions for distance measures used in cosmology, described in Sec \ref{ch1:distance_measure}, two of them are more frequently utilized. 
These are the luminosity distance $d_L$ and the angular diameter distance $d_A$. They are connected by equation \eqref{ch1:cddr} which first given by Etherington\cite{etherin1993} 
in the context of an FLRW metric, called the cosmic distance-duality relation (CDDR). Often referred to as Etherington's reciprocity 
theorem, the CDDR is considered to be correct for any general metric theories of gravity, in any background, based on two fundamental hypotheses. Firstly, the 
number of photons is conserved during cosmic evolution \cite{ellis1971, ellis2007}, and secondly, gravity is described by a metric theory with photons travelling 
along unique null geodesics in Riemannian geometry. The coupling of photons with unknown particles \cite{bassett}, the extinction of photons by intergalactic dust 
\cite{corasaniti}, the variation of fundamental constants \cite{ellis2013} could lead towards a violation of CDDR. 

As cosmography is strongly dependent on the validity of CDDR, any sizeable deviation indicates the possibility of some exotic physics beyond the standard cosmological 
model or the presence of systematic errors in observations \cite{bassett,bassett2}. Therefore, with the increasing quality and quantity of observational data, evaluating 
the reliability of CDDR has received significant attention lately. This chapter is devoted to the assessment of CDDR from observational data. 

The CDDR can straightway be put to test utilizing the luminosity distance $d_L (z)$ and the angular diameter distance $d_A (z)$ measurements at the same redshift 
$z$. The type-Ia supernovae \cite{suzuki, betoule, pan} observations generally serve as major sources for estimating $d_L$ with high precision, as described in 
chapter \ref{ch1}. But there is no simple way to measure the angular diameter distance $d_A$. The possible methods for determining $d_A$ include the combined 
data of the X-rays and Sunyaev-Zeldovich (SZ) effect of galaxy clusters \cite{filipps2005, bonamente2006}, measuring the BAO signals in the galaxy power spectrum 
\cite{beutler2011, blake2012, kazin2014}, the angular size of ultra-compact radio sources based on the approximately consistent linear size \cite{kellermann1993, 
gurvits1994, gurvits1999, jackson2004}, the images of quasars that are strongly lensed by foreground galaxies \cite{cao2015, liao2016}. Li and Lin\cite{li2018} 
and Lin, Li and Li\cite{lin2018} provide concise discussions on these methods for computing $d_A$, with their respective merits and limitations. 

While testing for validity of CDDR, the prime difficulty experienced is that both $d_L$ and $d_A$ are not computed at the same redshift $z$. Several approaches viz. the 
nearest neighbourhood method \cite{holanda2010, liao2016}, the interpolation method \cite{liang2013}, and the Gaussian Process Regression \cite{nair3,rana2017, li2018} 
have been proposed to resolve this drawback. It deserves mention that only data points in the overlapping redshift range are available for CDDR verification. 

Reconstruction of the CDDR has previously been addressed in the literature.  Until now, no evidence for CDDR violation has been recorded from the reconstruction 
techniques. These can be categorized into the cosmological model-dependent tests \cite{uzan2004,avgous2010,holanda2011,lima2011,nair,jhingan2014,holanda2016b,hu2018}, 
and the cosmological model-independent analysis \cite{holanda2010,li2011,goncalves2011,nair2,meng2012,chen2015,holanda2016,liao2016,rana2016,holanda2017,holanda2019,
xu2020,zheng2020}. A majority of these works involve a functional form for the CDDR chosen at the outset, followed by an estimation of parameters. This is undoubtedly 
biased, as a specific parametric form for the CDDR is already chosen. 

In this chapter, the Supernova distance modulus data, Cosmic Chronometer measurements of the observational Hubble data and the Baryon Acoustic Oscillation data 
have been utilized to examine the validity of CDDR in a non-parametric way. We adopted the GP regression for the reconstruction, avoiding any 
fiducial bias on the cosmological parameters included in the data sets.

\section{Reconstruction}
\label{ch2:reconst}

To test the validity of the cosmic distance duality relation, one analyses the following redshift dependence of CDDR, given by, 

\begin{equation} \label{ch2:eta}
\eta(z) = \frac{d_L}{d_A(1+z)^2}.
\end{equation}

In the case of $\eta = 1$ one can recover equation \eqref{ch1:cddr}, which implies the correctness of the standard CDDR. Any deviation of $\eta$ from unity 
indicates a non-validation of CDDR. We undertake a non-parametric GP reconstruction of the function $\eta$.

The uncertainty associated with the reconstructed function $\eta$ can be calculated by the standard error propagation formula, 
\begin{equation} \label{ch2:sig_eta}
\sigma_\eta = \eta \sqrt{\left( \frac{\sigma_{d_L}}{d_L}\right)^2 + \left( \frac{\sigma_{d_A}}{d_A} \right)^2} .
\end{equation}

The Squared Exponential \eqref{ch1:sqexp} as well as three orders of the Mat\'{e}rn $\nu$ \eqref{ch1:matern} covariance functions are taken into account 
for the GP analysis. For the Mat\'{e}rn covariance function, the orders of the polynomials are taken as $p=2,3,4$, consequently the parameter $\nu$ has 
values $\frac{5}{2},\frac{7}{2},\frac{9}{2}$ respectively.

The recent Pantheon \cite{pan} SNIa data is utilized for obtaining $d_L$, and $d_A$ are derived considering the volume-averaged BAO  
\cite{beutler2011,blake2012,ross2015,gilmarin2015,bautista2018,ata2018,agathe,blomqvist} data in combination with the 31 CC $H(z)$ 
\cite{cc_101,cc_102,cc_103,cc_104,cc_106,cc_105} measurements (obtained from a compilation of the 20 CCB and 11 CCH $H(z)$ values), 
in the same domain of redshift $0<z<2$.

\section{Methodology} \label{ch2:methodology}

\begin{figure*}[t!]
	\begin{center}
		\includegraphics[angle=0, width=0.45\textwidth]{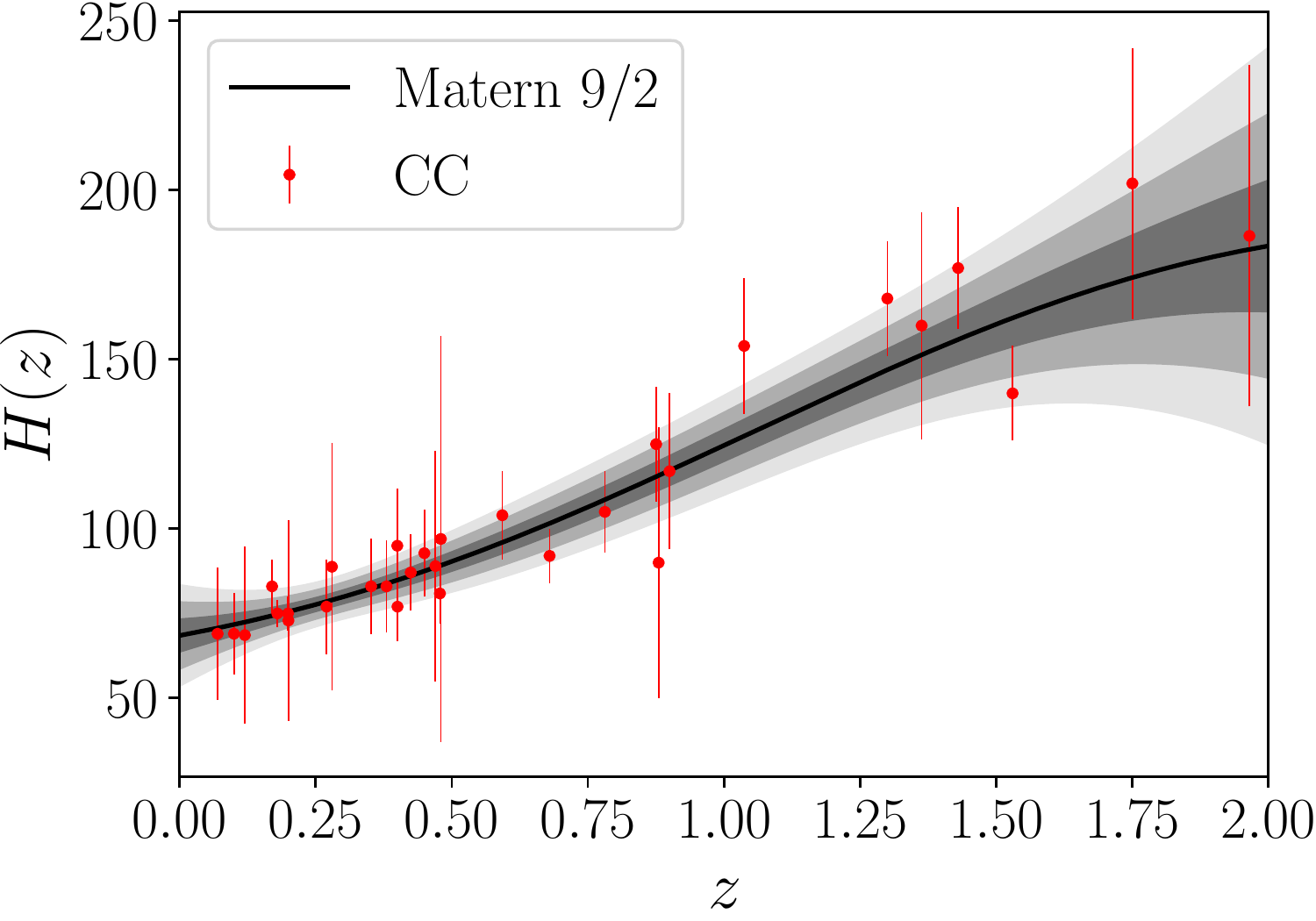}
		\includegraphics[angle=0, width=0.45\textwidth]{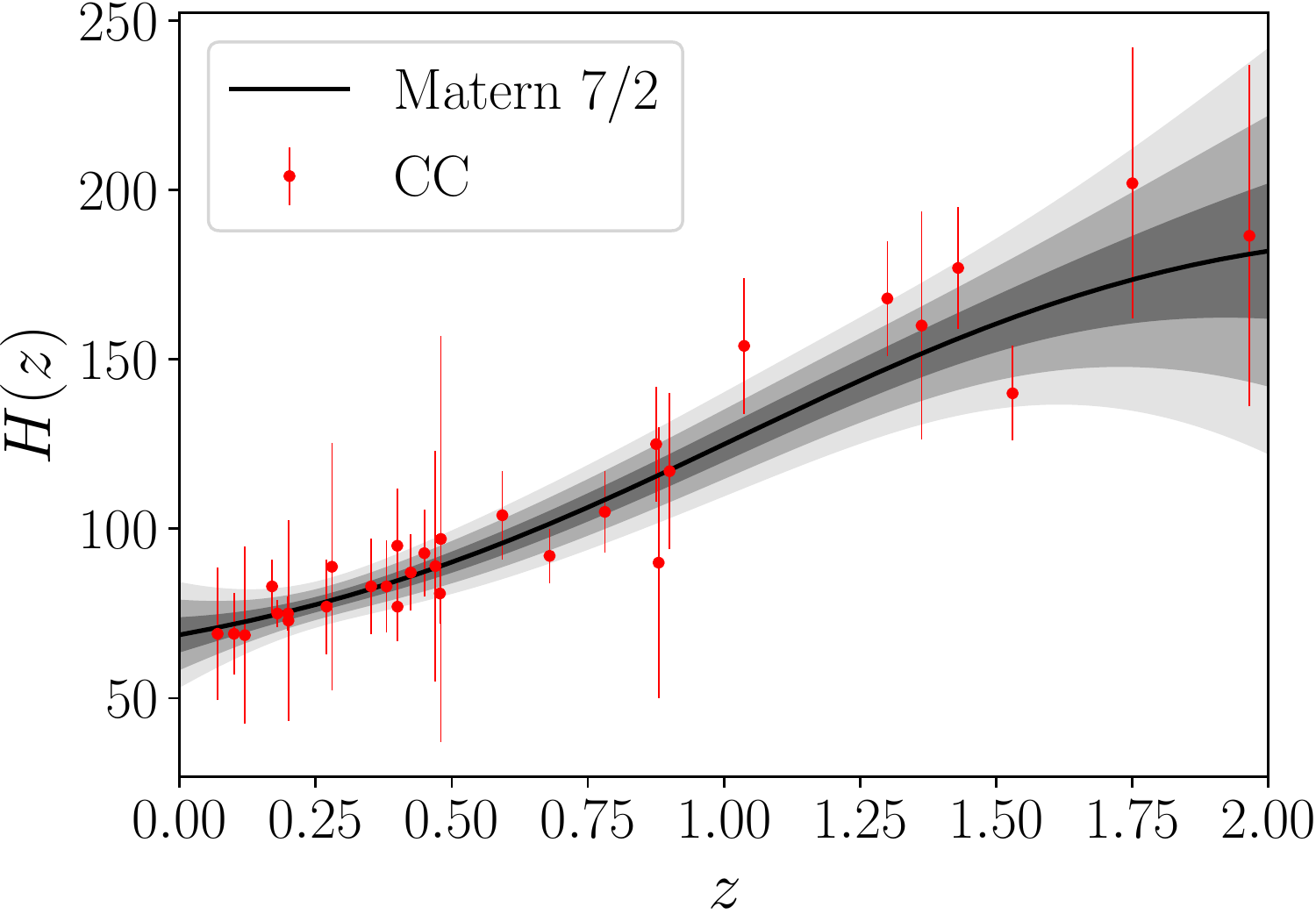}\\
		\includegraphics[angle=0, width=0.45\textwidth]{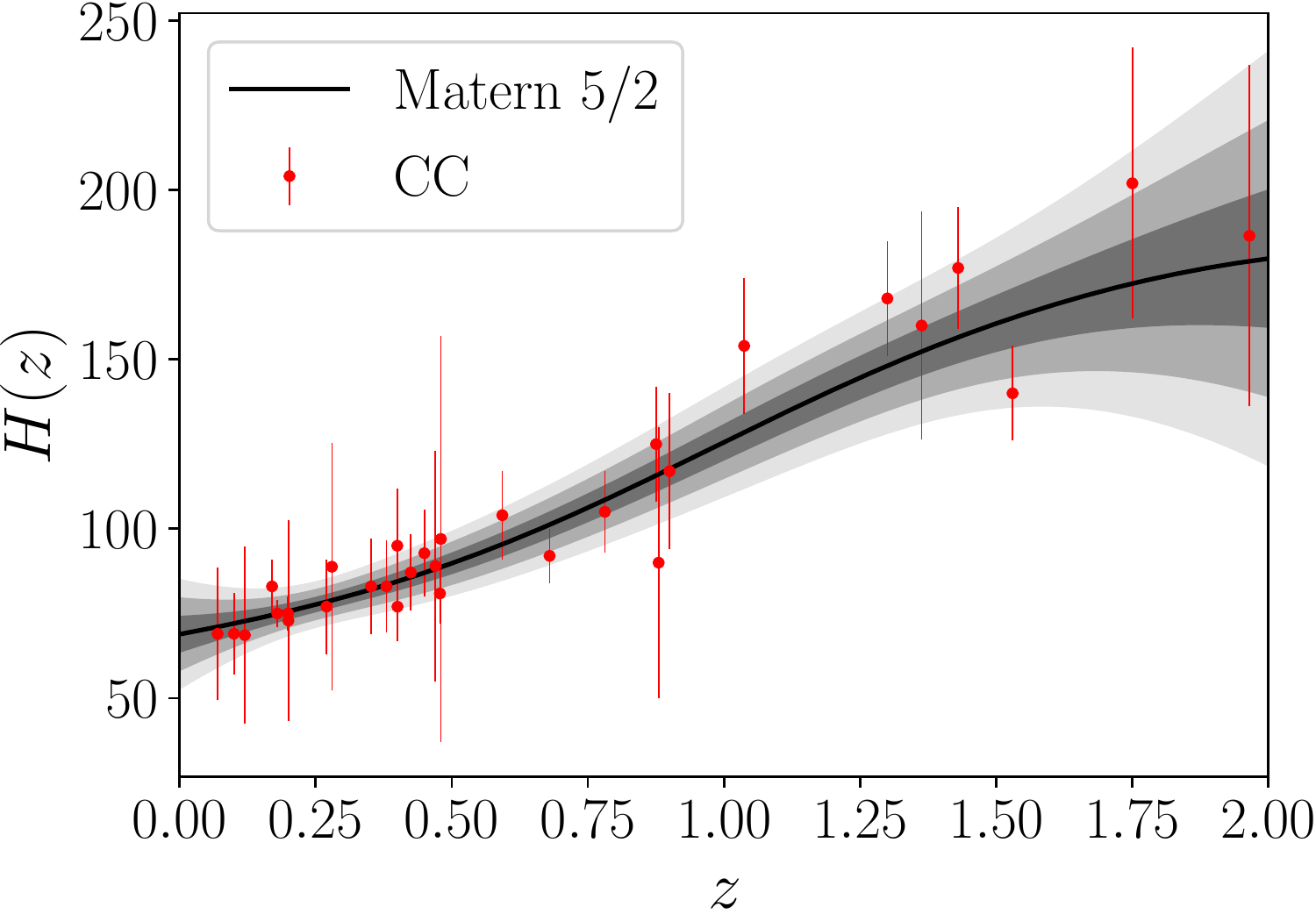}
		\includegraphics[angle=0, width=0.45\textwidth]{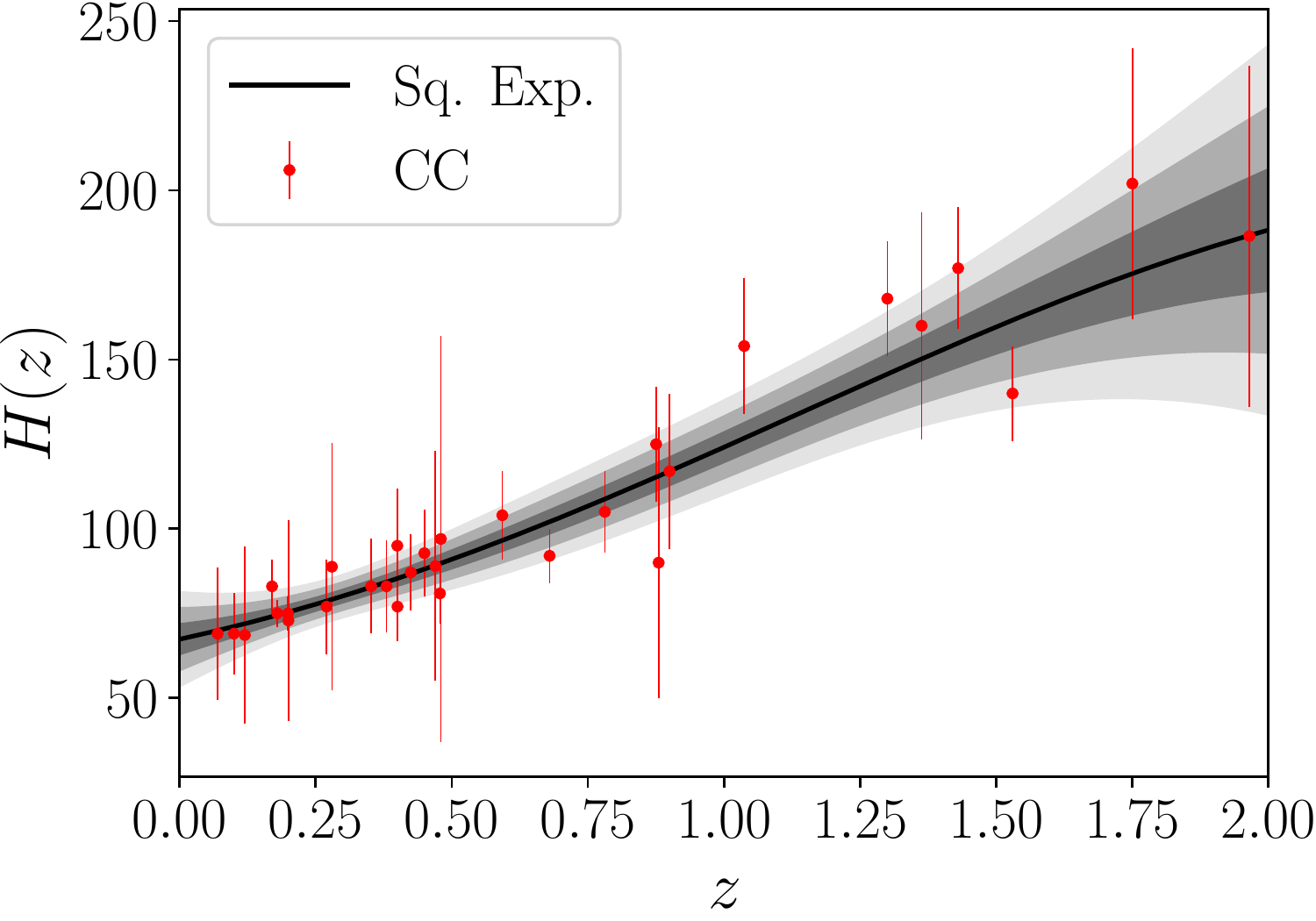}
	\end{center}
	\caption{{\small Plots for $H(z)$ (in units of km Mpc$^{-1}$ s$^{-1}$) reconstructed from CC  data using the Mat\'{e}rn 9/2, Mat\'{e}rn 7/2, Mat\'{e}rn 5/2 
			and Squared Exponential covariance function respectively. The solid black line is the best fit curve and the associated 
			1$\sigma$, 2$\sigma$ and 3$\sigma$ confidence regions are shown in lighter shades.}}
	\label{ch2:Hz_recon}
\end{figure*}

The Hubble parameter $H(z)$ is reconstructed using the GP regression and the results are shown in Fig. \ref{ch2:Hz_recon}. The present value of the Hubble 
parameter $H_0$, obtained from this reconstruction, is shown in Table \ref{ch2:Hz_res}. The reconstructed $H(z)$ data is normalized 
to obtain the reduced Hubble parameter $E(z) = H(z)/H_0$. The uncertainty in $E(z)$ is estimated by the standard technique of error 
propagation, as
\begin{equation} \label{ch2:sig_h}
{\sigma_{E}}^2 = \frac{{\sigma_H}^2}{ {H_0}^2} + \frac{H^2}{{H_0}^4}{\sigma_{H_0}}^2,
\end{equation} 
where $\sigma_{H_0}$ is the error associated with $H_0$.

\begin{table*}[t!]
	\caption{{\small Table showing the reconstructed value of $H_0$ (in units of km Mpc$^{-1}$ s$^{-1}$) for different choices of the covariance function, from 
			the CC data.}}
	\begin{center}
		\resizebox{0.98\textwidth}{!}{\renewcommand{\arraystretch}{1.3} \setlength{\tabcolsep}{15pt} \centering  
			\begin{tabular}{c c c c c} 
\hline
				$k(z,\tilde{z})$ & Mat\'{e}rn 9/2 & Mat\'{e}rn 7/2 & Mat\'{e}rn 5/2 &  Squared Exponential\\ 
				\hline
				\hline
				$H_0$  &  $68.471 \pm 5.081$  & $68.684 \pm 5.204$ & $68.858 \pm 5.466$	& $67.356 \pm 4.765$\\ 
				\hline
			\end{tabular}
		}
	\end{center}
	\label{ch2:Hz_res}
\end{table*}

Utilizing the reconstructed function $E(z)$, the normalized transverse comoving distance $D(z)$ for a flat spacetime is evaluated numerically using the composite 
trapezoidal integration rule \cite{trapez},
\begin{equation} \label{ch2:D_recon}
D(z) = \int_{0}^{z} \frac{\dif z'}{E(z')} \approx \frac{1}{2} \mathlarger{\mathlarger{\sum}}_{i=0}^{n-1} (z_{i+1} - z_i) \left[\frac{1}{E(z_{i+1})}+\frac{1}{E(z_{i})}\right].
\end{equation} 

The numerical error associated with $D(z)$ is of order $10^{-6}$ and does not adversely affect our analysis. The statistical uncertainty $\sigma_{D}$, associated 
with $D$, is obtained by the error propagation formula 
\begin{equation}
\sigma^2_{D} = \sum_{i=0}^{n} s_i^2 ,
\end{equation}
where, \begin{equation}
s_i = \frac{1}{2} (z_{i+1} - z_i) \left[\frac{\sigma^2_{E_{i+1}}}{E^4_{i+1}}+\frac{\sigma^2_{E_{i}}}{E^4_{i}}\right]^\frac{1}{2}.
\end{equation}

The reconstructed normalized comoving distance $D$ from the Hubble data, from equation \eqref{ch2:D_recon}, is utilized to calculate the distance modulus $\mu_{\mbox{\tiny H}}$ 
from the reconstructed Hubble data as 
\begin{equation}
\mu_{\mbox{\tiny H}} = 5 \log_{10} \left[D(1+z)\right] + 25.
\end{equation}

The 1$\sigma$ uncertainty $\sigma _{\mu_{\mbox{\tiny H}}}$, associated with $\mu_{\mbox{\tiny H}}$, is given by
\begin{equation}
\sigma_{\mu_{\mbox{\tiny H}}} = \frac{5}{\ln 10} \frac{\sigma_{D}}{D}.
\end{equation}

We reconstruct the observed apparent magnitudes $m_B = \mu_{\mbox{\tiny SN}} + M_B$ of the SNIa data, at the same redshift $z$ as that of the CC Hubble 
data employing another GP. As the absolute magnitude $M_B$ of SNIa is degenerate with the Hubble parameter $H_0$, the marginalized constraints on $M_B$ are 
obtained by minimizing the $\chi^2$ function   
\begin{equation} \label{ch2:chiSN}
\chi_{\mbox{\small SN}}^2 = \sum \Delta \mu^{\mbox{\small T}} \cdot \Sigma^{-1} \cdot \Delta \mu,
\end{equation}
considering a uniform prior $M_B$ $\in [-35,-5]$, where $\Delta \mu = (\mu_{\mbox{\tiny SN }}- \mu_{\mbox{\tiny H}})$ and $\Sigma = \Sigma_{\mu_{\mbox{\tiny SN }}} 
+ \sigma_{\mu_{\mbox{\tiny H}}}^2$ respectively.

\begin{figure*}[t!]\begin{center}
		\includegraphics[angle=0, width=0.24\textwidth]{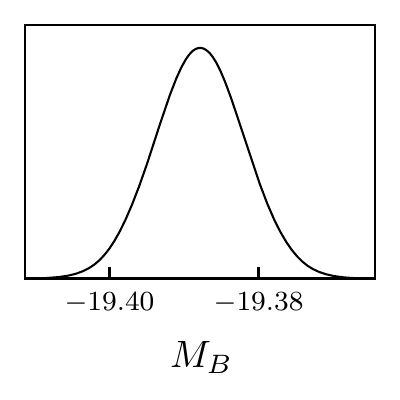}
		\includegraphics[angle=0, width=0.24\textwidth]{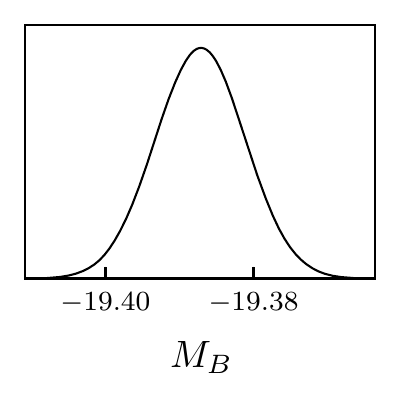}
		\includegraphics[angle=0, width=0.24\textwidth]{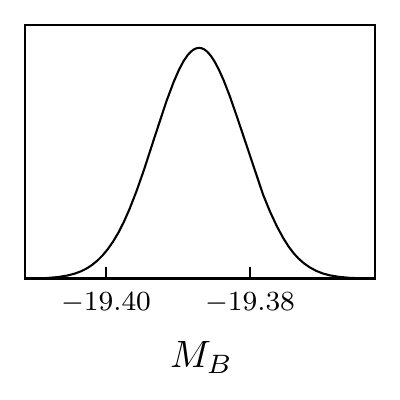}
		\includegraphics[angle=0, width=0.24\textwidth]{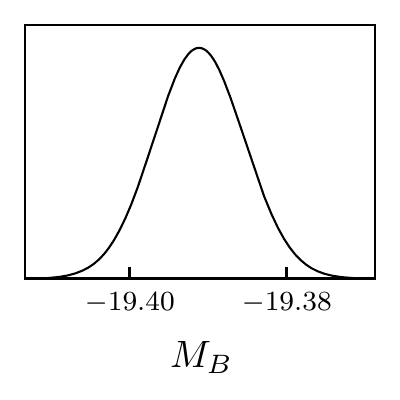}
	\end{center}
	\caption{{\small Plots for marginalized likelihood of absolute magnitude $M_B$ using the Mat\'{e}rn 9/2, Mat\'{e}rn 7/2, Mat\'{e}rn 5/2 and Squared Exponential 
			covariance function (from left to right) respectively.}}
	\label{ch2:MB_plot}
\end{figure*}

Further, the volume-averaged $\frac{D_V}{r_d}$ measurements from the BAO data compilation is reconstructed via another GP in the same redshift range $0<z<2$. 
The comoving sound horizon at the drag epoch $r_d$ (defined in eq. \eqref{ch1:rdrag_def}) is considered as a free parameter in this particular analysis and 
is estimated via two different strategies. 

The WriggleZ DES \cite{blake2012} data measures the acoustic scale parameter $A(z)$ given by,
\begin{equation}
A(z) = D_V \frac{\sqrt{\Omega_{m0} ~ H_0^2}}{c z}.
\end{equation} 

As the matter density parameter, $\Omega_{m0}$, is correlated with $H_0$, we constrain $\Omega_{m0}$ assuming a fiducial $\Lambda$CDM model, 
\begin{equation}\label{ch2:chiCC}
\chi_{\mbox{\small H}}^2 = \mathlarger{\mathlarger{\sum}}_i \frac{\left[ E(z_i) -  \sqrt{\Omega_{m0}(1+z_i)^3 + 1-\Omega_{m0}} \right]^2}{\sigma_{E}^2(z_i)},
\end{equation}
with uniform prior $\Omega_{m0} \in [0.01,0.7]$ using the reconstructed $E(z)$ data, for the same $H_0$ as given in Table \ref{ch2:Hz_res}.

\begin{figure*}[t!]\begin{center}
		\includegraphics[angle=0, width=0.24\textwidth]{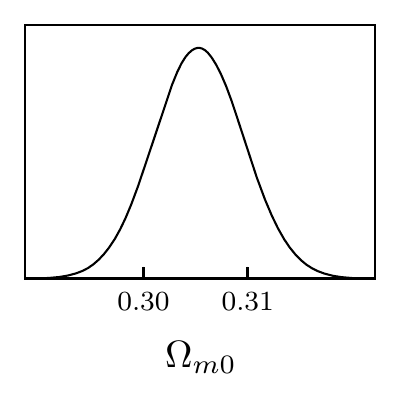}
		\includegraphics[angle=0, width=0.24\textwidth]{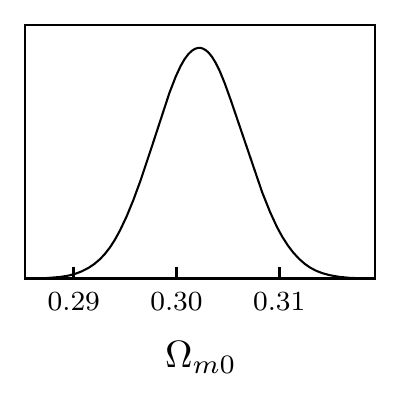}
		\includegraphics[angle=0, width=0.24\textwidth]{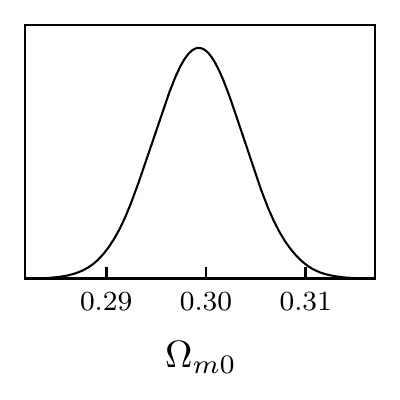}
		\includegraphics[angle=0, width=0.24\textwidth]{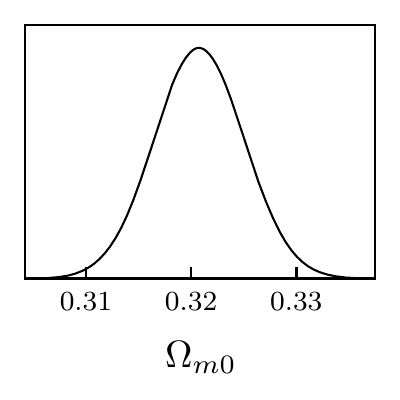}
	\end{center}
	\caption{{\small Plots for marginalized likelihood of matter density parameter  $\Omega_{m0}$ using the Mat\'{e}rn 9/2, Mat\'{e}rn 7/2, Mat\'{e}rn 5/2 and 
			Squared Exponential covariance function (from left to right) respectively.}}
	\label{ch2:Om_plot}
\end{figure*}

For the remaining BAO data sets, we reconstruct $D_V/r_d$ by dividing with $r_{\mbox{\tiny d, fid}} = 147.49$ wherever applicable. Eq. \eqref{ch1:dilation_d} 
can be written in terms of the reconstructed $D(z)$ and $H(z)$ from the CC data, as
\begin{equation}
\left. D_V \right\vert_{\mbox{\tiny H}} = \left[\frac{c^2  D^2 (z) }{H_0^2} \frac{ c z}{H(z)}\right]^{\frac{1}{3}} = \frac{c}{H_0} \left[ \frac{D^2(z) z}
{E(z)}\right]^{\frac{1}{3}}.
\end{equation} 

Finally, the marginalized constraints on $r_d$ are obtained by minimizing the $\chi^2$ function, 
\begin{equation} \label{ch2:chiBAO}
\chi_{\mbox{\small BAO}}^2 = \mathlarger{\mathlarger{\sum}} \frac{\left( \left.\frac{D_V}{r_d}\right\vert_{\mbox{\tiny BAO}} - \left.\frac{D_V}{r_d}\right\vert_{\mbox{\tiny H}}
	\right)^2}{{\sigma_{\left. \frac{D_V}{r_d}\right\vert_{\mbox{\tiny BAO}}}^2} +{\sigma_{\left. \frac{D_V}{r_d}\right\vert_{\mbox{\tiny H}}}^2}} ,
\end{equation} 
for a uniform prior assumption $r_d \in [130, 160]$.

\begin{figure*}[t!]\begin{center}
		\includegraphics[angle=0, width=0.24\textwidth]{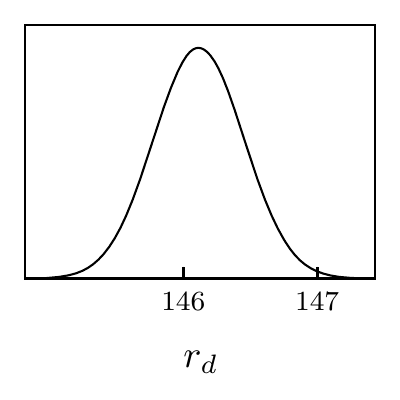}
		\includegraphics[angle=0, width=0.24\textwidth]{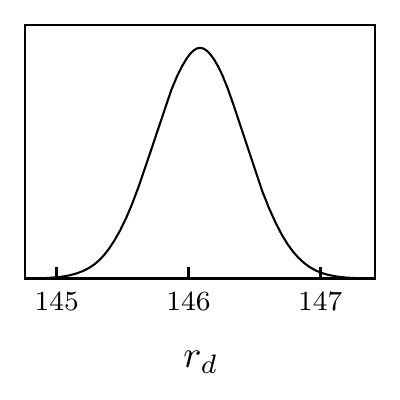}
		\includegraphics[angle=0, width=0.24\textwidth]{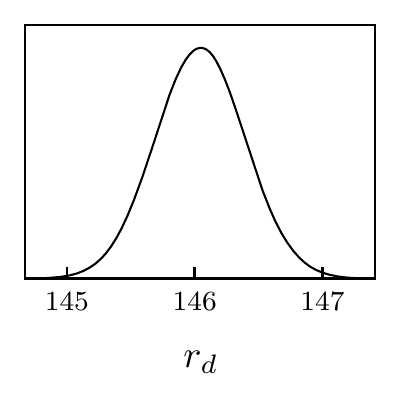}
		\includegraphics[angle=0, width=0.24\textwidth]{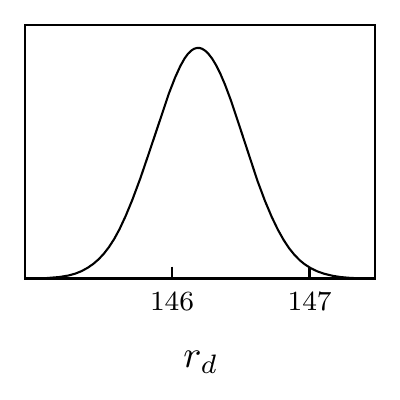}
	\end{center}
	\caption{{\small Plots for marginalized likelihood of comoving sound horizon at drag epoch $r_d$ (in units of Mpc) using the Mat\'{e}rn 9/2, Mat\'{e}rn 
			7/2, Mat\'{e}rn 5/2 and Squared Exponential covariance function (from left to right) respectively.}}
	\label{ch2:rd_plot}
\end{figure*}

Uncertainty in the parameters $M_B$, $\Omega_{m0}$ and $r_d$ are obtained by a Markov Chain Monte Carlo (MCMC) analysis. Plots for the marginalized $M_B$, 
$\Omega_{m0}$ and $r_d$ constraints are shown in Figures \ref{ch2:MB_plot}, \ref{ch2:Om_plot} and \ref{ch2:rd_plot}. The best-fit values are given 
in Table \ref{ch2:MOr_res}. 

\begin{table*}[t!]
	\caption{{\small Table showing the marginalized constraints on $M_B$, $\Omega_{m0}$ and $r_d$ (in units of Mpc) for different choices of the covariance 
			function using equations \eqref{ch2:chiSN}, \eqref{ch2:chiCC} and \eqref{ch2:chiBAO} respectively.}}
	\begin{center}
		\resizebox{0.98\textwidth}{!}{\renewcommand{\arraystretch}{1.3} \setlength{\tabcolsep}{15pt} \centering  
			\begin{tabular}{c c c c c} 
\hline
				$k(z,\tilde{z})$  &  Mat\'{e}rn 9/2 & Mat\'{e}rn 7/2 & Mat\'{e}rn 5/2 & Squared Exponential\\ 
				\hline
				\hline
				$M_B$ & $-19.388 \pm 0.006$ &  $-19.387 \pm 0.006$  & $-19.387 \pm 0.006$ & $-19.391 \pm 0.006$	\\ 
				\hline
				$\Omega_{m0}$ & $0.305 \pm 0.004$ &  $0.302 \pm 0.004 $  & $0.299 \pm 0.005$ & $0.321 \pm 0.004 $	\\ 
				\hline
				$r_d$ & $146.116 \pm 0.336$ &  $146.086f \pm 0.340$  & $146.044 \pm 0.350 $ & $146.193 \pm 0.326$	\\ 
				\hline
			\end{tabular}
		}
	\end{center}
	\label{ch2:MOr_res}
\end{table*}

Eisenstein and Hu\cite{rdrag_def} arrived at an approximation for the comoving sound horizon at drag epoch $r_d$, given by
\begin{equation}
r_d  \approx  \frac{44.5 \log \left(\frac{9.83}{\Omega_{m0} h^2}\right)}{\sqrt{1+10\left(\Omega_{b0}h^2\right)^{3/4}}}.  \label{ch1:rdrag_approx}
\end{equation} 

As an alternative method, we consider equation \eqref{ch1:rdrag_approx} for estimating the approximate value of $r_d$. Using the reconstructed value of $H_0$ 
as given in Table \ref{ch2:Hz_res}, the marginalized $\Omega_{m0}$ constraints from Table \ref{ch2:MOr_res}, and the value of $\Omega_{b0}h_0^2 = 0.022383$ from 
Planck 2018 \cite{planck} probe in equation \eqref{ch1:rdrag_approx}, one can evaluate the approximate value of $r_d$ along with its associated 1$\sigma$ uncertainty. 
The calculated values of $r_d$ using the aforementioned procedure is shown in Table \ref{ch2:rdrag_tab}.

\begin{table*}[t!]
	\caption{{\small Table showing the value of $r_d$ (in units of Mpc) for different choices of the covariance function, considering the approximated definition 
			from equation \eqref{ch1:rdrag_approx}.}}
	\begin{center}
		\resizebox{0.98\textwidth}{!}{\renewcommand{\arraystretch}{1.3} \setlength{\tabcolsep}{15pt} \centering  
			\begin{tabular}{c c c c c} 
\hline
				$k(z,\tilde{z})$ & Mat\'{e}rn 9/2 & Mat\'{e}rn 7/2 & Mat\'{e}rn 5/2 &  Squared Exponential\\ 
				\hline
				\hline
				$r_d$  &  $149.828 \pm 12.059$  & $149.959 \pm 12.313$ & $150.134 \pm 12.921$	& $149.181 \pm 11.496$\\ 
				\hline
			\end{tabular}
		}
	\end{center}
	\label{ch2:rdrag_tab}
\end{table*}

\section{Result}

\begin{figure*}[t!]
	\begin{center}
		\includegraphics[angle=0, width=0.45\textwidth]{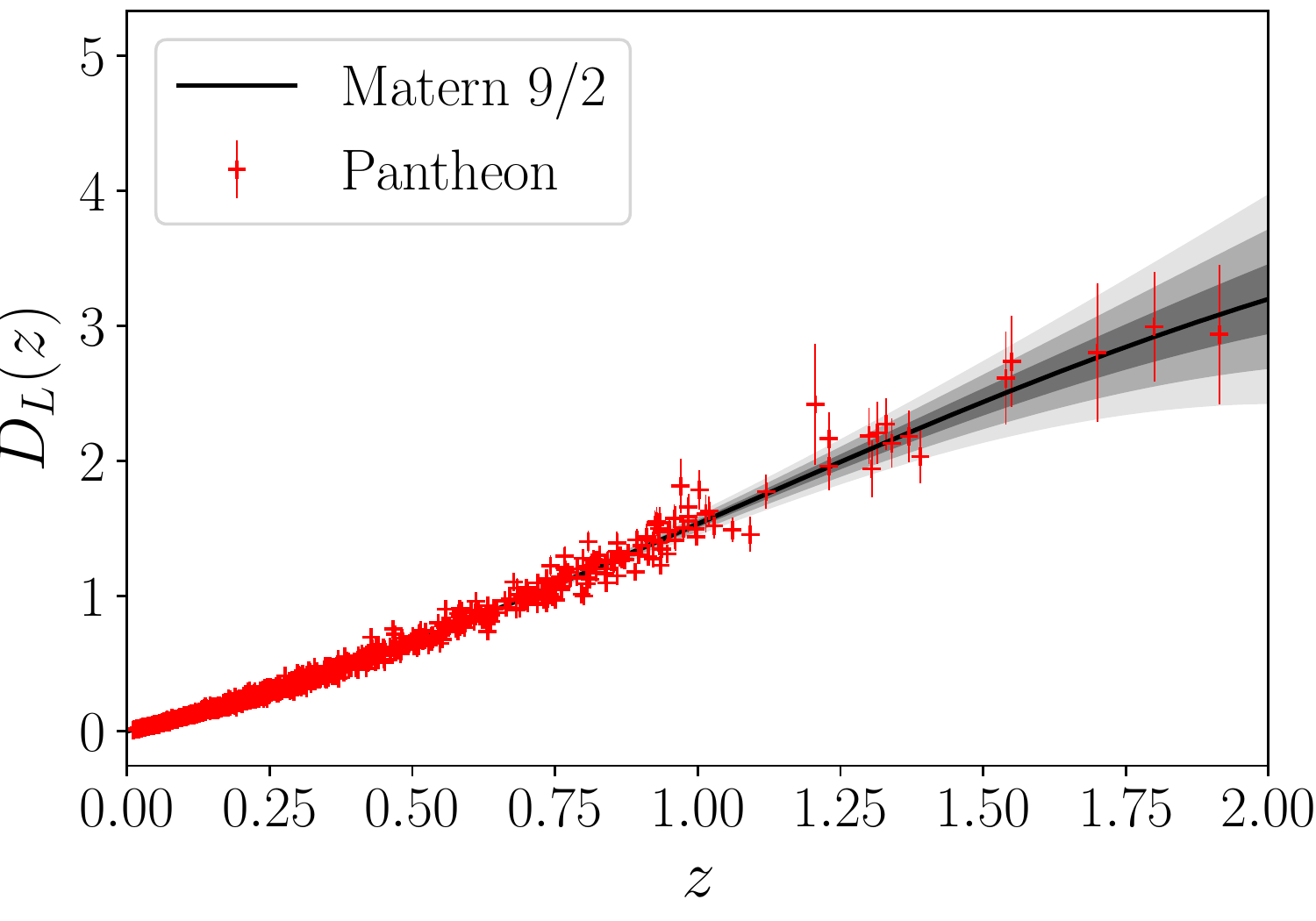}
		\includegraphics[angle=0, width=0.45\textwidth]{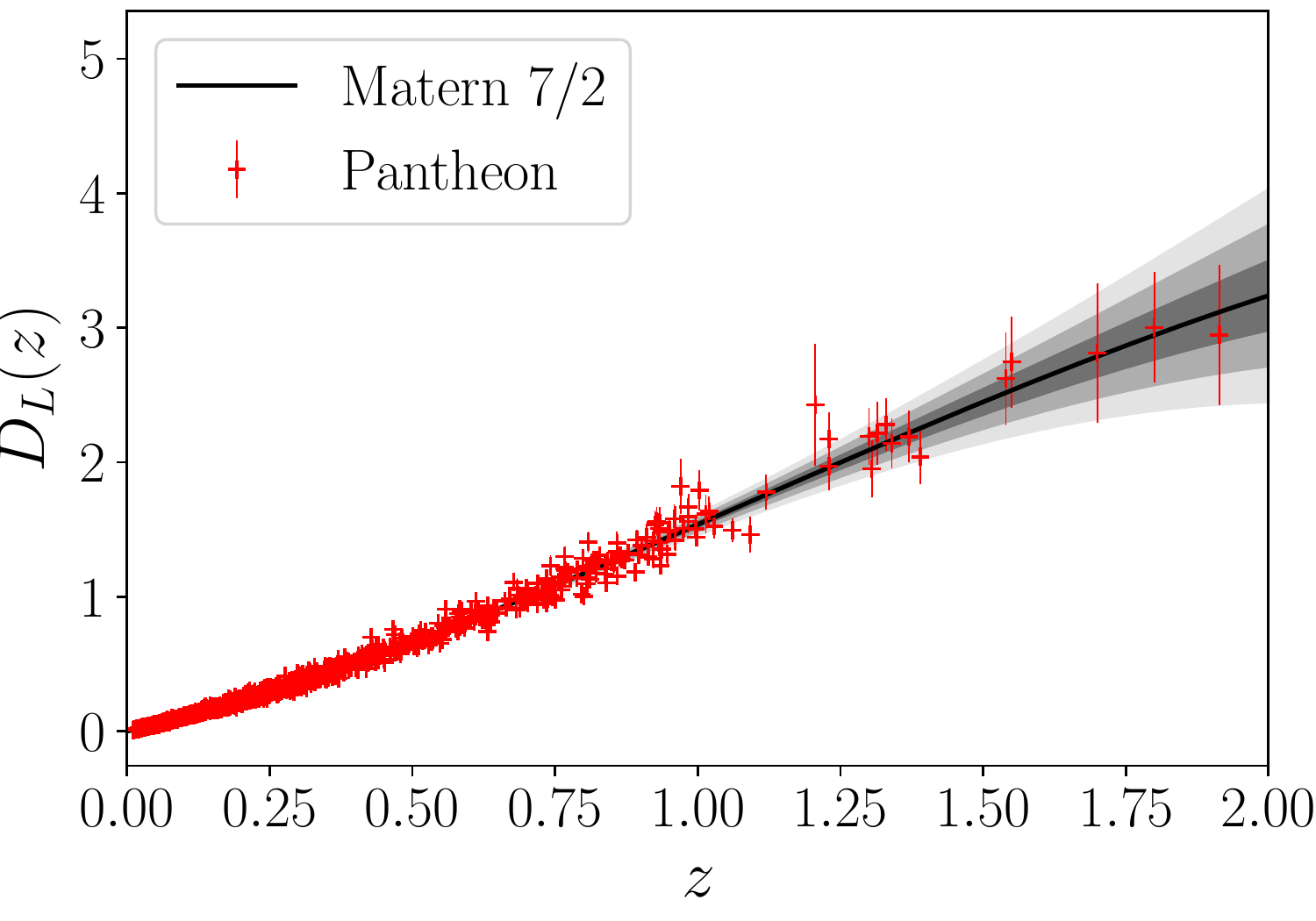}\\
		\includegraphics[angle=0, width=0.45\textwidth]{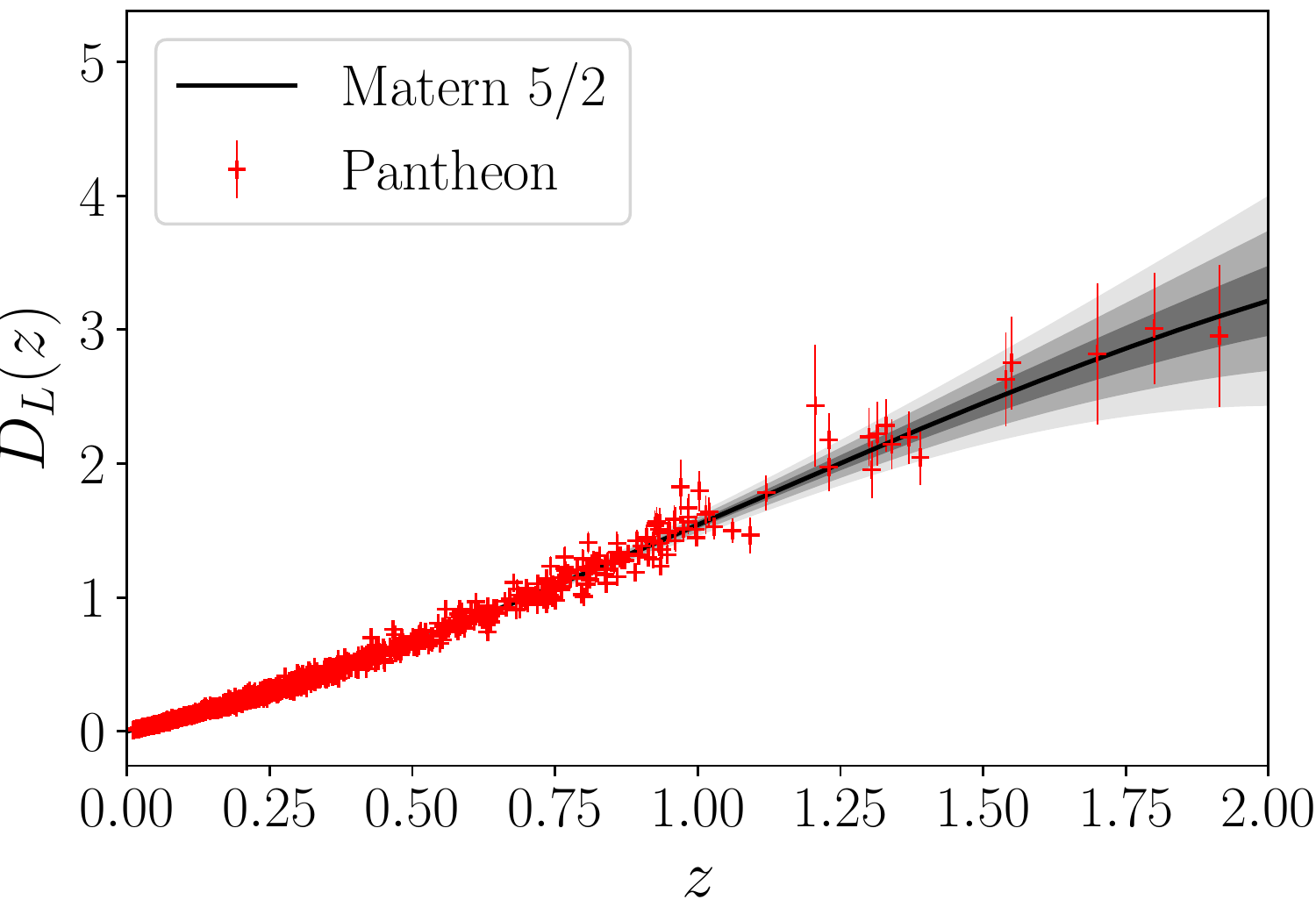}
		\includegraphics[angle=0, width=0.45\textwidth]{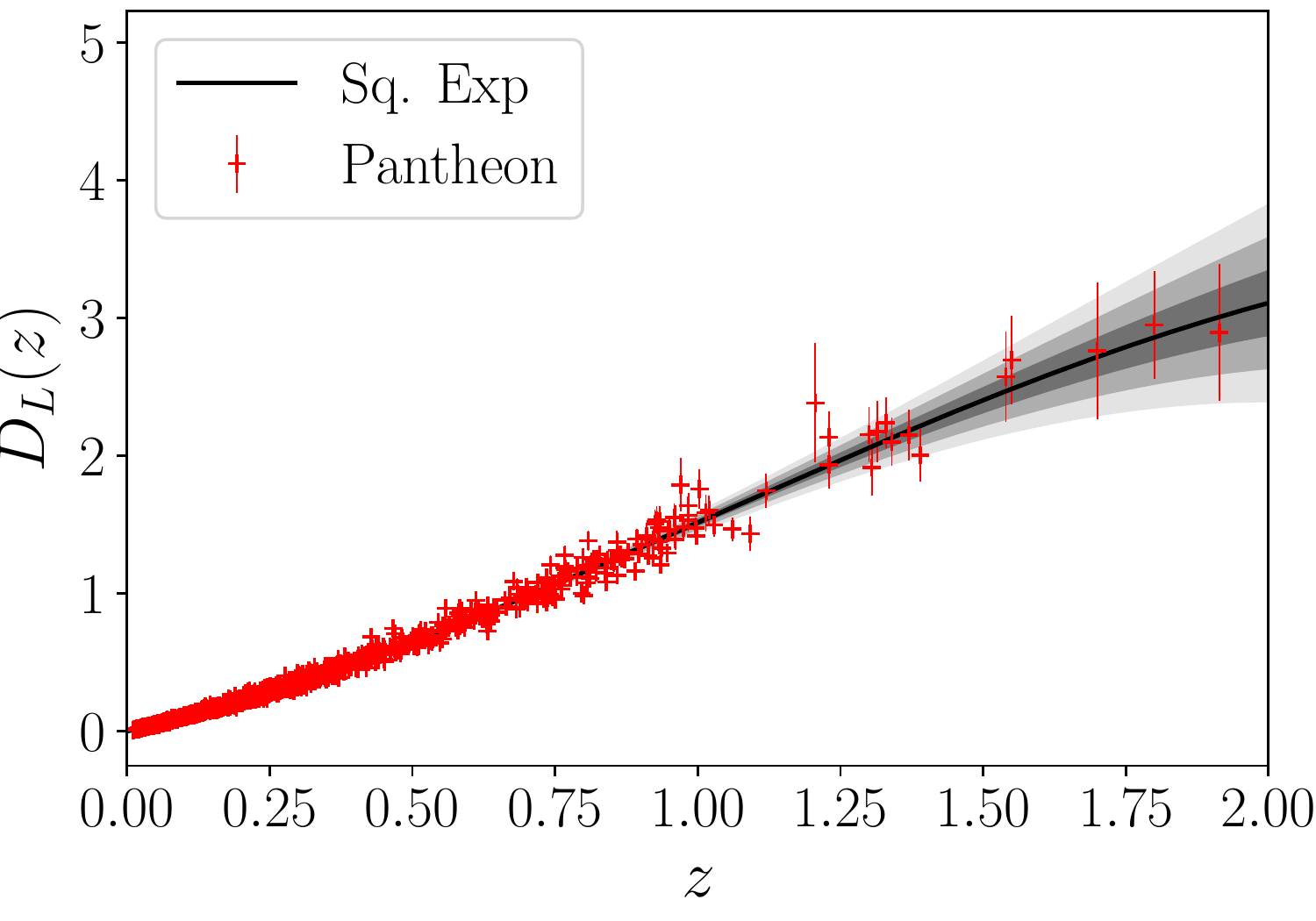}
	\end{center}
	\caption{{\small Plots for the reconstructed dimensionless or normalized luminosity distance $D_L$ considering the Mat\'{e}rn 9/2, Mat\'{e}rn 7/2, 
			Mat\'{e}rn 5/2 and Squared Exponential covariance function. The black solid lines represent the best fit curves. 
			The associated 1$\sigma$, 2$\sigma$ and 3$\sigma$ confidence levels are shown by the shaded regions.}}
	\label{ch2:Dl_plot}
\end{figure*}

\begin{figure*}[t!]
	\begin{center}
		\includegraphics[angle=0, width=0.45\textwidth]{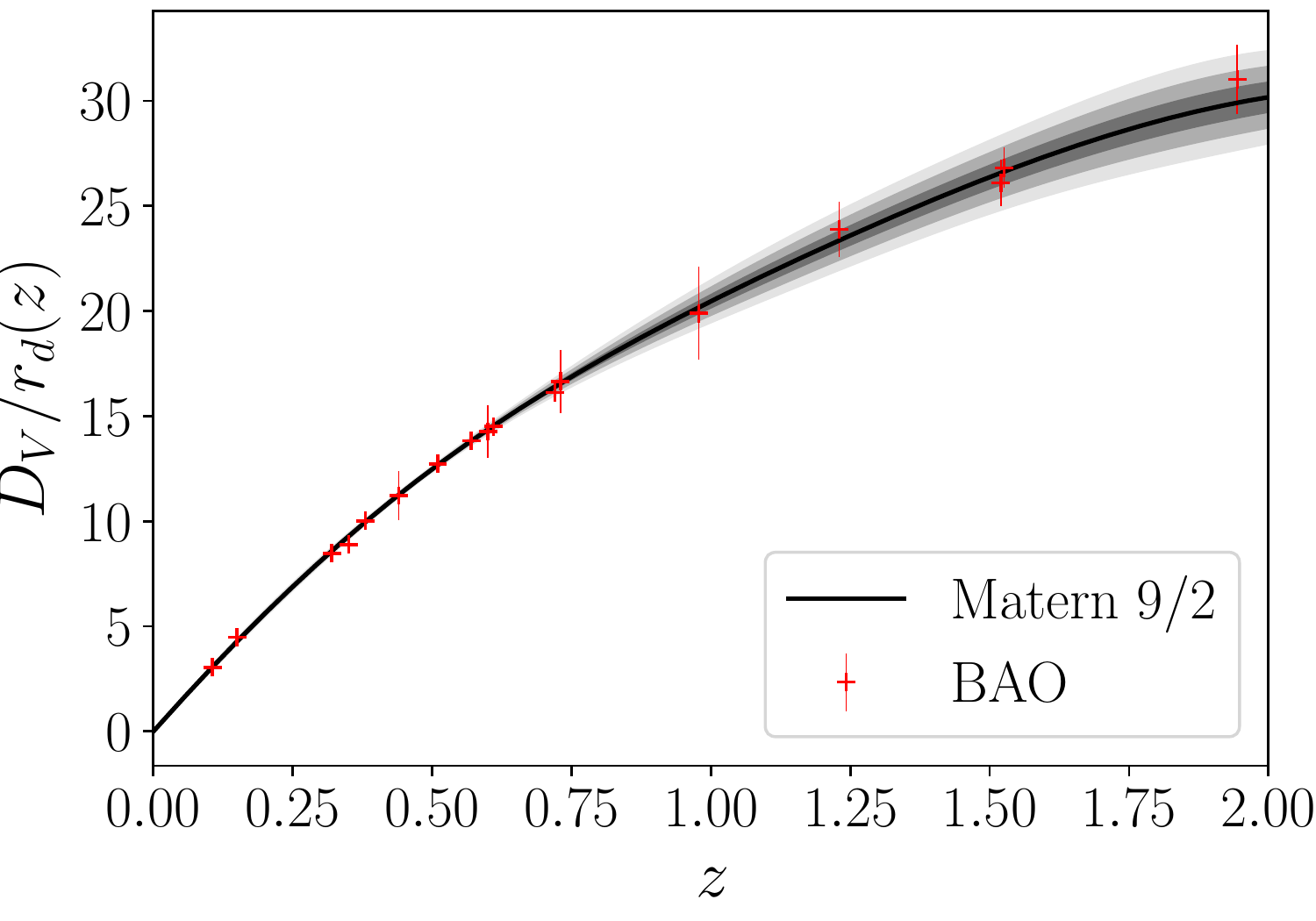}
		\includegraphics[angle=0, width=0.45\textwidth]{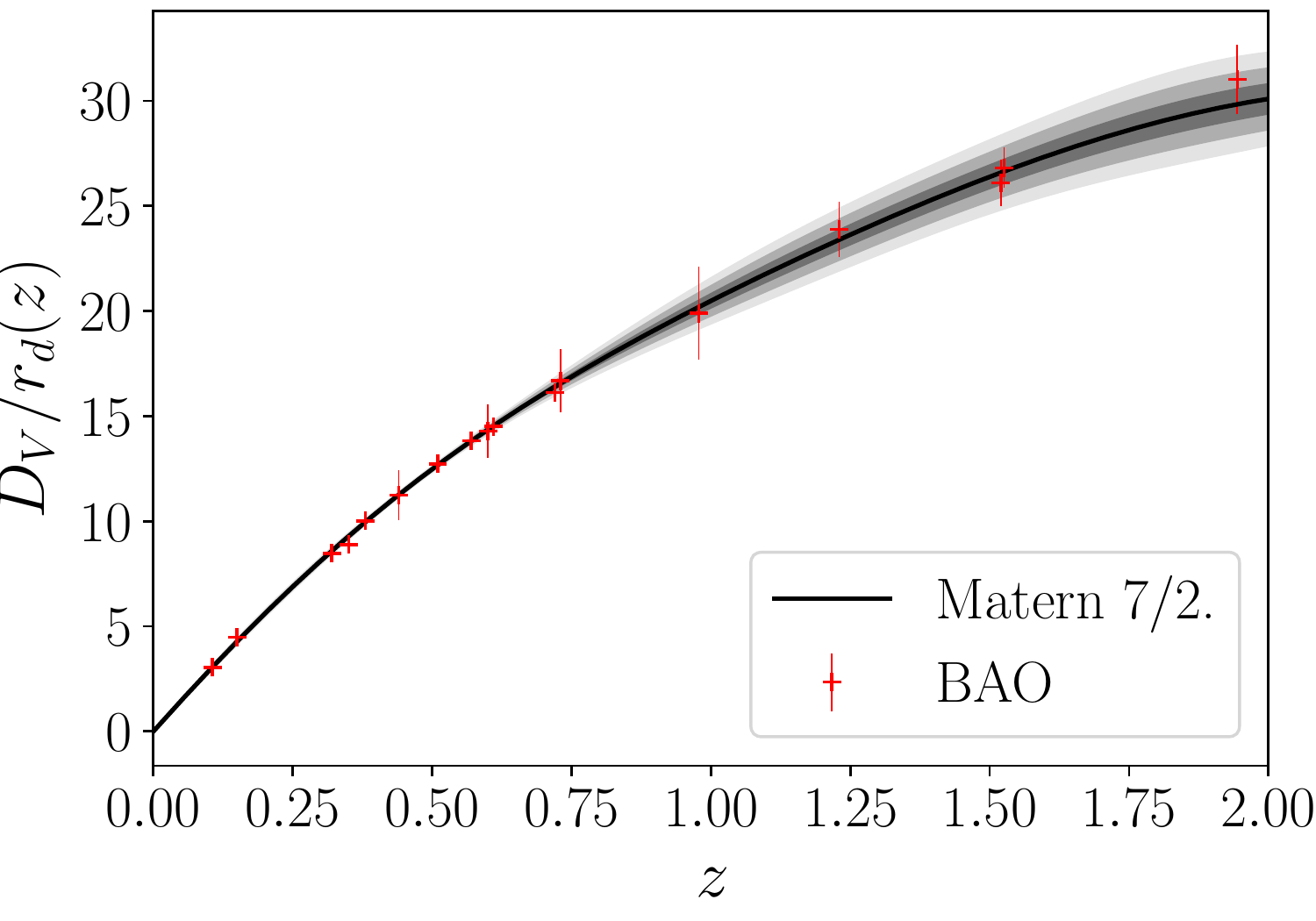}\\
		\includegraphics[angle=0, width=0.45\textwidth]{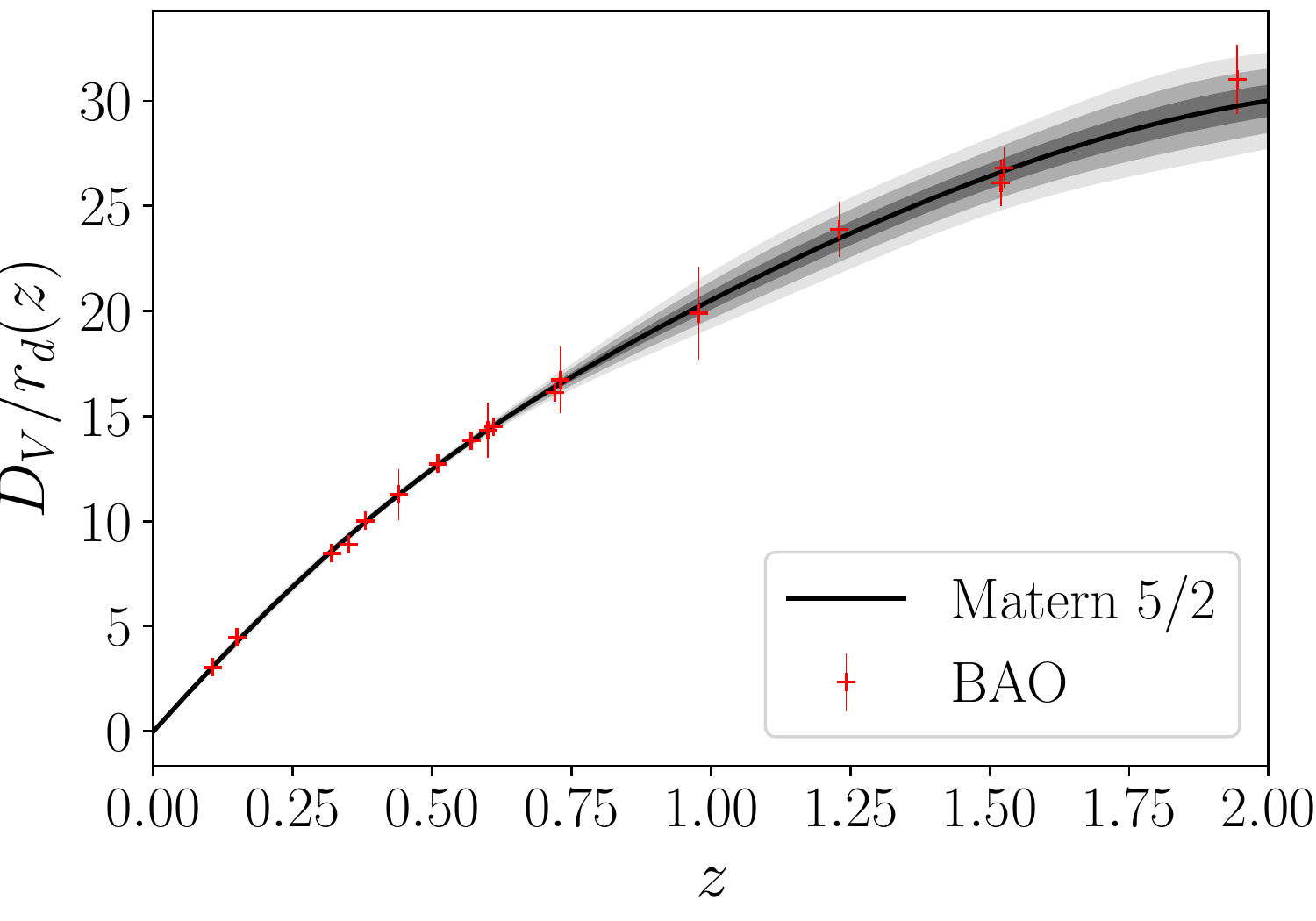}
		\includegraphics[angle=0, width=0.45\textwidth]{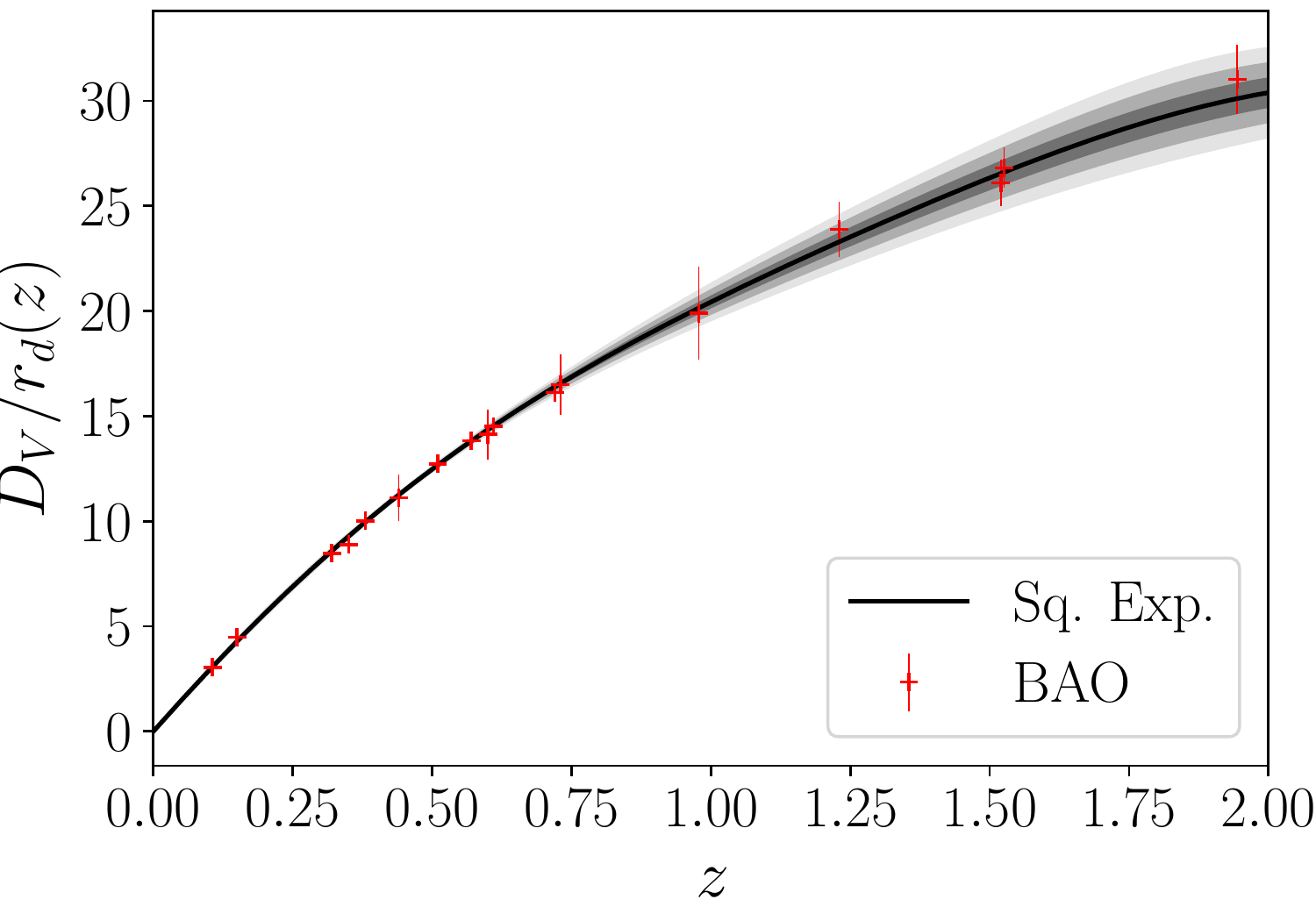}
	\end{center}
	\caption{{\small Plots for the reconstructed dimensionless ratio of volume-averaged distance to the comoving sound horizon at drag epoch $\frac{D_V}{r_d}$ 
			considering the Mat\'{e}rn 9/2, Mat\'{e}rn 7/2, Mat\'{e}rn 5/2 and Squared Exponential covariance function. The black solid 
			lines represent the best fit curves. The associated 1$\sigma$, 2$\sigma$ and 3$\sigma$ confidence levels are shown by the shaded regions.}}
	\label{ch2:Dv_plot}
\end{figure*}

Plots for the reconstructed dimensionless functions $D_L(z)$ (the normalized luminosity distance) and $\frac{D_V}{r_d}(z)$ in the redshift range $0 < z < 2$ 
as that of reconstructed $H(z)$, for different choices of the covariance function, are shown in Fig \ref{ch2:Dl_plot} and \ref{ch2:Dv_plot} respectively. 
The specific points (in the $H$, $D_L$ and $\frac{D_V}{r_d}$ plots) with error bars represent the observational data used in the reconstruction. The 
hyperparameters ($\sigma_f, l$) are optimized by maximizing log marginal likelihood while proceeding with the GP reconstruction.

\begin{figure*}
	\begin{center}
		\includegraphics[angle=0, width=0.45\textwidth]{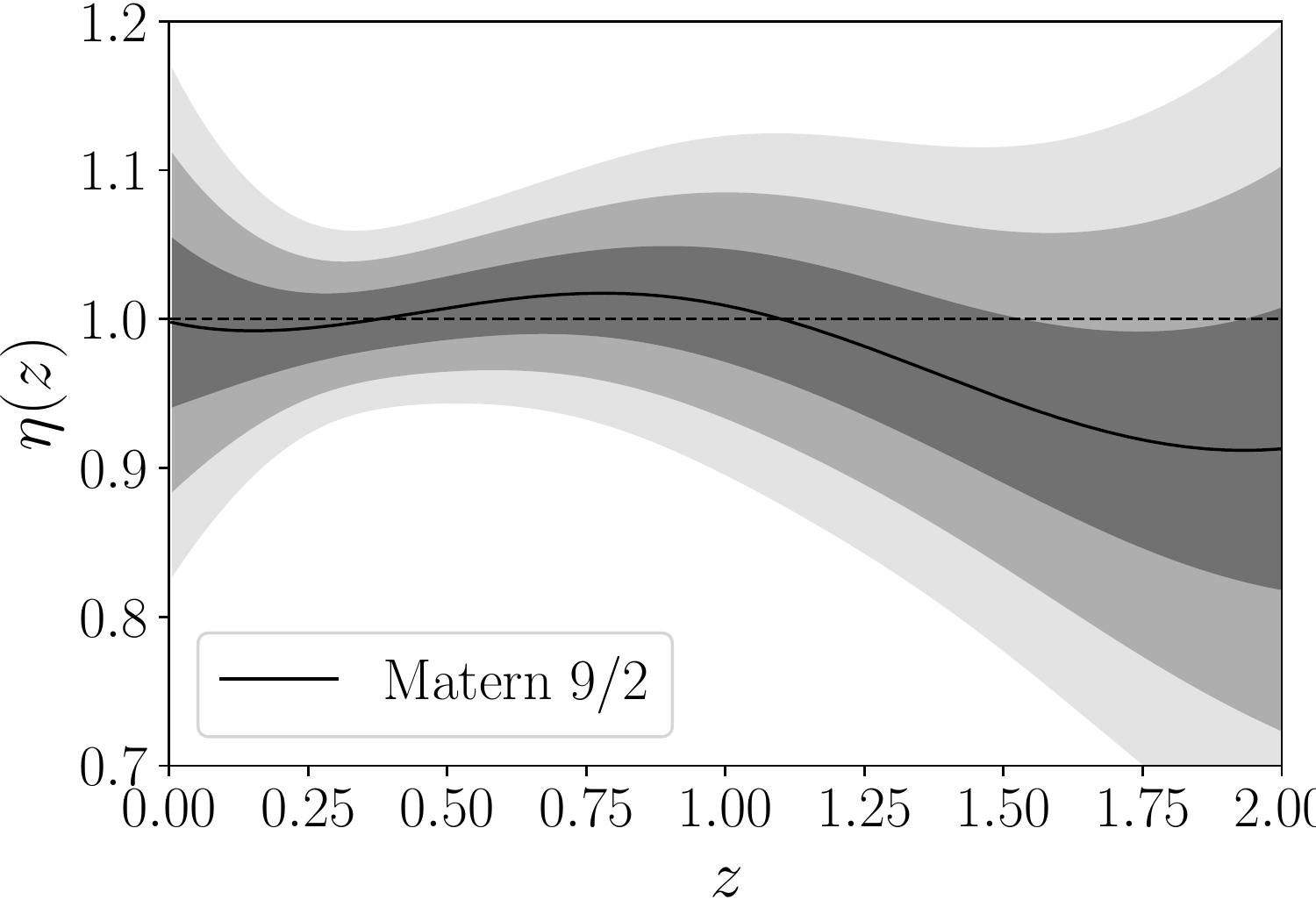}
		\includegraphics[angle=0, width=0.45\textwidth]{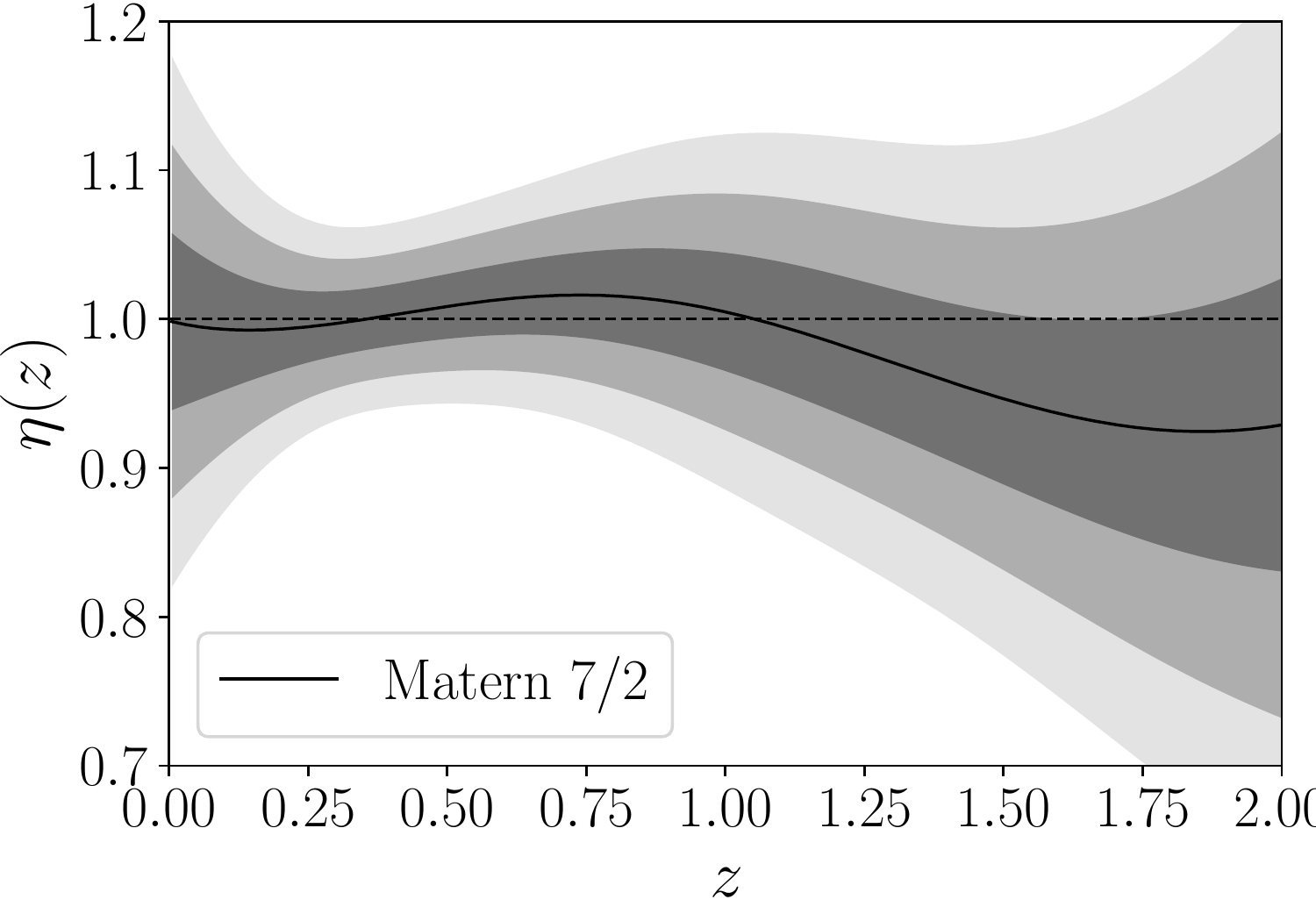}\\
		\includegraphics[angle=0, width=0.45\textwidth]{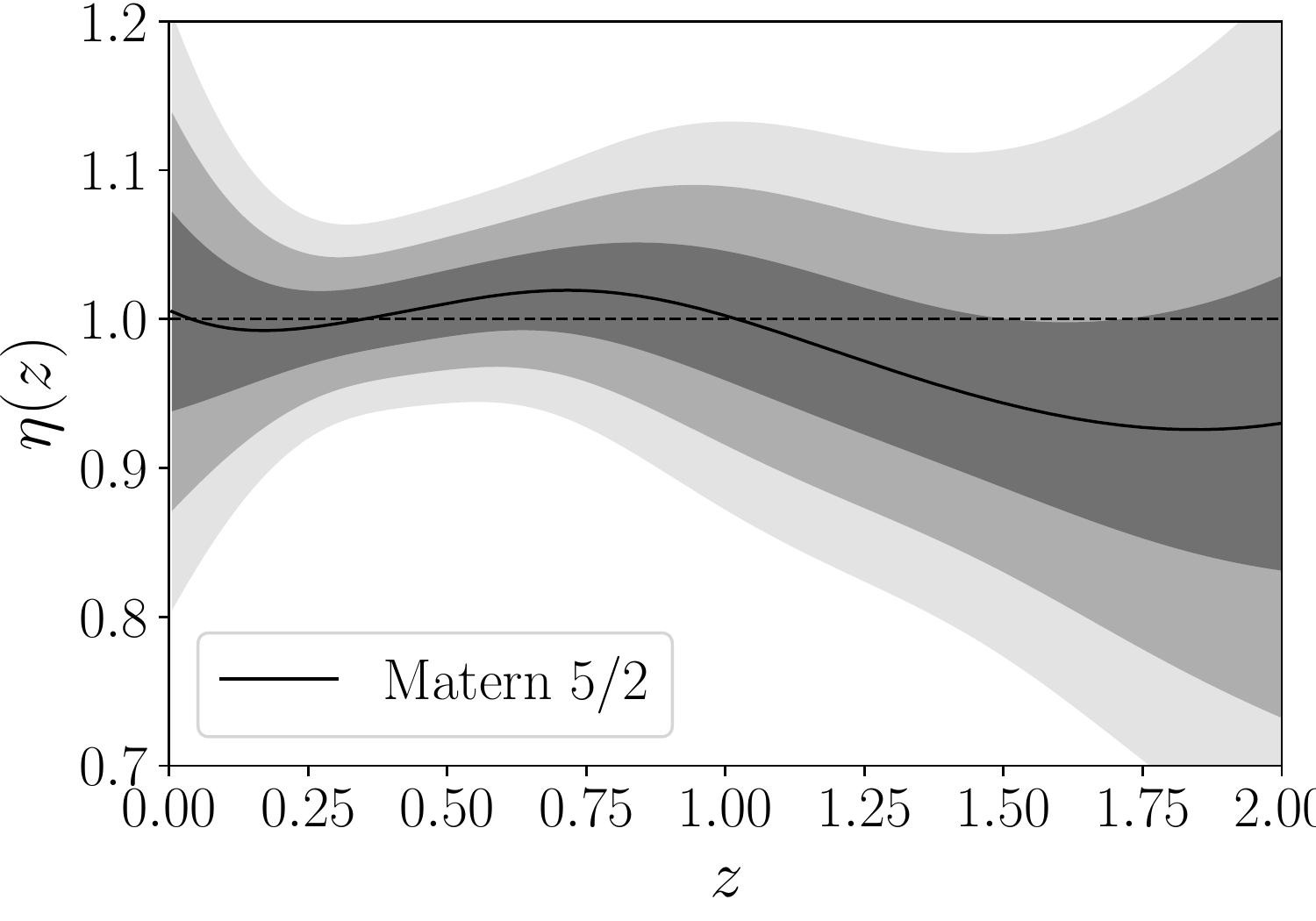}
		\includegraphics[angle=0, width=0.45\textwidth]{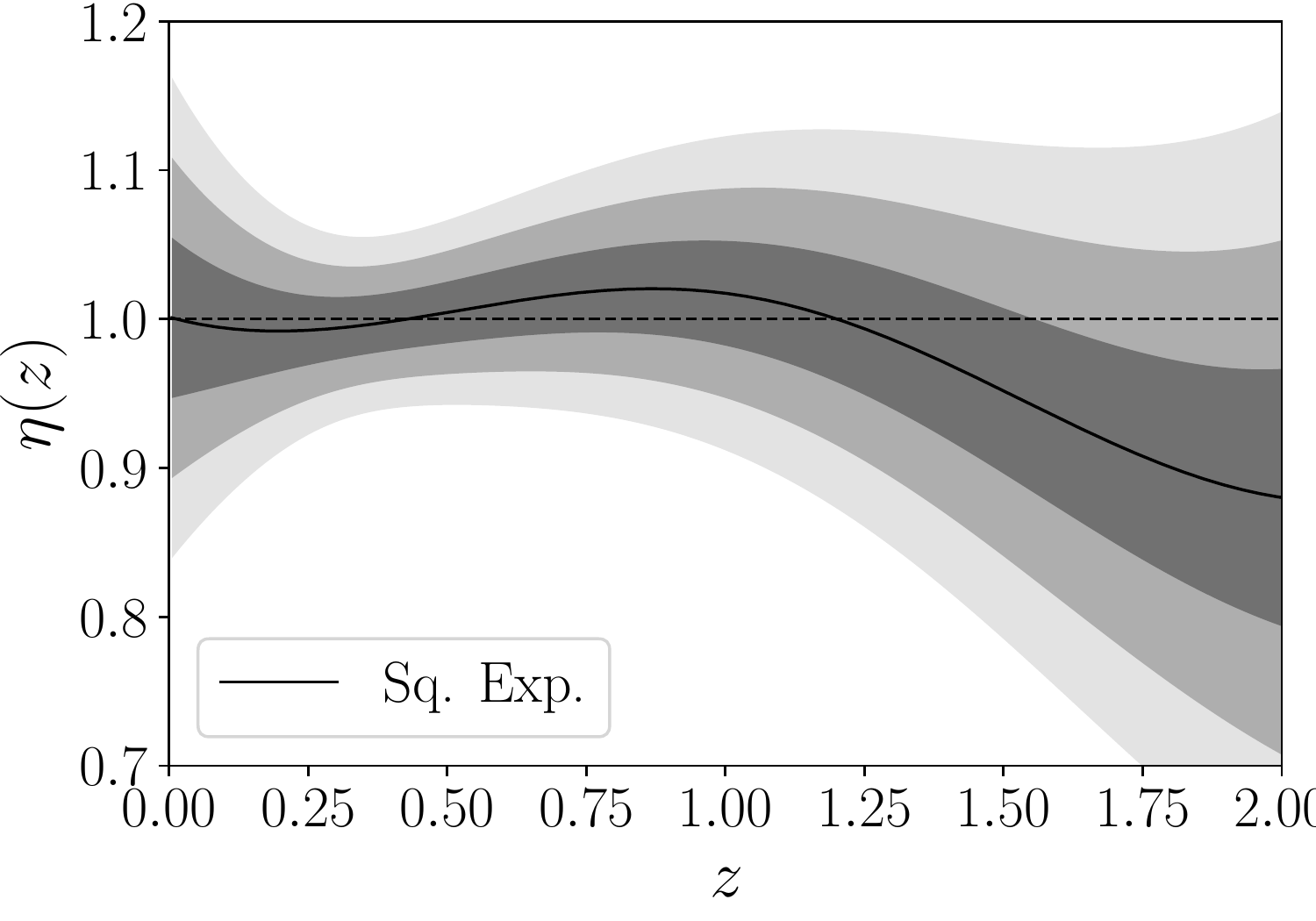}
	\end{center}
	\caption{{\small Plots for the reconstructed cosmic distance duality ratio $\eta$ considering the Mat\'{e}rn 9/2, Mat\'{e}rn 7/2, Mat\'{e}rn 5/2 and 
			Squared Exponential covariance function using $r_d$ from Table \ref{ch2:MOr_res}. The black solid lines represent the best 
			fit curves. The associated 1$\sigma$, 2$\sigma$ and 3$\sigma$ confidence levels are shown by the shaded regions.}}
	\label{ch2:eta_np_plot}
\end{figure*}

\begin{figure*}\begin{center}
		\includegraphics[angle=0, width=0.45\textwidth]{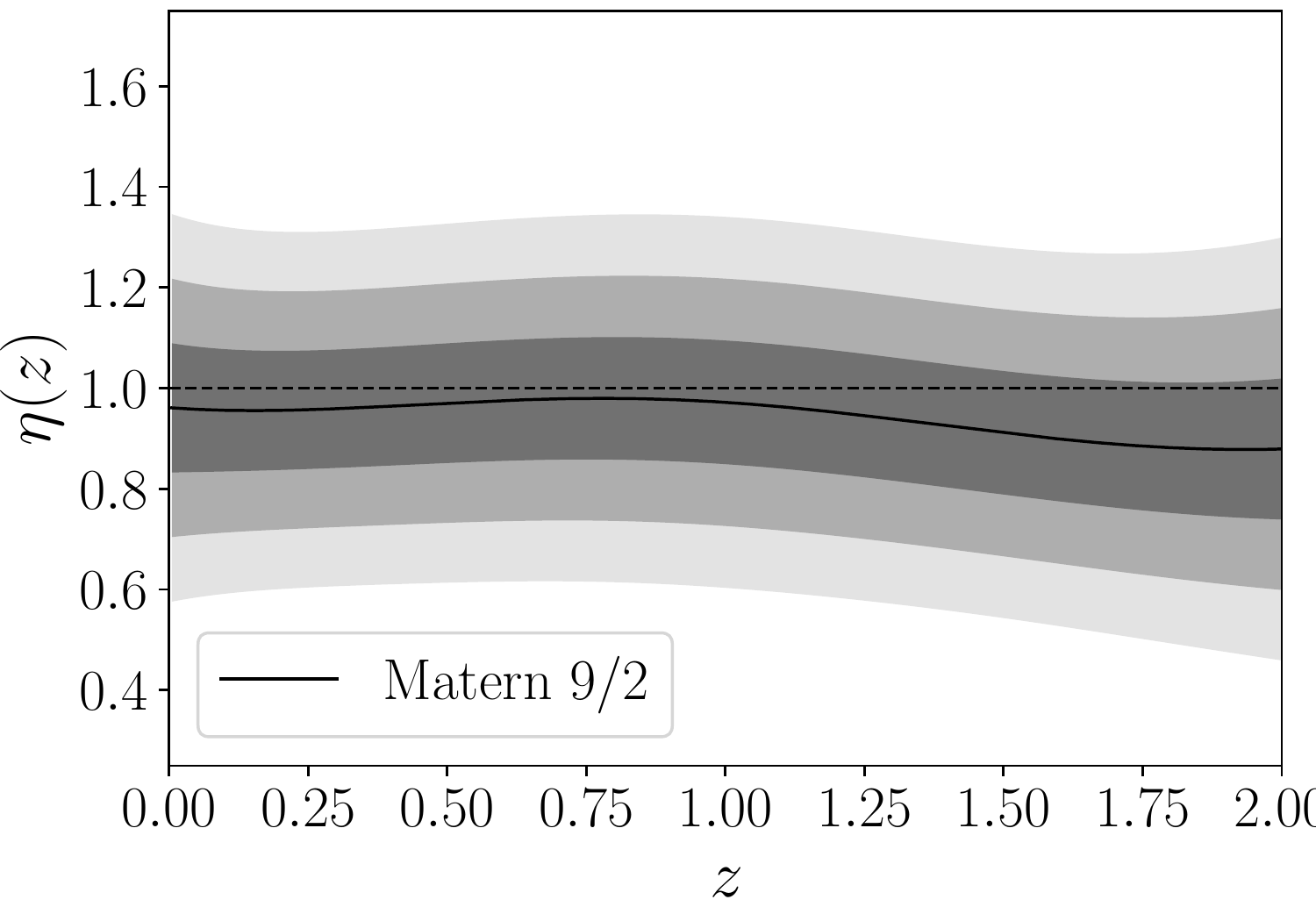}
		\includegraphics[angle=0, width=0.45\textwidth]{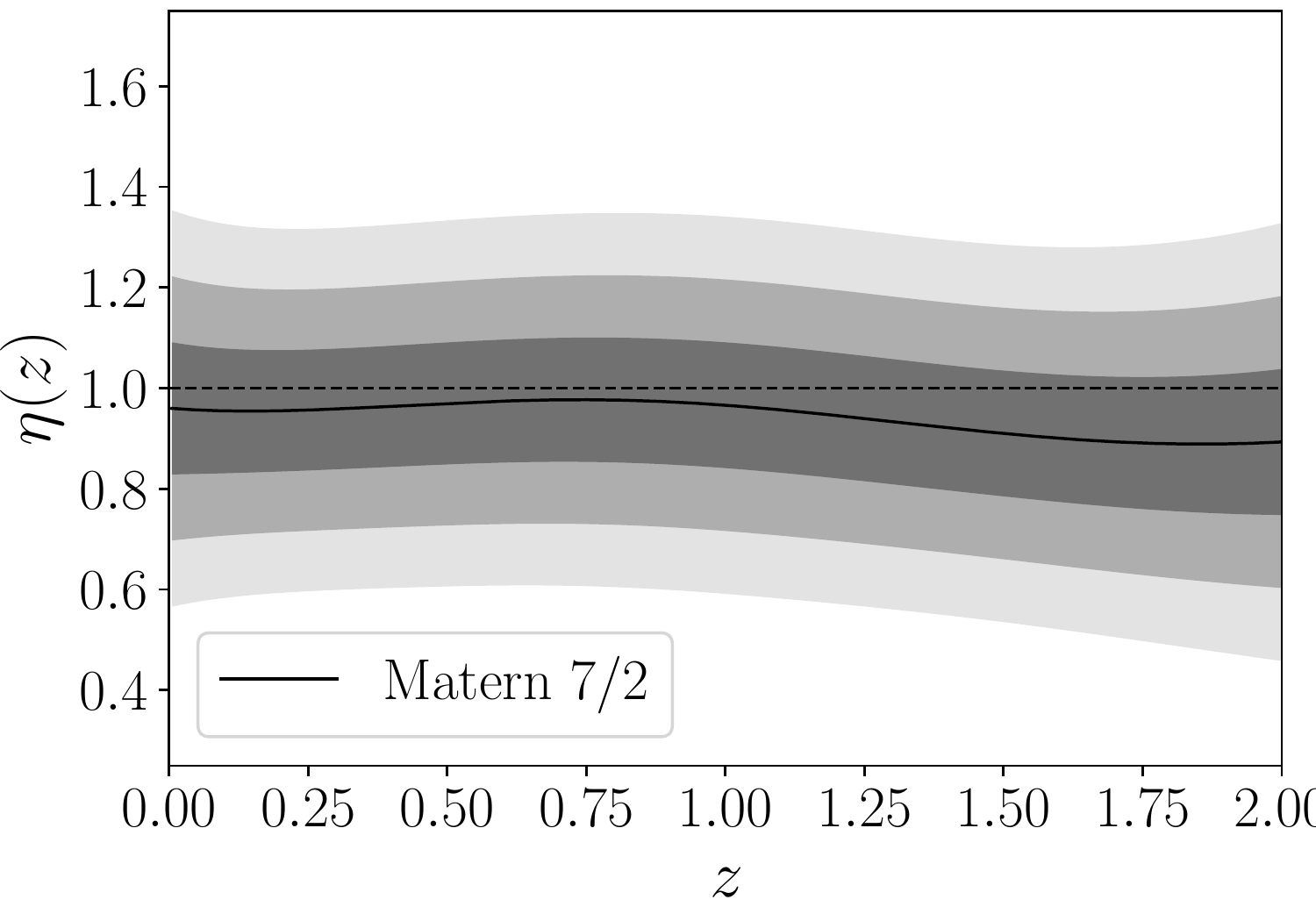}\\
		\includegraphics[angle=0, width=0.45\textwidth]{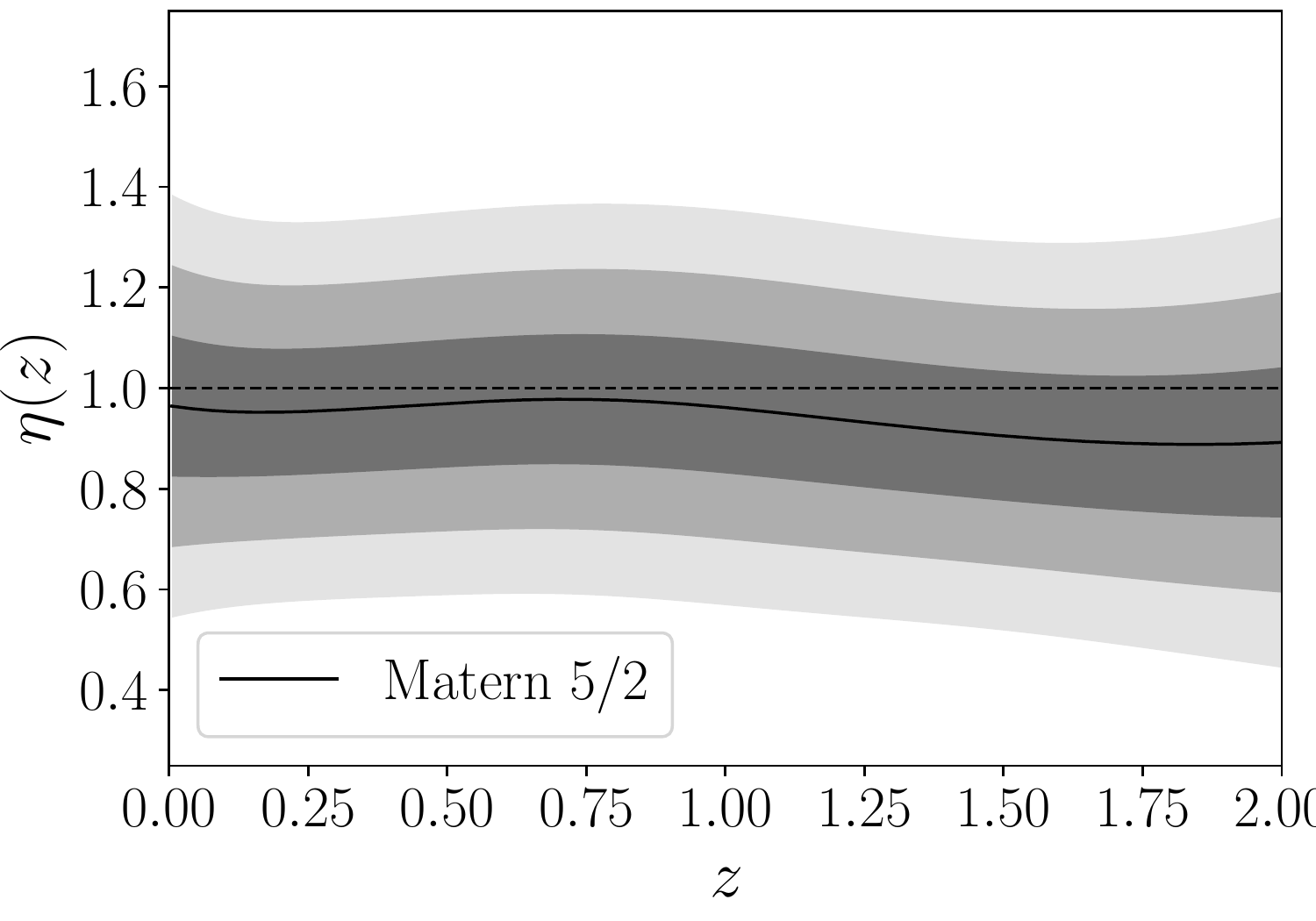}
		\includegraphics[angle=0, width=0.45\textwidth]{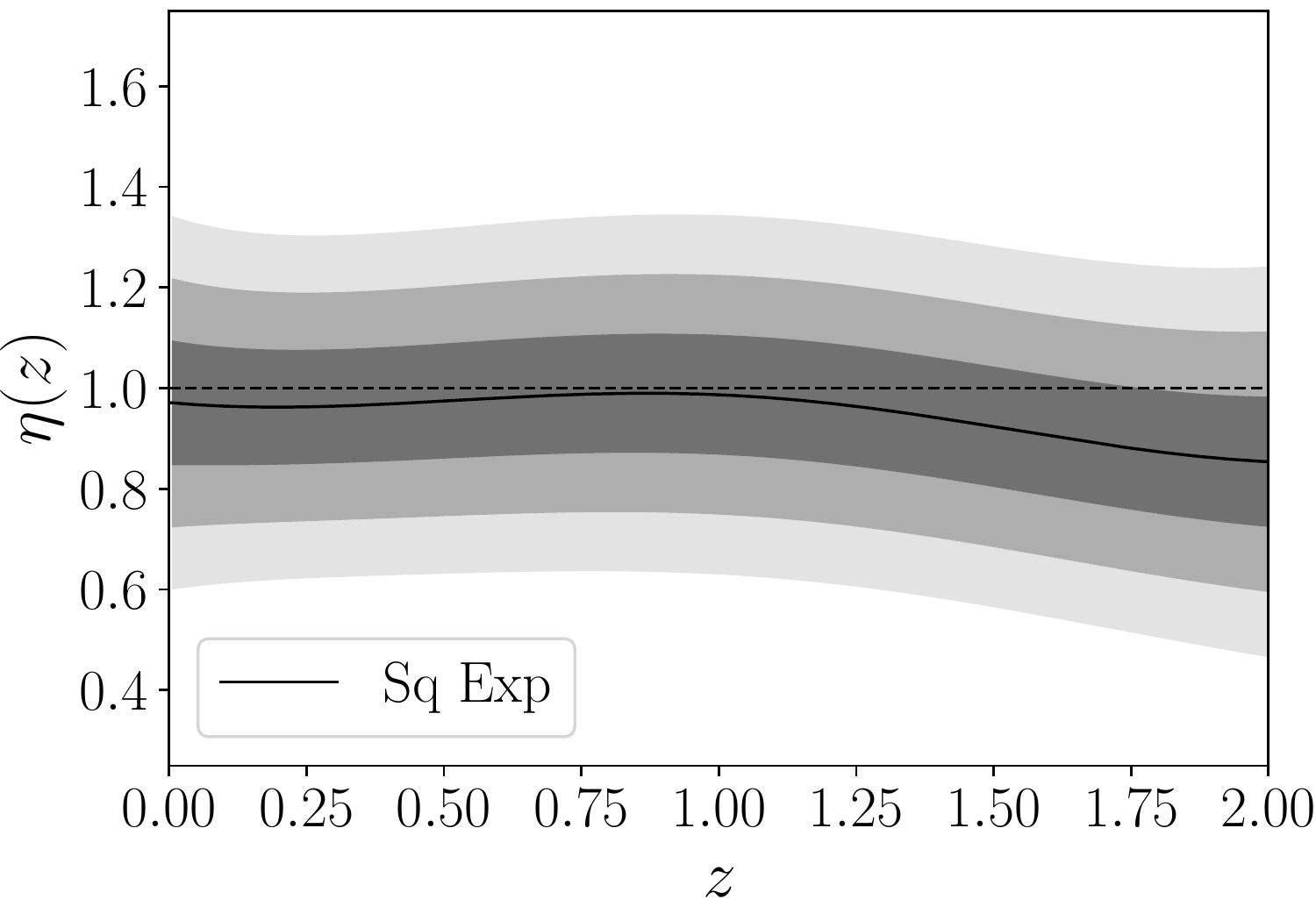}
	\end{center}
	\caption{{\small Plots for the reconstructed cosmic distance duality ratio $\eta$ considering the Mat\'{e}rn 9/2, Mat\'{e}rn 7/2, Mat\'{e}rn 5/2 and 
			Squared Exponential covariance function using $r_d$ from Table \ref{ch2:rdrag_tab}. The black solid lines represent the best 
			fit curves. The associated 1$\sigma$, 2$\sigma$ and 3$\sigma$ confidence levels are shown by the shaded regions.}}
	\label{ch2:eta_np_plot2}
\end{figure*}

Finally, the non-parametric reconstruction of $\eta$ is achieved using the reconstructed $H(z)$, $d_L(z)$ and $D_V(z)$, following the relation
\begin{equation}
\eta(z) = \frac{d_L ~\sqrt{c ~z}}{D_V^{\frac{3}{2}}~ H^{\frac{1}{2}} ~(1+z)}.
\end{equation} 
Plots for the reconstructed $\eta(z)$ are shown in Fig \ref{ch2:eta_np_plot} and \ref{ch2:eta_np_plot2} considering different choices of the covariance function. 
In the case of Fig \ref{ch2:eta_np_plot}, the marginalized constraints on $r_d$ are obtained via equation \eqref{ch2:chiBAO}, whereas Fig. \ref{ch2:eta_np_plot2} 
is the plot considering the approximated value of $r_d$ derived from equation \eqref{ch1:rdrag_approx}. It appears in Fig. \ref{ch2:eta_np_plot2}, that the uncertainty 
does not increase significantly at higher redshift, as compared to Fig \ref{ch2:eta_np_plot}. However, comparing the Y-axes range of both sets, it can be clearly 
seen that Fig \ref{ch2:eta_np_plot} is better constrained than that of \ref{ch2:eta_np_plot2}. This could be the effect of 1$\sigma$ uncertainty in $r_d$ which is 
quite large in the case of Fig \ref{ch2:eta_np_plot2} as compared to Fig \ref{ch2:eta_np_plot}, given in Tables \ref{ch2:rdrag_tab} and \ref{ch2:MOr_res} respectively.

The black solid line represents the best fit values of $\eta$. The shaded regions correspond to the 68\% ($1\sigma$), 95\% ($2\sigma$) and 99.7\% ($3\sigma$) confidence 
levels (CL). Plots reveal that the reconstructed values of $\eta(z)$ in the low redshift range $0<z<1$ remains very close to unity. Thus, CDDR is always allowed within 
the 1$\sigma$ uncertainty for low redshift. 

At higher redshift $z>1.5$, in cases of the Mat\'{e}rn 5/2 and 7/2 covariance functions, CDDR is always allowed in 1$\sigma$ 
for Fig \ref{ch2:eta_np_plot2} and almost in the case of Fig \ref{ch2:eta_np_plot}. For the Mat\'{e}rn 9/2 and Squared Exponential covariance functions, CDDR is almost 
always allowed within the $1\sigma$ uncertainty in Fig \ref{ch2:eta_np_plot2} for $z>1.5$. Nonetheless, CDDR is always within a 2$\sigma$ uncertainty which indicates a 
non-violation of CDDR in the late-time Universe for all cases studied.

\section{Discussion}
\label{ch2:conclusion}

In the present study, the validity of the cosmic distance duality relation is investigated by the non-parametric GP method. The distance modulus measurements 
of type Ia supernovae from the latest Pantheon sample, the cosmic chronometer measurements of the Hubble parameter, and the recent measurements of volume-averaged 
Baryon Acoustic Oscillation data are utilized in the analysis. Firstly, a reconstruction of the luminosity distance $d_L$ is done from the Pantheon SNIa 
compilation. A reconstruction of the Hubble parameter $H(z)$ from the CC data, followed by another reconstruction of $D_V$ from the isotropic BAO data, have 
been carried out using the GP method. The analysis has been done for four choices of the covariance function, namely the Squared exponential, Mat\'{e}rn 9/2, 
Mat\'{e}rn 7/2 and Mat\'{e}rn 5/2 covariance, in the same domain of redshift $0<z<2$. The angular distance $d_A$ are obtained via combining the reconstructed 
$H$ and $D_V$ from the CC  and BAO data, respectively.

Finally, the reconstructed functions viz. luminosity distance $d_L(z)$, Hubble parameter $H(z)$ and the volume-averaged distance $D_V(z)$ have been combined 
to get the cosmic distance-duality ratio $\eta(z)$. The reconstructed $\eta$ curve is obtained directly from the data without assuming any functional form 
\textit{a priori}. A common feature is that the best-fit curve for $\eta$, shown in Fig \ref{ch2:eta_np_plot} and \ref{ch2:eta_np_plot2}, is close to $\eta=1$ 
in the range $0<z<1$. Also, $\eta = 1$ is always included in 1$\sigma$ for the low redshift range $0<z<1.5$ and always included in 2$\sigma$ for all choices of 
the covariance functions at high redshift, in case of both Fig \ref{ch2:eta_np_plot} and \ref{ch2:eta_np_plot2}. As the BAO data points are mostly concentrated 
up to $z \approx 1$, the reconstructed $\eta$ is well constrained up to $z\approx 1$. The uncertainties in the reconstructed $\eta(z)$ curves increase with 
increasing redshift. 

It deserves mention that marginalized constraints on $M_B$ are obtained by keeping the nuisance parameters $\alpha$, $\beta$, colour, stretch and bias 
corrections fixed using the BBC framework for the Pantheon SNIa compilation. Fixing the value of $M_B$ from the global $\Lambda$CDM fits from Scolnic 
\textit{et al}\cite{pan}, may result in inconsistencies as $M_B$ is degenerate with $H_0$. In case of the BAO $\frac{D_V}{r_d}$ data, the comoving sound 
horizon at photon drag epoch, $r_d$ is constrained considering a fiducial measure on $r_{\mbox{\tiny d, fid}}$ equal to $147.49$. We have tried to keep our 
analysis model-independent, as far as possible. Since the BAO measurements from WiggleZ DES in Blake \textit{et al}\cite{blake2012} are dependent on the 
cosmological parameters $H_0$ and $\Omega_{m0}$, so precise constraints on $\Omega_{m0}$ are derived assuming a fiducial $\Lambda$CDM model, which dampens 
the spirit of a model-independent reconstruction to a certain extent. The uncertainties on these parameters are propagated properly for the evaluation of 
error in the distance measurements. As there may be correlations between these parameters $H_0$, $M_B$, $\Omega_{m0}$ and  $r_d$, keeping these values fixed 
may lead to inconsistent results, if proper care is not taken.

The GP method has previously been used for a non-parametric reconstruction of $\eta$. Nair \textit{et al}\cite{nair3} utilized Union 2.1 SNIa data compilation, 
BAO measurements from SDSS, 6dFGS, WiggleZ, BOSS, and observational Hubble data compilation to reconstruct the luminosity distance, the volume-averaged 
distance and the Hubble rate using the GP regression technique in the redshift range $0.1 < z <0.73$. Our work is similar to the work by Nair \textit{et 
al}\cite{nair3} but there are quite a few differences to list. We have utilized the latest updated Pantheon SNIa compilation, which spans up to a redshift range $ 
z = 2.26$, instead of the Union 2.1 sample with available data points limited up to $z = 1.41$. Here, constraints have been obtained on a wider range of overlapping 
redshift $0 < z < 2$, and tighter constraints are obtained in the low redshift regime due to availability of more number of BAO data. Another non-parametric 
reconstruction of the CDDR by Rana \textit{et al}\cite{rana2017} using different dynamic and geometric properties of strong gravitational 
lensing (SGL) along with the JLA SNIa observations, do not indicate any deviation from CDDR and are in concordance with the standard value of unity within a 
2$\sigma$ confidence region. The difference between our work and that by Rana \textit{et al}\cite{rana2017} lies in terms of the data sets involved and the 
redshift range considered for reconstruction. In case of Rana \textit{et al}\cite{rana2017}, the reconstruction was mainly focused on the redshift range of $0<z<1$. 
Zhou and Li\cite{zhou2019} reconstructed the distance-redshift relation from observations of the Dark Energy Survey SNIa with simulated fiducial $H(z)$ data and 
obtained that, except for the very low redshift range $z<0.2$, there is no significant deviation from the theoretical CDDR. The prime objective of work by Zhou and  
Li\cite{zhou2019} was to test the fidelity of Gaussian processes for cosmography where CDDR was reconstructed as a consistency check. The present work shows that the GP 
successfully reproduces the CDDR even at higher redshifts.

The results obtained in the present analysis are in agreement with those from the existing literature, discussed above. We have extended the analysis 
to higher redshifts. But due to the scarcity of observational data at high redshift, the uncertainty increases with higher $z$. We can 
conclude that all the recent studies of cosmic distance measurements indicate a non-violation of the CDDR at low redshift ($z<1.5$). Future high redshift 
observations of BAO, SNIa and other observables would be able to provide tighter constraints on CDDR at higher redshift. Similar analysis with future 
observations would be useful to decide whether CDDR is equally valid at high redshift, or redshift dependent higher order correction terms are essentially 
required.

\clearpage{}
\clearpage{}\chapter{Reconstruction of the cosmological \textit{jerk} parameter}\blfootnote{\begin{flushleft} The work presented in this chapter is based on ``Non-parametric reconstruction of the cosmological \textit{jerk} parameter", \textbf{Purba Mukherjee} and Narayan Banerjee, Euro.\ J.\ Phys.\ C \textbf{81}, 36 (2021).\end{flushleft}}  \label{ch4:chap4}
\chaptermark{Reconstruction of the cosmological \textit{jerk} parameter}
\section{Introduction}

The current chapter presents a non-parametric reconstruction of the cosmological jerk parameter from observational data. At the outset, we do not 
assume any fiducial model of the Universe, except that it is spatially flat, homogeneous and isotropic, thus described by the FLRW metric. Ignoring the Einstein 
equations, we pick up only the kinematical quantities, that are defined as the time derivatives of the scale factor $a(t)$. The first order derivative of the 
scale factor, the Hubble parameter $H$ that measures the rate of cosmic expansion, is found to evolve with time. So, the natural choice as a relevant parameter is 
the next order derivative of $a$, namely the \textit{deceleration} parameter, defined as 
\begin{equation} \label{ch4:qdef}
q = -\frac{1}{aH^2} \frac{d^2 a}{dt^2}. 
\end{equation} 
Now that $q$ can be estimated and is found to be evolving, the third order derivative of the scale factor $a(t)$, called the {\it jerk}, is now a quantity of interest. 
The jerk parameter $j$ is defined in a dimensionless way as
\begin{equation} \label{ch4:jerkdef}
j =  \frac{1}{aH^3} \frac{d^3 a}{dt^3}. 
\end{equation} 

Reconstruction of cosmological models through the deceleration parameter $q$ can be found in the work of Gong and Wang\cite{gong1, gong2}, Wang, Xu, Lu and 
Gui\cite{wang}, Lobo, Mimoso and Visser\cite{jose}, Mamon and Das\cite{sudipta}, Mamon\cite{mamon}, Cardenas and Motta\cite{motta}, Jesus, Holanda and 
Pereira\cite{jesus}, Yang and Gong\cite{yangong}. The list is surely not quite exhaustive. Reconstruction through the jerk parameter has been carried out by 
Luongo \cite{luongo}, Rapetti {\it et al}\cite{rapetti}, Zhai {\it et al}\cite{zz}, Mukherjee and Banerjee\cite{ankan1, ankan2}, Mamon and Bamba\cite{mamon}, 
Mukherjee, Paul and Jassal\cite{ankan3}. Density perturbations also have been investigated for models reconstructed through the jerk parameter by Sinha and 
Banerjee\cite{srijita}. The $\Lambda$CDM model has been recovered from a reconstruction of cosmographic parameters like $q$ and $j$ by Amirhashchi and 
Amirhashchi\cite{amirhashchi}. The jerk parameter and a combination of deceleration and jerk parameters together have been identified as the statefinder 
parameters in the investigations by Sahni \textit{et al}\cite{jsahni} and Alam \textit{et al}\cite{jalam}. Although the possible importance of the jerk 
parameter in the game of reconstruction was pointed out long back\cite{jalam}, not much work has been done to utilize its full potential. These investigations 
mostly rely on assuming a functional form of $j$ as ansatz and estimating the parameters from observational data. This is necessarily restrictive as 
the functional form for $j$ is already chosen.  

A more unbiased way is to attempt a non-parametric reconstruction, where the evolution of the relevant quantity is determined directly from observational 
data without any ansatz \textit{a priori}. There have already been efforts towards a non-parametric reconstruction of $q$, that can be found in the works of 
Bilicki and Seikel\cite{bilicki}, Zhang and Xia\cite{xia},  Lin, Li and Tang\cite{lin}, Nunes \textit{et al}\cite{nunes},  Bengaly\cite{carlos}, Jesus, Valentim, 
Escobal and Pereira\cite{jesus_nonpara}, Arjona and Nesseris\cite{arjona}, Velten, Gomes and Busti\cite{busti}, G\'{o}mez-Valent\cite{adria} and Haridasu 
\textit{et al}\cite{haridasu}. However, there is hardly any attempt to model the dark energy by reconstructing the jerk parameter $j$ in a non-parametric 
way. Although there is no convincing reason that a reconstruction of kinematic parameters like $q$ or $j$ is more useful than that of a physical quantity like the 
dark energy equation of state parameter, this indeed provides an alternative route towards the understanding of dark energy in the absence of a convincing 
physical theory. 

We have utilized various combinations of the CC \cite{cc_101,cc_102,cc_103,cc_104,cc_106,cc_105} measurements of the Hubble parameter (obtained from a compilation 
between the 15 CCM and 11 CCH $H(z)$ values), the Pantheon \cite{pan} SNIa distance modulus data sample consisting of 1048 supernovae, the 30 $r$BAO $H(z)$ \cite{bao_61,bao_1,alam2017,bao_107,bao_108,bautista2017,bao_2,bao_4,blake2012,bao_77,anderson2014,bao_9,bao_110} data (given in Table \ref{ch1:tabbao}), and also the 
CMB Shift parameter data to examine their effect on the reconstruction. One may find that some of the $r$BAO $H(z)$ data from clustering measurements are correlated 
since they either belong to the same analysis or there is an overlap between the galaxy samples. Here in this paper, we mainly consider the central value and standard 
deviation of the data into consideration. Therefore, they are assumed to be independent measurements as in Li {\it et al}\cite{li_rsd} and Geng {\it et al}\cite{geng}. 
Indeed there are apprehensions that the CMB Shift parameter data depends crucially on a fiducial cosmological model\cite{elgaroy} and so does the BAO data\cite{carter}. 
Nevertheless, we do not ignore them. Our reconstruction is based on the combinations both including and excluding the model-dependent datasets. Finally, we extract the 
physical information regarding the effective equation of state parameter $w_{eff}$ for various combinations of the datasets. 

Different strategies for determining the value of $H_0$ is well known in the recent literature \cite{cosmoletter2}. Locally, the Hubble parameter has been 
measured to be $H_0 = 74.03 \pm 1.42 $ km Mpc$^{-1}$ s$^{-1}$\cite{riess} obtained from HST observations of 70 long-period Cepheids in the Large Magellanic 
Clouds by the SH0ES team (hereafter referred to as R19). Another strategy involves an extrapolation of data on the early Universe from the CMB, which yields 
the value $H_0 = 67.27 \pm 0.60$ km Mpc$^{-1}$ s$^{-1}$\cite{planck} from Planck (TT, TE, EE+lowE) 2018 survey assuming a base $\Lambda$CDM model (hereafter 
referred to as P18). In view of the $4.4 \sigma$ tension between the P18 and R19 values of $H_0$, reconstruction using both of them have been carried out 
separately. This exercise is undertaken to check if there is any qualitative change in the results affected by these priors.

At the very beginning, the model-independent datasets viz. CC and Pantheon, are taken into consideration. The reconstructed values of $j$ are found to be consistent with 
the standard $\Lambda$CDM model within $2\sigma$ CL. The model-dependent datasets like BAO and the CMB Shift are also included thereafter, which does not significantly 
help in improving or deteriorating the confidence level in favour of $\Lambda$CDM. The deceleration parameter $q$ can also be reconstructed from the same datasets. This 
is further used to find the effective equation of state parameter for the model-independent datasets only. Results depict that the $\Lambda$CDM model is excluded for some 
part of the evolution in $1\sigma$, but is definitely included within $2\sigma$ in the domain $0 \leq z \leq 2.36$ of all the reconstructions.

\section{Reconstruction Methodology}

We consider a spatially homogeneous and isotropic Universe described by the FLRW metric. We shall be dealing with a spatially flat Universe for which the curvature index $k=0$.  
In this particular case, the dimensionless transverse comoving distance of luminous objects, like SNIa, is given by
\begin{equation} \label{ch4:D}
D(z)=  \int_{0}^{z} \frac{\dif z'}{E(z')} ,
\end{equation} 
where $E(z) = \frac{H(z)}{H_0}$ is the reduced Hubble parameter.

Equation \eqref{ch4:D} gives a relation between the reduced Hubble parameter $E(z)$ and the comoving distance $D(z)$ as
\begin{equation} \label{ch4:D'}
D'(z)=\frac{1}{E(z)}.
\end{equation} 

The uncertainty $\sigma_{D'}$ associated with $D'$ is propagated from the uncertainty $\sigma_E$, the uncertainty in $E$, via the error propagation rule 
\begin{equation}
\sigma_{D'} = \frac{\vert \sigma_{E} \vert}{E^2} . \label{ch4:sigD'} 
\end{equation}

The deceleration parameter $q$ and jerk parameter $j$, defined in equations \eqref{ch1:qdef} and \eqref{ch1:jdef} can be written as a function of the reduced Hubble 
parameter $E(z)$ along with its derivatives at redshift $z$ as
\begin{equation} 
q(z) = -1 + (1+z) \frac{E'}{E},	 \label{ch4:qz}
\end{equation}
\begin{equation}
j(z) =   1 - 2(1+z)\frac {E'}{E} + (1+z)^2 \frac{\left(E'^2 + E E''\right)}{E^2} . \label{ch4:jz}
\end{equation} 

These expression for $q(z)$ and $j(z)$ (Eq. \eqref{ch4:qz} and \eqref{ch4:jz}) can be expressed in terms of the comoving luminosity distance $D$ (defined in 
Eq. \eqref{ch4:D}) and its derivatives as,
\begin{equation} 
q(z) = -1 - (1+z) \frac{D''}{D'},	 \label{ch4:qz2}
\end{equation}
\begin{equation} 
j(z) = 1  + 2 (1+z) \frac{D''}{D'} +  (1+z)^2 \frac{\left(3 D''^2 - D' D'''\right)}{D'^2}.  \label{ch4:jz2}
\end{equation}

The uncertainty associated with $j(z)$, $\sigma_j$ are obtained from Eq. \eqref{ch4:jz} and \eqref{ch4:jz2} via the standard rule of error propagation. To implement 
the reconstruction, the widely used GP method has been adopted. Throughout this work, we consider a zero mean function $\mu(z)=0$ as a prior on the GP. We employ only 
the Mat\'{e}rn ($\nu = \frac{9}{2}$, $p=4$) covariance function for this analysis. The reconstruction of $j$ in the present work involves a two-step analysis. In the 
first step, the marginalized $M_B$ constraints are obtained for the Pantheon SN data. In the second step, these constraints are utilized in reconstructing $E(z)$, 
$D(z)$, and their higher derivatives for different combinations of datasets. Finally, the jerk parameter $j(z)$ is derived by using the reconstructed $E(z)$, $D(z)$, 
their derivatives as in equations \eqref{ch4:jz} and \eqref{ch4:jz2} respectively.

\begin{figure*}[t!]
	\begin{center}
		\includegraphics[angle=0, width=0.32\textwidth]{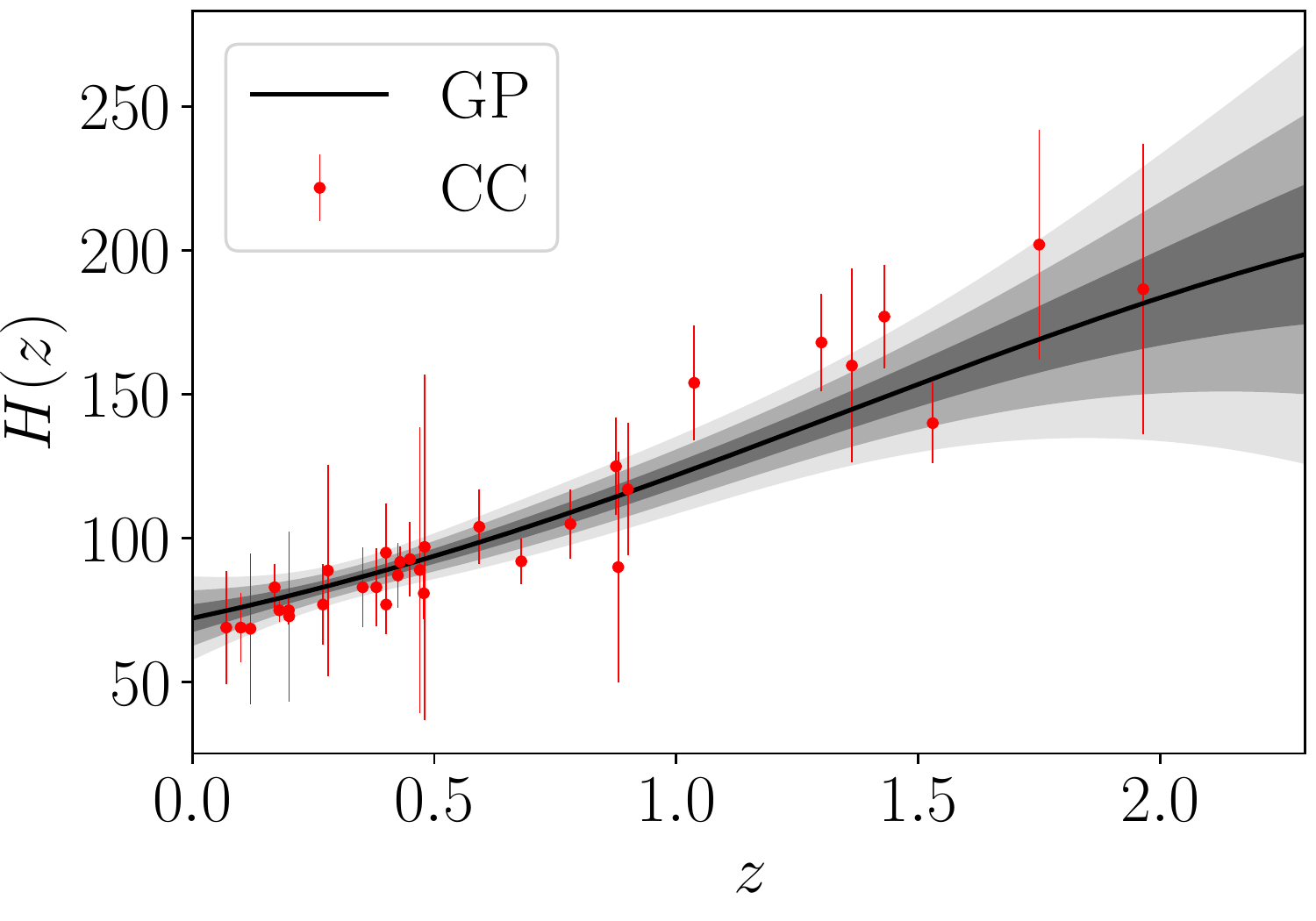}
		\includegraphics[angle=0, width=0.32\textwidth]{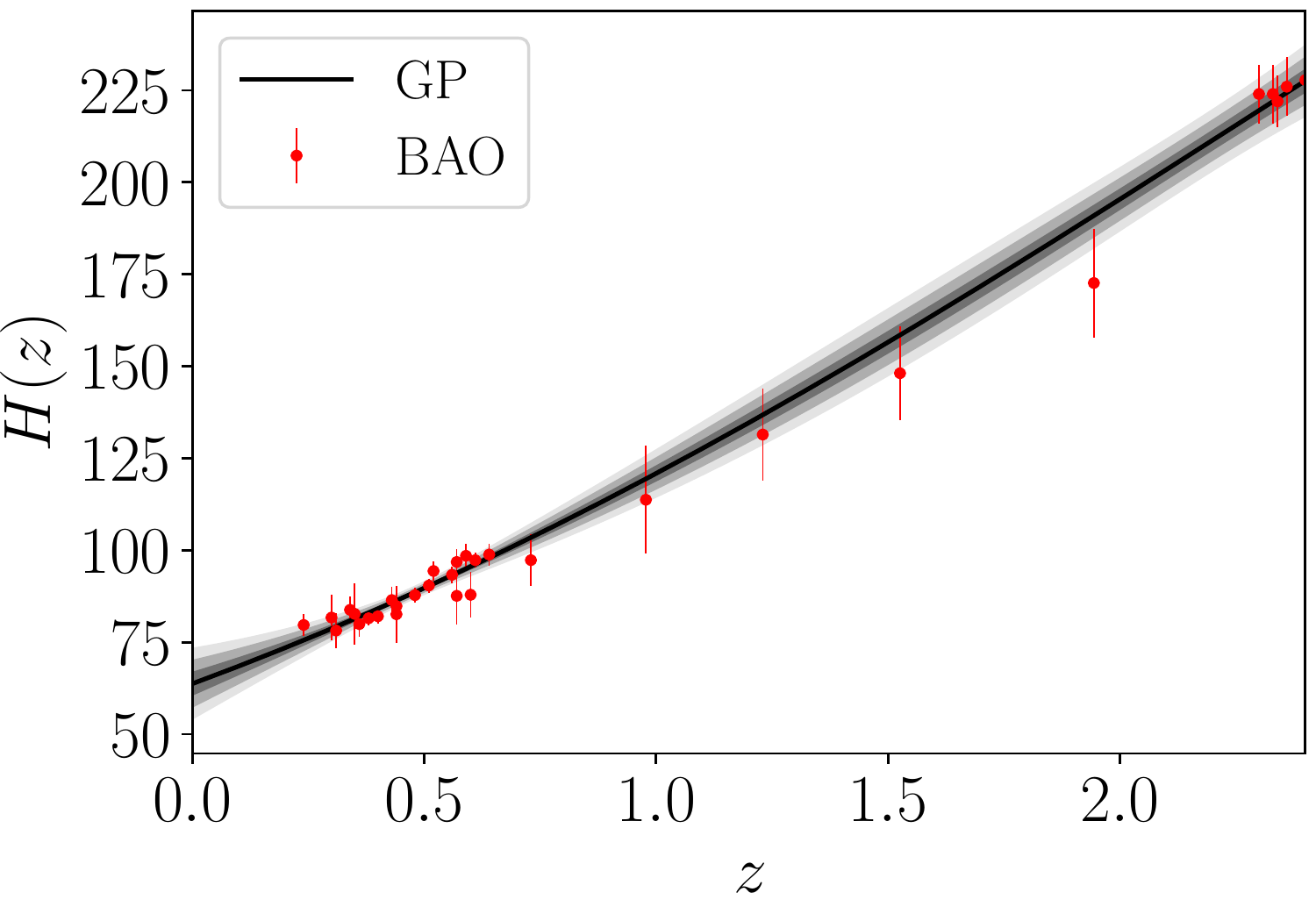}
		\includegraphics[angle=0, width=0.32\textwidth]{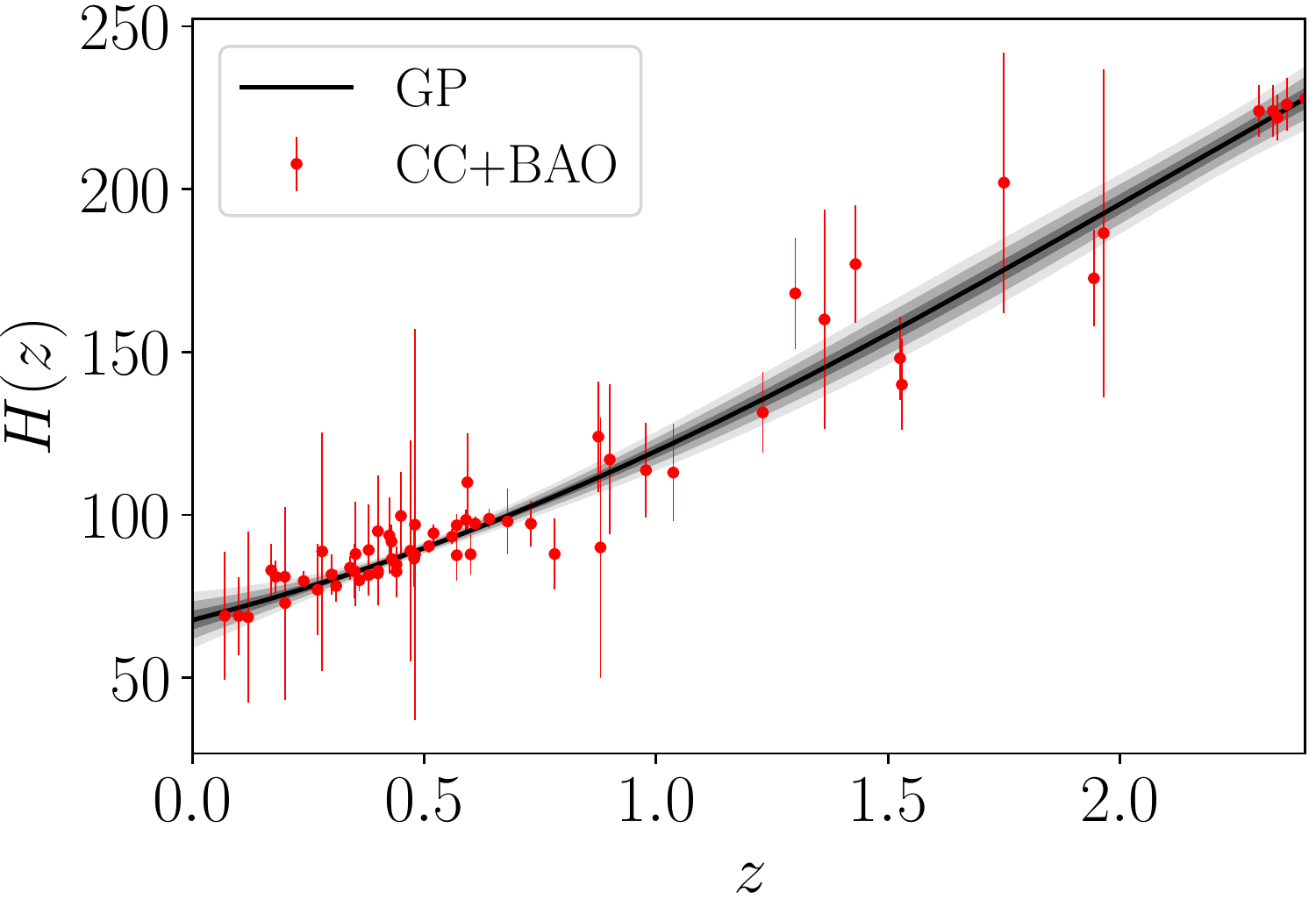}
	\end{center}
	\caption{{\small Plots for $H(z)$ reconstructed from CC data (left),  $r$BAO data (middle), and combined CC+$r$BAO data (right). The solid black line is the ``best fit'' and the associated 1$\sigma$, 2$\sigma$ and 3$\sigma$ confidence regions are shown in lighter shades.}}
	\label{ch4:Hz_recon}
\end{figure*}

\begin{table}[t!] 
	\caption{{\small Table showing the reconstructed value of $H_0$ and best fit values of $M_B$ for the Pantheon SNIa data corresponding 
			to the different $H(z)$ datasets used in reconstruction.}}
	\begin{center}
		\resizebox{0.99\textwidth}{!}{\renewcommand{\arraystretch}{1.3} \setlength{\tabcolsep}{30pt} \centering  
			\begin{tabular}{ c c c c } 
\hline
				\textbf{Dataset} & CC &  $r$BAO & CC + $r$BAO  \\ 
				\hline
				\hline
				$H_0$ & $71.249 \pm 4.569$ &  $64.721 \pm 3.264$  & $67.862 \pm 2.894$	\\ 
				\hline				
				$M_B$ & $-19.360$ &  $-19.453$  & $-19.398$	\\ 
				\hline
			\end{tabular}
		}
	\end{center}
	\label{ch4:Hz_res}
\end{table}

\begin{figure*}[t!]
	\centering
	\begin{minipage}{0.24\textwidth}[a]
		\centering
		\includegraphics[width=\linewidth]{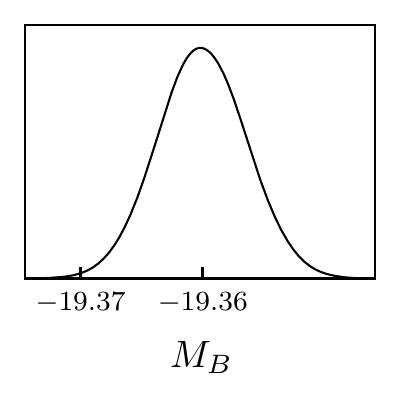}
	\end{minipage}
	\begin{minipage}{0.24\textwidth}[b]
		\centering
		\includegraphics[width=\linewidth]{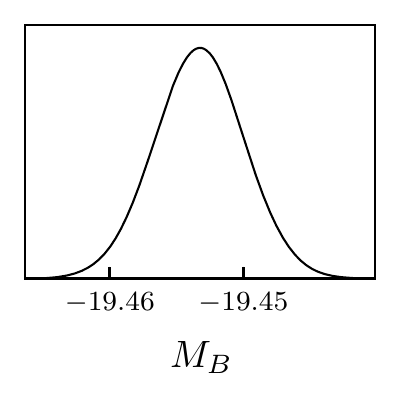}
	\end{minipage}
	\begin{minipage}{0.24\textwidth}[c]
		\centering
		\includegraphics[width=\linewidth]{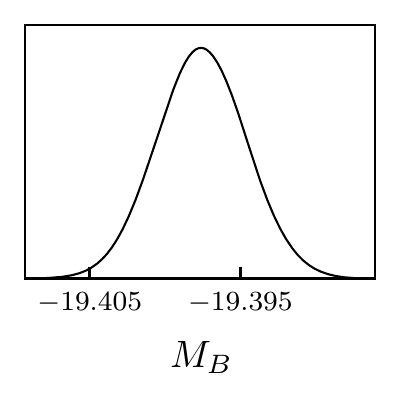}
	\end{minipage}
	\begin{minipage}{0.24\textwidth}[d]
		\centering
		\includegraphics[width=\linewidth]{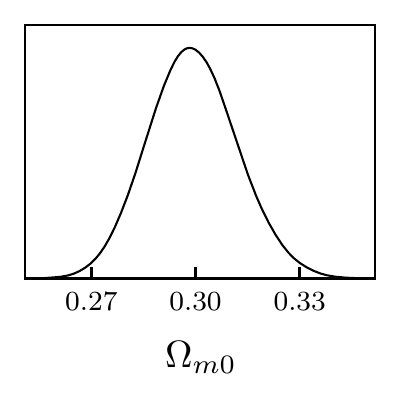}
	\end{minipage}
	\caption{{\small Plots for marginalized likelihood of absolute magnitude $M_B$  for the Pantheon SNIa data using reconstructed $H(z)$ 
			from CC data ([a]),  $r$BAO data ([b]), and combined CC+$r$BAO data ([c]). Plot for the marginalized matter density parameter $\Omega_{m0}$ 
			([d]) from the CMB Shift parameter data considering a fiducial $\Lambda$CDM model.}}
	\label{ch4:fit_plot}
\end{figure*}

After the preparation of Hubble data, we utilize the GP method to reconstruct the Hubble parameter $H(z)$ as shown in Fig. \ref{ch4:Hz_recon}. 
This reconstructed function $H(z)$ is normalized to obtain a smooth function of the reduced Hubble parameter $E(z) = H(z)/H_0$. The 1$\sigma$ uncertainty associated 
with $E$, $\sigma_E$, is evaluated from equation \eqref{ch2:sig_h}. With the smooth reconstructed function $E(z)$, we use a composite trapezoidal rule \cite{trapez} 
to obtain the normalized comoving distance $D$ from the Hubble data via Eq. \eqref{ch2:D_recon}. This reconstructed normalized comoving distance $D$ from the Hubble 
data is utilized to calculate the distance modulus $\mu_{\mbox{\tiny H}}$ and its associated 1$\sigma$ uncertainty $\sigma_{\mu_{\mbox{\tiny H}}}$. The apparent 
magnitude $m_{\mbox{\tiny SN}}$ is reconstructed via another GP using the observational data by Scolnic \textit{et al}\cite{pan} and the marginalized $M_B$ constraints 
are obtained  by minimizing the $\chi^2$ function, given in equation \eqref{ch2:chiSN}, by following a similar procedure as section \ref{ch2:methodology}.

For the CMB shift parameter data, we utilize Eq. \eqref{ch1:cmbeqn}. We use the value of CMB shift parameter $\mathcal{R} = 1.7488 \pm 0.0074$ \cite{planck_cmb} and 
the matter density parameter $\Omega_{m0}$ is marginalized assuming a fiducial $\Lambda$CDM model. The $\chi^2$ for CMB Shift parameter data is given by, 
\begin{equation}
	\chi^2_\text{CMB} = \left[\frac{1.7488 - \mathcal{R}(\Omega_{m0},~ 1089)}{0.0074}\right]^2. \label{ch4:chiCMB}
\end{equation}

The present values of the Hubble parameter $H_0$ obtained from a GP reconstruction with the CC or combined CC+$r$BAO data is 
shown in Table \ref{ch4:Hz_res}. Plots for the marginalized $M_B$ constraints are shown in the first three columns ([a], [b], [c]) of Figure 
\ref{ch4:fit_plot} and the best fit $M_B$ values are given in Table \ref{ch4:Hz_res}. Plot for the marginalized $\Omega_{m0}$ is shown in the last 
column [d] of Figure \ref{ch4:fit_plot} and the best-fit result is $\Omega_{m0} \approx 0.299 \pm 0.013.$

\begin{table}[t!] 
	\caption{{\small Table showing the effect of local (R19) and global (P18) measurements of Hubble parameter on reconstructed value of $H_0$, and 
			the corresponding marginalized constraints on the absolute magnitude $M_B$ of SNIa.}}
	\begin{center}
		\resizebox{0.99\textwidth}{!}{\renewcommand{\arraystretch}{1.3} \setlength{\tabcolsep}{20pt} \centering  
			\begin{tabular}{c c c c c } 
\hline
				\textbf{Dataset} & CC+P18 &  CC+R19 & CC+$r$BAO+P18 & CC+$r$BAO+R19 \\ 
				\hline
				\hline
				$H_0$ & $67.324 \pm 0.593$ &  $73.834\pm 1.357$  & $67.640 \pm  0.414$ & $73.516 \pm 1.314$	\\ 
				\hline				
				$M_B$ & $-19.415$ &  $-19.319$  & $-19.408$	& $-19.328$ \\ 
				\hline
			\end{tabular}
		}
	\end{center}
	\label{ch4:H0_res}
\end{table}

\begin{figure}[t!]
	\begin{center}
		\includegraphics[angle=0, width=\textwidth]{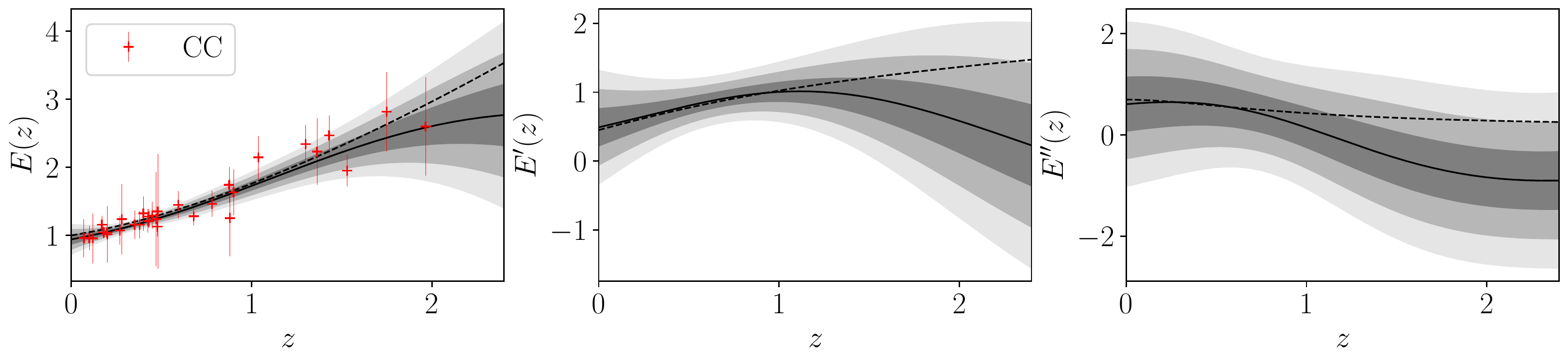}
	\end{center}
	\caption{{\small Plots for the reconstructed reduced Hubble parameter $E(z)$, its derivatives $E'(z)$ and $E''(z)$ using CC data. The black 
			solid line is the best fit curve and the associated 1$\sigma$, 2$\sigma$ and 3$\sigma$ confidence regions are shown in lighter shades. 
			The specific points (in the left figure) with error bars represent the observational data. The black dashed line is for the $\Lambda$CDM 
			model with $\Omega_{m0} = 0.3$.}}
	\label{ch4:h_HS}
\end{figure}

\begin{figure}[t!]
	\begin{center}
		\includegraphics[angle=0, width=\textwidth]{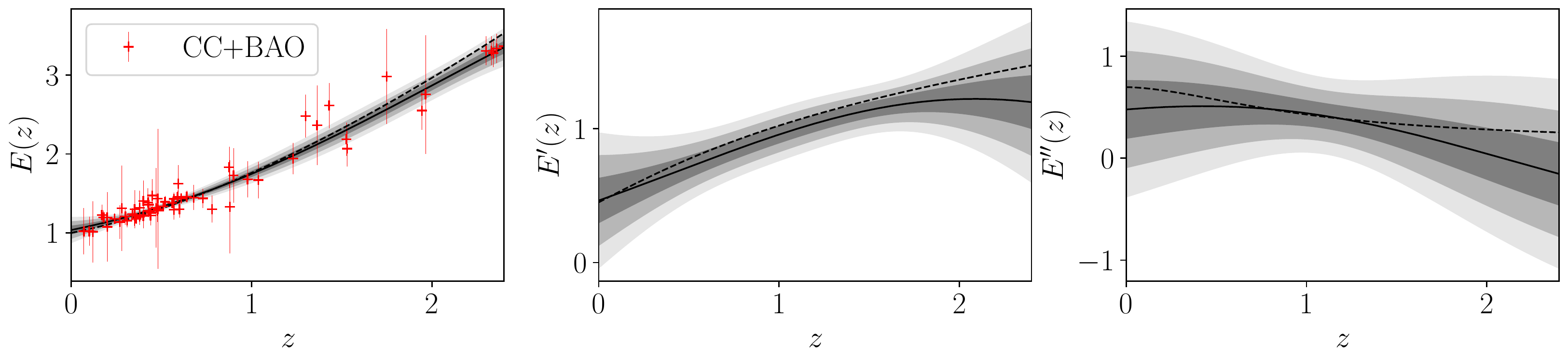}
	\end{center}
	\caption{{\small Plots for the reconstructed reduced Hubble parameter $E(z)$, its derivatives $E'(z)$ and $E''(z)$ using CC+$r$BAO data. The black 
			solid line is the best fit curve and the associated 1$\sigma$, 2$\sigma$ and 3$\sigma$ confidence regions are shown in lighter shades. The 
			specific points (in the left figure) with error bars represent the observational data. The black dashed line is for the $\Lambda$CDM model 
			with $\Omega_{m0} = 0.3$.}}
	\label{ch4:h_HSC}
\end{figure}

\begin{figure}[t!]
	\begin{center}
		\includegraphics[angle=0, width=\textwidth]{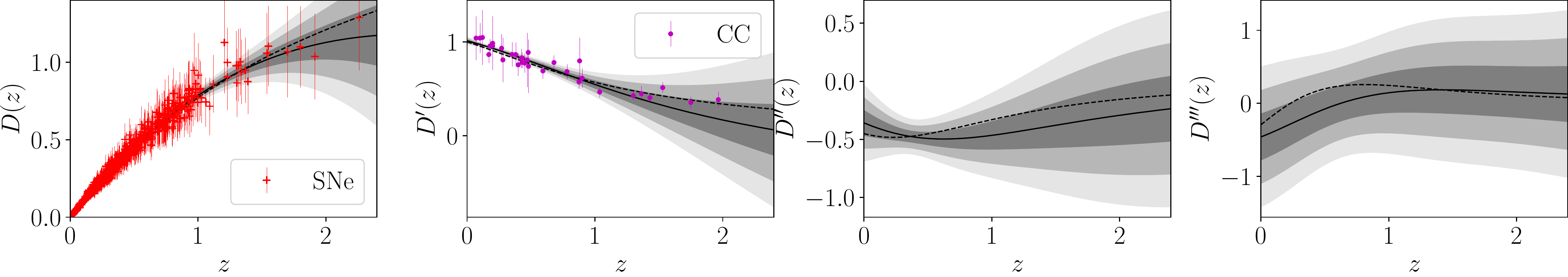}
	\end{center}
	\caption{{\small Plots for the reconstructed dimensionless comoving distance $D(z)$, its derivatives $D'(z)$, $D''(z)$ and $D'''(z)$ 
			using the Pantheon data. The black solid line is the best fit curve. The associated 1$\sigma$, 2$\sigma$ and 3$\sigma$ confidence regions 
			are shown in lighter shades. The specific points (in the top two figures) with error bars represent the observational data. The black dashed 
			line is for the $\Lambda$CDM model: $\Omega_{m0} =0.3$.}}
	\label{ch4:D_HS}
\end{figure}

\begin{figure}[t!]
	\begin{center}
		\includegraphics[angle=0, width=\textwidth]{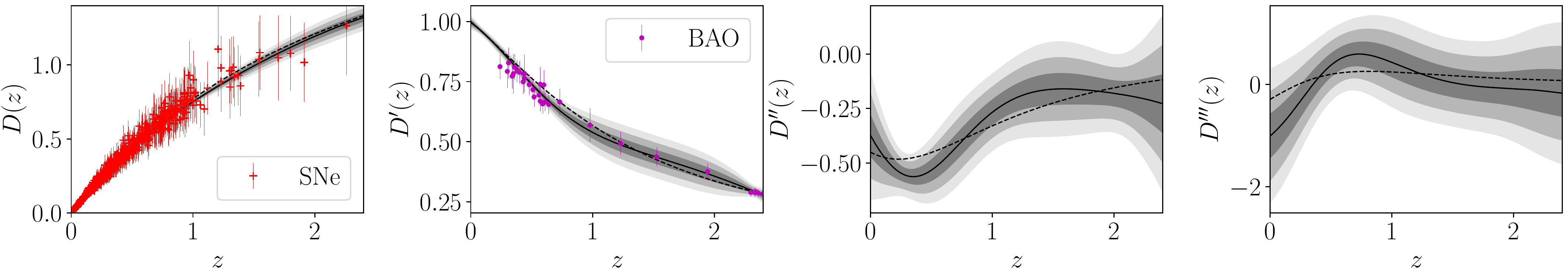}
	\end{center}
	\caption{{\small Plots for the reconstructed dimensionless comoving distance $D(z)$, its derivatives $D'(z)$, $D''(z)$ and $D'''(z)$ 
			using combined Pantheon+$r$BAO data. The black solid line is the best fit curve. The associated 1$\sigma$, 2$\sigma$ and 3$\sigma$ confidence 
			regions are shown in lighter shades. The specific points (in the top two figures) with error bars represent the observational data. The black 
			dashed line is for the $\Lambda$CDM model: $\Omega_{m0} =0.3$.}}
	\label{ch4:D_HSC}
\end{figure}

\begin{figure}[t!]
	\begin{center}
		\includegraphics[angle=0, width=\textwidth]{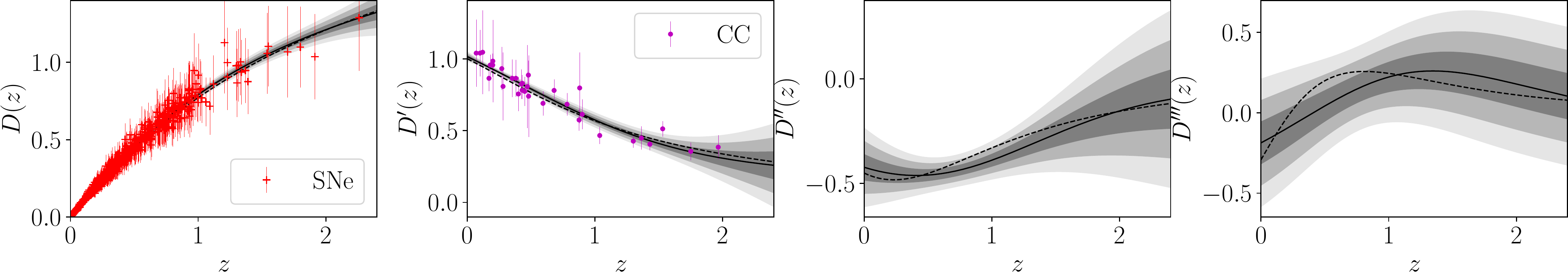}
	\end{center}
	\caption{{\small Plots for the reconstructed dimensionless comoving distance $D(z)$, its derivatives $D'(z)$, $D''(z)$ and $D'''(z)$ 
			using combined Pantheon+CC data. The black solid line is the best fit curve. The associated 1$\sigma$, 2$\sigma$ and 3$\sigma$ confidence 
			regions are shown in grey. The specific points (in the top two figures) with error bars represent the observational data. The black dashed 
			line is for the $\Lambda$CDM model: $\Omega_{m0} =0.3$.}}
	\label{ch4:all_HS}
\end{figure}

\begin{figure}[t!]
	\begin{center}
		\includegraphics[angle=0, width=\textwidth]{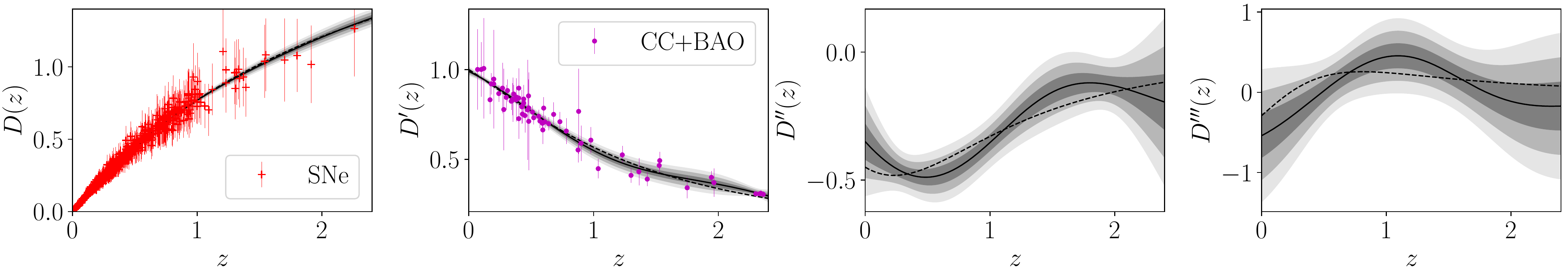}
	\end{center}
	\caption{{\small Plots for the reconstructed dimensionless comoving distance $D(z)$, its derivatives $D'(z)$, $D''(z)$ and $D'''(z)$ 
			using combined Pantheon+CC+$r$BAO+CMB data. The black solid line is the best fit curve. The associated 1$\sigma$, 2$\sigma$ and 3$\sigma$ 
			confidence regions are shown in lighter shades. The specific points (in the top two figures) with error bars represent the observational data. 
			The black dashed line is for the $\Lambda$CDM model: $\Omega_{m0} =0.3$.}}
	\label{ch4:all_HSC}
\end{figure}

These constraints are utilized in obtaining the functions $D$ and $E$ from combined datasets for the reconstruction of $j$. The reconstructed functions 
$E(z)$, $D(z)$, and their respective derivatives are plotted against $z$ for different sets of the data, and shown in Fig. \ref{ch4:h_HS}, \ref{ch4:h_HSC}, 
\ref{ch4:D_HS}, \ref{ch4:D_HSC}, \ref{ch4:all_HS} and \ref{ch4:all_HSC}. The black solid line is the best fit curve, and the black dashed line represents 
the $\Lambda$CDM model with $\Omega_{m0}=0.3$. The shaded regions correspond to the $68\%$, $95\%$ and $99.7\%$ CLs. The specific points marked with error 
bars represent the observational data used in the reconstruction.  We find that incorporating the reduced Hubble parameter measurements from the CC and 
$r$BAO Hubble datasets provides additional constraints on the first-order derivatives of $D(z)$ in our analysis, which reduces the uncertainties associated 
with the reconstructed functions $D(z)$ and $D'(z)$ at high redshifts.

The Pantheon data is used to estimate the $D$ data points and the uncertainty $\mathbf{\Sigma}_D$ from the observed $\mu$ and $\mathbf{\Sigma}_{\mu}$ data 
compilation respectively. For the CC and $r$BAO data, the $H$-$\sigma_H$ data is converted to $E$-$\sigma_{E}$ data set utilizing equation \eqref{ch2:sig_h} 
considering the GP reconstructed $H_0$ given in Table \ref{ch4:Hz_res}. From \eqref{ch4:D'} we can clearly see that $D'(z)$ is related to $E(z)$. So, we can 
take into account the $E$ dataset, its associated $\sigma_{E}$ uncertainties, and represent them using equations \eqref{ch4:D'} and \eqref{ch4:sigD'} 
respectively.

It is further examined whether the two different strategies for determining the value of $H_0$ affect our reconstruction differently. We proceed with 
the analysis similar to that above, except for adding the P18 or R19 data to the CC $H(z)$ dataset. For a comparison, one can refer to Table \ref{ch4:Hz_res} 
and Table \ref{ch4:H0_res} to get an insight as to how the inclusion of $H_0$ measurement affects our reconstruction. While considering the CC+P18 and 
CC+R19 combinations, normalization of the $H-\sigma_H$ dataset is carried out using the respective P18 and R19 $H_0$ values, as given in Table \ref{ch4:H0_res}.

\subsection{Reconstruction of $j$}

\begin{figure*}[t!] 
	\begin{center}
		\includegraphics[angle=0, width=\textwidth]{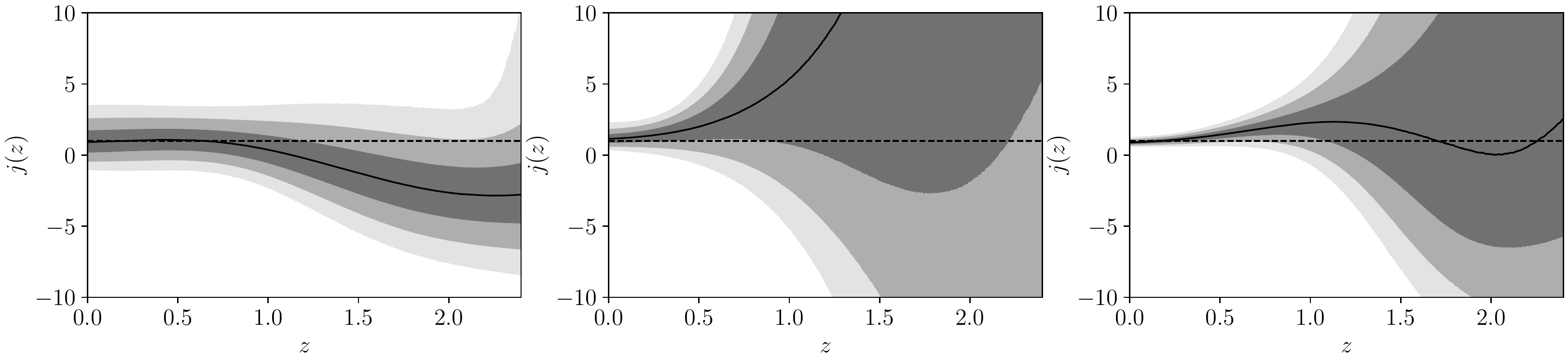}\\
		\includegraphics[angle=0, width=\textwidth]{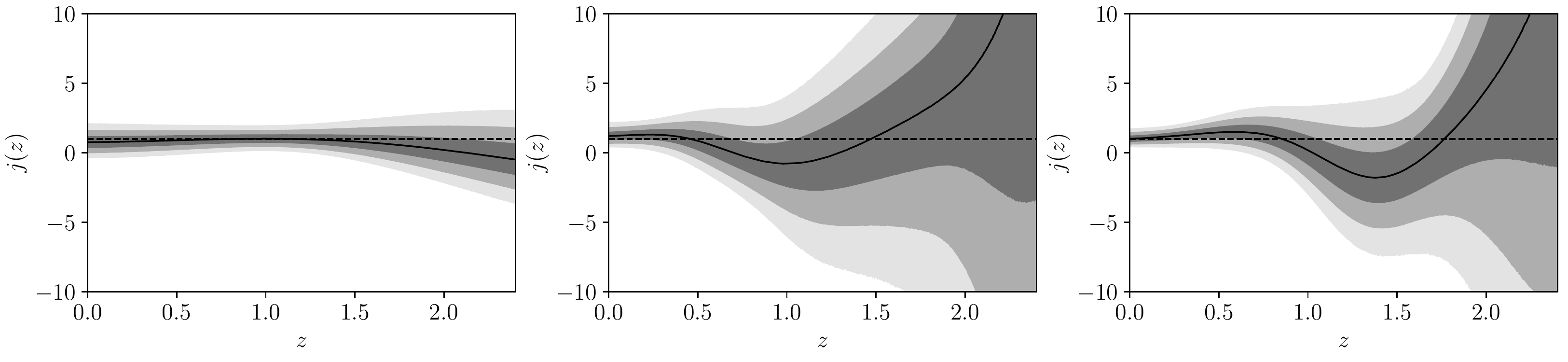}
	\end{center}
	\caption{{\small Plots for $j(z)$ reconstructed from CC data (top left),  Pantheon data (top middle), Pantheon+CC data (top right), CC+$r$BAO data 
			(bottom left),  Pantheon+$r$BAO+CMB data (bottom middle), and Pantheon+CC+$r$BAO+CMB data (bottom right). The solid black line is the ``best fit''. 
			The black dashed line represents the $\Lambda$CDM model with $\Omega_{m0}=0.3$ where $j=1$.}}
	\label{ch4:jerkplot}
\end{figure*}

The cosmological jerk parameter $j$ is derived from the reconstructed functions $E(z)$, $D(z)$ and their higher-order derivatives using equations \eqref{ch4:jz} 
and \eqref{ch4:jz2}. If one assumes the standard Einstein equations with a cold dark matter and a cosmological constant $\Lambda$, $j$ is a constant whose value 
is unity. Results for the reconstructed jerk is given in Fig. \ref{ch4:jerkplot}. The shaded regions correspond to the $68\%$, $95\%$ and $99.7\%$ CLs. The black 
solid lines show the best-fit values of the reconstructed function, and the black dashed lines corresponds to the $\Lambda$CDM model with $\Omega_{m0}=0.3$ 
and $j=1$. Plots for the best-fit values of the reconstructed jerk parameter clearly indicates that $j$ has an evolution, that may well be non-monotonic. The 
reconstructed best fit $j_0$ at the present epoch is shown in Table \ref{ch4:res_table}.  However, the plot shows that the $\Lambda$CDM model is allowed within 
$2\sigma$ uncertainty. 

\begin{table}[t!] 
	\caption{{\small Table showing the reconstructed value of $j_0$ corresponding to the different datasets used along with the 1$\sigma$ uncertainty.}}
	\begin{center}
		\resizebox{0.98\textwidth}{!}{\renewcommand{\arraystretch}{1.5} \setlength{\tabcolsep}{6pt} \centering  
			\begin{tabular}{c c c c c c c } 
\hline
				\textbf{Datasets} & CC & SN & SN+CC & CC+$r$BAO & SN+$r$BAO & SN+CC+$r$BAO+CMB  \\ 
				\hline
				\hline
				$j_0$ & ${0.911}^{+0.736}_{-0.818}$ &  ${1.139}^{+0.289}_{-0.330}$  & ${0.909}^{+0.149}_{-0.158}$  & ${0.893}^{+0.420}_{-0.444}$ &  ${1.211}^{+0.287}_{-0.316}$  & $ {1.035}^{+0.219}_{-0.227}$	\\ 
				\hline
			\end{tabular}
		}
	\end{center}
	\label{ch4:res_table}
\end{table}

\begin{figure*}[t!]
	\begin{center}
		\includegraphics[angle=0, width=0.32\textwidth]{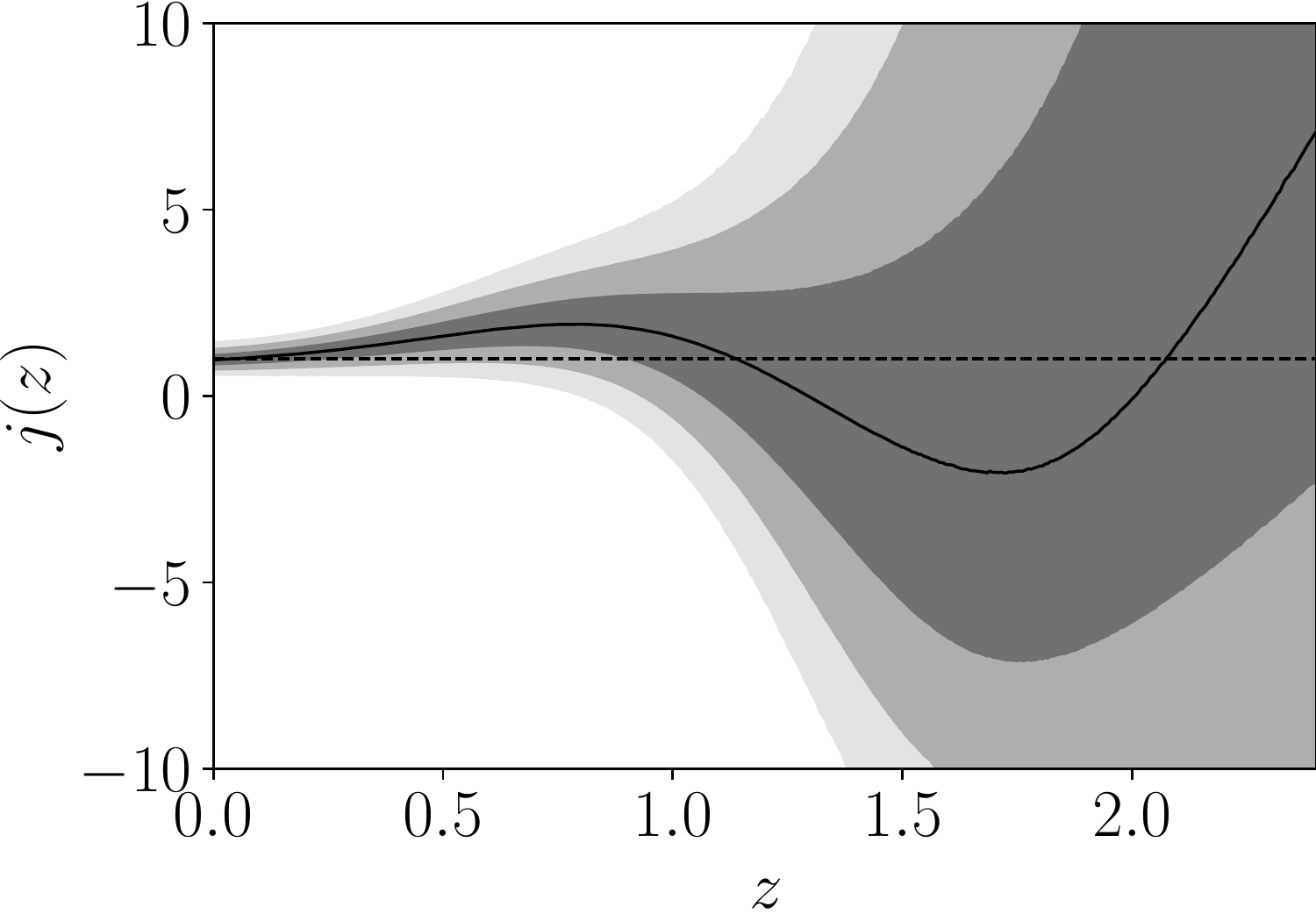}
		\includegraphics[angle=0, width=0.32\textwidth]{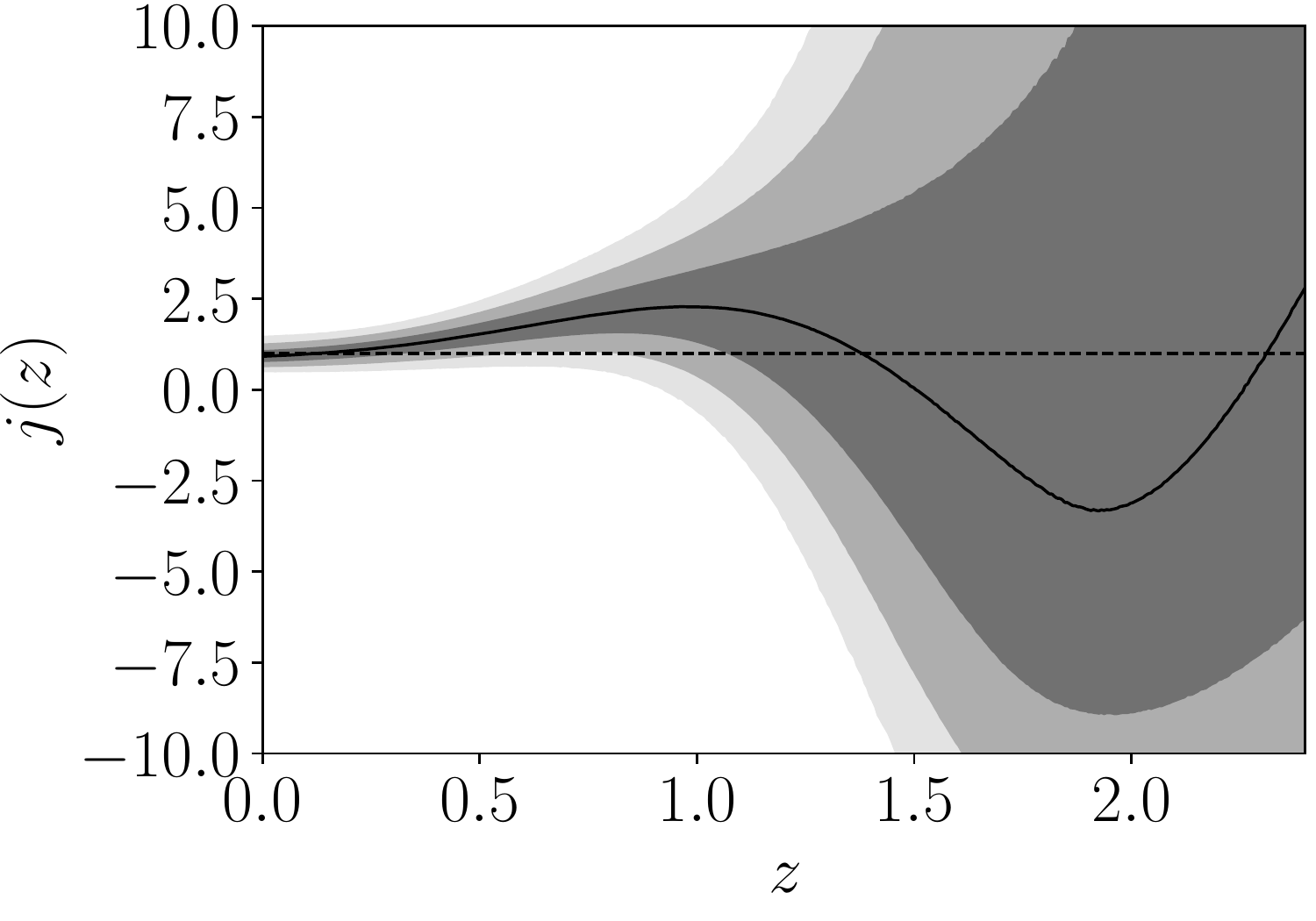}
		\includegraphics[angle=0, width=0.32\textwidth]{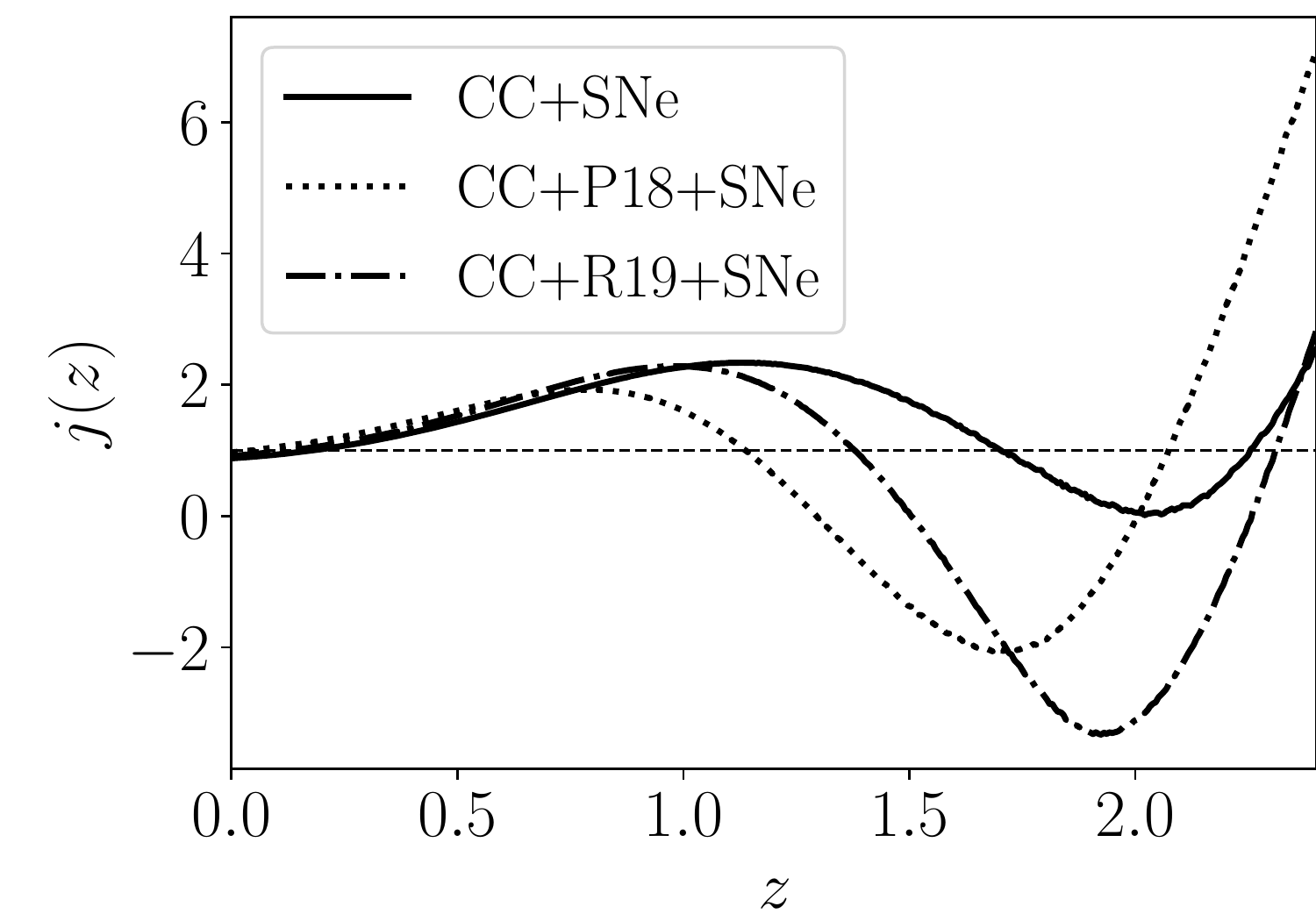}
	\end{center}
	\caption{{\small Plots for the reconstructed jerk $j$, using the  Pantheon+CC+P18 data (left) and from Pantheon+CC+R19 data (middle). The 
			black dashed line corresponds to the $\Lambda$CDM model, with $\Omega_{m0} = 0.3$ where $j=1$. A comparison among the three cases is shown 
			in the right column.}}
	\label{ch4:H0_effect}
\end{figure*}

In Fig. \ref{ch4:jerkplot}, we used the GP reconstructed values of $H_0$ from Table \ref{ch4:Hz_res}, without considering any prior $H_0$ measurement. For checking 
if the two different strategies for determining the value of $H_0$ affect the reconstruction of $j$, the P18 and R19 $H_0$ values from Table \ref{ch4:H0_res} are 
considered for the purpose. So, we have three cases when reconstructing the function $E(z)$, (i) with no prior of $H_0$, (ii) with the P18 prior of $H_0$ and (iii) 
with the R19 prior of $H_0$. Finally, we compare the results obtained for cases (ii) and (iii) with that of case (i). Plots for the reconstructed $j(z)$ along 
with a comparison for the model-independent Pantheon+CC data is shown in Fig \ref{ch4:H0_effect}. The black solid, dotted and dashed lines represent the results 
obtained for the (i) Pantheon+CC, (ii)  Pantheon+CC+P18, and (iii)  Pantheon+CC+R19 combinations, respectively.

\subsection{Reconstructing the $w_{\text{eff}}$}

A reconstruction of the dynamical or physical quantities, like the equation of state, requires prior knowledge regarding the theory of gravity. The standard 
cosmological model assumes that the Universe is described by Einstein's theory of General Relativity, which is the correct description for gravity. So, for 
this particular exercise, we have introduced the Einstein equations, 
\begin{eqnarray} \label{ch4:friedmann}
3H^2 &=& 8\pi G \rho ,\\
2\dot{H} + 3H^2 &=& -8\pi G p ,
\end{eqnarray} 
where $\rho$ and $p$ are the total energy density and pressure contribution from all components constituting the Universe. 

The definition of the deceleration parameter
\begin{equation} \label{ch4:q}
\frac{\dot{H}}{H^2} = -(1+q),
\end{equation}
is substituted in Einstein equations. 

Therefore, the effective equation of state parameter is
\begin{equation}  \label{w_eff}
w_{\text{eff}} = \frac{p}{\rho} = -\frac{2\dot{H} + 3H^2 }{3H^2 }  = \frac{-1 + 2q}{3}.
\end{equation}

\begin{figure*}[t!] 
	\begin{center}
		\includegraphics[angle=0, width=\textwidth]{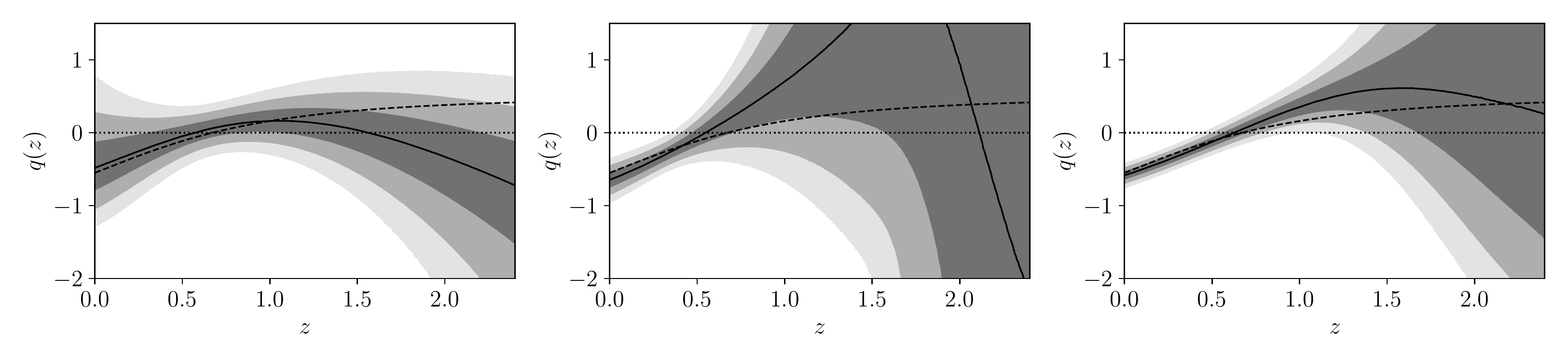}\\
		\includegraphics[angle=0, width=\textwidth]{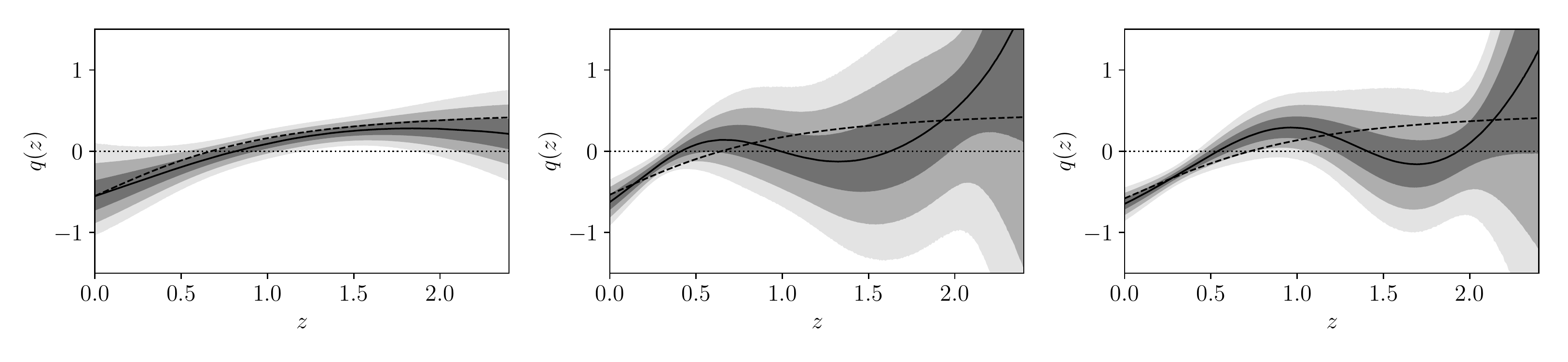}
	\end{center}
	\caption{{\small Plots for $q(z)$ reconstructed from CC data (top left),  Pantheon data (top middle), Pantheon+CC data (top right), CC+$r$BAO data 
			(bottom left),  Pantheon+$r$BAO+CMB data (bottom middle), and Pantheon+CC+$r$BAO+CMB data (bottom right). The solid black line is the ``best fit'' 
			and the black dashed line represents the $\Lambda$CDM model with $\Omega_{m0}=0.3$.}}
	\label{ch4:qplot}
\end{figure*}

\begin{table}[t!] 
	\caption{{\small Table showing the reconstructed value of $q_0$ corresponding to the different datasets used along with the 1$\sigma$ uncertainty.}}
	\begin{center}
		\resizebox{0.98\textwidth}{!}{\renewcommand{\arraystretch}{1.5} \setlength{\tabcolsep}{6pt} \centering  
			\begin{tabular}{c c c c c c c} 
\hline
				\textbf{Datasets} & CC & SN & CC+SN & CC+$r$BAO & SN+$r$BAO & CC+SN+$r$BAO+CMB  \\ 
				\hline 
				\hline
				$q_0$ & ${-0.482}^{+0.307}_{-0.360}$ & ${-0.646}^{+0.106}_{-0.103}$ & ${-0.584}^{+0.059}_{-0.058}$ &  ${-0.552}^{+0.177}_{-0.195}$  
				& ${-0.625}^{+0.097}_{-0.094}$ &	${-0.647}^{+0.070}_{-0.069}$ \\ 
				\hline
			\end{tabular}
		}
	\end{center}
	\label{ch4:res_table2}
\end{table}

Using the reconstructed values of $E(z)$, $D(z)$ and their derivatives in equations \eqref{ch4:qz} and \eqref{ch4:qz2}, the deceleration parameter 
$q$ can also be reconstructed for the same combinations of datasets. The plots are shown in Fig \ref{ch4:qplot}. The shaded regions correspond to 
the $68\%$, $95\%$ and $99.7\%$ CLs. The black solid lines represent the best-fit values of the reconstructed function, and the black dashed lines 
corresponds to the $\Lambda$CDM model with $\Omega_{m0}=0.3$. For a comparison, one can note that the expected value of $q_{\Lambda CDM}$ at $z=0$ 
is ${q_0}_{\Lambda CDM} = \frac{3}{2}\Omega_{m0} - 1 = -0.55$. Although the best fit curve appears to deviate from a monotonic behaviour, the 
deceleration parameter corresponding to the $\Lambda$CDM model is included generally in $1\sigma$, and at most in the $2\sigma$ CL.

\begin{figure*}[t!]
	\begin{center}
		\includegraphics[angle=0, width=0.24\textwidth]{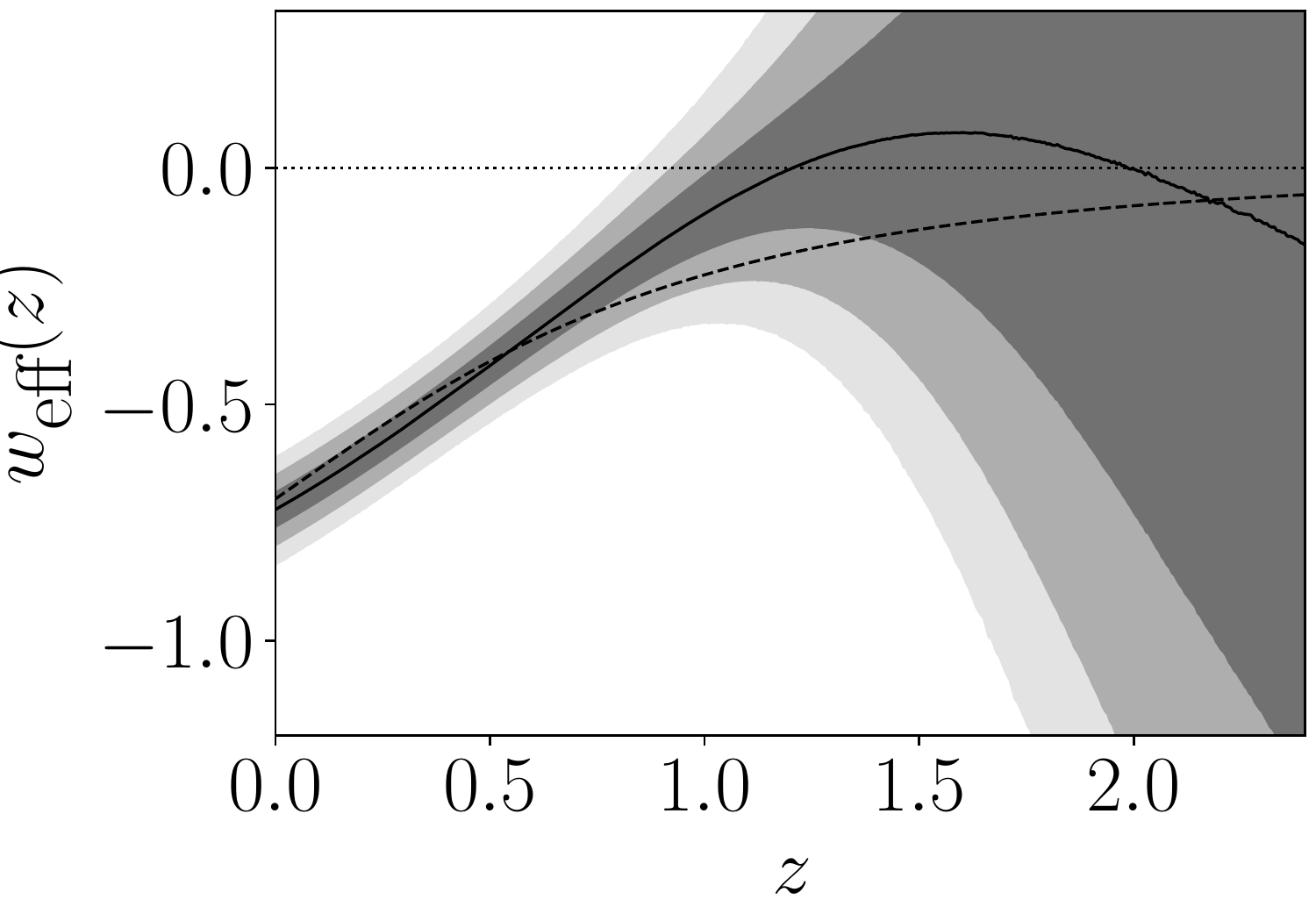}
		\includegraphics[angle=0, width=0.24\textwidth]{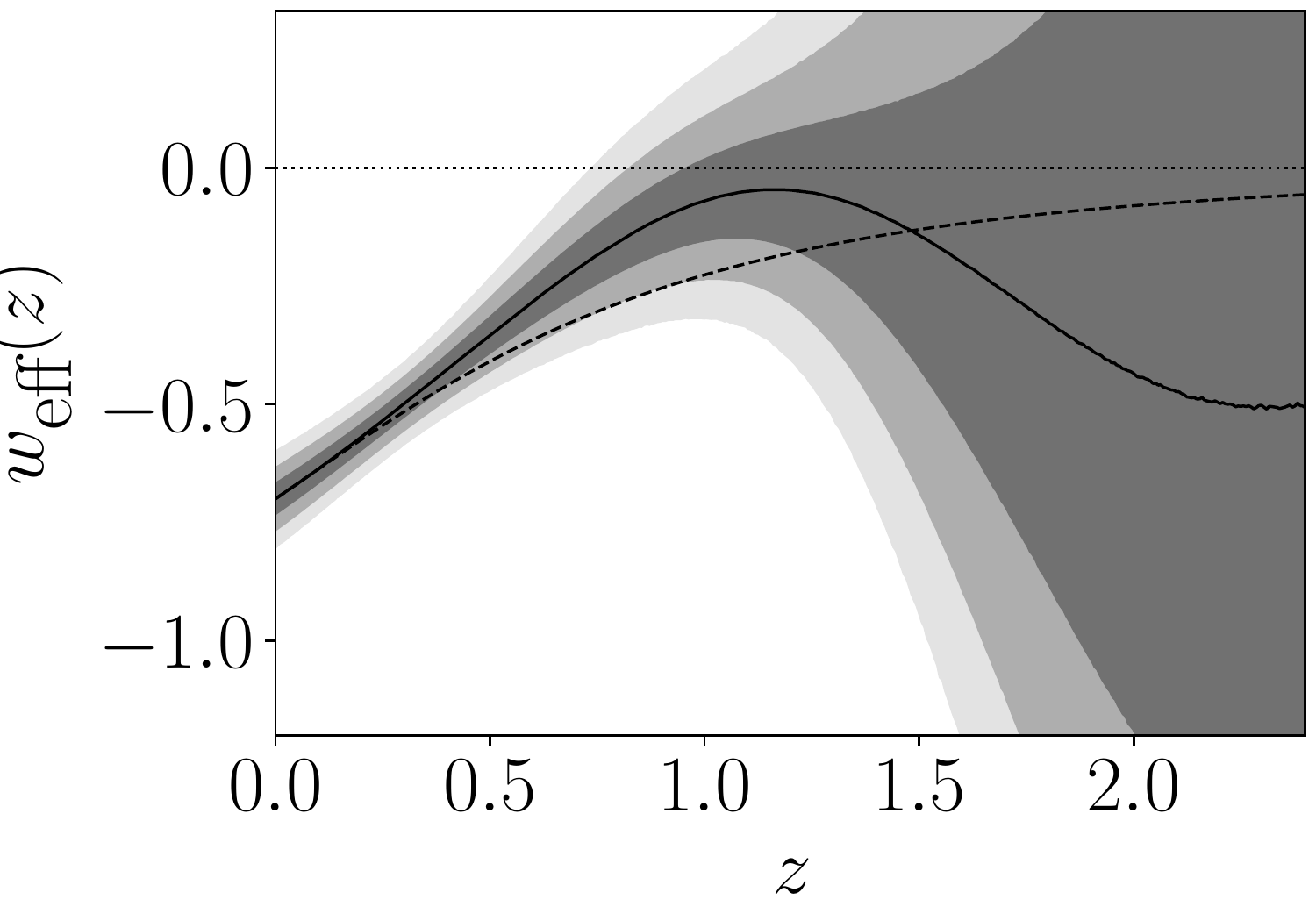}
		\includegraphics[angle=0, width=0.24\textwidth]{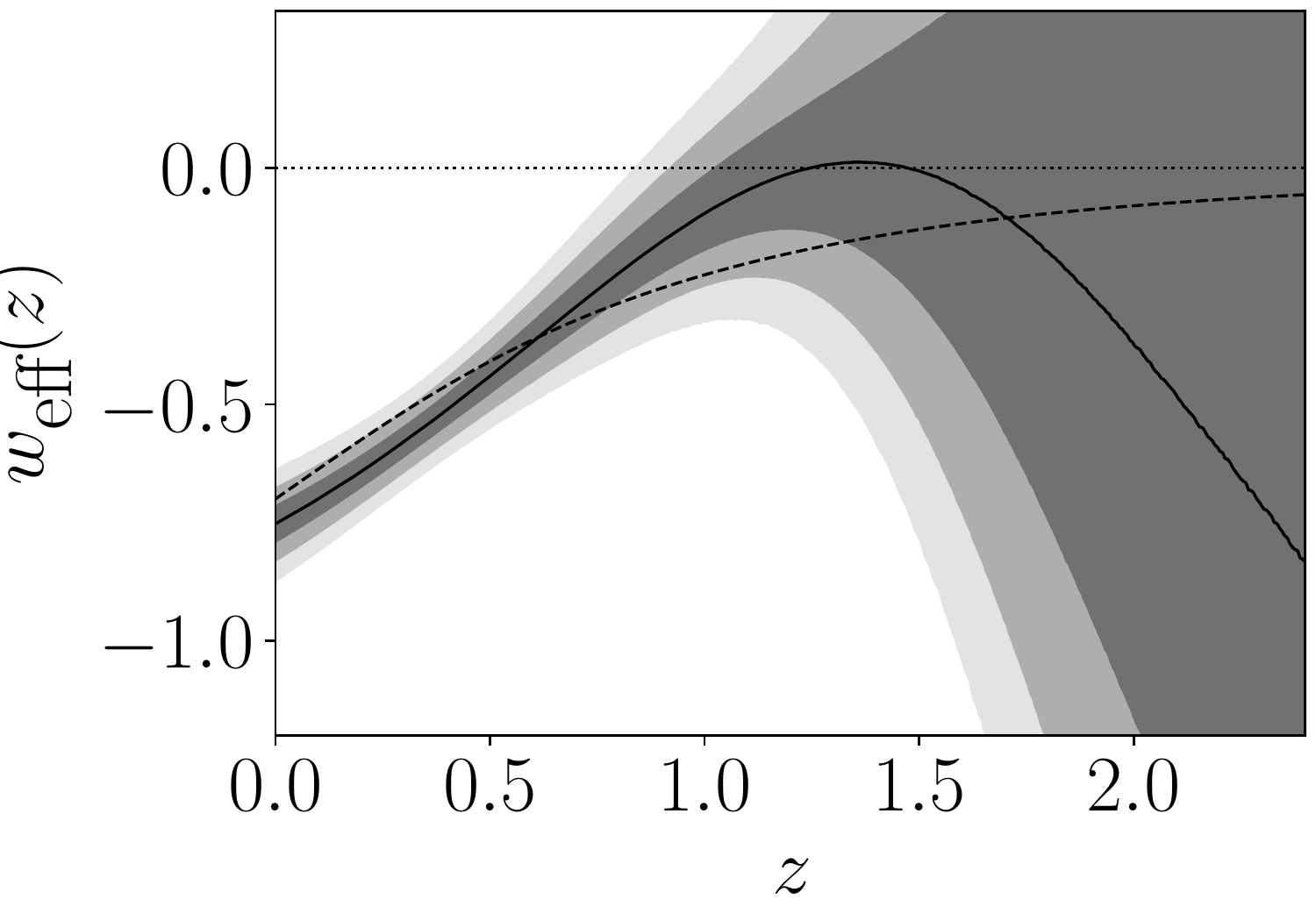}
		\includegraphics[angle=0, width=0.24\textwidth]{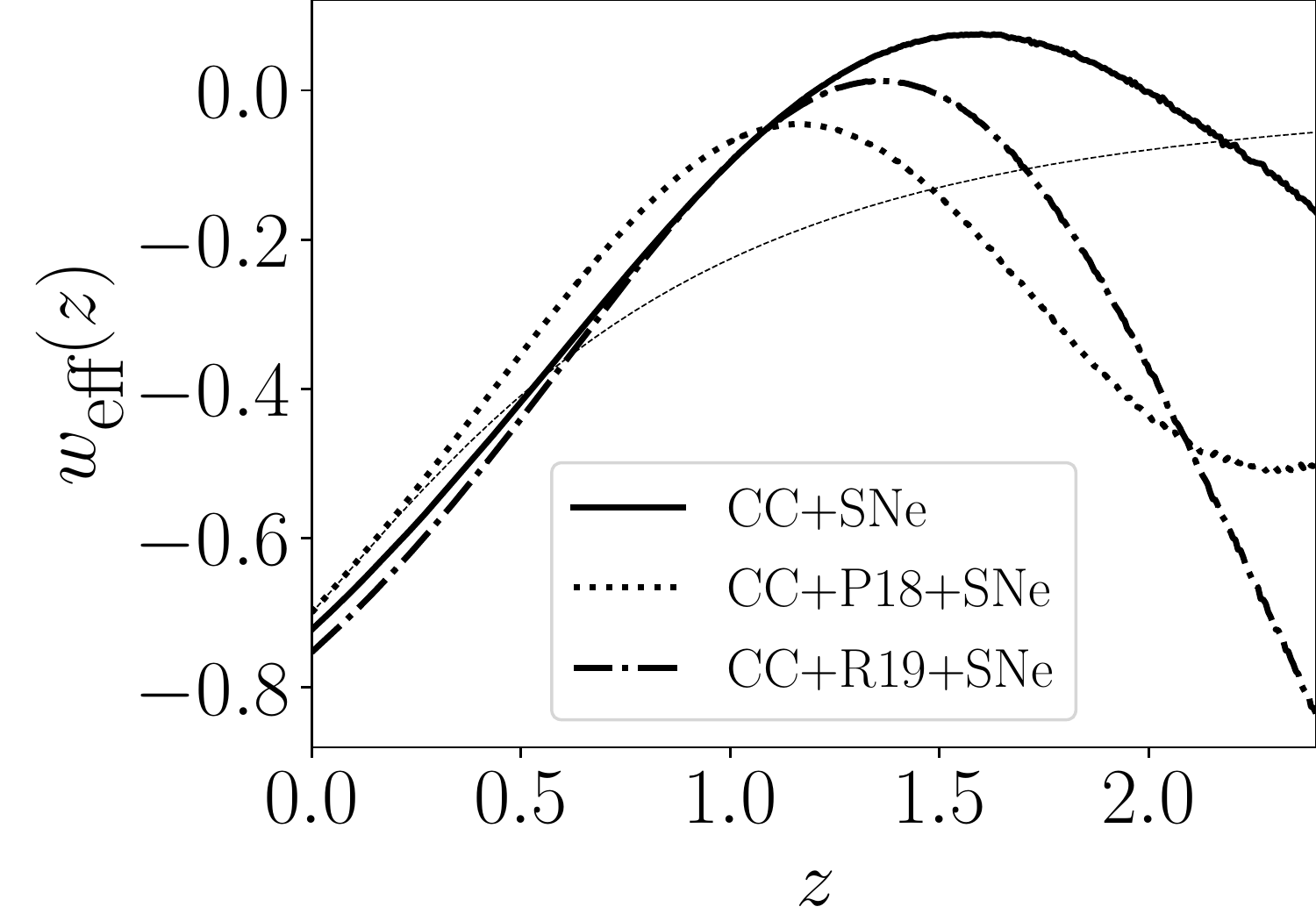}
	\end{center}
	\caption{{\small Plots for the effective equation of state parameter, from the reconstructed deceleration $q$, for combined Pantheon+CC data (left), 
			Pantheon+CC+P18 data (second from left) and Pantheon+CC+R19 data (third from left). The black dashed line represents the effective EoS for 
			$\Lambda$CDM model, considering $\Omega_{m0} = 0.3$. A comparison among the three results is given in the extreme right.}}
	\label{ch4:weffplot}
\end{figure*}

Using the 
reconstructed $q(z)$ for the different data combinations, one arrives at the effective EoS parameter $w_{\text{eff}}$ non-parametrically. 
The evolution of $w_{\text{eff}}$ is shown in Fig. \ref{ch4:weffplot}. The black solid line represents the best-fit curve. The shaded regions show the 
uncertainty associated with $w_{\text{eff}}$ corresponding to the 1$\sigma$, 2$\sigma$ and 3$\sigma$ CLs. The reconstructed values of $q$ and 
$w_{\text{eff}}$ at $z=0$ are shown in Tables \ref{ch4:res_table2} and \ref{ch4:weff_res}.

\begin{table}[t!] 
	\caption{{\small Table showing the reconstructed value of $w_{\text{eff}}(0)$ corresponding to the different datasets used along with the 1$\sigma$, 2$\sigma$ and 3$\sigma$ uncertainty.}}
	\begin{center}
		\resizebox{0.98\textwidth}{!}{\renewcommand{\arraystretch}{1.5} \setlength{\tabcolsep}{15pt} \centering  
			\begin{tabular}{ c c c c } 
\hline
				\textbf{Datasets} & CC+SN & CC+SN+P18 & CC+SN+R19  \\ 
				\hline
				\hline
				$w_{\text{eff}}(z=0)$ & ${-0.723}^{+0.039}_{-0.038}~^{+0.078}_{-0.075}~^{+0.120}_{-0.113}$ 
		& ${-0.700}^{+0.035}_{-0.035}~^{+0.070}_{-0.068}~^{+0.107}_{-0.102} $ & ${-0.752}^{+0.041}_{-0.040}~^{+0.080}_{-0.078}~^{+0.123}_{-0.117} $\\ 
				\hline
			\end{tabular}
		}
	\end{center}
	\label{ch4:weff_res}
\end{table}

For the $\Lambda$CDM model, the cold dark matter contributes only to the energy density while the cosmological constant $\Lambda$ contributes to both 
the energy density and pressure. The effective EoS \eqref{w_eff} thus takes the following form, 

\begin{equation}
w_{{\text{eff}},\Lambda\text{CDM}} =  \frac{p_\Lambda}{\rho_\Lambda + \rho_m} = -\frac{1}{1 + \frac{\Omega_{m0}}{1-\Omega_{m0}}(1+z)^3}.
\end{equation} 

Considering the value of $\Omega_{m0} = 0.299 \pm 0.013$ from the CMB Shift parameter marginalization, we calculate the value of the effective EoS for 
the $\Lambda$CDM model to be $-0.701 \pm 0.013$ at $z=0$ using the standard error propagation rule. 

For higher redshifts ($z>1.5$), the reconstructed $w_{\text{eff}}$ in the present work indicates a non-monotonic behaviour. However, the corresponding 
$w_{{\text{eff}}}$ for the $\Lambda$CDM model is included definitely in the $2\sigma$ CL.

\subsection{Fitting function for $j(z)$}

In this section, we attempt to write an approximate fitting function for the reconstructed jerk parameter in the low redshift range $0< z < 1$ for the model 
independent Pantheon+CC,  Pantheon+CC+R19 and Pantheon+CC+P18 combinations. A polynomial function for $j(z)$ is considered, where
\begin{equation} \label{ch4:fit}
j_{\text{fit}} (z) = \sum_{i=0}^{n} j_i z^i.
\end{equation}

This polynomial is non-linear in $z$ but linear in $j_i$'s. Thus estimating the above equation by the method of least squares or $\chi^2$ minimization 
holds.

We define the $\chi^2$ function as,
\begin{equation}
	\chi^2 =  \mathlarger{\mathlarger{\sum}}_{s} \dfrac{\left[j(z_s)- j_{\text{fit}}(z_s)\right]^2}{\sigma(z_s)^2}.
\end{equation}

We work out the fitting using a trial and error estimation for different orders of $i$ in equation \eqref{ch4:fit}. To check the goodness 
of the fit, the minimized $\chi^2$ for every $i^{\mbox{\small th}}$ order fitting are calculated. The value of reduced $\chi^2_\nu = \frac{\chi^2}{\nu}$, where 
$\nu$ signifies the degrees of freedom, is estimated. This procedure entails going from order to order in the polynomial and getting the best-fitting $\chi^2$ 
and truncating once $\chi_\nu^2$ falls below one. 

\begin{figure*}[t!]
	\begin{center}
		\includegraphics[angle=0, width=0.32\textwidth]{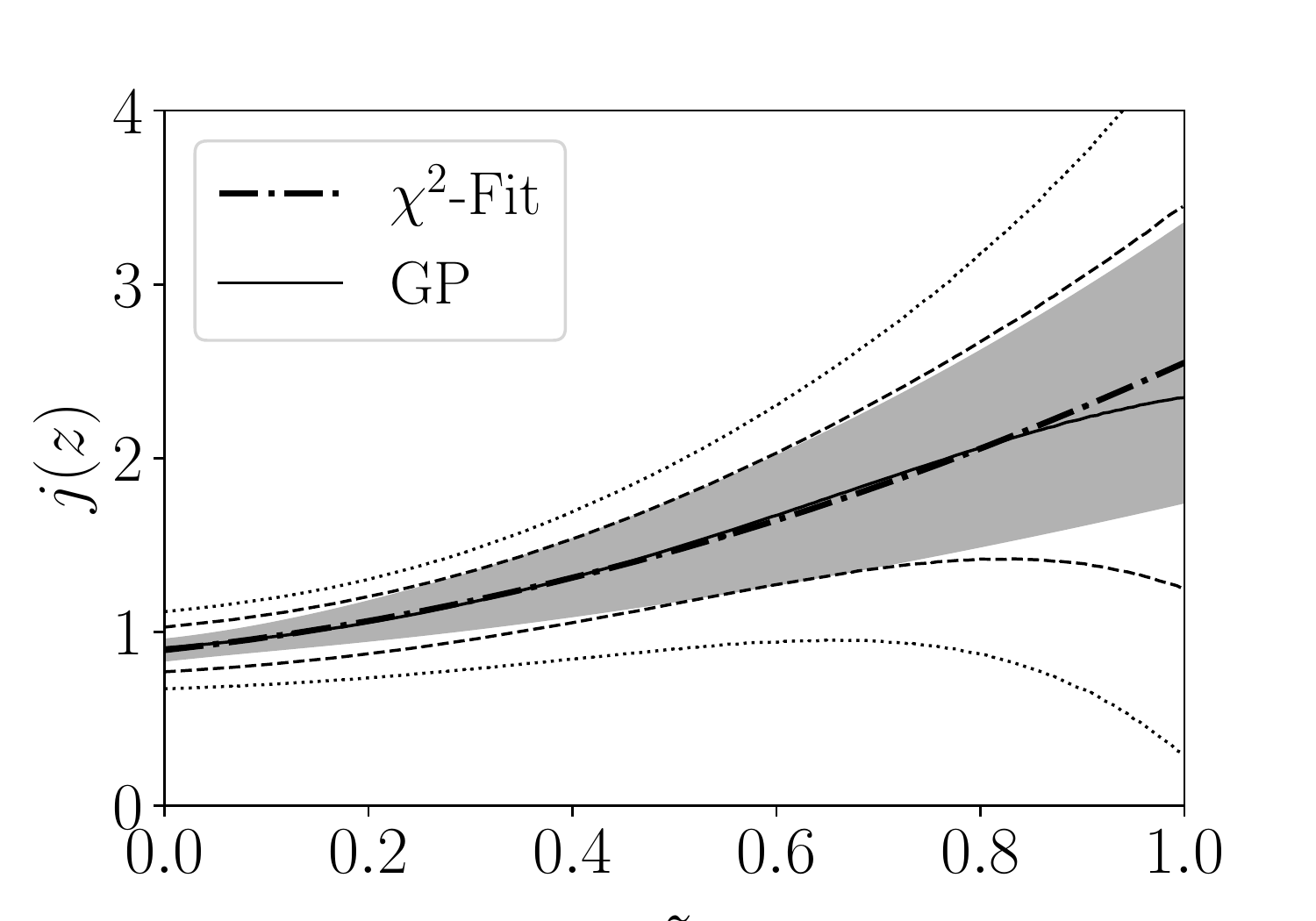}
		\includegraphics[angle=0, width=0.32\textwidth]{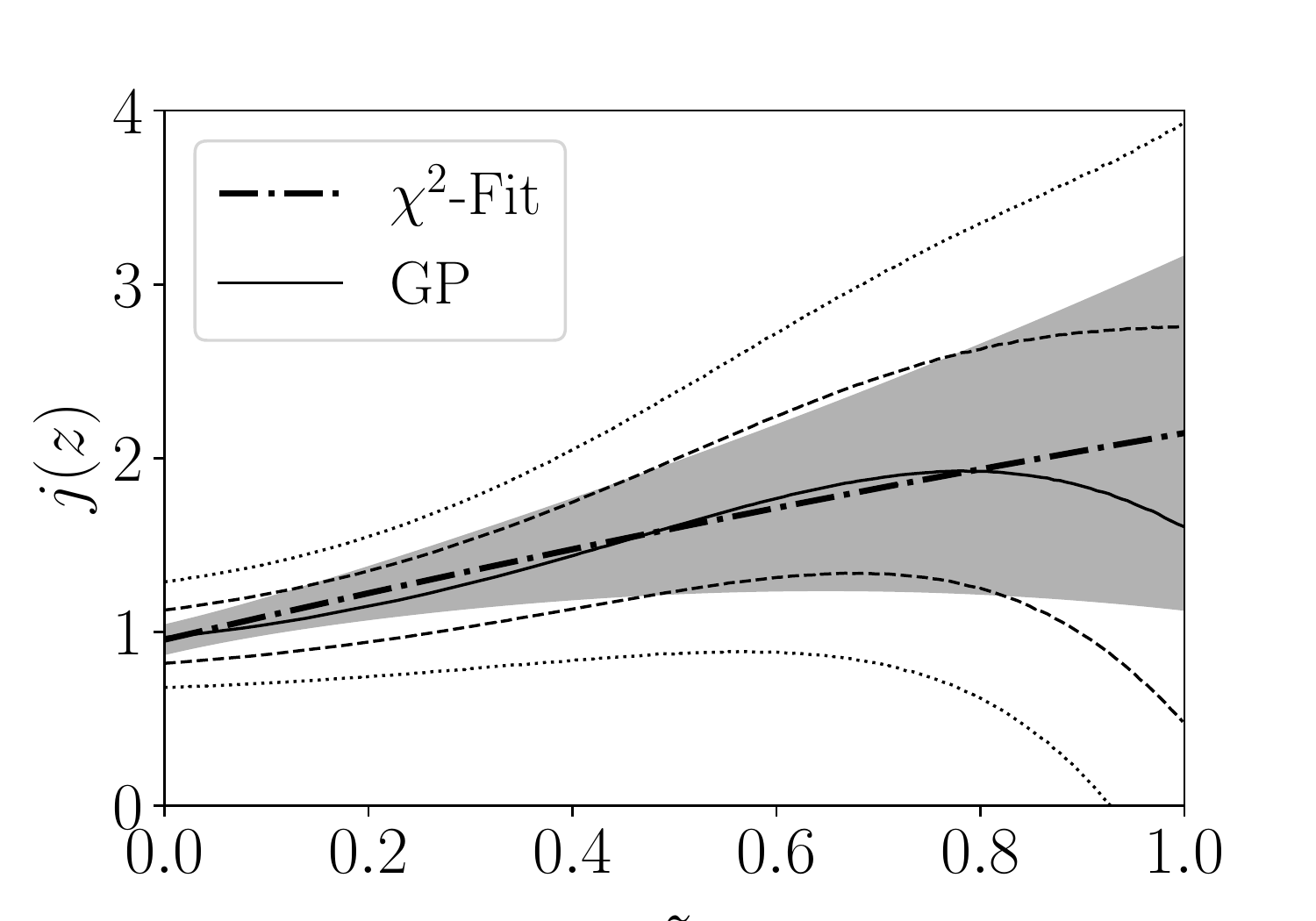}
		\includegraphics[angle=0, width=0.32\textwidth]{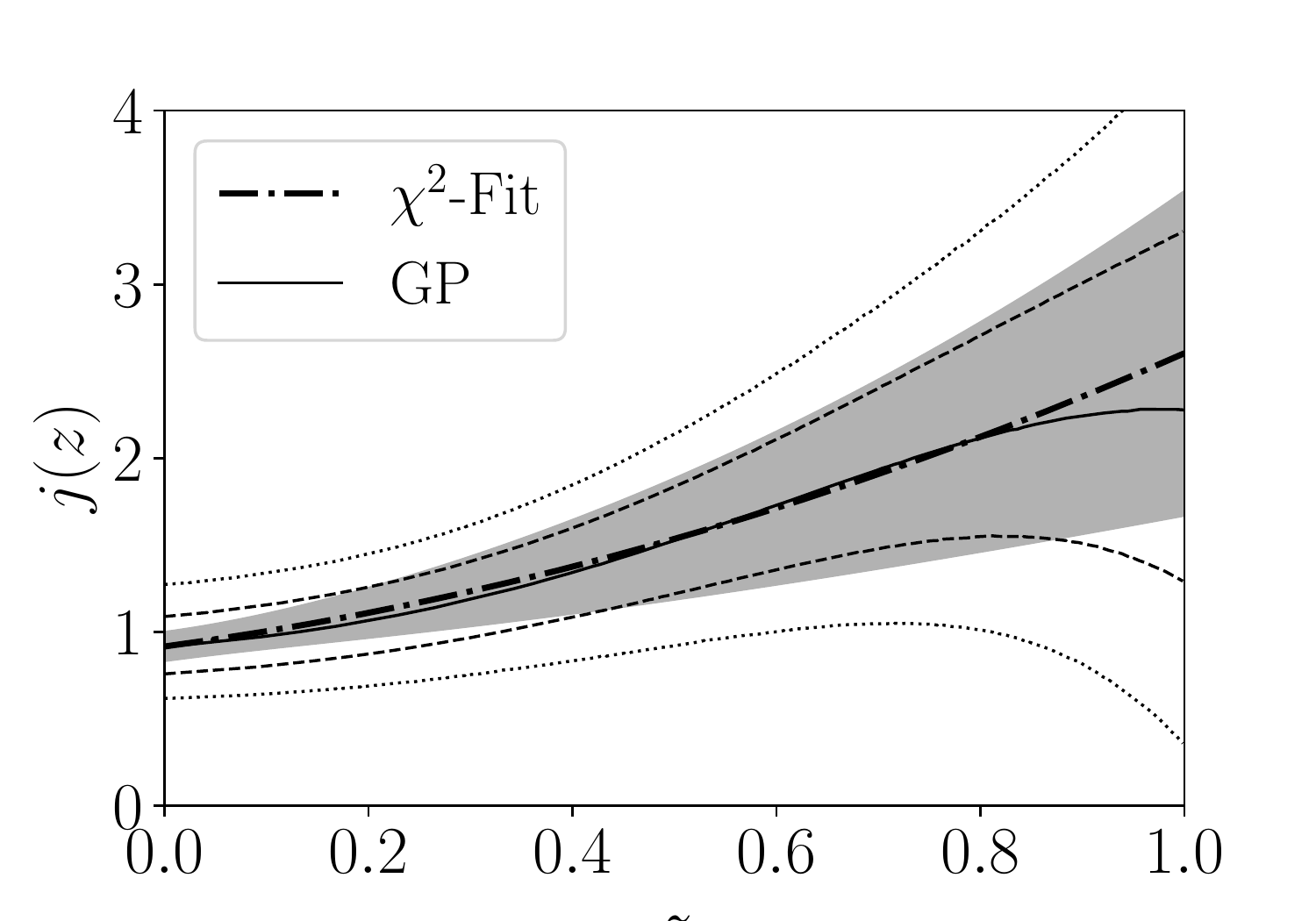}
	\end{center}
	\caption{{\small Plots showing a comparison between the reconstructed jerk $j(z)$ and the estimated $j_{\text{fit}}(z)$ using the combined 
			Pantheon+CC data (left), Pantheon+CC+P18 data (middle) and the Pantheon+CC+R19 data (right). The black solid line is the best fit curve 
			from GP reconstruction. The 1$\sigma$ and 2$\sigma$ C.L. are shown in dashed and dotted lines respectively. The bold dot-dashed line represents 
			the best fit function from $\chi^2$-minimization. The associated 1$\sigma$ uncertainty is shown by the shaded region.}}
	\label{ch4:jfit_plot}
\end{figure*}

We start with $n = 1$ followed by $n=2$ and so on and check for which order $n$ the value of $\chi^2_\nu < 1$. 
The measure of  $\chi^2_\nu $ obtained for the three cases studied are mentioned below. The estimated $j_i$'s and their respective $1\sigma$ uncertainties are also provided. A comparison between the reconstructed $j(z)$ and estimated $j_{\text{fit}}$ are shown in Fig. \ref{ch4:jfit_plot}. 

We note that the process fails 
for $z>1$, but we can do a reasonable estimate for $z<1$. The plots shown are only in the domain $0 \leq z \leq 1$.

For Pantheon+CC data,
\begin{equation}
j_{\text{fit}}(z) = 0.901~ + 0.611~ z + 0.987~ z^2.
\end{equation} 
The set of parameters $j_i$'s and their 1$\sigma$ uncertainties are,

$j_0  = 0.901^{+0.032}_{-0.032}$~, $j_1 = 0.611^{+0.231}_{-0.230}$ ~and~ $j_2 = 0.987^{+0.314}_{-0.314}$ ~with~ $\chi^2_\nu = 0.670$.

Similarly, for the Pantheon+CC+P18 data,
\begin{equation}
j_{\text{fit}}(z) = 0.917 + 0.788~ z + 0.899~ z^2.
\end{equation} 
The set of parameters $j_i$'s and their 1$\sigma$ uncertainties are,

$j_0  = 0.917^{+0.045}_{-0.045}$~, $j_1 = 0.788^{+0.291}_{-0.291}$ ~and~ $j_2 = 0.899^{+0.366}_{-0.367}$ ~with~ $\chi^2_\nu = 0.868$.

And finally for the Pantheon+CC+R19 data,
\begin{equation}
j_{\text{fit}}(z) = 0.956 + 1.376~ z - 0.188~ z^2.
\end{equation}
The set of parameters $j_i$'s and their 1$\sigma$ uncertainties are,

$j_0  = 0.956^{+0.044}_{-0.044}$~, $j_1 = 1.376^{+0.311}_{-0.311}$ ~and~ $j_2 = -0.188^{+0.403}_{-0.403}$ ~with~ $\chi^2_\nu = 0.817$.

In case we proceed with the fitting considering any higher-order polynomial, the $1\sigma$ uncertainty for the fitted function will not be contained within the $1\sigma$ 
error margin of $j(z)$ reconstructed by GP.

\section{Discussion}

The primary aim of the present work is a reconstruction of the cosmological jerk parameter $j$ from diverse observational datasets. The reconstruction is 
non-parametric, so $j$ is unbiased of any particular functional form to start with. Also, it does not depend on the theory of gravity. The basic assumption 
is that the Universe is described by the spatially flat, homogeneous and isotropic FLRW metric. The CC and Pantheon datasets are completely model-independent 
estimates used to reconstruct $j$. As the $r$BAOs in galaxy surveys and CMB Shift parameter measurement make use of a fiducial cosmological model, 
we reconstruct $j$ from the combined CC, SN, BAO and CMB datasets, rather as an additional exercise, to examine their possible effect on the reconstruction.

Although a non-parametric reconstruction is there in the literature for quite some time now for reconstructing physical quantities like 
the equation of state parameter or the quintessence potential and the cosmographic quantity like the deceleration parameter $q$, it has hardly been used to reconstruct 
the jerk parameter. As the deceleration parameter $q$ is now an observed quantity and is found to evolve, the following higher-order derivative, the jerk parameter, is 
a natural focus of attention. Indeed the parameters made out of even higher derivatives like snap (4th order derivative of $a$), crack (5th order derivative) etc., 
could well be evolving\cite{capoz}. But we focus on $j$ which is the evolution of $q$, the highest order derivative that is an observationally estimated quantity. 

For the reconstructed $j$, the plots reveal that $j$ has an evolution that is not necessarily monotonic. This non-monotonicity is not apparent when only CC or 
Pantheon data are individually employed, and $j$ decreases slowly for the former and increases rapidly for the latter (Fig. \ref{ch4:jerkplot}). When these two 
data sets are combined, the non-monotonicity appears and this nature is preserved even when the model-dependent datasets like CMB Shift and BAO data are included. 
The jerk parameter corresponding to the $\Lambda$CDM model is included in the $2\sigma$ CL for various combinations of datasets. It is found that the exclusion 
of the CMB Shift and $r$BAO data does not seriously affect the agreement with $\Lambda$CDM within the $2\sigma$ uncertainty.

The effective equation of state parameter $w_{\text{eff}}$ is linear in $q$, so the plots for both of them will look similar. We use the reconstruction of $q$ 
to plot $w_{\text{eff}}$ against the redshift $z$. The plots reveal that $w_{\text{eff}}$ also has a non-monotonic evolution. The plots also indicate that the 
Universe might have another stint of accelerated expansion in the recent past before entering into a decelerated phase and finally giving way to the current 
accelerated expansion. For the reconstruction of $w_{\text{eff}}$, the model-dependent data sets like $r$BAO and CMB Shift data are not included. 

It may be noted that we obtained the marginalized constraints on $M_B$ and $\Omega_{m0}$ by keeping the nuisance parameters, $\alpha$ and $\beta$, fixed using 
the BBC framework. As there may be correlations between these parameters keeping them fixed may adversely affect the model-independent nature of the reconstruction 
to an extent. 

A very recent work by Bengaly\cite{carlos} shows that the accelerated expansion of the Universe is correct even in $7\sigma$. So the observational constraints on 
the kinematic parameters find even more importance. The present work attempts to reconstruct the jerk parameter in a non-parametric way. Some of the recent 
parametric reconstructions of $j$ show that the present value of $j$ indicates that the $\Lambda$CDM model is not included in the $1\sigma$ 
CL \cite{mamon,ankan3}. The present work also shows that the evolution of $j$ may exclude the corresponding $\Lambda$CDM value for some part of the evolution in 
$1\sigma$, but at a $2\sigma$ level $j=1$ is indeed included. It deserves mention that, a recent study by Mehrabi and Basilakos\cite{mehrabi} shows that if the 
Gamma Ray Burst (GRB) data are included, the current value of $j$ is quite different from the standard $\Lambda$CDM value of $j=1$.

A recent work by Steinhardt, Sneppen and Sen\cite{bidisha} points out some errors in the quoted values of the redshift $z$ in the Pantheon dataset. We have worked out 
the reconstruction of $j$ with the corrected values of $z$ given in Ref \cite{bidisha}. There is hardly any qualitative difference in the plots. The only noticeable 
difference is found in the lower middle panel of Fig. \ref{ch4:jerkplot} for the best-fit curve where Pantheon+$r$BAO data are combined. However, even in $1\sigma$, 
there are no changes to note in conclusion.


\clearpage
\clearpage{}\chapter{Reconstruction of interaction in the cosmic dark sector}\blfootnote{\begin{flushleft} The work presented in this chapter is based on ``Nonparametric reconstruction of interaction in the cosmic dark sector", \textbf{Purba Mukherjee} and Narayan Banerjee, Phys.\ Rev.\ D \textbf{103}, 123530 (2021).\end{flushleft}}  \label{ch5:chap5}
\chaptermark{Reconstruction of interaction in the cosmic dark sector}
\section{Introduction}

For the different classes of dark energy models mentioned in section \ref{ch1:models}, the standard practice assumes that the exotic dark energy 
$\rho_D$ and the familiar cold dark matter $\rho_m$ evolve independently. Observational data reveal that at the present epoch, $\rho_{D}$ and 
$\rho_{m}$ are of the same order of magnitude! i.e., $\frac{\rho_{D0}}{\rho_{m0}} \sim \mathcal{O}(1)$. This fact is an indication towards a special 
period of the cosmic history we are currently living in. The corresponding question ``why now" constitutes the cosmological 
{\it coincidence problem} \cite{quint2}, which inspired a search for any non-gravitational interaction between $\rho_{D}$ and $\rho_m$. 

A possible coupling of the vacuum energy and the pressureless matter was investigated long back by Henriksen\cite{henrik}, and Olson and 
Jordan\cite{olson}. In context of the late-time cosmic acceleration, one argument is that the dark matter and the dark energy may not evolve 
independently. There is rather a transfer of energy between them. As a result, they do not satisfy individual conservation equations, but 
the total energy is conserved via the equation
\begin{equation} \label{ch1:Tmunutot}
\nabla_\mu \left( T^{\mu \nu}_m +  T^{\mu \nu}_D \right) = 0 ,
\end{equation}
where $T^{\mu \nu}_m$ and $T^{\mu \nu}_D$ denote the energy-momentum tensors for the dark matter and the dark energy, respectively.

Some investigations \cite{farrar, caiwang, amendola, guo, he, caldera, cai_su, honorez, besseda, yang14} in this direction have already been 
carried out. A novel function $Q$ is introduced such that, $ \nabla_\mu T^{\mu \nu}_m = - \nabla_\mu T^{\mu \nu}_D \propto Q$. The function 
$Q$ determines the rate of energy transfer  from one sector to the other. 

In the present chapter, the possibility of a non-gravitational interaction between the dark matter and the dark energy has been reconstructed 
using some recent observational datasets. As the origin and nature of this non-gravitational interaction are not known, $Q$ is phenomenologically 
chosen in various forms. Usually, it is assumed to be proportional to the Hubble parameter $H$ and one of the densities that of the dark matter or 
the dark energy. More general forms, where a linear combination of both the densities or even their derivatives are involved, can also be found in 
the literature. Various choices and their consequences are discussed in references \cite{r31,r32,r33,r34,holo_ankan,r35,r36,ankan_pan,holo_purba,
r37,r38,r39,r40,r41,r42,r45,r47,r48,r49,r50,r52}. It was shown that an interacting scenario can also potentially resolve the issues connected to 
the local value of the Hubble parameter \cite{r40,r50,r53}. We refer to the work of Wang \textit{et al}\cite{Q_review} 
for a comprehensive review on $Q$.

A reconstruction of the interaction from observations normally depends on the parametric form of $Q$ and the estimation of these parameters. Another approach is the non-parametric reconstruction, where attempts are made to ascertain the quantity directly from data without assuming any particular functional form. So this is clearly more unbiased. As there is no \textit{a priori} reason to rule out an interaction in the dark sector on the one hand, and also for the lack of any compulsive theoretical model for that on the other, certainly a reconstruction of the interaction $Q$, in an unbiased way without assuming any functional form of $Q$ deserves a lot more attention than that is available in the literature.

Cai and Su\cite{cai_su} investigated the possible interaction, independent of any specific form, by dividing the whole range of redshift into a few 
bins and setting the interaction term a constant in each bin. The result indicated that there could be an oscillatory interaction. Wang 
\textit{et al}\cite{wang_pca} adopted a non-parametric Bayesian approach and suggested that an interacting vacuum is not preferred over the standard 
$\Lambda$CDM. Yang, Guo and Cai\cite{yangguocai} presented a non-parametric reconstruction of the interaction between dark energy and dark matter 
directly from SNIa Union 2.1 data using the GP method. It was found that unless the EoS parameter $w$ for the dark energy deviates significantly 
from $-1$, the interaction is not evident. If $w$ is widely different from $-1$, the interaction cannot be ruled out at a $95\%$ CL. Another recent 
analysis by Cai, Tamanini and Yang\cite{Qlisa} indicates an interesting possibility that a gravity wave signal might carry signatures of this interaction 
in the dark sector encoded in the wave signal. This work is based on LISA space-based interferometer. It was shown that a 10-year survey could unveil the 
interaction in a wide redshift domain between $1<z<10$. 

This work aims in undertaking a non-parametric reconstruction of the interaction term $Q$ as a function of the redshift $z$ from recent 
observational data. The motivation is to find the nature of deviation from the zero interaction scenario. The Cosmic Chronometer Hubble data, 
the Pantheon Supernova compilation of CANDELS and CLASH MCT programs obtained by the HST, and the Baryon Acoustic Oscillation Hubble data are 
utilized for this purpose. Three cases for the dark energy EoS are considered. These are the decaying vacuum energy $\Lambda$ with $w = -1$, 
the $w$CDM model and the CPL parametrization of dark energy. For various combinations of datasets and different choices of dark energy, the most 
important common feature found is that ``no interaction'' is almost always included in $2\sigma$ and definitely in $3\sigma$. 

\section{The Model}

In a flat FLRW Universe composed of dark matter and dark energy, the Einstein equations are given by 
\begin{eqnarray}  
H^2 &=& \frac{8 \pi G}{3}(\rho_m + \rho_D) \label{ch5:friedmann},\\
\dot{H} + H^2 &=& -\frac{8 \pi G}{6}(\rho_m + \rho_D + 3 p_D ),
\end{eqnarray} 
where $\rho_m$ denotes the energy density of dark matter, $\rho_D$ the energy density of dark energy component, $p_D$ signifies the pressure component from 
the dark energy sector and $p_m = 0$ for pressureless dust. In what follows, we have considered $8 \pi G= 1$ for simplicity.

The energy conservation equation is given by the contracted Bianchi identity
\begin{equation}
\dot{\rho} + 3H (1+w_{\mbox{\tiny eff}})\rho  = 0,
\end{equation} 
where, $\rho = (\rho_m + \rho_D)$ the total energy density and $w_{\mbox{\tiny eff}}$ is the effective equation of state, defined as 
\begin{equation}
w_{\mbox{\tiny eff}} = \frac{p}{\rho} = \frac{p_D}{\rho_m + \rho_D}.
\end{equation} 

This conservation equation can be separated into two parts,
\begin{eqnarray} 
\dot{\rho_m}+ 3H\rho_m &=& -Q ,\\
\dot{\rho_D} + 3H (1+w)\rho_D &=& Q,
\end{eqnarray} 
where $w = \frac{p_D}{\rho_D}$ is the dark energy equation of state and $Q$ describes the rate of transfer of energy between dark matter and dark energy. One recovers the standard $\Lambda$CDM model for $Q = 0$ and $w = -1$. Unlike the usual practice of parametrizing the interaction term $Q$ using 
any parametric form proportional to $H\rho$, we focus on a non-parametric reconstruction of $Q$ from observational data.

The conservation equations can be rewritten with redshift $z$ as the argument, in the form
\begin{eqnarray} 
-H (1+z) \rho_m^\prime + 3 H \rho_m &=& -Q (z), \label{ch5:conservation} \\
-H (1+z) \rho_D^\prime + 3 H (1+w) \rho_D &=& Q (z), \label{ch5:conservation2}
\end{eqnarray}
where a 'prime' represents a differentiation w.r.t. the redshift $z$.

Equation \eqref{ch5:friedmann} can be written in terms of the reduced Hubble parameter, $E = \frac{H}{H_0}$, as
\begin{equation} \label{ch5:fried_reduced}
E^2(z) = \tilde{\rho}_m + \tilde{\rho}_D,
\end{equation} 
where $\tilde{\rho}_m$ and $ \tilde{\rho}_D$ are $\rho_m$ and $\rho_D$ respectively scaled by a factor of $\frac{1}{3 H_0^2}$. 

On differentiating Eq. \eqref{ch5:fried_reduced} w.r.t. $z$, we get
\begin{equation} \label{ch5:fried_diff}
2 E E'(z) = \tilde{\rho}_m^\prime + \tilde{\rho}_D^\prime .
\end{equation} 

The conservation equations \eqref{ch5:conservation} and \eqref{ch5:conservation2} can be reduced to the following forms,
\begin{eqnarray} 
-E (1+z) \tilde{\rho}_m^\prime + 3 E \tilde{\rho}_m &=& -\tilde{Q} (z), \label{ch5:cons_reduced_m}\\
-E (1+z) \tilde{\rho}_D^\prime + 3 E (1+w) \tilde{\rho}_D &=& \tilde{Q} (z). \label{ch5:cons_reduced_d}
\end{eqnarray} 
The dimensionless $\tilde{Q}$ characterizes the interaction, where $\tilde{Q} = \frac{1}{ 3 H_0^3} Q$.

By combining equation \eqref{ch5:fried_diff} with equations \eqref{ch5:cons_reduced_m} and \eqref{ch5:cons_reduced_d}, one can obtain,
\begin{equation} \label{ch5:Q_h}
\begin{split}
\tilde{Q} = \left( \frac{E^2(1+w)}{w} + \frac{(1+z)E^2 w'}{3 w^2} \right) \left[ 2(1+z)E' - 3E \right] + ~~~~~~~~~~~\\ 
~~~~~~~~~~~~~~~~~~~~~~~~- \frac{(1+z)E}{3 w}\left[2(1+z)(E'^2 + E E'') - 4 EE' \right].
\end{split}
\end{equation} 
This will be the key equation for the reconstruction of $\tilde{Q}$. On utilizing the observed dimensionless Hubble parameter $E(z)$, one can 
reconstruct the interaction $\tilde{Q}$, once the equation of state $w(z)$ of dark energy is known.

\begin{table*}[h!] 
	\caption{{\small Table showing the reconstructed value of $H_0$ (in units of km Mpc$^{-1}$ s$^{-1}$) using samples from Set A and B.}}
	\begin{center}
		\resizebox{\textwidth}{!}{\renewcommand{\arraystretch}{1.5} \setlength{\tabcolsep}{20pt} \centering  
			\begin{tabular}{l c c c c} 
				\hline
				$\kappa(z,\tilde{z}$)& \textbf{A 1} & \textbf{A 2} & \textbf{B 1} & \textbf{B 2} \\
				\hline
				\hline
				Sq. Exp & $67.36\pm4.77  $ & $ 72.13\pm4.85$ & $65.19\pm2.63$ & $67.66\pm2.79$\\ 
				\hline				
				Mat\'{e}rn 9/2 & $68.47\pm5.08$ & $72.76\pm5.00$ & $65.15\pm2.72$ & $67.57\pm2.90$\\ 
				\hline
			\end{tabular}
		}
	\end{center}
	\label{ch5:Hz_res}
\end{table*}

\section{The Reconstruction}

To reconstruct the interaction $\tilde{Q}$ we need a non-parametric method to obtain $E(z)$, and its derivatives $E'(z)$ and $E''(z)$.  
The GP method is adopted as a numerical tool. We have used both the squared exponential and the Mat\'{e}rn $\frac{9}{2}$ covariance functions. 
The observational Hubble data from CCs \cite{cc_101,cc_102,cc_103,cc_104,cc_106,cc_105} (shown in Table \ref{ch1:tabcc}) and $r$BAOs in 
galaxies and galaxy clusters \cite{bao_61,bao_1,alam2017,bao_107,bao_108,bautista2017,bao_2,bao_4,bao_3,blake2012,bao_77,anderson2014,bao_9,bao_110} 
(shown in Table \ref{ch1:tabbao} consisting of $30 H(z)$ measurements), along with the reduced Hubble data from the Pantheon supernova compilation 
of CANDELS and CLASH MCT programs \cite{mct} (shown in Table \ref{ch1:tabmct}) are utilized to obtain the target function $E(z)$, and its derivatives 
$E'(z)$ and $E''(z)$ respectively. For the CC data, we have taken into consideration two different compilations, hereafter referred to as the CC$_1$ 
and CC$_2$ samples respectively. The CC$_1$ sample has a total of 31 $H(z)$ values, obtained from a combination of the CCB and CCH samples. Similarly, 
the CC$_2$ sample consists of 26 $H(z)$ values, obtained from another compilation between the CCM and CCH samples.

\subsection{Methodology} \label{ch5:methodology}

Three choices are considered for the dark energy equation of state $w = \frac{p_D}{{\rho}_D}$. Firstly the decaying vacuum energy case, 
followed by the $w$CDM model and finally the CPL parametrization, given as \begin{eqnarray}
\Lambda\mbox{CDM}	: w(z) &=& -1 ,\\
w\mbox{CDM}	:  w(z) &=& w  ,\\
\mbox{CPL}	 :  w(z) &=& w_0 + w_a \left(\frac{z}{1+z}\right) .
\end{eqnarray}
For the dark energy EoS, considering the $w$CDM model, we take the best-fit value of $w = -1.006 \pm 0.045$ from the Planck 2015 \cite{planck_cmb} survey, 
and for the CPL parametrization, we take $w_0 = -1.046^{+0.179}_{-0.170}$ and $w_a = 0.14^{+0.60}_{-0.76}$ 
respectively from HST Cluster Supernova Survey 2011  \cite{suzuki}.

We attempt to reconstruct $\tilde{Q}$ directly for the following combination of datasets,
\begin{itemize}
	\item \textbf{Set A}
	\begin{enumerate}
		\item CC$_1$+Pantheon+MCT
		\item CC$_2$+Pantheon+MCT
	\end{enumerate}
	\item \textbf{Set B}
	\begin{enumerate}
		\item CC$_1$+Pantheon+MCT+BAO
		\item CC$_2$+Pantheon+MCT+BAO
	\end{enumerate}	
\end{itemize}

Set A comprises of model-independent data combinations like CC and Pantheon. Set B includes the $r$BAOs in combination with the CC and Pantheon datasets. 
But these $r$BAOs in galaxy surveys assume some fiducial cosmological model for acquiring these measurements. This makes Set B model-dependent.

We start with constraining the Hubble parameter in the present epoch $H_0$.  The GP method is utilized to reconstruct $H(z)$ from the Hubble data sets.  
The value of $H_0$ obtained for the CC and CC+$r$BAO combination is shown in Table \ref{ch5:Hz_res}. Further, the reconstructed dataset is normalized 
to obtain the reduced Hubble parameter $E(z) = \frac{H(z)}{H_0}$. The uncertainty associated with $E$, i.e. $\sigma_{E}$ is obtained via equation 
\eqref{ch2:sig_h}. 

For Set A, the CC Hubble data is normalized with the reconstructed $H_0$, as given in Table \ref{ch5:Hz_res} to obtain the reduced Hubble parameter $E(z)$. 
Similarly, the CC+$r$BAO Hubble data has been normalized for Set B with the reconstructed $H_0$ from Table \ref{ch5:Hz_res}. One should note that, $E(z=0) = 
1$ with $0$ uncertainty. 

The $E(z)$ datasets reconstructed from the CC and CC+$r$BAO combination are now combined with the $E$ dataset obtained from the Pantheon+MCT compilation. The 
error uncertainties and the covariance matrix associated with individual data sets have been combined and considered for the analysis. Assuming that these 
composite $E$ datasets, A and B, obey a Gaussian distribution with a mean and variance, the posterior distribution of the reconstructed function $E(z)$ and 
its derivatives can be expressed as a joint Gaussian distribution of individual datasets considered.

Thus, given a set of observational data, we have used the GP to construct the most probable underlying continuous function $E(z)$ describing this data, 
along with its derivatives, and have also obtained the associated confidence levels. From the reconstructed values of $E(z)$, $E'(z)$ and $E''(z)$ in Eq. 
\eqref{ch5:Q_h}, the interaction $\tilde{Q}(z)$ is reconstructed.

\begin{table*}[t!] 
	\caption{{\small Table showing the reconstructed value of $\tilde{Q}(z=0)$ using samples from Set A and B for the decaying dark energy EoS given by $w = -1$.}}
	\begin{center}
		\resizebox{\textwidth}{!}{\renewcommand{\arraystretch}{1.5} \setlength{\tabcolsep}{5pt} \centering  
			\begin{tabular}{l c c c c } 
				\hline
				$\kappa(z,\tilde{z})$& \textbf{A 1} & \textbf{A 2} & \textbf{B 1} & \textbf{B 2} \\
				\hline
				\hline
				Sq. Exp & $-0.133^{+0.126~+0.343~+0.686}_{-0.096~-0.190~-0.296}$ & $-0.162^{+0.119~+0.301~+0.581}_{-0.088~-0.170~-0.261}$ & 
				$-0.129^{+0.102~+0.234~+0.417}_{-0.076~-0.138~-0.198}$ & $-0.063^{+0.146~+0.331~+0.571}_{-0.109~-0.193~-0.269}$ \\ 
				\hline				
				Mat\'{e}rn 9/2 & $-0.085^{+0.204~+0.529~+1.016}_{-0.145~-0.280~-0.429}$ & $-0.127^{+0.174~+0.433~+0.812}_{-0.122~-0.230~-0.347}$ & 
				$-0.141^{+0.121~+0.280~+0.492}_{-0.092~-0.168~-0.243}$ & $-0.085^{+0.168~+0.381~+0.671}_{-0.126~-0.224~-0.316}$ \\ 
				\hline
			\end{tabular}
		}
	\end{center}
	\label{ch5:Q_lcdm_res}
\end{table*}

\begin{table*}[t!] 
	\caption{{\small Table showing the reconstructed value of $\tilde{Q}(z=0)$ using samples from Set A and B for the $w$CDM model with EoS given by 
			$w = -1.006 \pm 0.045$ \cite{planck_cmb}.}}
	\begin{center}
		\resizebox{\textwidth}{!}{\renewcommand{\arraystretch}{1.5} \setlength{\tabcolsep}{5pt} \centering  
			\begin{tabular}{l c c c c} 
				\hline
				$\kappa(z,\tilde{z})$& \textbf{A 1} & \textbf{A 2} & \textbf{B 1} & \textbf{B 2} \\
				\hline
				\hline
				Sq. Exp & $-0.135^{+0.124~+0.335~+0.670}_{-0.096~-0.190~-0.296}$ & $-0.165^{+0.116~+0.296~+0.567}_{-0.086~-0.168~-0.257}$ & 
				$-0.133^{+0.099~+0.228~+0.409}_{-0.075~-0.136~-0.197}$ & $-0.069^{+0.143~+0.324~+0.564}_{-0.107~-0.189~-0.266}$ \\ 
				\hline				
				Mat\'{e}rn 9/2 & $-0.088^{+0.202~+0.518~+1.008}_{-0.143~-0.278~-0.425}$ & $-0.130^{+0.170~+0.425~+0.804}_{-0.120~-0.229~-0.345}$ & 
				$-0.146^{+0.118~+0.273~+0.488}_{-0.091~-0.167~-0.244}$ & $-0.091^{+0.165~+0.376~+0.653}_{-0.124~-0.219~-0.311}$ \\ 
				\hline
			\end{tabular}
		}
	\end{center}
	\label{ch5:Q_wcdm_res}
\end{table*}

\begin{table*}[t!] 
\caption{{\small Table showing the reconstructed value of $\tilde{Q}(z=0)$ using samples from Set A and B for the CPL parametrization of dark energy 
		with EoS given by $w(z) = w_0 + w_a (\frac{z}{1+z})$, $w_0 = -1.046^{+0.179}_{-0.170}$ and $w_a = 0.14^{+0.60}_{-0.76}$ \cite{suzuki}.}}
	\begin{center}
		\resizebox{\textwidth}{!}{\renewcommand{\arraystretch}{1.5} \setlength{\tabcolsep}{5pt} \centering  
			\begin{tabular}{l c c c c} 
				\hline
				$\kappa(z,\tilde{z})$& \textbf{A 1} & \textbf{A 2} & \textbf{B 1} & \textbf{B 2} \\
				\hline
				\hline
				Sq. Exp & $-0.175^{+0.097~+0.249~+0.520}_{-0.091~-0.185~-0.293}$ & $-0.214^{+0.087~+0.217~+0.429}_{-0.080~-0.162~-0.253}$ & 
				$-0.202^{+0.073~+0.169~+0.306}_{-0.061~-0.117~-0.177}$ & $-0.159^{+0.111~+0.253~+0.448}_{-0.085~-0.155~-0.255}$ \\ 
				\hline				
				Mat\'{e}rn 9/2 & $-0.136^{+0.157~+0.410~+0.814}_{-0.132~-0.263~-0.412}$ & $-0.187^{+0.131~+0.330~+0.638}_{-0.108~-0.214~-0.331}$ & 
				$-0.213^{+0.091~+0.209~+0.374}_{-0.076~-0.146~-0.221}$ & $-0.178^{+0.131~+0.299~+0.530}_{-0.100~-0.186~-0.272}$ \\ 
				\hline
			\end{tabular}
		}
	\end{center}
	\label{ch5:Q_cpl_res}
\end{table*}

\begin{figure*}[t!]
	\begin{center}
		\includegraphics[angle=0, width=\textwidth]{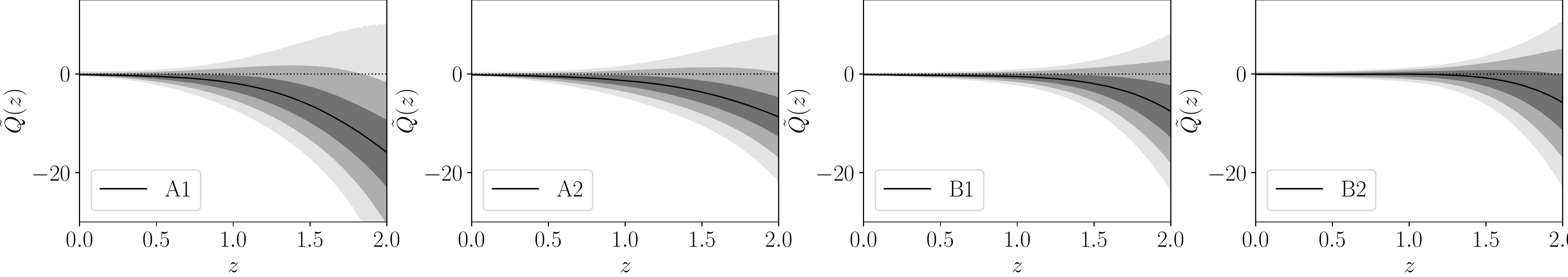}\\
		\includegraphics[angle=0, width=\textwidth]{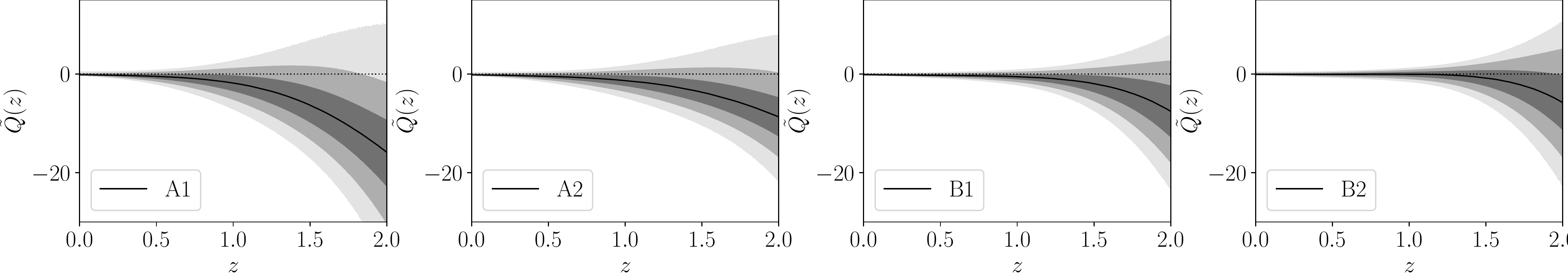}\\
		\includegraphics[angle=0, width=\textwidth]{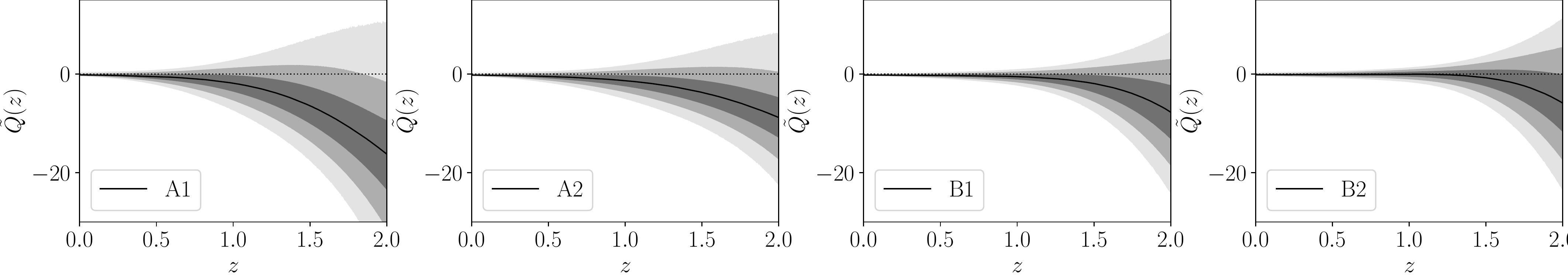}
	\end{center}
	\caption{{\small Plots for the reconstructed interaction term $\tilde{Q}$ from the dataset samples of Set A and B using a squared 
			exponential covariance function considering the decaying dark energy EoS given by $w = -1$ (top row), the $w$CDM model of dark energy with 
			EoS given by $w = -1.006 \pm 0.045$ \cite{planck_cmb} (middle row), and the CPL parametrization of dark energy with EoS 
			given by $w(z) = w_0 + w_a (\frac{z}{1+z})$, $w_0 = -1.046^{+0.179}_{-0.170}$ and $w_a = 0.14^{+0.60}_{-0.76}$ \cite{suzuki} 
			(bottom row). The black solid curve shows the best fit values and the shaded regions correspond to the 1$\sigma$, 2$\sigma$ 
			and 3$\sigma$ uncertainty.}}
	\label{ch5:Q_sqexp}
\end{figure*}

\begin{figure*}[t!]
	\begin{center}
		\includegraphics[angle=0, width=\textwidth]{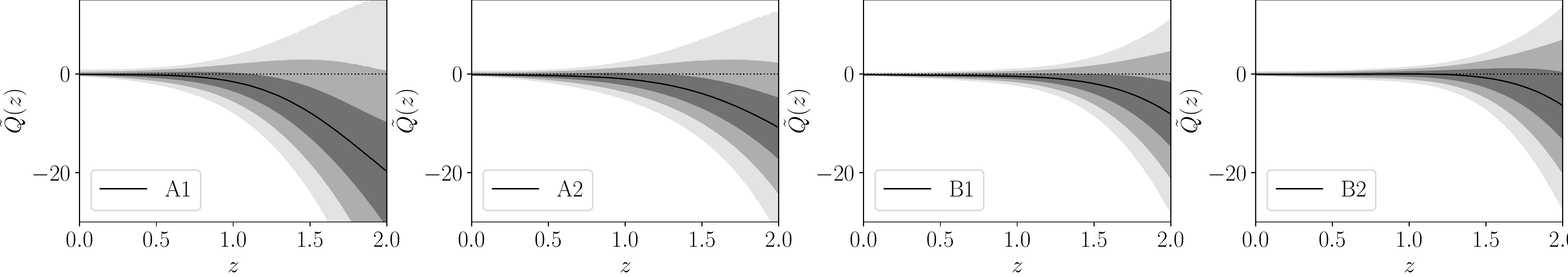}\\
		\includegraphics[angle=0, width=\textwidth]{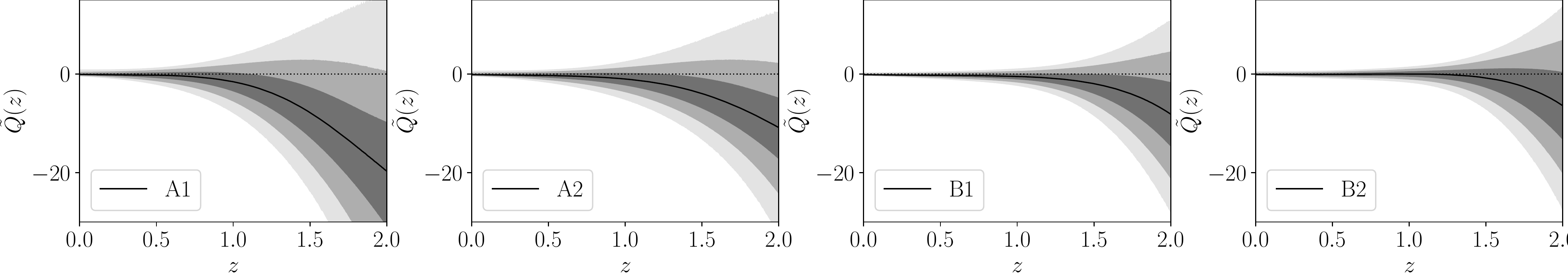}\\
		\includegraphics[angle=0, width=\textwidth]{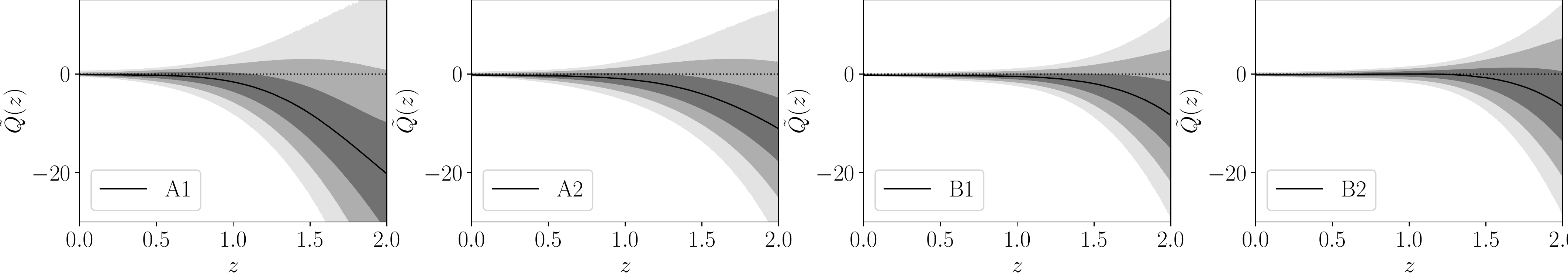}
	\end{center}
	\caption{{\small Plots for the reconstructed interaction term $\tilde{Q}$ from the dataset samples of Set A and B using a Mat\'{e}rn 
			9/2 covariance function considering the decaying dark energy EoS given by $w = -1$ (top row), the $w$CDM model of dark energy with EoS 
			given by $w = -1.006 \pm 0.045$ \cite{planck_cmb} (middle row), and the CPL parametrization of dark energy with EoS given by 
			$w(z) = w_0 + w_a (\frac{z}{1+z})$, $w_0 = -1.046^{+0.179}_{-0.170}$ and $w_a = 0.14^{+0.60}_{-0.76}$ \cite{suzuki} (bottom row). 
			The black solid curve shows the best fit values and the shaded regions correspond to the 1$\sigma$, 2$\sigma$ and 3$\sigma$ 
			uncertainty.}}
	\label{ch5:Q_mat92}
\end{figure*}

\begin{figure*}[t!]
	\begin{center}
		\includegraphics[angle=0, width=\textwidth]{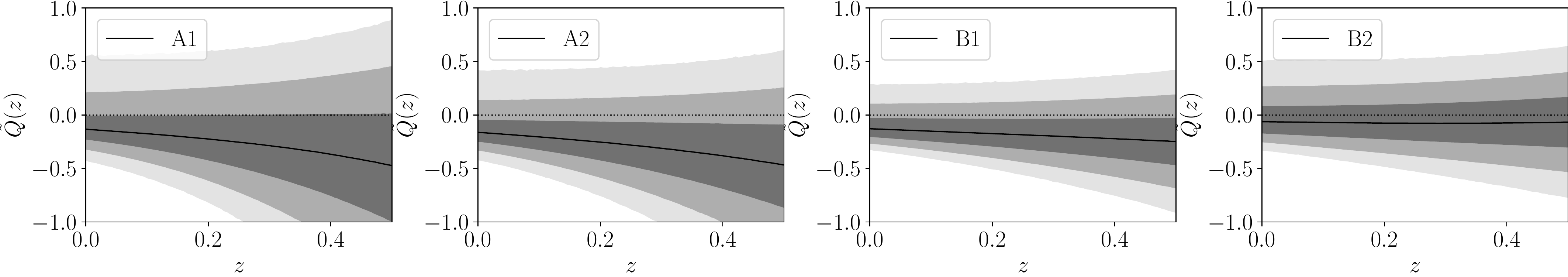}\\
		\includegraphics[angle=0, width=\textwidth]{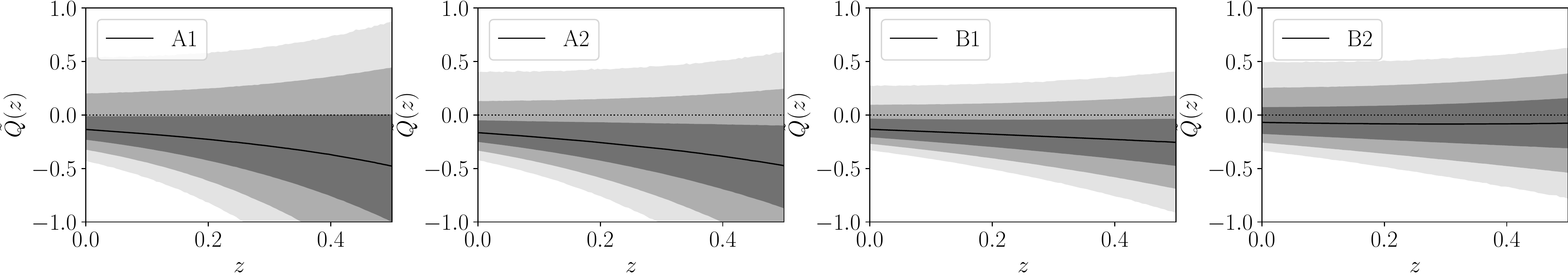}\\
		\includegraphics[angle=0, width=\textwidth]{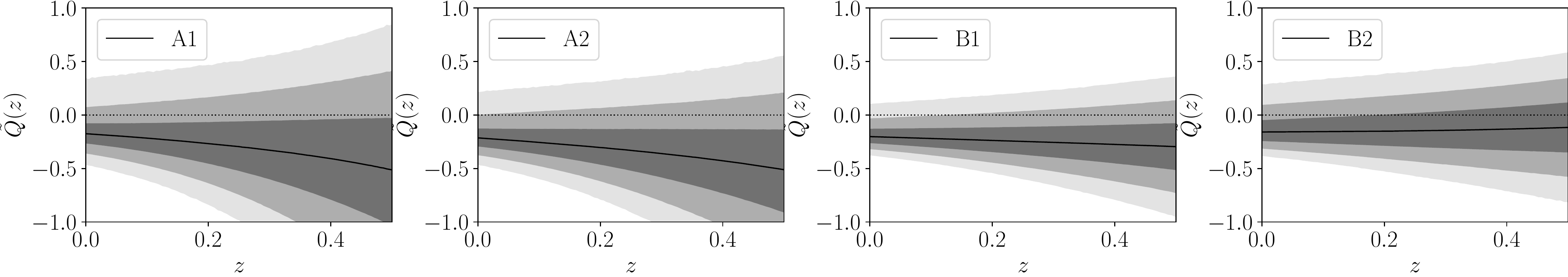}
	\end{center}
	\caption{{\small Plots for the reconstructed $\tilde{Q}$ function in the low redshift range $0<z<0.5$ from the dataset samples of Set A 
			and B using a squared exponential covariance function considering the decaying dark energy EoS given by $w = -1$ (top row), the 
			$w$CDM model with DE EoS given by $w = -1.006 \pm 0.045$ \cite{planck_cmb} (middle row), and the CPL parametrization of dark energy 
			with EoS given by $w(z) = w_0 + w_a (\frac{z}{1+z})$, $w_0 = -1.046^{+0.179}_{-0.170}$ and $w_a = 0.14^{+0.60}_{-0.76}$ \cite{suzuki} 
			(bottom row). The black solid curve shows the best fit values and the shaded regions correspond to the 1$\sigma$, 2$\sigma$ and 
			3$\sigma$ uncertainty.}}
	\label{ch5:Qscaled_sqexp}
\end{figure*}

\begin{figure*} [t!]
	\begin{center}
		\includegraphics[angle=0, width=\textwidth]{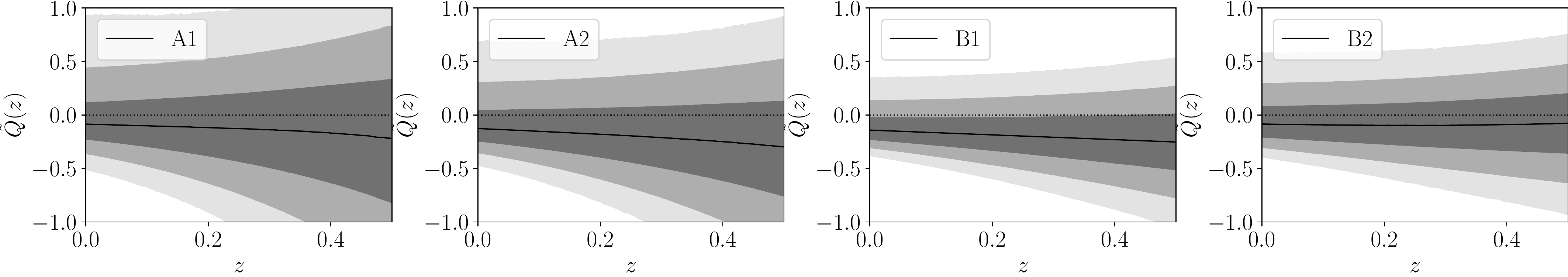}\\
		\includegraphics[angle=0, width=\textwidth]{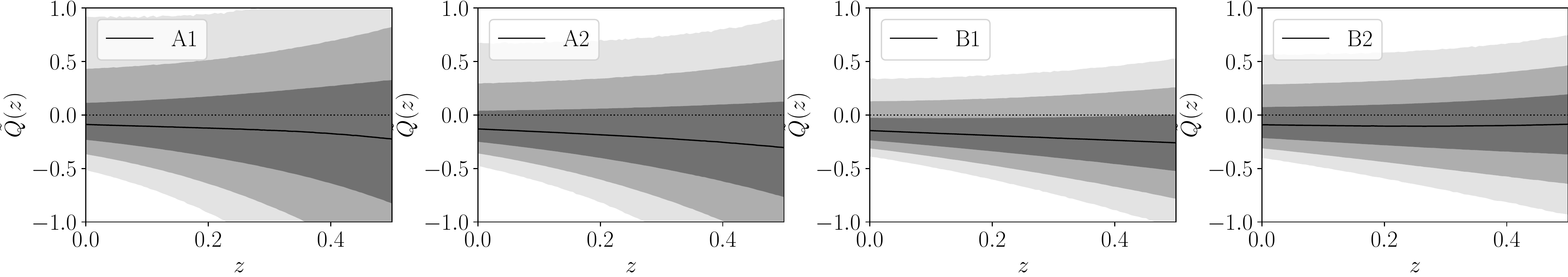}\\
		\includegraphics[angle=0, width=\textwidth]{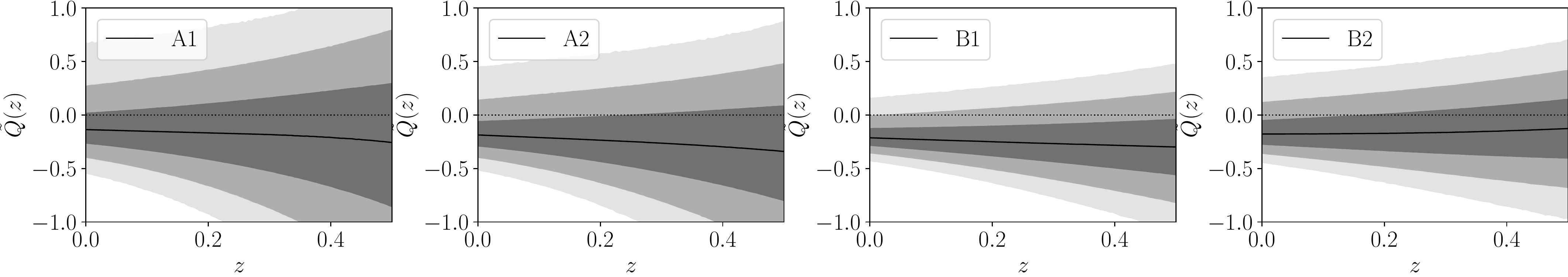}
	\end{center}
	\caption{{\small Plots for the reconstructed $\tilde{Q}$ function in the low redshift range $0<z<0.5$ from the dataset samples of Set A 
			and B using a Mat\'{e}rn 9/2 covariance function considering the decaying dark energy EoS given by $w = -1$ (top row), the $w$CDM 
			model with DE EoS given by $w = -1.006 \pm 0.045$ \cite{planck_cmb} (middle row), and the CPL parametrization of dark energy with 
			EoS given by $w(z) = w_0 + w_a (\frac{z}{1+z})$, $w_0 = -1.046^{+0.179}_{-0.170}$ and $w_a = 0.14^{+0.60}_{-0.76}$ \cite{suzuki} 
			(bottom row). The black solid curve shows the best fit values and the shaded regions correspond to the 1$\sigma$, 2$\sigma$ and 
			3$\sigma$ uncertainty.}}
	\label{ch5:Qscaled_mat92}
\end{figure*}

\subsection{Results}

The reconstructed interaction $\tilde{Q}$ for various combinations of datasets are shown in Fig. \ref{ch5:Q_sqexp} and 
\ref{ch5:Q_mat92}. The shaded regions correspond to the $68\%$, $95\%$ and $99.7\%$ CLs respectively from darker to lighter shades. The black 
solid line shows the curve with best-fit values of $\tilde{Q}$. Tables \ref{ch5:Q_lcdm_res}, \ref{ch5:Q_wcdm_res} and \ref{ch5:Q_cpl_res} show 
the best fit results for $\tilde{Q}(z=0)$ along with the $1\sigma$, 2$\sigma$ and 3$\sigma$ uncertainties for all the combinations. In Fig. 
\ref{ch5:Qscaled_sqexp} and \ref{ch5:Qscaled_mat92}, the plots for $\tilde{Q}$ are zoomed in for the range $0<z<0.5$, to examine its behaviour 
at very low redshift more closely. 

From equation \eqref{ch5:conservation} one can understand that a negative $Q$ indicates the energy flow from dark energy to the dark matter sector, and 
a positive $Q$ indicates the reverse. The plots show that the reconstructed $\tilde{Q}$ remains close to $0$, indicating no appreciable interaction for 
low redshift ranges. The best fit curve shows a small deviation towards negative values, but the zero interaction scenario is always included in 2$\sigma$ 
for most of these combinations. So, the energy gets transferred from the dark energy to the dark matter sector, if it happens at all.

Pav\'{o}n and Wang\cite{diego} formulated the Le Ch\^{a}telier-Braun principle in cosmological physics which predicts that for an approach towards 
thermodynamic equilibrium, the transfer of energy between the dark energy and dark matter sectors, must be such that the latter gains energy from the 
former and not the other way around. This direction of flow of energy is consistent with the thermodynamic requirement \cite{diego}. It is interesting to 
note that the case of $\tilde{Q} < 0$ guarantees that the ratio $\frac{\rho_{m}}{\rho_D}$ asymptotically tends to a constant \cite{pavon}, thus 
alleviating the coincidence problem.

\subsection{Fitting function for $\tilde{Q}$}\label{ch5:fitting}

An approximate fitting formula for the reconstructed interaction has been derived. This exercise is done in the low redshift range $0 < z < 1$ 
using the combined datasets A1, A2, B1 and B2. The goal is to find a simple analytic form of $\tilde{Q}$. As both the covariance functions yield similar 
results, we pick up only the Mat\'{e}rn 9/2 covariance as an example. 

We consider a polynomial for $\tilde{Q}(z)$ as a function of redshift $z$ as,
\begin{equation} \label{Qfit}
\tilde{Q}_{\mbox{\tiny fit}} (z) = \sum_{i=0}^{n} \tilde{Q}_i z^i.
\end{equation}

The coefficients $\tilde{Q}_i$'s of the above equation are estimated by the $\chi^2$ minimization, where we define the $\chi^2$ function as 
\begin{equation}
\chi^2 =  \mathlarger{\mathlarger{\sum}}_{s}\dfrac{\left[\tilde{Q}(z_s)- \tilde{Q}_{\mbox{\tiny fit}}(z_s)\right]^2}{\sigma^2(z_s)}.
\end{equation}

\begin{figure*}[h!]
	\begin{center}
		\includegraphics[angle=0, width=0.325\textwidth]{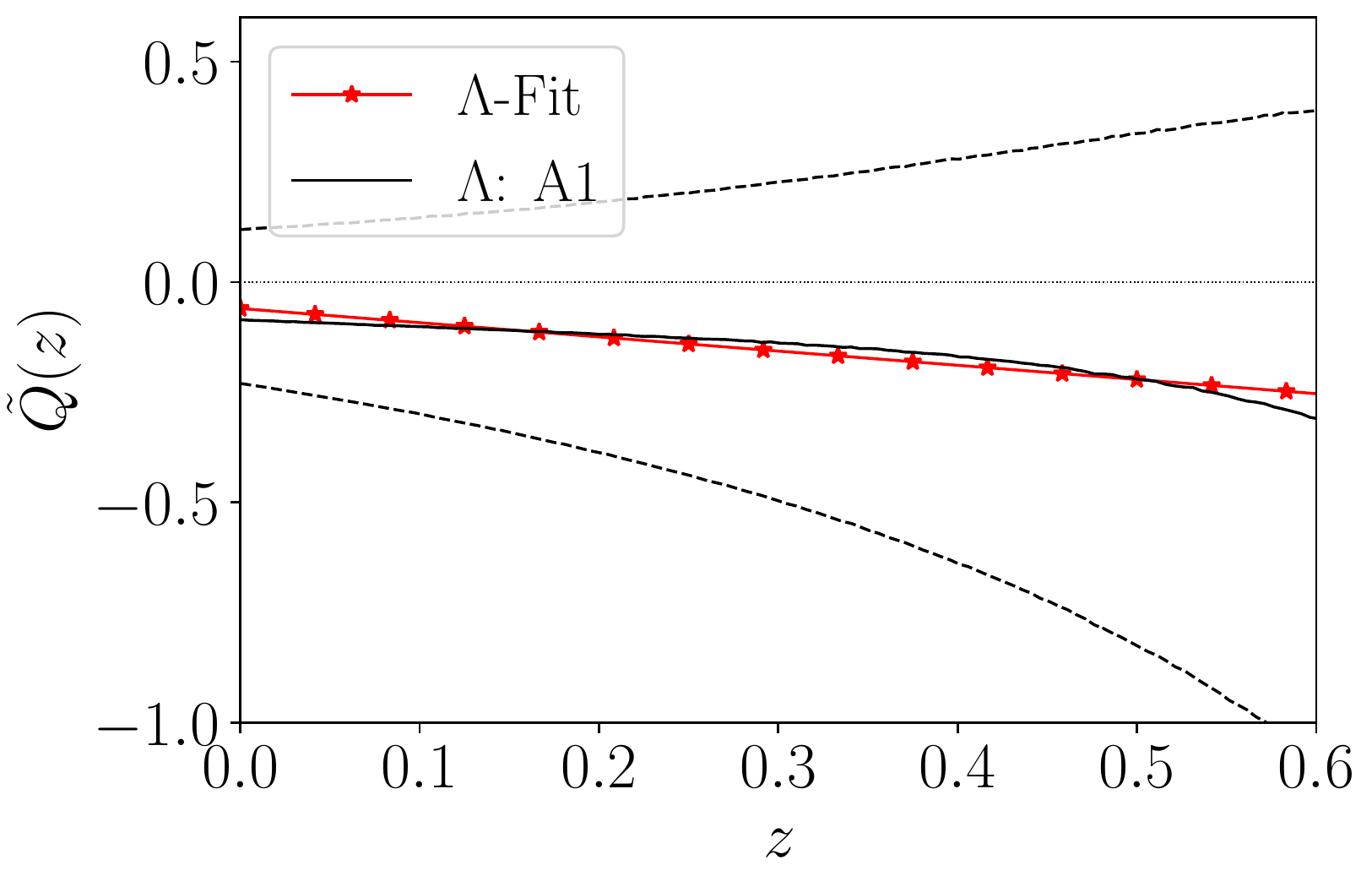}
		\includegraphics[angle=0, width=0.325\textwidth]{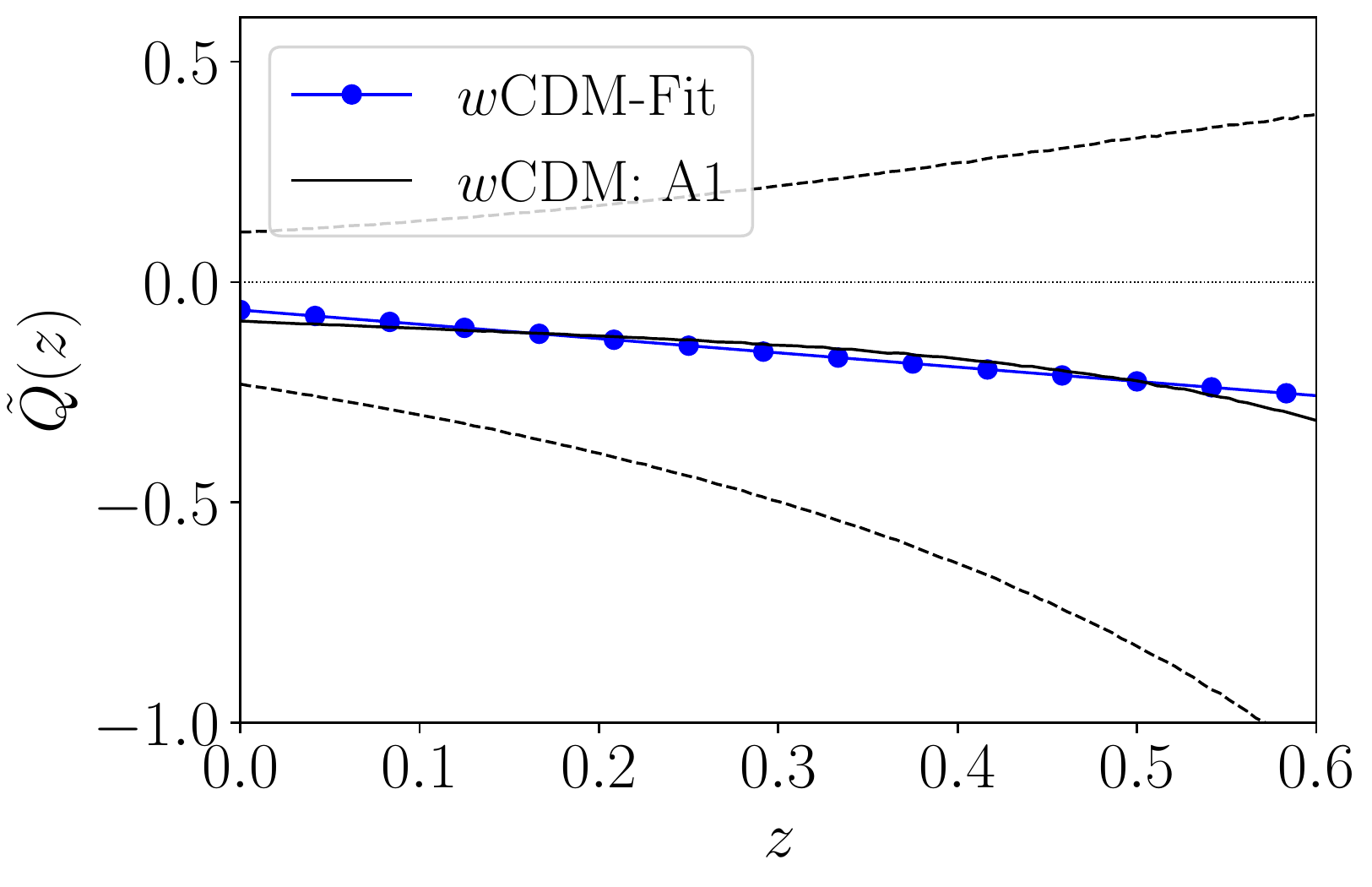}
		\includegraphics[angle=0, width=0.325\textwidth]{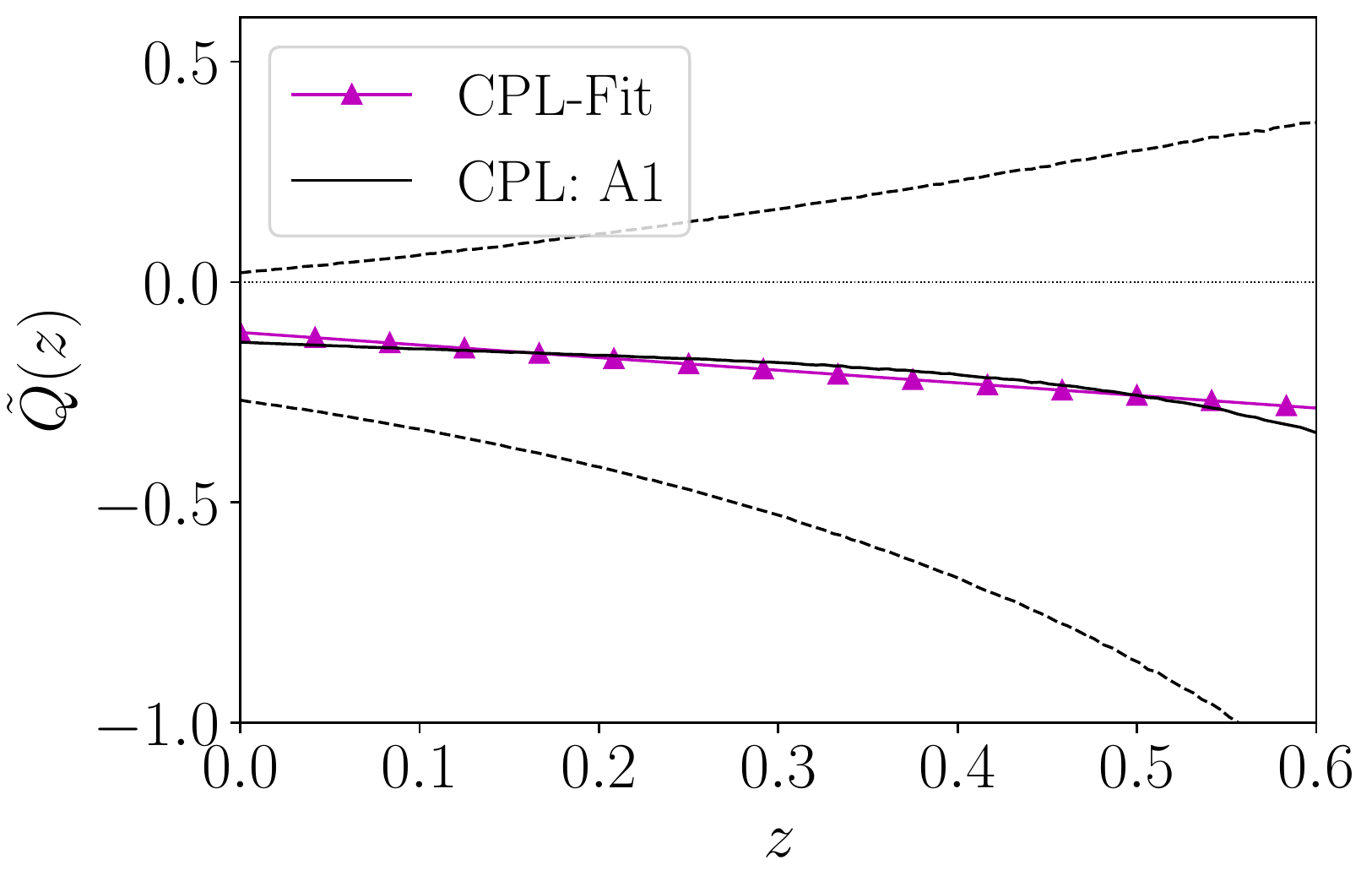}\\
		\includegraphics[angle=0, width=0.325\textwidth]{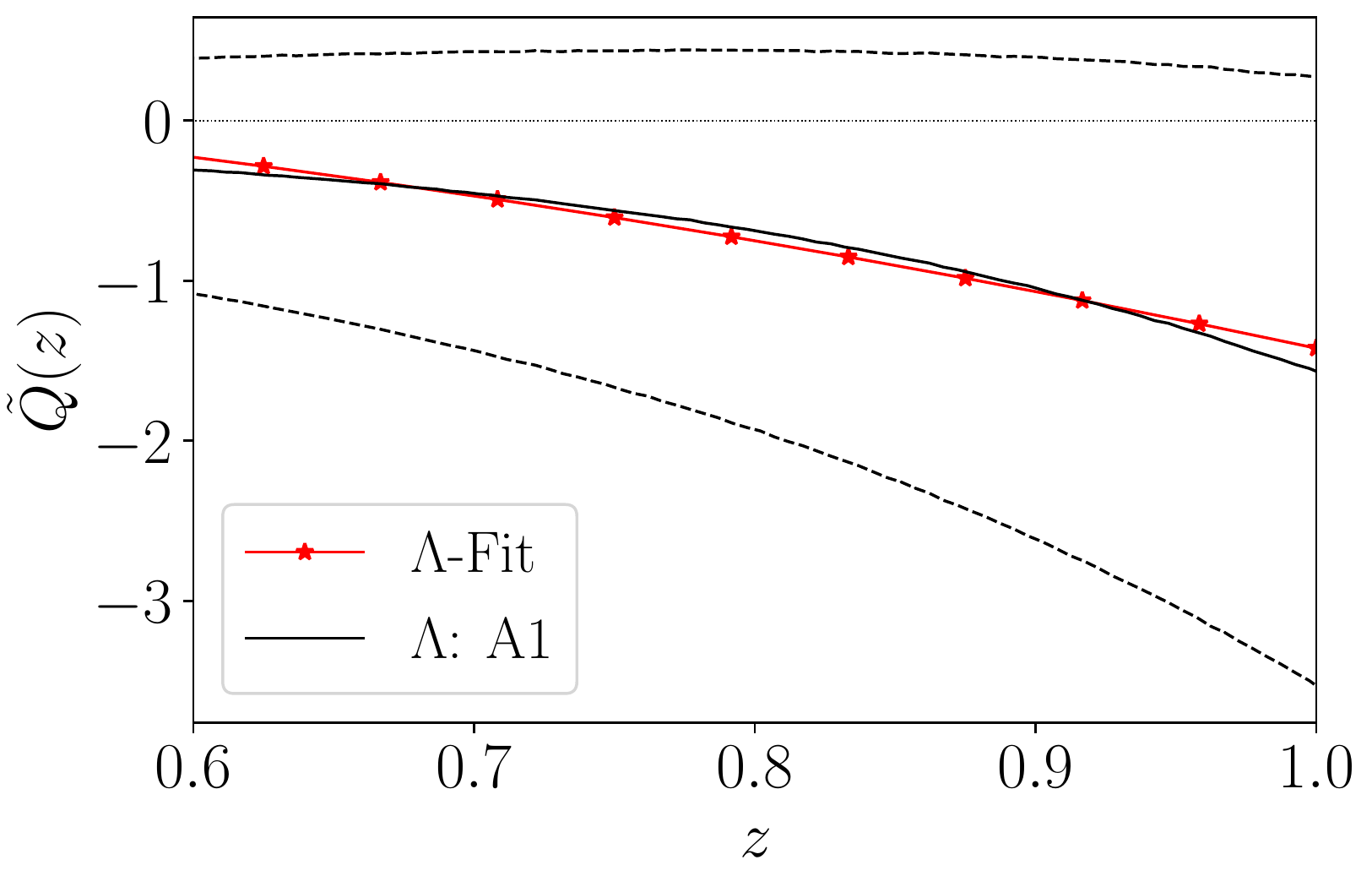}
		\includegraphics[angle=0, width=0.325\textwidth]{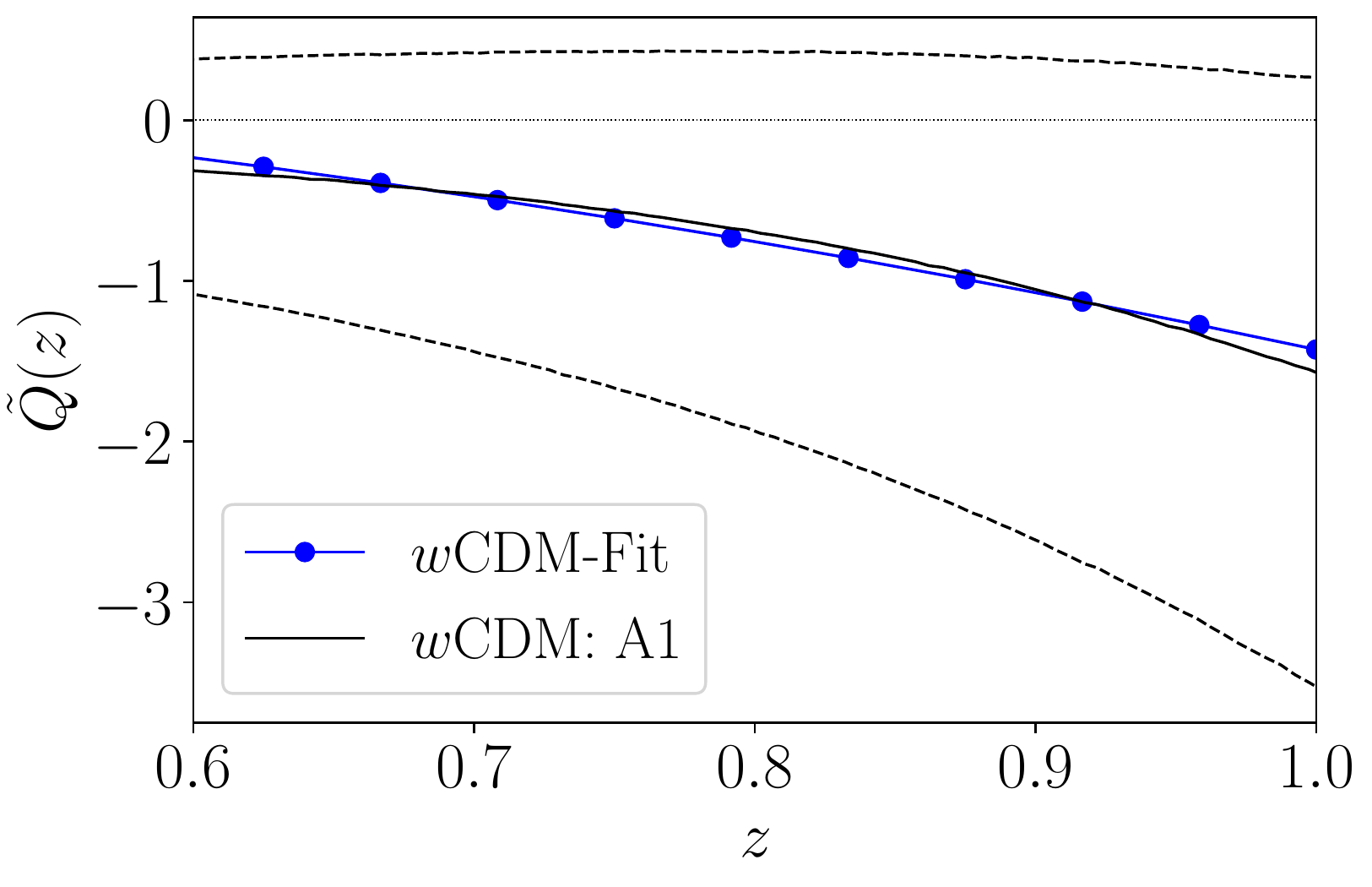}
		\includegraphics[angle=0, width=0.325\textwidth]{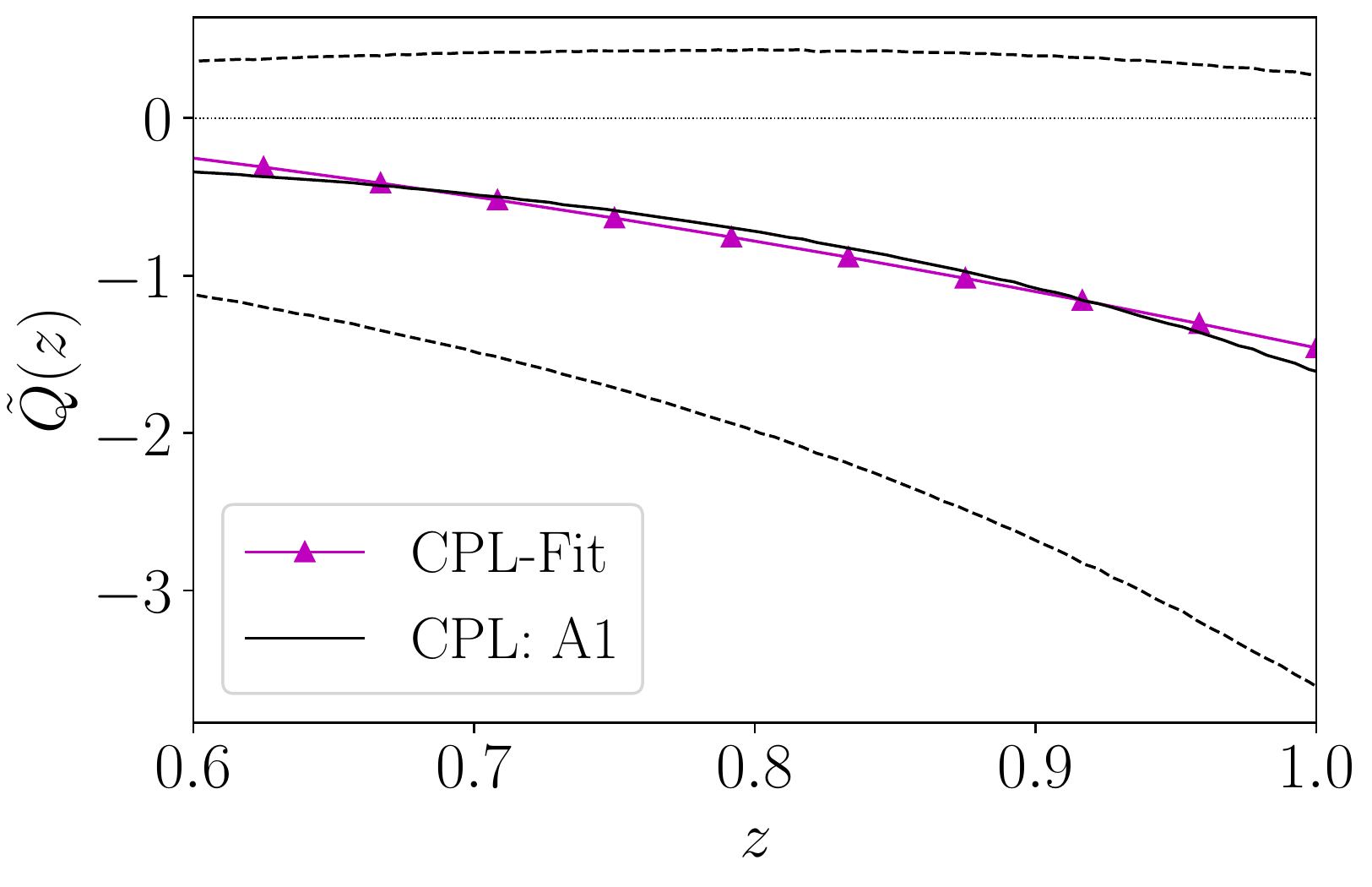}
	\end{center}
	\caption{{\small Plots showing a comparison between the reconstructed interaction $\tilde{Q}(z)$ and the estimated $\tilde{Q}_{\mbox{\tiny fit}}(z)$ 
			using the combined dataset A1, for EoS given by $w=-1$(left), $w$CDM model (middle) and the CPL parametrization (right). The black solid 
			line is the reconstructed function. The line with marker represents the best fit result from $\chi^2$-minimization. The 1$\sigma$ C.L.s 
			are shown in dashed lines.}}
	\label{ch5:Qfit_plot1}
\end{figure*}

\begin{figure*}[h!]
	\begin{center}
		\includegraphics[angle=0, width=0.325\textwidth]{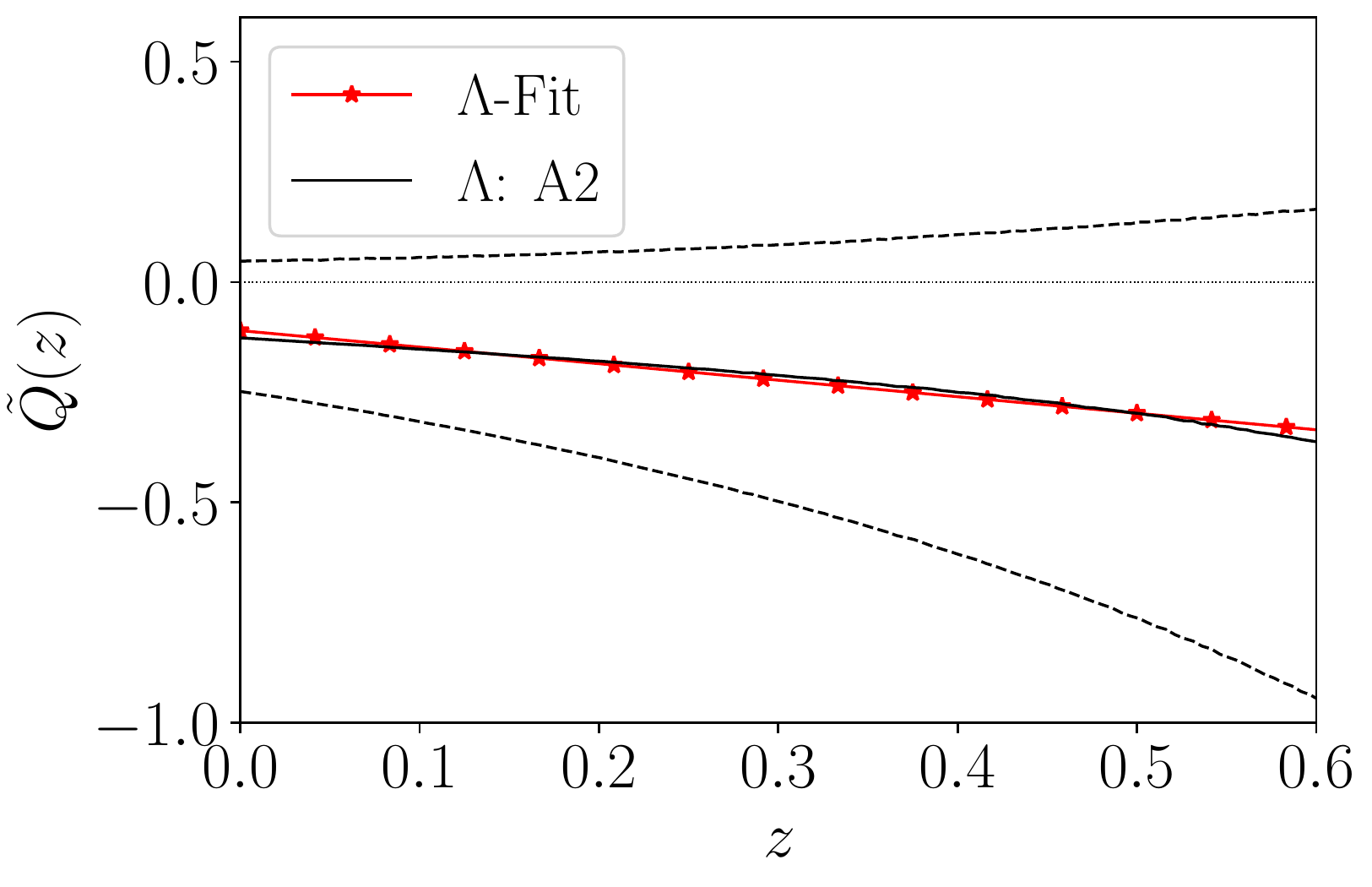}
		\includegraphics[angle=0, width=0.325\textwidth]{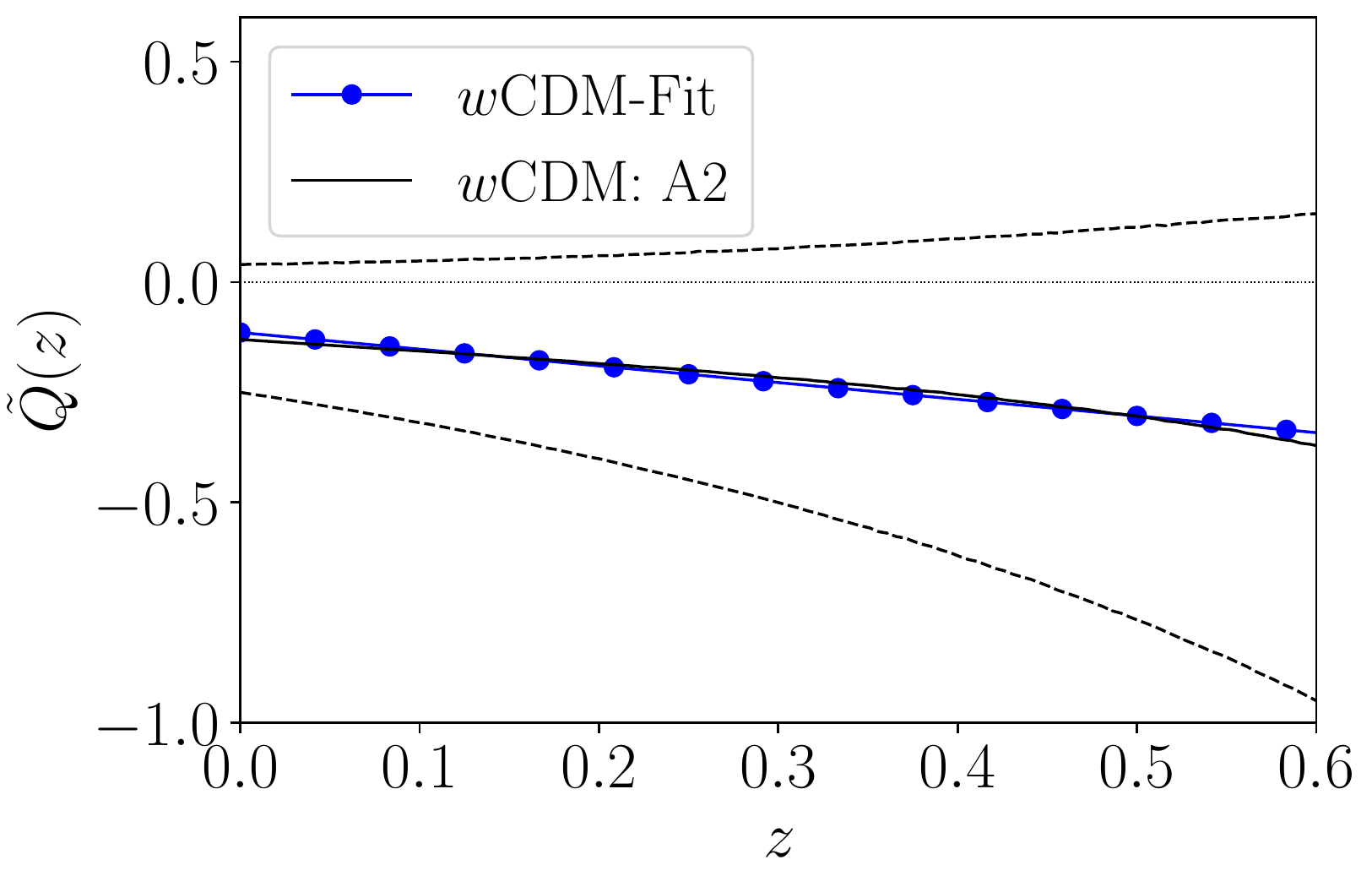}
		\includegraphics[angle=0, width=0.325\textwidth]{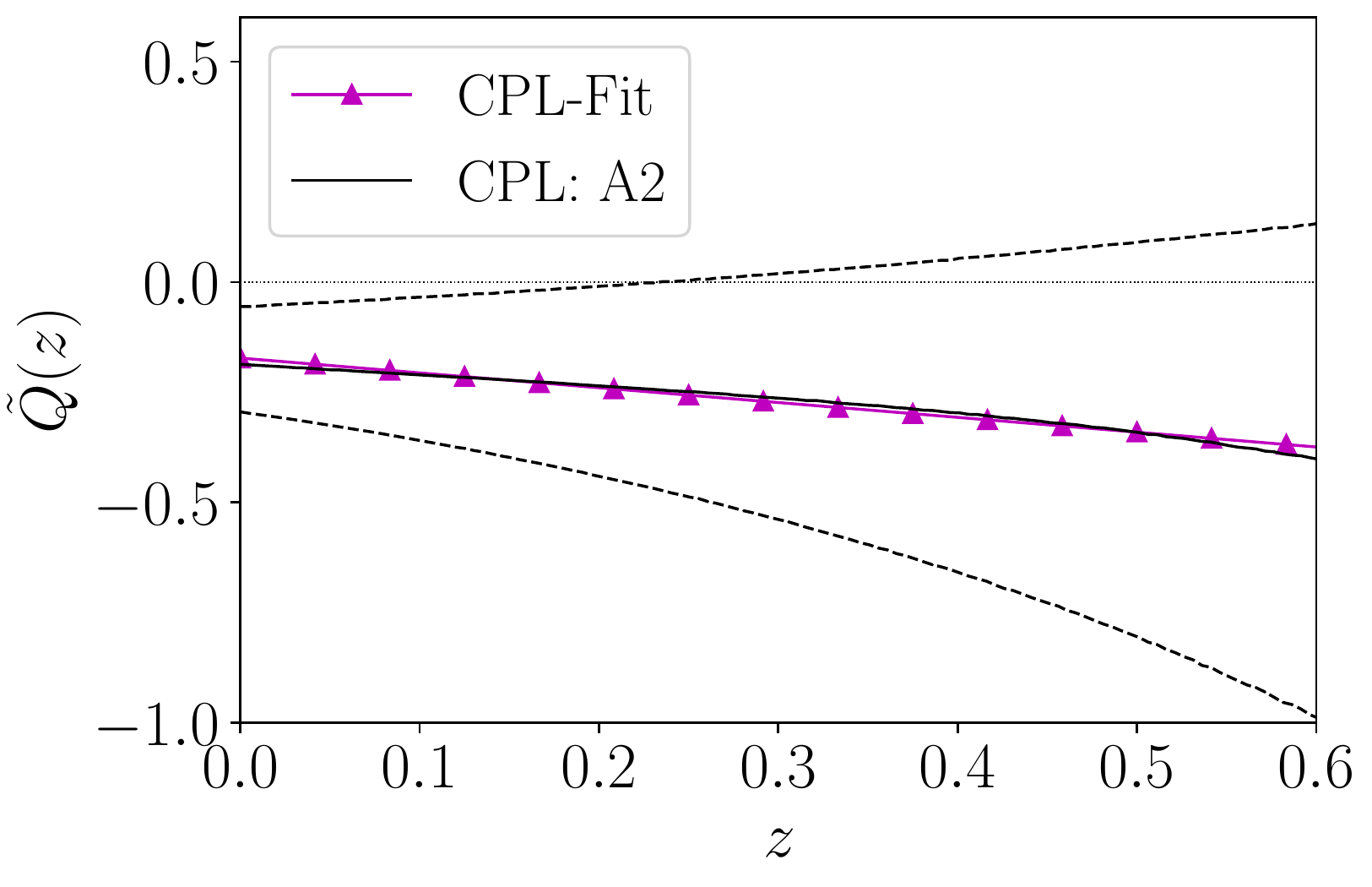}\\
		\includegraphics[angle=0, width=0.325\textwidth]{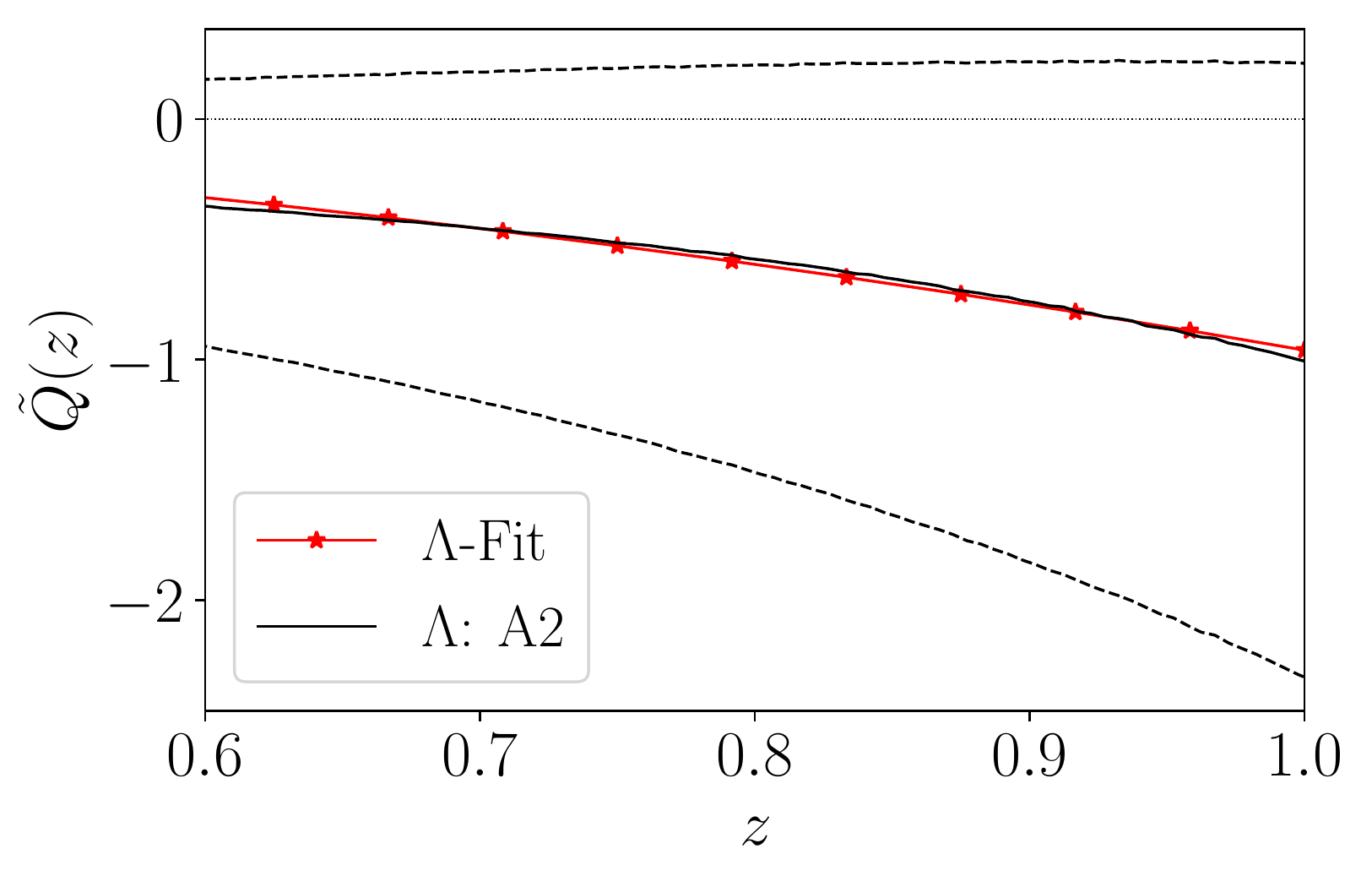}
		\includegraphics[angle=0, width=0.325\textwidth]{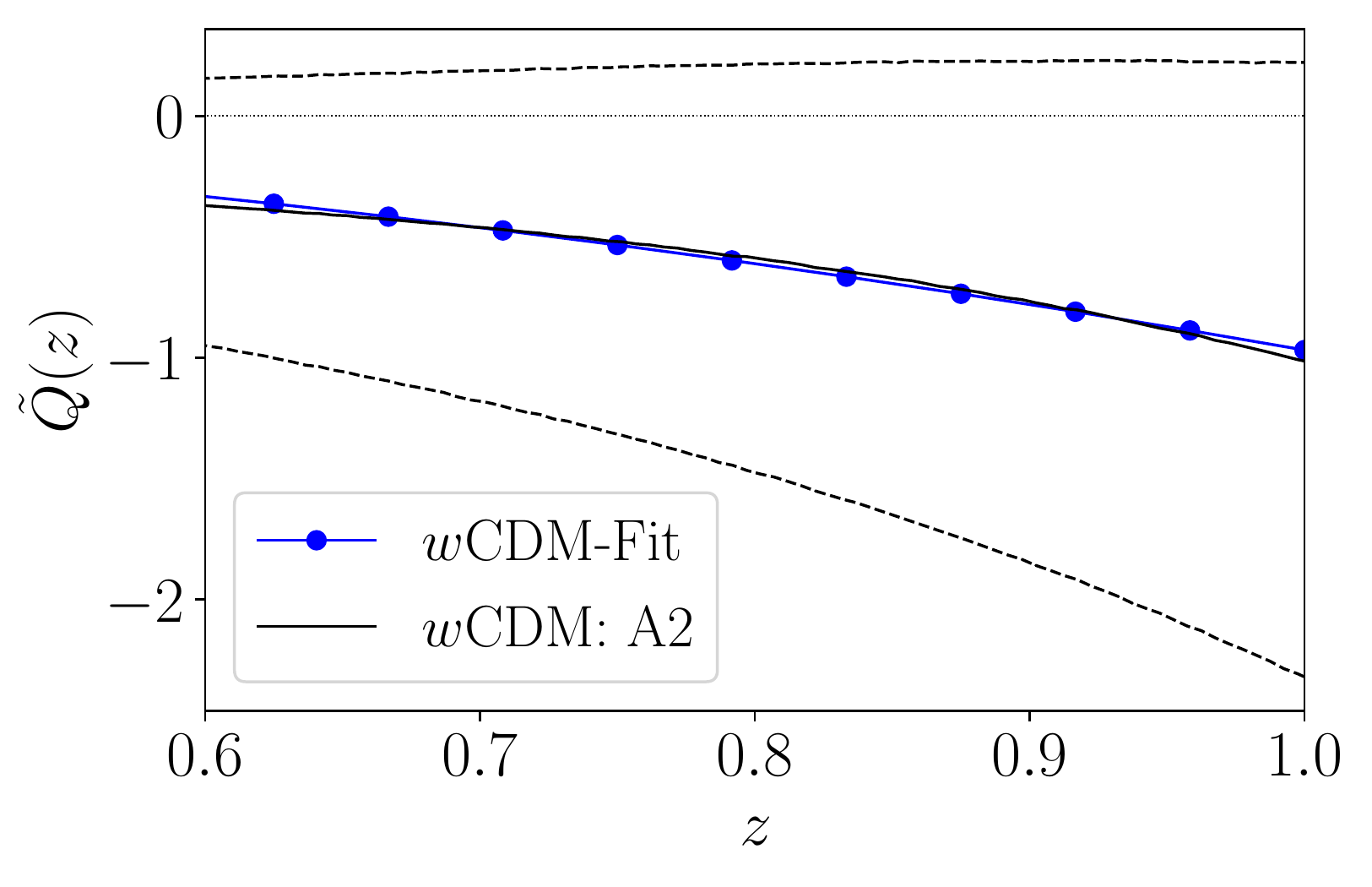}
		\includegraphics[angle=0, width=0.325\textwidth]{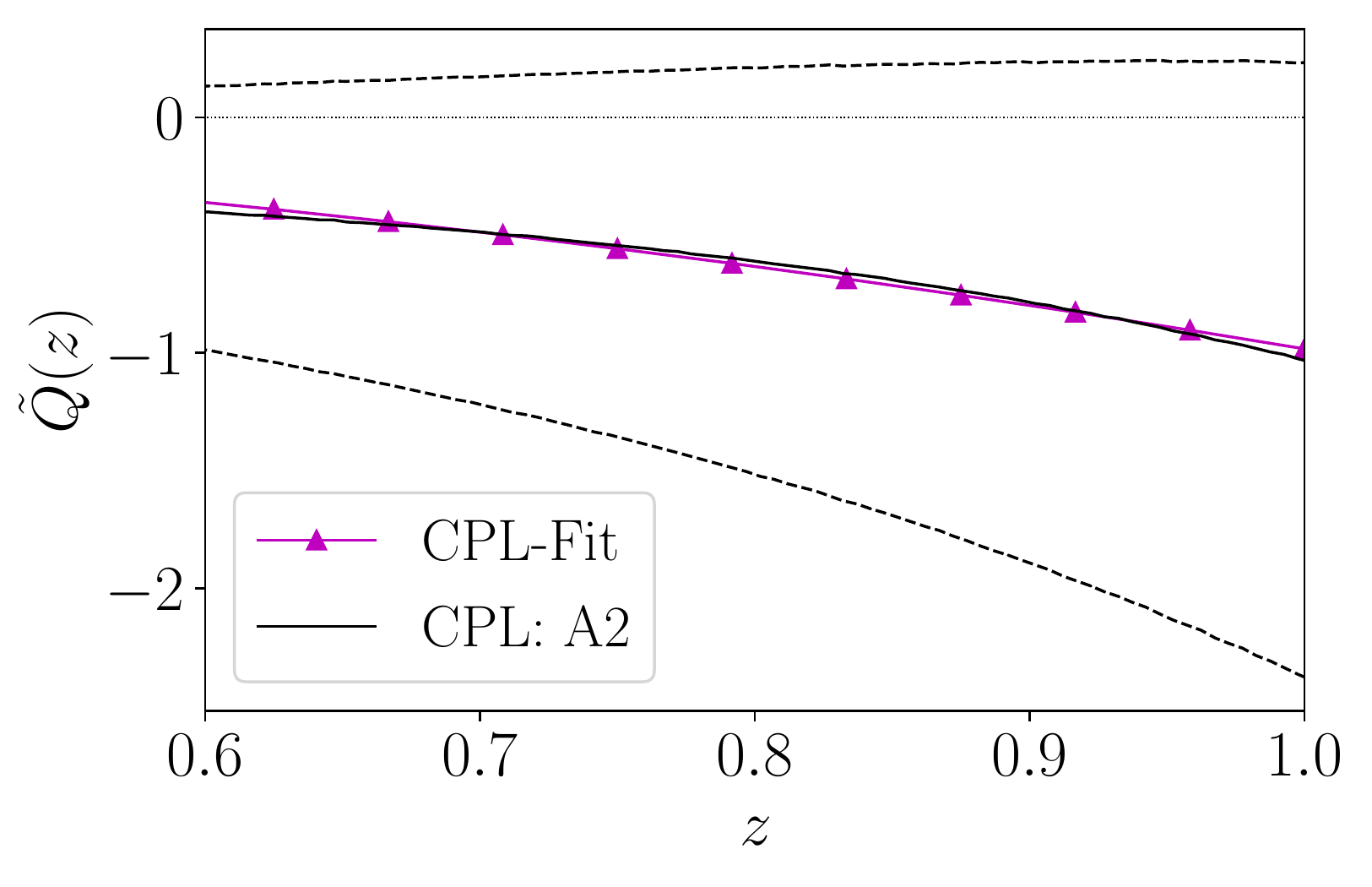}
	\end{center}
	\caption{{\small Plots showing a comparison between the reconstructed interaction $\tilde{Q}(z)$ and the estimated $\tilde{Q}_{\mbox{\tiny fit}}(z)$ 
			using the combined dataset A2, for EoS given by $w=-1$(left), $w$CDM model (middle) and the CPL parametrization (right). The black solid line 
			is the reconstructed function. The line with marker represents the best fit result from $\chi^2$-minimization. The 1$\sigma$ C.L.s are shown 
			in dashed lines.}}
	\label{ch5:Qfit_plot2}
\end{figure*}

\begin{figure*}[t!]
	\begin{center}
		\includegraphics[angle=0, width=0.325\textwidth]{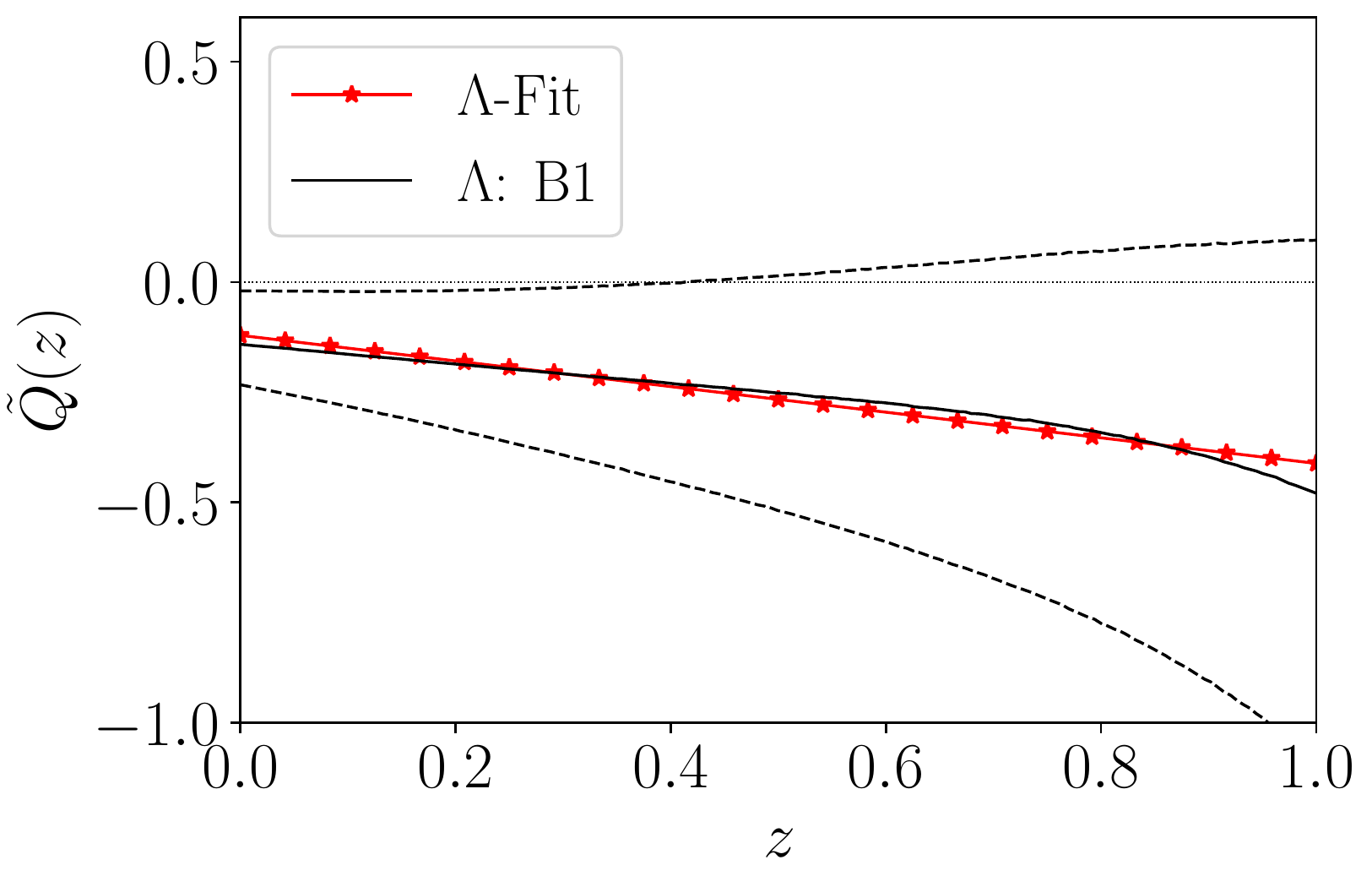}
		\includegraphics[angle=0, width=0.325\textwidth]{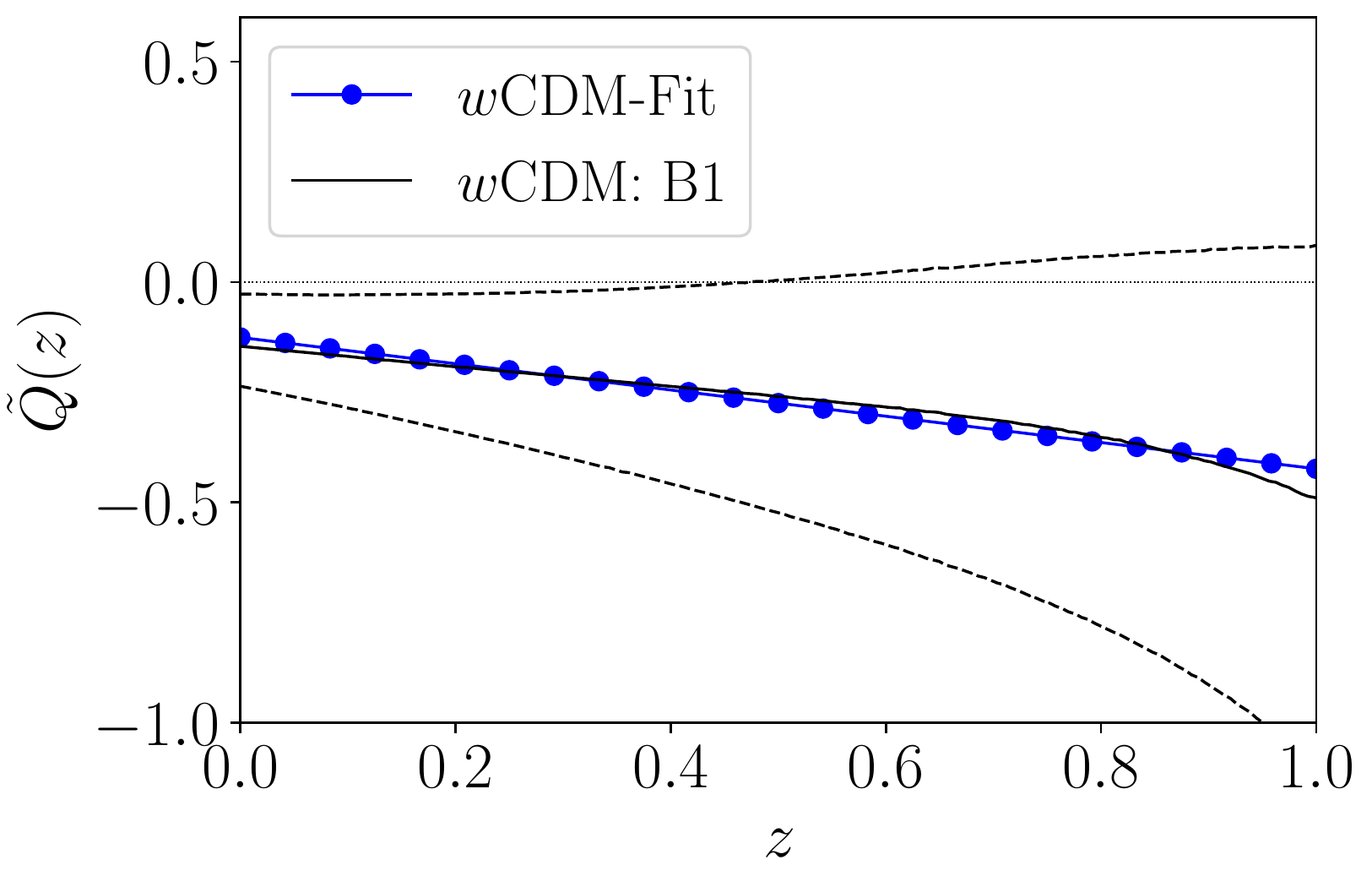}
		\includegraphics[angle=0, width=0.325\textwidth]{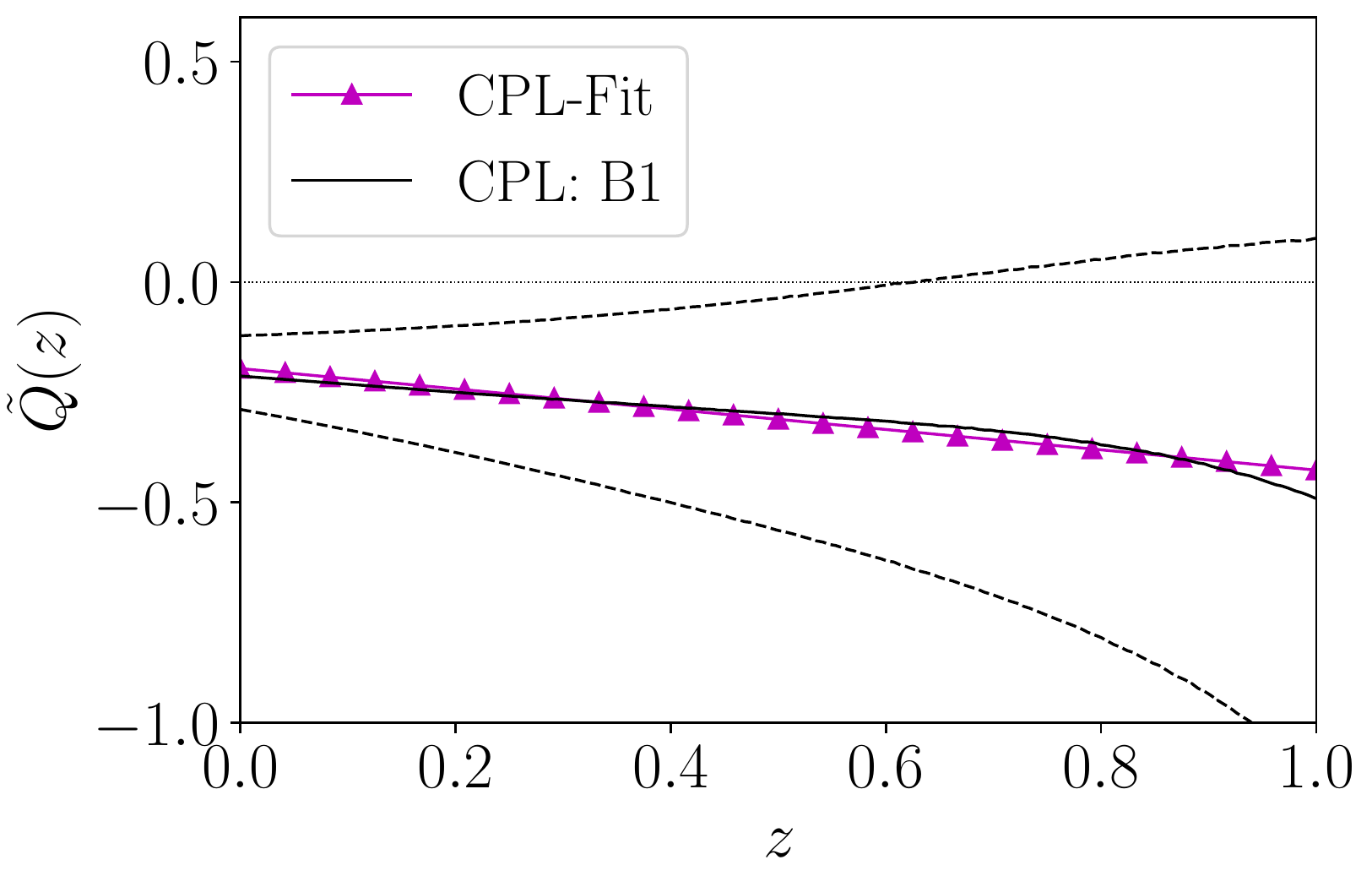}
	\end{center}
	\caption{{\small Plots showing a comparison between the reconstructed interaction $\tilde{Q}(z)$ and the estimated $\tilde{Q}_{\mbox{\tiny fit}}(z)$ 
			using the combined dataset B1, for EoS given by $w=-1$(left), $w$CDM model (middle) and the CPL parametrization (right). The black solid line 
			is the reconstructed function. The line with marker represents the best fit result from $\chi^2$-minimization. The 1$\sigma$ C.L.s are shown 
			in dashed lines.}}
	\label{ch5:Qfit_plot3}
\end{figure*}

\begin{figure*}[t!]
	\begin{center}
		\includegraphics[angle=0, width=0.325\textwidth]{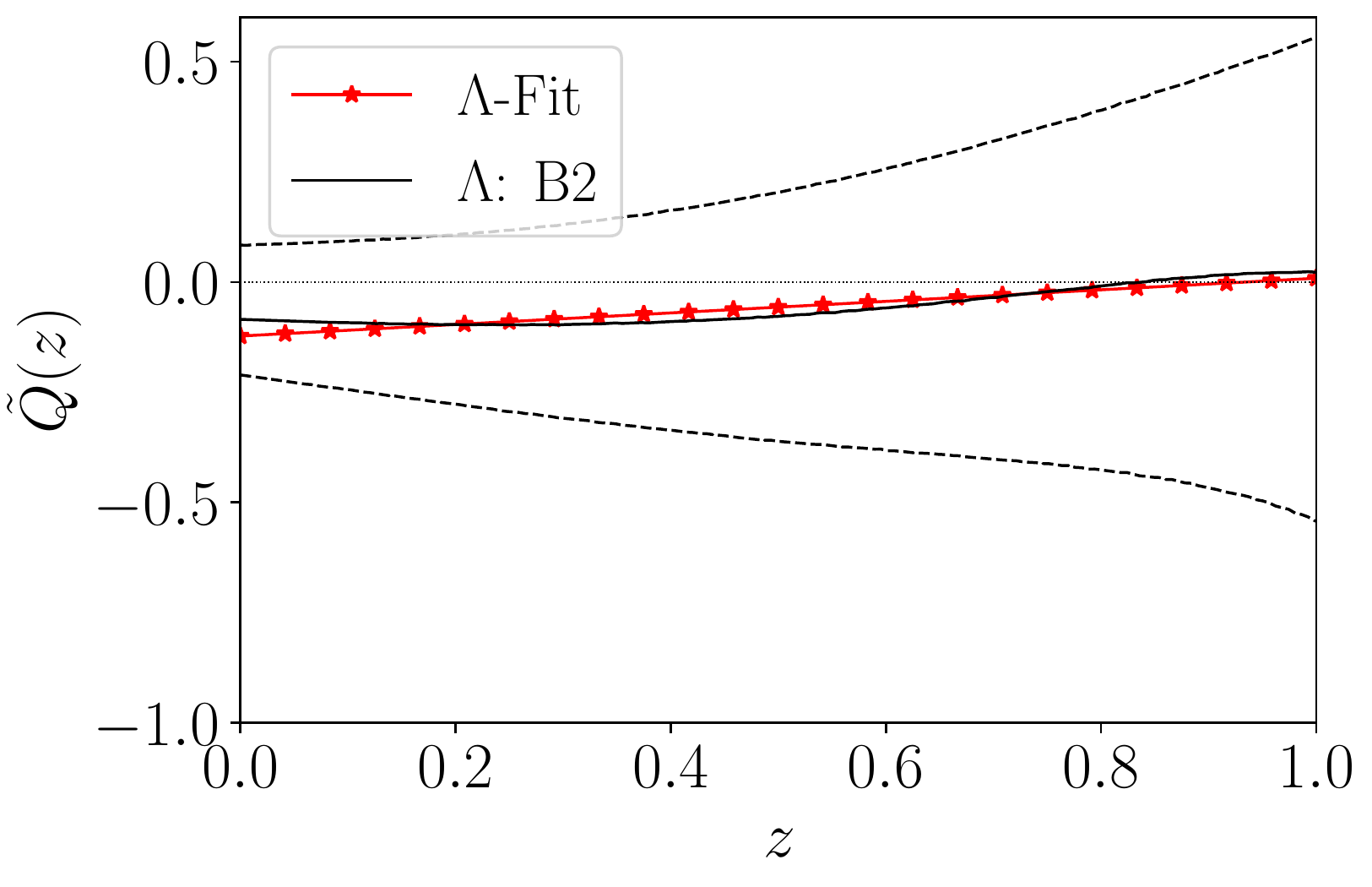}
		\includegraphics[angle=0, width=0.325\textwidth]{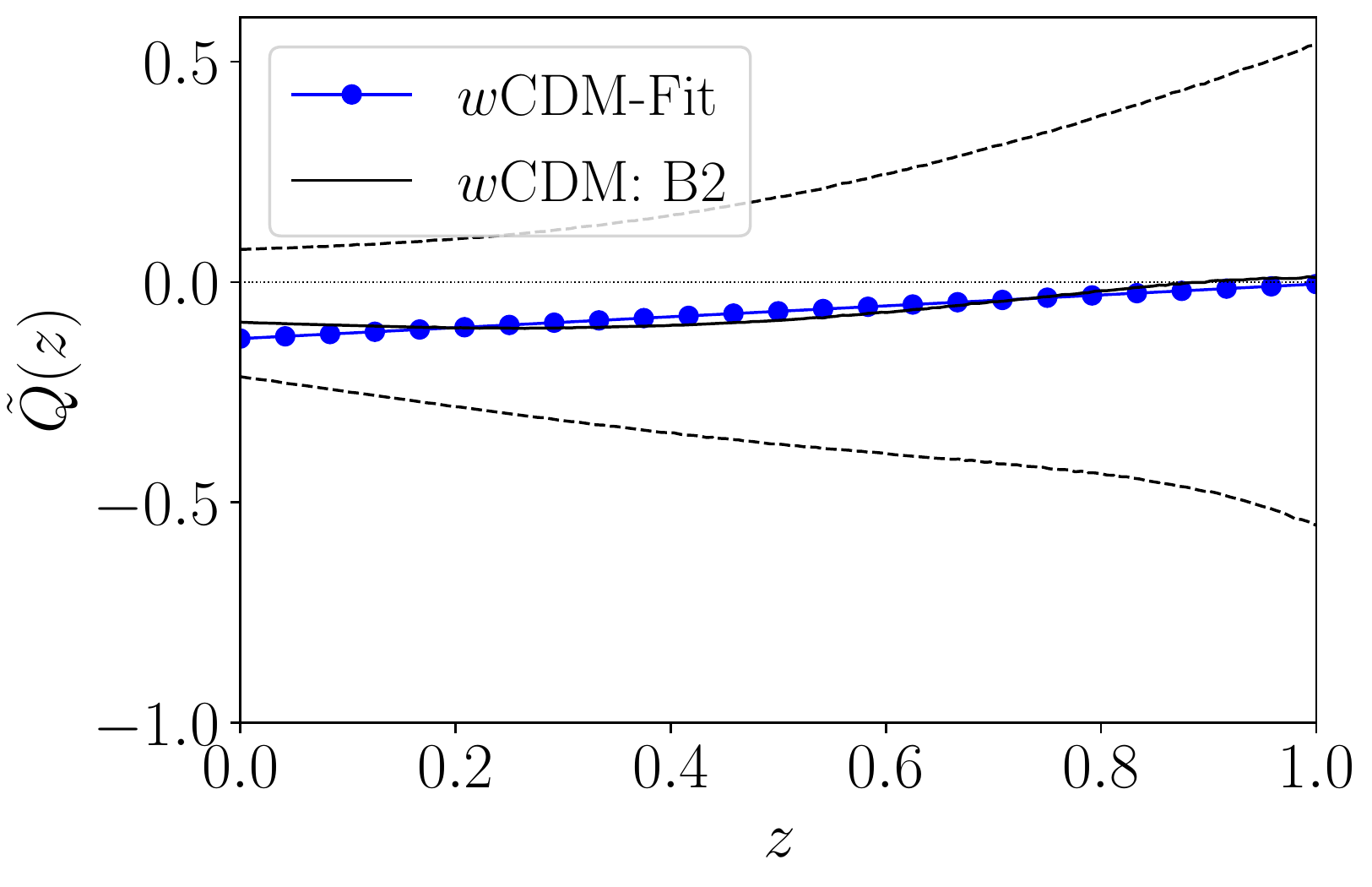}
		\includegraphics[angle=0, width=0.325\textwidth]{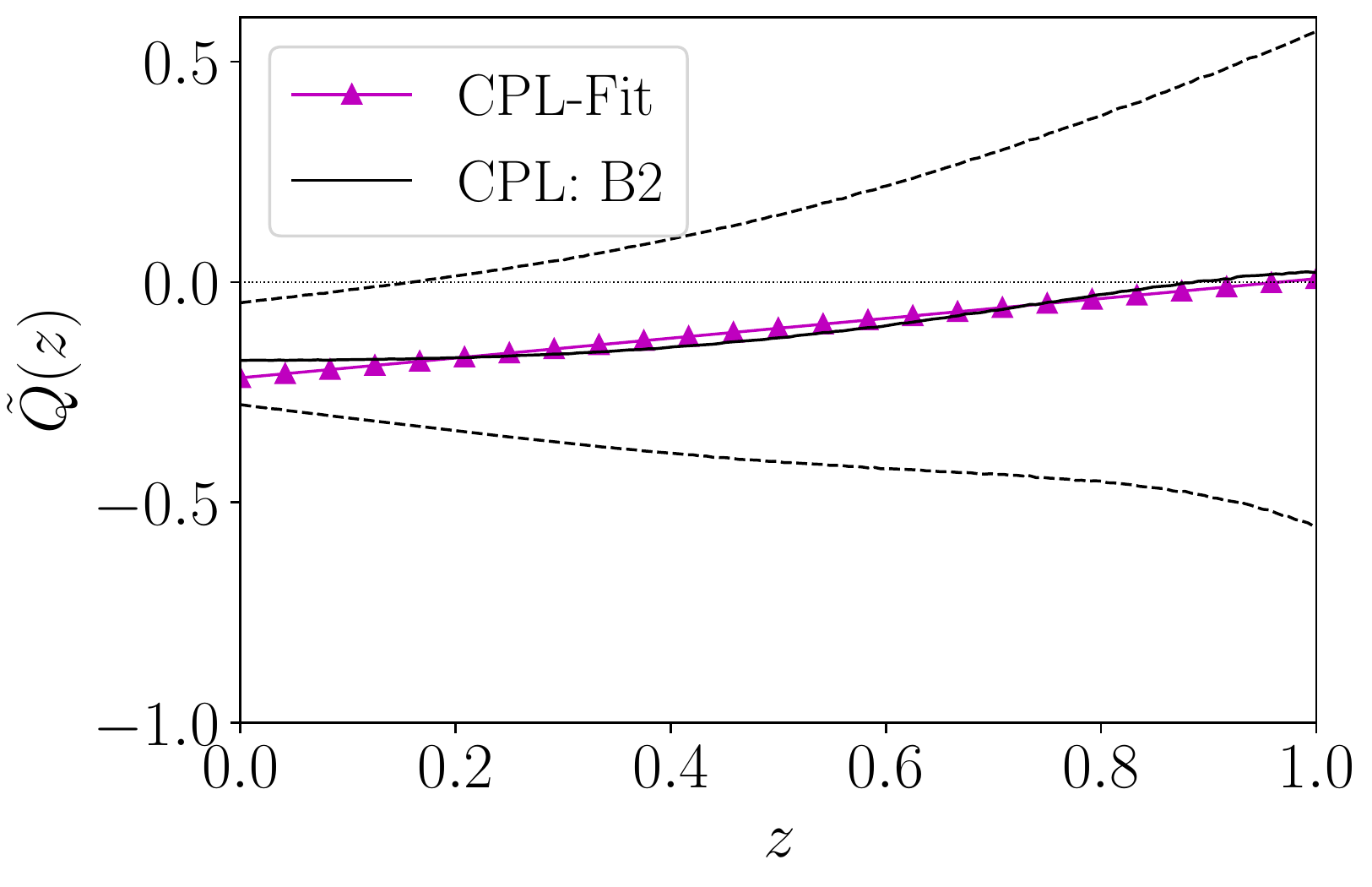}
	\end{center}
	\caption{{\small Plots showing a comparison between the reconstructed interaction $\tilde{Q}(z)$ and the estimated $\tilde{Q}_{\mbox{\tiny fit}}(z)$ 
			using the combined dataset B2, for EoS given by $w=-1$(left), $w$CDM model (middle) and the CPL parametrization (right). The black solid line 
			is the reconstructed function. The line with markers represents the best fit result from $\chi^2$-minimization. The 1$\sigma$ C.L.s are shown 
			in dashed lines.}}
	\label{ch5:Qfit_plot4}
\end{figure*}

The fitting is done using a trial and error estimation for different orders of $n$ in equation \eqref{Qfit}. The reduced $\chi_\nu^2= \frac{\chi^2}{\nu}$, where 
$\nu$ signifies the degrees of freedom, values are estimated. This procedure entails to go from order to order in the polynomial and getting the best-fitting 
$\chi_\nu^2$, and truncating once $\chi_\nu^2$ falls below unity to prevent any over fitting. The estimated values of the best fit $\tilde{Q}_i$'s along with 
their $1\sigma$ uncertainties are given in Tables \ref{ch5:qfit_tabA} and \ref{ch5:qfit_tabB}. A comparison between the reconstructed $\tilde{Q}(z)$ and estimated 
$\tilde{Q}_{\mbox{\tiny fit}}$, for various combinations of datasets are shown in Figures \ref{ch5:Qfit_plot1}, \ref{ch5:Qfit_plot2}, \ref{ch5:Qfit_plot3} and 
\ref{ch5:Qfit_plot4}.

For A1 dataset, in the redshift range $0<z<0.6$,
\begin{eqnarray}
\tilde{Q}_{\mbox{\tiny fit}}(z) &=& -0.060 -0.322~z   \mbox{ for $w=-1$} .\\
&=& -0.063 -0.324~z   \mbox{ for $w$CDM} .\\
&=& -0.114 -0.286~z  \mbox{ for CPL} .
\end{eqnarray} 

For A1 dataset, in the redshift range $0.6<z<1$,
\begin{eqnarray}
\tilde{Q}_{\mbox{\tiny fit}}(z) &=& 0.443 -1.865~z^2  \mbox{ for $w=-1$} .\\
&=& 0.438 -1.866~z^2  \mbox{ for $w$CDM} .\\
&=& 0.424 -1.882~z^2  \mbox{ for CPL} .
\end{eqnarray}

For A2 dataset, in the redshift range $0<z<0.6$,
\begin{eqnarray}
\tilde{Q}_{\mbox{\tiny fit}}(z) &=& -0.110 -0.375~z  \mbox{ for $w=-1$} .\\
&=& -0.115 -0.379~z  \mbox{ for $w$CDM} .\\
&=& -0.173 -0.336~z \mbox{ for CPL} .
\end{eqnarray} 

For A2 dataset, in the redshift range $0.6<z<1$,
\begin{eqnarray}
\tilde{Q}_{\mbox{\tiny fit}}(z) &=& 0.031 -0.992~z^2  \mbox{ for $w=-1$} .\\
&=& 0.025 -0.993~z^2  \mbox{ for $w$CDM} .\\
&=& -0.011 -0.973~z^2  \mbox{ for CPL} .
\end{eqnarray}

\begin{table*}[t!] 
	\caption{{\small Table showing the coefficient $\tilde{Q}_i$'s for best fit $\tilde{Q}_{\mbox{\tiny fit}} = \tilde{Q}_0 + \tilde{Q}_1 z $ in the redshift 
			range $0<z<0.6$ and $\tilde{Q}_{\mbox{\tiny fit}} = \tilde{Q}_0 + \tilde{Q}_2 z^2 $ in the redshift range $0.6<z<1$ for datasets A1 and A2.}}
	\begin{center}
		\resizebox{\textwidth}{!}{\renewcommand{\arraystretch}{1.3} \setlength{\tabcolsep}{12pt} \centering  
			\begin{tabular}{c c c c c c c } 
				\hline
				EoS & \textbf{Datasets} & $\tilde{Q}_0$ & $\tilde{Q}_1$ & \textbf{Datasets} & $\tilde{Q}_0$ & $\tilde{Q}_2$ \\
				\hline
				\hline
				$w=-1$ & A1 $(0<z<0.6)$ &  $-0.060^{+0.180}_{-0.180}$ & $-0.322^{+0.516}_{-0.514}$ & A1 $(0.6<z<1)$ &  $0.443^{+0.329}_{-0.383}$ &  $-1.865^{+0.558}_{-0.481}$  \\ 
				\hline				
				$w$CDM & A1 $(0<z<0.6)$ & $-0.063^{+0.180}_{-0.180}$ & $-0.324^{+0.516}_{-0.516}$ & A1 $(0.6<z<1)$ & $0.438^{+0.332}_{-0.383}$ &  $-1.866^{+0.558}_{-0.484}$   \\ 
				\hline
				CPL & A1 $(0<z<0.6)$ & $-0.114^{+0.180}_{-0.180}$ & $-0.286^{+0.515}_{-0.518}$ & A1 $(0.6<z<1)$ & $0.434^{+0.336}_{-0.385}$  &  $-1.883^{+0.561}_{-0.491}$  \\ 
				\hline
				$w=-1$ & A2 $(0<z<0.6)$ & $-0.110^{+0.180}_{-0.180}$ & $-0.375^{+0.517}_{-0.516}$ & A2 $(0.6<z<1)$ &  $0.031^{+0.400}_{-0.401}$ &  $-0.992^{+0.586}_{-0.583}$  \\ 
				\hline				
				$w$CDM & A2 $(0<z<0.6)$ & $-0.115^{+0.179}_{-0.180}$ & $-0.379^{+0.517}_{-0.516}$ & A2 $(0.6<z<1)$ & $0.025^{+0.401}_{-0.401}$ &  $-0.993^{+0.586}_{-0.583}$ \\ 
				\hline
				CPL & A2 $(0<z<0.6)$ & $-0.173^{+0.180}_{-0.179}$ & $-0.336^{+0.517}_{-0.517}$ & A2 $(0.6<z<1)$ & $-0.011^{+0.401}_{-0.400}$  &  $-0.973^{+0.582}_{-0.584}$  \\ 
				\hline
		\end{tabular}}
	\end{center}
	\label{ch5:qfit_tabA}
\end{table*}

For B1 dataset, in the redshift range $0<z<1$,
\begin{eqnarray}
\tilde{Q}_{\mbox{\tiny fit}}(z) &=& -0.121 -0.291~z \mbox{ for $w=-1$} .\\
&=& -0.126 -0.298~z \mbox{ for $w$CDM} .\\
&=& -0.196 -0.231~z \mbox{ for CPL} .
\end{eqnarray}

For B2 dataset, in the redshift range $0<z<1$,
\begin{eqnarray}
\tilde{Q}_{\mbox{\tiny fit}}(z) &=& -0.122 +0.131~z \mbox{ for $w=-1$} .\\
&=& -0.128 +0.124~z \mbox{ for $w$CDM} .\\
&=& -0.217 -0.225~z \mbox{ for CPL} .
\end{eqnarray}

\begin{table*}[t!] 
\caption{{\small Table showing the coefficient $\tilde{Q}_i$'s for best fit $\tilde{Q}_{\mbox{\tiny fit}} = \tilde{Q}_0 + \tilde{Q}_1 z $ in the redshift 
			range $0<z<1$ for datasets B1 and B2.}}
	\begin{center}
		\resizebox{\textwidth}{!}{\renewcommand{\arraystretch}{1.3} \setlength{\tabcolsep}{12pt} \centering  
			\begin{tabular}{c c c c c c c } 
				\hline
				EoS & \textbf{Datasets} & $\tilde{Q}_0$ & $\tilde{Q}_1$ & \textbf{Datasets} & $\tilde{Q}_0$ & $\tilde{Q}_1$ \\
				\hline
				\hline
				$w=-1$ & B1 $(0<z<1)$ &  $-0.121^{+0.153}_{-0.153}$ &  $-0.291^{+0.262}_{-0.263}$ & B2 $(0<z<1)$ &  $-0.122^{+0.153}_{-0.153}$ &  $0.131^{+0.263}_{-0.263}$  \\ 
				\hline				
				$w$CDM & B1 $(0<z<1)$ & $-0.126^{+0.152}_{-0.153}$ &  $-0.298^{+0.264}_{-0.262}$ & B2 $(0<z<1)$ & $-0.128^{+0.153}_{-0.153}$ &  $0.124^{+0.264}_{-0.263}$ \\ 
				\hline
				CPL & B1 $(0<z<1)$ & $-0.196^{+0.153}_{-0.153}$ &  $-0.231^{+0.263}_{-0.263}$ & B2 $(0<z<1)$ & $-0.217^{+0.153}_{-0.153}$  &  $0.225^{+0.263}_{-0.263}$ \\ 
				\hline
			\end{tabular}
		}
	\end{center}
	\label{ch5:qfit_tabB}
\end{table*}

On proceeding with any higher order polynomial, it is found that the fitted function is no longer contained within the $1\sigma$ error margin of 
$\tilde{Q}(z)$ reconstructed by GP.

\section{Evolution of the cosmological density parameters}

With the nature of the interaction function reconstructed, one can obtain the evolution of energy density parameters as well. The model is a 
spatially flat, homogenous and isotropic Universe where the total energy density is composed of only pressureless matter and dark energy. We 
define the density parameters $\Omega_i$'s as
\begin{eqnarray} 
\Omega_{m} &=& \frac{\tilde{\rho}_m}{E^2}, \label{ch5:omega_m}\\
\Omega_{D} &=& \frac{\tilde{\rho}_D}{E^2}, \label{ch5:omega_d}
\end{eqnarray} 
such that $\Omega_{m} + \Omega_{D} = 1$.

We make use of the equation \eqref{ch5:cons_reduced_m} and rewrite it as, 
\begin{eqnarray} 	
\frac{\dif \tilde{\rho}_m}{\dif z} - \frac{3 \tilde{\rho}_m}{1+z} &=& \frac{\tilde{Q}}{E (1+z)}. \label{ch5:rho_m}
\end{eqnarray}

One can see that \eqref{ch5:rho_m} is a linear first-order non-homogeneous differential equation of the form
\begin{equation}\label{ch5:lin1DE}
\frac{\dif y}{\dif z} + A(z)y = B(z) ,
\end{equation} 
with $A(z) = -\frac{3}{1+z}$ and $B(z) = \frac{\tilde{Q}}{E (1+z)}$. The integrating factor for \eqref{ch5:lin1DE} is given by $e^{\int A(z)\dif z}$, and the general solution is 
\begin{equation}\label{ch5:lin1DEsol}
y = e^{-\int A \dif z} \int\left(B e^{\int A \dif z} \right) \dif z+ C,
\end{equation} 
where $C$ is the constant of integration.

Thus, the solution for $\tilde{\rho}_m$ can be written as
\begin{eqnarray} \label{ch5:rho_msol}
\tilde{\rho}_m = \tilde{\rho}_{m0}(1+z)^{3} + {(1+z)^{3}}\int_{0}^{z}\frac{\tilde{Q}}{E } (1+z)^{-4} \dif z.
\end{eqnarray} 

If $\tilde{Q} = 0$, equation \eqref{ch5:rho_msol} reduces to the relation $\tilde{\rho}_{m} = \tilde{\rho}_{m0}(1+z)^3$ (where $\tilde{\rho}_{m0} 
= \tilde{\rho}_{m} (z=0)$) which is the standard evolution scenario for a pressureless matter that evolves independently. Thus, one can rewrite the 
density parameters with the help of the equations \eqref{ch5:omega_m} and \eqref{ch5:omega_d} as
\begin{eqnarray} \label{ch5:omega_msol}
\Omega_{m} &=& 	\frac{\tilde{\rho}_{m0}(1+z)^{3}}{E^2} + \frac{(1+z)^{3}}{E^2}\int_{0}^{z}\frac{\tilde{Q}}{E (1+z)^{4}} ~ \dif z,~~~~~\\
\Omega_{D} &=& 1 - \Omega_{m}.
\end{eqnarray}

With the smooth functions of $E(z)$ and $\tilde{Q}(z)$ reconstructed from the combined datasets, we use the trapezoidal rule \cite{trapez} to 
calculate the integral
\begin{eqnarray} \label{ch5:f_gp}
f(z) &=& \int_{0}^{z}\frac{\tilde{Q}}{E } (1+z)^{-4} ~\dif z\nonumber \\
&=& \int_{0}^{z} g(z)~ \dif z\nonumber \\
&\approx& \frac{1}{2} \mathlarger{\mathlarger{\sum}}_{i=0}^{n} (z_{i+1} - z_i) \left[{g(z_{i+1})}+{g(z_{i})}\right],
\end{eqnarray} 
where $g(z) = \frac{\tilde{Q}}{E } (1+z)^{-4}$. 

The uncertainty in $f(z)$ is obtained by the error propagation formula,
\begin{equation} \label{ch5:sigf_gp}
\sigma^2_f =  \frac{1}{4}  \mathlarger{\mathlarger{\sum}}_{i=0}^{n}\left(z_{i+1} - z_i\right)^2 \left[\sigma^2_{g_{i+1}}+\sigma^2_{g_{i}}\right],
\end{equation} 
where contribution from uncertainties in $\tilde{Q}$ and $E$ have been included.

\begin{figure*}[t!]
	\begin{center}
		\includegraphics[angle=0, width=\textwidth]{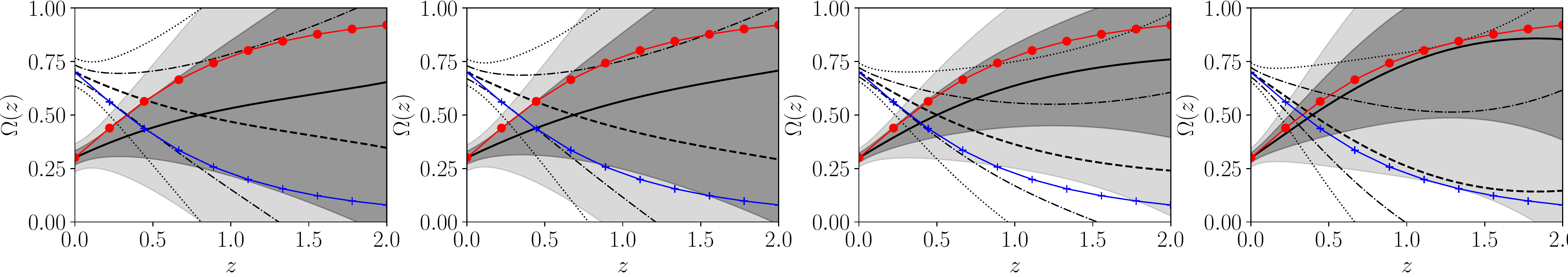}\\
		\includegraphics[angle=0, width=\textwidth]{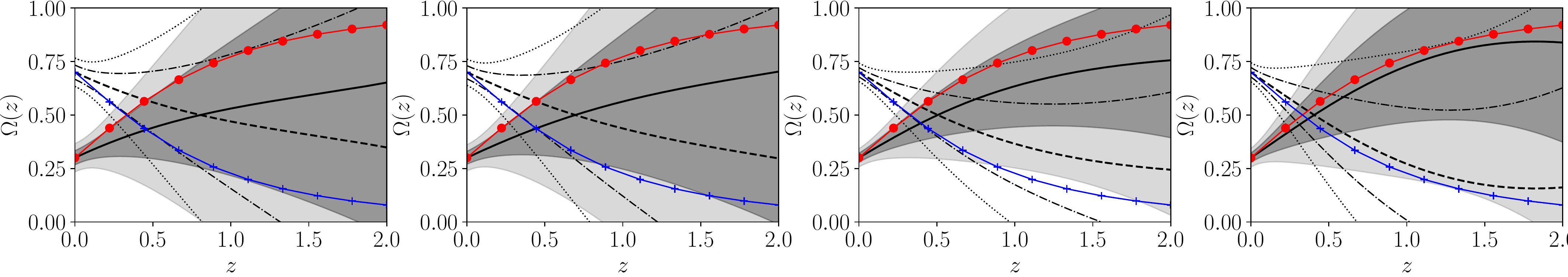}\\
		\includegraphics[angle=0, width=\textwidth]{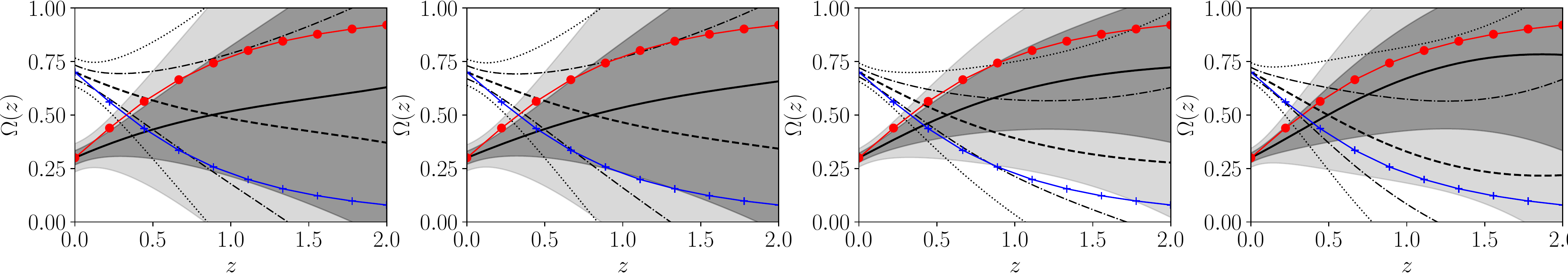}
	\end{center}
	\caption{{\small Plots for the dark energy density $\Omega_D$ and the matter density $\Omega_{m}$ from the dataset samples of Set A1 (column 1), 
			Set A2 (column 2), Set B1 (column 3) and Set B2 (column 4) using a squared exponential covariance function considering the decaying 
			dark energy EoS given by $w = -1$ (top row), the $w$CDM model with DE EoS given by $w = -1.006 \pm 0.045$ \cite{planck_cmb} (middle row), 
			and the CPL parametrization of dark energy with EoS given by $w(z) = w_0 + w_a (\frac{z}{1+z})$, $w_0 = -1.046^{+0.179}_{-0.170}$ and 
			$w_a = 0.14^{+0.60}_{-0.76}$ \cite{suzuki} (bottom row). The black solid curve corresponds to $\Omega_m$ while the black dashed line 
			represents $\Omega_{D}$. The 1$\sigma$ and 2$\sigma$ uncertainties in $\Omega_{m}$ is shown by the dark and light shaded regions, and 
			those of $\Omega_{D}$ is given by the region bounded with dashed-dotted and dotted lines respectively. The line drawn with circles 
			represents $\Omega_{m}$ and the line with cross markers is that of $\Omega_{D}$, for the $\Lambda$CDM model.}}
	\label{ch5:omega_sqexp}
\end{figure*}

\begin{figure*}[t!]
	\begin{center}
		\includegraphics[angle=0, width=\textwidth]{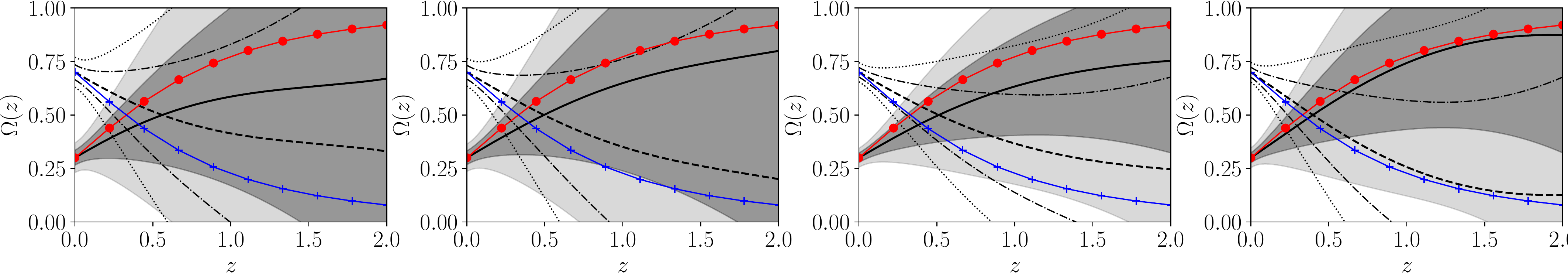}\\
		\includegraphics[angle=0, width=\textwidth]{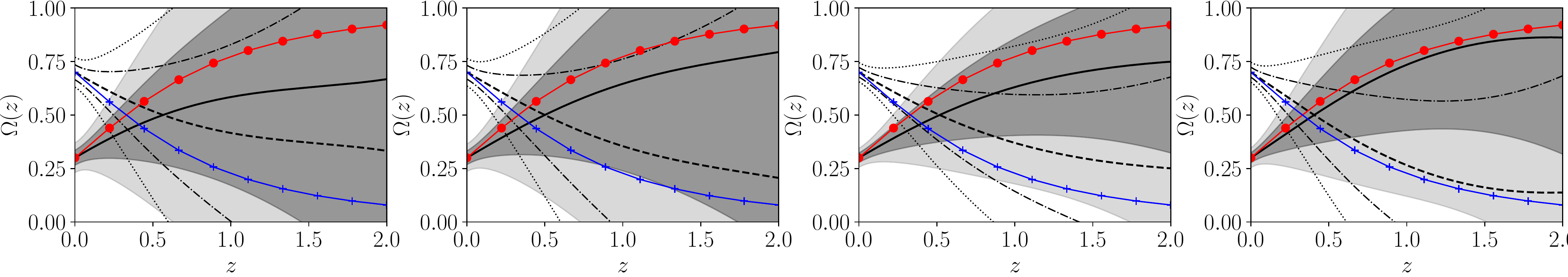}\\
		\includegraphics[angle=0, width=\textwidth]{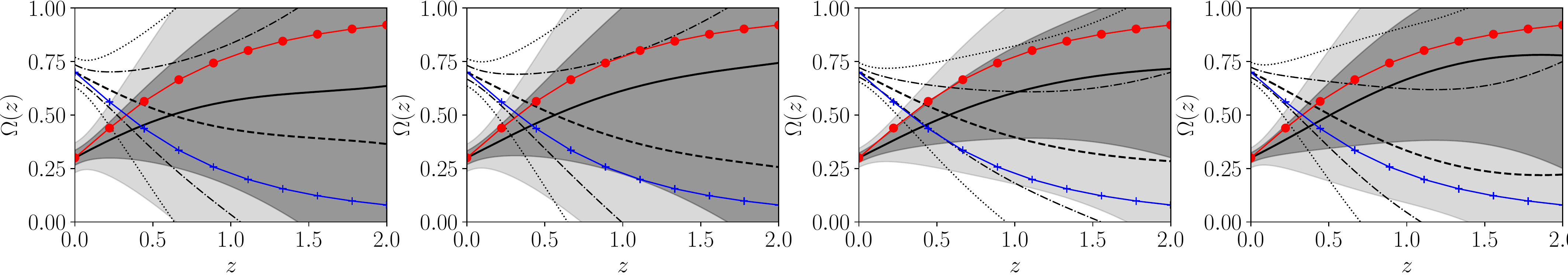}
	\end{center}
	\caption{{\small Plots for the dark energy density $\Omega_D$ and the matter density $\Omega_{m}$ from the dataset samples of Set A1 (column 1), Set 
			A2 (column 2), Set B1 (column 3) and Set B2 (column 4) using a Mat\'{e}rn 9/2 covariance function considering the decaying dark energy EoS 
			given by $w = -1$ (top row), the $w$CDM model with DE EoS given by $w = -1.006 \pm 0.045$ \cite{planck_cmb} (middle row), and the CPL 
			parametrization of dark energy with EoS given by $w(z) = w_0 + w_a (\frac{z}{1+z})$, $w_0 = -1.046^{+0.179}_{-0.170}$ and 
			$w_a = 0.14^{+0.60}_{-0.76}$ \cite{suzuki} (bottom row). The black solid curve corresponds to $\Omega_m$ while the black dashed line represents 
			$\Omega_{D}$. The 1$\sigma$ and 2$\sigma$ uncertainties in $\Omega_{m}$ is shown by the dark and light shaded regions, and that of $\Omega_{D}$ 
			is given by the region bounded with dashed-dotted and dotted lines respectively. The line drawn with circles represents $\Omega_{m}$ and the 
			line with cross markers is that of $\Omega_{D}$, for the $\Lambda$CDM model.}}
	\label{ch5:omega_mat92}
\end{figure*}


We plot the density parameters $\Omega_{m}$ and $\Omega_{D}$ using the equations \eqref{ch5:omega_m} and \eqref{ch5:omega_d}. We choose the value of 
$\tilde{\rho}_{m0} = 0.3$, i.e., $\Omega_{m0} = \frac{0.3}{E(0)^2}$. The plots are shown in Fig. \ref{ch5:omega_sqexp} and \ref{ch5:omega_mat92} for 
both choices of the covariance function. 

For the three different choices of the interacting dark energy models, the evolution of the density parameters is qualitatively similar and not too 
sensitive to the choice of the datasets. This feature hardly depends on the choice of the covariance function, only except the fact 
the use of Mat\'{e}rn $9/2$ covariance function brings the transition to dark energy dominance a bit closer to $z=0.5$. One intriguing common feature 
to note is that for the interacting models, $\Omega_D$ takes over as the dominant role over $\Omega_m$ later in the evolution (closer to $z=0$) 
compared to the corresponding $\Lambda$CDM model. 

One can note that the transition from a matter-dominated phase to a dark energy dominated phase occurs within the redshift range $0.5<z<1$. 

\section{Thermodynamics of the Interaction}

For studying the thermodynamic properties of the model, we consider the Universe as a system  bounded by some cosmological horizon and the 
matter content of the Universe is enclosed within a volume defined by a radius not bigger than the horizon. This idea primarily originated from the 
consideration of black hole thermodynamics, which is equally valid for a cosmological horizon\cite{gibbons, jacob, padma}. However, in an evolving 
scenario like cosmology, an apparent horizon is more relevant than an event horizon. An apparent horizon is determined by 
the equation  
\begin{equation}
	g^{\mu\nu} R_{,\mu} R_{,\nu} = 0, \label{ch5:apph}
\end{equation}
where, $R = a(t)r$ is the proper radius of the 2-sphere and $r$ is the comoving radius. 

For a spatially flat FLRW Universe, equation \eqref{ch5:apph} tells us that the apparent horizon ($r_h$) is in fact the Hubble horizon, 
\begin{equation}
r_h = \frac{1}{H}.
\end{equation}
This serves the purpose for recovering the first law of thermodynamics. For a comprehensive description, we refer to the work of Ferreira and 
Pav\'{o}n\cite{diego2}, and the monograph by Faraoni\cite{valerio}. 

For the second law to be valid, the entropy $S$ should be non-decreasing with respect to the expansion of the Universe. If $S_f$, $S_h$ stand for entropy of the 
fluid and that of the horizon containing the fluid respectively, then the total entropy of the system, given by $S = S_f + S_h$, should satisfy the relation 
\begin{equation}\label{ch5:cond1}
\frac{\dif S}{\dif x} \geq 0,
\end{equation} 
where $x = \ln a$, $a$ being the scale factor of the Universe. 

For an approach to equilibrium, the condition is 
\begin{equation}\label{ch5:cond2}
\frac{\dif ^2 S}{\dif x^2} < 0. 
\end{equation}

With the apparent horizon as the cosmological horizon, the entropy of the horizon $S_h$ can be written as \cite{bak}
\begin{equation} \label{ch5:Sh}
S_h =  8 \pi^2 r_h^2 = \frac{8 \pi^2}{H^2}.
\end{equation}

Further, the temperature of the dynamical apparent horizon is related to the horizon radius by,
\begin{equation} \label{ch5:HKtemp}
T_h = \frac{1}{2 \pi r_h} \left[ 1 - \frac{\dot{r_h}}{2 H r_h}\right] = \frac{2 H^2 + \dot{H}}{4 \pi H}~,
\end{equation} 
known as the Hayward-Kodama temperature \cite{hktemp1, hktemp2}. As the cosmological horizon is evolving, Hawking temperature is replaced 
by Hayward-Kodama temperature \cite{valerio}. 

We consider $S_m$, $S_D$ as the entropies of the matter sector and the dark energy sector, such that  $S_f = S_m + S_D$. If $T$ is the temperature of composite matter distribution inside the horizon, then the first law of thermodynamics, $T\dif S = \dif E + p~\dif V$, can be recast for 
the individual sectors in the following forms 
\begin{eqnarray} 
T \dif S_m &=& \dif E_m + p_m \dif V = \dif E_m , \label{ch5:tdsm} \\
T \dif S_D &=& \dif E_D + p_D \dif V, \label{ch5:tdsd}
\end{eqnarray} 
where $V = \frac{4}{3} \pi r_h^3 = \frac{4 \pi}{3 H^2}$, is the fluid volume. $E_m$, $E_D$ represent the internal energies of the dark matter 
and energy components given by $E_m = \frac{4}{3}\pi r_h^3 \rho_{m} = \rho_{m} V$ and $E_D = \frac{4}{3}\pi r_h^3 \rho_{D} = \rho_{D} V$ respectively. 
Now, differentiating equations \eqref{ch5:Sh}, \eqref{ch5:tdsm} and \eqref{ch5:tdsd} with respect to the cosmic time $t$ along with the assumption $T$ should 
be equal to $T_h$ (equation \eqref{ch5:HKtemp}), we get
\begin{eqnarray}
\dot{S_m} + \dot{S_D} &=& 16 \pi^2 \frac{\dot{H}}{H^3} \left(1+\frac{\dot{H}}{2H^2+\dot{H}}\right), \\
\dot{S_h} &=& -16 \pi^2 \frac{\dot{H}}{H^3}.
\end{eqnarray}

Therefore
\begin{equation} \label{ch5:Sdot}
\dot{S} = \dot{S_m} + \dot{S_D} +\dot{S_h} = 16 \pi^2 \frac{\dot{H}^2}{H^3} \left(\frac{1}{2H^2+\dot{H}}\right).
\end{equation}

It requires mention that it may not always be justified to assume the fluid temperature equal to the horizon temperature. This assumption is particularly unjustified for a radiation distribution that obeys Stefan's law. However, for a pressureless matter, the equality of $T$ and $T_h$ is valid, and this equality is approximately correct for dark energy. Thus in the present context, our assumption is not at all drastic. For an account of this justification, we refer to the work of Mimoso and Pav\'{o}n\cite{mimoso}. 

The relation \eqref{ch5:Sdot} can be written with $x$ as the argument, where $x = \ln ~a = -\ln (1+z)$, as 
\begin{equation} \label{ch5:Sdiffx}
\frac{\dif S}{\dif x} = \frac{16 \pi^2}{H^4} \left(\frac{\dif H}{\dif x}\right)^2 \Psi(x),
\end{equation} 
where 
\begin{equation}
\Psi(x) = \left[2 + \frac{1}{H} \frac{\dif H}{\dif x}\right]^{-1}.
\end{equation}

Again on differentiating equation \eqref{ch5:Sdiffx} w.r.t. $x$, one obtains
\begin{equation} \label{ch5:Sdiff2x}
\frac{\dif^2 S}{\dif x^2} = \frac{16 \pi^2 \Psi^2}{H^4} \left( \frac{\dif H}{\dif x}\right)^2 \Phi(x),
\end{equation} 
where
\begin{equation}
\Phi = \frac{1}{H}\frac{{\dif}^2 H}{\dif x^2} - \frac{3}{H^2}\left(\frac{\dif H}{\dif x}\right)^2 + \frac{4}{\frac{\dif H}{\dif x}} \frac{{\dif}^2 H}{\dif x^2} - \frac{8}{H} \frac{\dif H}{\dif x} .
\end{equation}

\begin{figure*}[t!] 
	\begin{center}
		\includegraphics[angle=0, width=\textwidth]{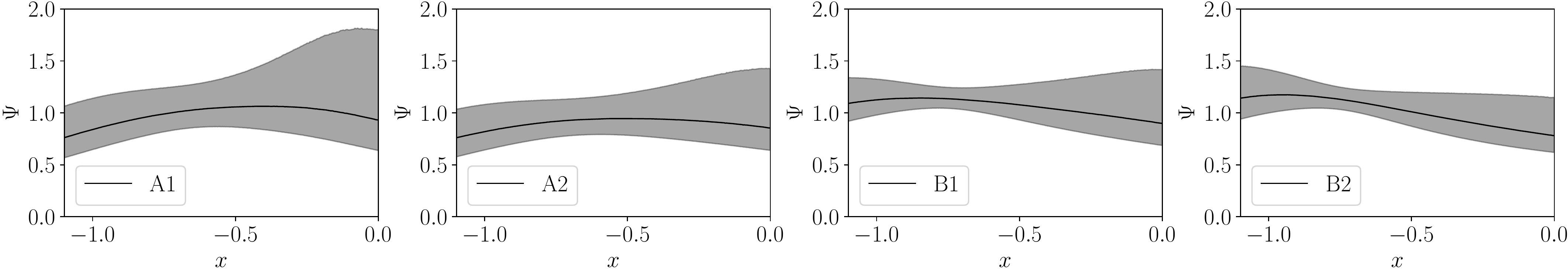}\\
		\includegraphics[angle=0, width=\textwidth]{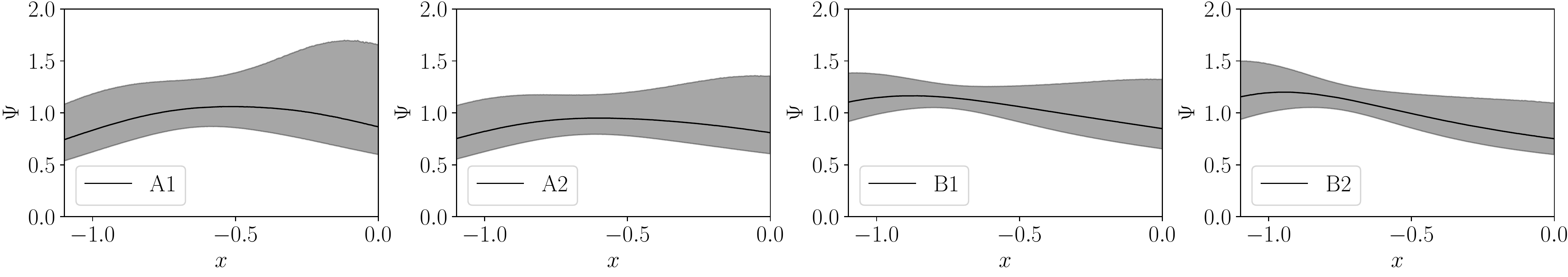}
	\end{center}
	\caption{{\small Plots for $\Psi$ from the dataset samples of Set A and B using a Squared Exponential covariance (top row) and the Mat\'{e}rn $9/2$ 
			covariance function. The solid black line gives the best fit values of $\Psi$. The shaded region correspond to the 1$\sigma$ uncertainty.}}
	\label{ch5:psi_plot}
\end{figure*}

\begin{figure*}[t!]
	\begin{center}
		\includegraphics[angle=0, width=\textwidth]{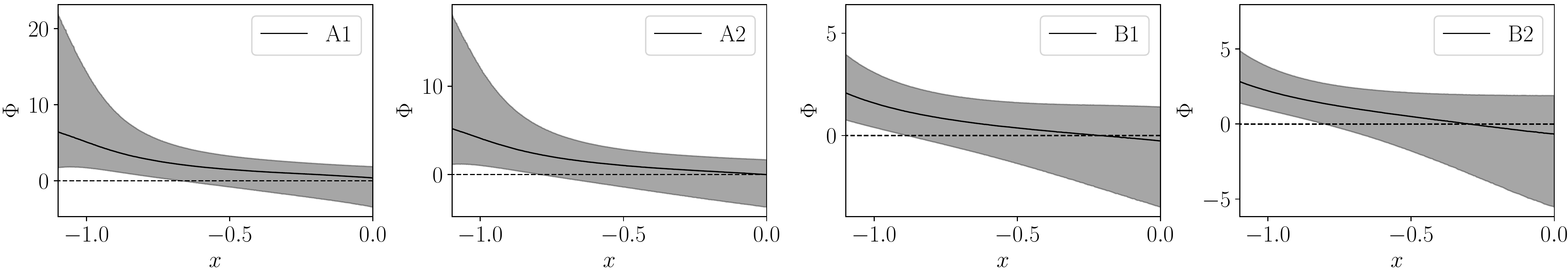}\\
		\includegraphics[angle=0, width=\textwidth]{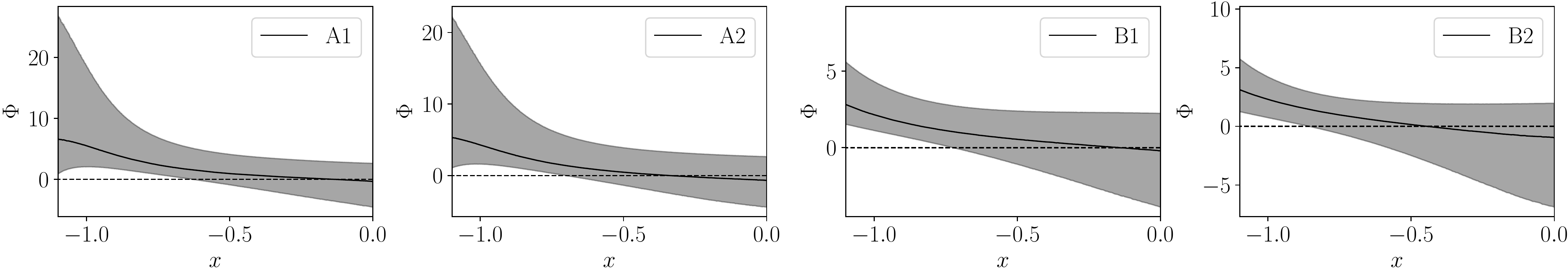}
	\end{center}
	\caption{{\small Plots for $\Phi$ from the dataset samples of Set A and B using a Squared Exponential covariance (top row) and the Mat\'{e}rn $9/2$ 
			covariance function. The solid black line gives the best fit values of $\Phi$. The shaded region correspond to the 1$\sigma$ uncertainty.}}
	\label{ch5:phi_plot}
\end{figure*}

From equation \eqref{ch5:Sdiffx}  we see that for the inequality (\ref{ch5:cond1}) to hold, the required condition is $\Psi \geq 0$. Equation 
\eqref{ch5:Sdiff2x} shows that condition (\ref{ch5:cond2}) shall be satisfied, provided $\Phi<0$. We find the behaviour of $\Psi$ and 
$\Phi$ by plotting them, in figure \ref{ch5:psi_plot} and \ref{ch5:phi_plot} respectively, as functions of $x$ where $x$ = $-\ln(1+z)$. 

The plots in figure \ref{ch5:psi_plot} show that $\Psi$ remains positive in 1$\sigma$ throughout the domain of reconstruction $0<z<2$. Therefore, the second 
law of thermodynamics is indeed satisfied for the reconstructed scenario. In figure \ref{ch5:phi_plot}, the plots reveal that $\Phi$ was positive in 
the past, but as we approach the present epoch, the value of $\Phi$ decreases gradually and changes its signature. The best fit value $\Phi$ becomes 
negative as $x$ increases. This hints towards a possibility that the Universe is undergoing a change from a thermodynamic non-equilibrium in the past 
towards an equilibrium state in the present epoch.

\section{Discussion}

It is often argued that the possibility of some non-gravitational interaction in the cosmic dark sector should not be ruled out \textit{a priori}. As the nature of dark energy is not known, it is impossible to model the cosmic interaction theoretically. The usual practice is to assume a transfer of energy $Q$ between the dark matter and the dark energy sectors. This $Q$ is parametrized as a function of the densities $\rho_D$, or $\rho_m$, or both, and even their derivatives \cite{yangnbpan}. The model parameters are then reconstructed using observational data. This approach is biased as some functional form of $Q$ is already chosen.

In the present work, an attempt is made to reconstruct the transfer of energy $Q$ in a dimensionless representation, defined as $\tilde{Q} = \frac{Q}{3H_{0}^{2}}$, directly from observational data without any parametrization. Various combinations of datasets are utilized, properly described in section \ref{ch5:methodology}. This investigation has been done for three different models of dark energy (i) an interacting vacuum with $w=-1$, (ii) a $w$CDM model where $w$ is close to $-1$ but not exactly equal to that and (iii) the CPL parametrization where $w(z) = w_0 + w_a \frac{z}{1+z}$. As a general feature, we find that for any of these cases and any combination of datasets, an interaction in the dark sector is not significant at the present epoch. The presence of such an interaction may not be ruled out in the past, beyond $z\geq 0.5$. But a zero interaction scenario is indeed a possibility normally in $2\sigma$ and at most in $3\sigma$. The results obtained are closer to that given by Wang \textit{et al}\cite{wang_pca} where a non-parametric Bayesian approach indicated that an interacting vacuum is not preferred but is quite different from the oscillatory behaviour as noted by Cai and Su\cite{cai_su}.

An analytic expression for the energy transfer rate $Q$ in the form of a polynomial in $z$ is given in section \ref{ch5:fitting}. The reduced ${\chi}^2$ 
test allowed up to second order in $z$ for the dataset combinations A1 and A2 while only up to first order in $z$ for the other two dataset combinations 
B1 and B2. 

The evolution of the density parameters, ${\Omega}_m$ and  ${\Omega}_D$, are checked in the presence of this interacting scenario. The common feature observed is that the dominance of dark energy is a bit delayed than that of the non-interacting model.

The thermodynamic considerations also reveal an interesting possibility. While the reconstructed interaction does not infringe upon the thermodynamic 
viability in terms of the increase in entropy, the Universe seems to be evolving towards a thermodynamic equilibrium only from a recent past $x\sim 
\approx 0.5$, i.e., close to $z\approx 0.6$ for all combinations of datasets. 

\clearpage{}
\clearpage{}\chapter{Revisiting a non-parametric reconstruction of the deceleration parameter}\blfootnote{\begin{flushleft} The work presented in this chapter is based on ``Revisiting a non-parametric reconstruction of the deceleration parameter from combined background and the growth rate data", \textbf{Purba Mukherjee} and Narayan Banerjee, arXiv:2007.15941 (2021).\end{flushleft}}  \label{ch7:chap7}
\chaptermark{Revisiting a reconstruction of the deceleration parameter}
\section{Introduction}

The present chapter is devoted to revisiting a non-parametric reconstruction of the cosmic deceleration parameter $q$. In chapter \ref{ch4:chap4}, a non-parametric reconstruction of the cosmographical quantities has already been carried out. Nevertheless, as new data are pouring in and new techniques are evolving, revisiting the nature of $q$ with newer datasets is quite imperative. In the absence of a universally accepted form of dark energy, this kind of revisit is an essential tool for refining the present understanding of the accelerated expansion of the Universe.

Different combinations of the background datasets like Pantheon SNIa distance modulus compilation \cite{pan}, Cosmic Chronometer Hubble parameter measurements \cite{cc_101, cc_102, cc_103, cc_104, cc_105, cc_106} including the full systematics as given by Moresco \textit{et al}\cite{sps2} and Baryon Acoustic Oscillation data have been utilized for the purpose.  In this chapter, we have focussed on a better model-independent treatment of the SNIa as well as the BAO data, and included all the recently updated systematic uncertainties in the CC dataset. Attempts are made to estimate the Hubble parameter at the present epoch, $H_0$, for a combination of datasets in a novel way, which serves as a normalization constant for the individual datasets in the final GP reconstruction. 
Since the growth of perturbations plays a promising role in distinguishing between diverse dark energy models, the growth rate measurements from the RSD data are utilized, which has commonly been ignored for a non-parametric reconstruction of $q$ in the literature \cite{bilicki, xia, lin, jesus_nonpara, adria, haridasu, arjona, nunes, purba_j, busti}.

Cosmological observations indicate that the Universe is undergoing an accelerated expansion in the present epoch. However, this acceleration has set in during the recent past and is not a permanent feature of the evolution. Transition from a decelerated to an accelerated phase of expansion is marked by a change in the signature of $q$, which occurs at some particular $z_t$, known as the deceleration-acceleration transition redshift. This $z_t$ has been estimated.

For the CC data, two different samples have been taken into consideration, hereafter referred to as the CC$_1$ and CC$_2$ samples respectively. The CC$_1$ sample has a total of 31 $H(z)$ values, obtained from combining the CCB and CCH compilation, whereas the CC$_2$ sample consists of 15 $H(z)$ values from the CCM compilation, as given in Table \ref{ch1:tabcc}. For the BAO data, we make use of the volume-averaged compilation \cite{beutler2011,kazin2014,ross2015,gilmarin2015,bautista2018,ata2018,agathe,blomqvist} and the BAO Hubble parameter measurements. The latest compilation of the 9 BAO $H(z)r_d$ measurements from different galaxy surveys, which includes the BOSS DR12 samples at 3 effective binned redshifts $z = 0.38, 0.51, 0.61$ \cite{alam2017}, eBOSS DR14 samples of LRG and quasars at 4 effective redshifts $z = 0.98, 1.23, 1.52, 1.94$ \cite{bao_4}, and the Ly$\alpha$ forest samples at $z=2.34$ \cite{agathe} and $z = 2.35$ \cite{blomqvist} respectively, are further taken into account for the reconstruction of $q(z)$. We consider the $H(z)\frac{r_d}{r_{d, fid}}$ measurements along with the full covariance matrix, where the subscript `$fid$' stands for the fiducial value assumed in the process of acquiring these measurements in the respective data samples.

In this work, we test the possible effect of spatial curvature, which has mostly been avoided for simplicity in the previous works, except for the work of Zhang and Xia\cite{xia}. A non-zero spatial curvature prior was considered by Zhang and Xia\cite{xia} when working with the Union 2.1 SNIa compilation only, but the authors ignored a combination of datasets for the reconstruction of $q$. Cosmological observations suggest that the spatial geometry of the Universe is very close to flat. This  prediction can be tested to high accuracy by a combination of the Planck temperature and polarization power spectra with the CMB lensing, which gives $\Omega_{k0} = -0.0106 \pm 0.0065$ (TT, TE, EE+lowE+lensing) \cite{planck}. We have investigated the effect of this non-zero $\Omega_{k0}$ prior on the reconstruction of $q$.

In light of the tension in measurement of the present value of Hubble parameter $H_0$ \cite{Q_review,dhawan,cosmoletter2,planck,riess21,wendy,aiola,macaulay}, a prior choice on the $H_0$ value for the reconstruction of $q$, has been investigated. The Hubble parameter has recently been measured to be $H_0 = 73.2 \pm 1.3$ km s$^{-1}$ Mpc$^{-1}$, obtained from the expanded sample of 75 Milky Way Cepheids with Hubble Space Telescope (HST) photometry and Gaia EDR3 parallaxes by the SH0ES team \cite{riess21} (hereafter referred to as R21). The Planck 2018 survey on the early Universe yields $H_0 = 67.27 \pm 0.60$ km Mpc$^{-1}$ s$^{-1}$\cite{planck} (TT, TE, EE+lowE) on assuming a base $\Lambda$CDM model (hereafter referred to as P18). The outcome of including these two different $H_0$ measurements, one from the Planck survey and other by the SH0ES team, having a maximum discrepancy at the $4.2\sigma$ level, has also been checked. This is in addition to the $H_0$ value reconstructed independently in the present case.

Results indicate that the $\Lambda$CDM model is well consistent at 2$\sigma$ CL. The use of any prior measurement for $H_0$, or the spatial curvature density parameter $\Omega_{k0}$ does not make any qualitative difference in this regard. The matter density parameter $\Omega_{m0}$ is observed to have a strong influence on the reconstruction of $q$ from the growth rate data. Lastly, we have compared our method and the results obtained with those of the existing literature in the final section. This comparison can also be used as an inventory of results.

\section{Reconstruction from Background data}

A spatially homogeneous and isotropic Universe, described by the FLRW metric is considered. The transverse comoving distance $d_C$ of luminous objects, like supernovae, are given by
\begin{equation} \label{ch7:D}
d_C(z)=  \frac{c}{H_0 \sqrt{\vert \Omega_{k0} \vert}}\sin\mbox{$n$} \left( \sqrt{\vert \Omega_{k0} \vert } \int_{0}^{z} \frac{\dif z'}{E(z')}\right) ,
\end{equation} 
in which the $\sin n$ function is a shorthand for
\[ \sin nx= \begin{cases}
\sinh x &  (\Omega_{k0}>0), \\
~~x & (\Omega_{k0} \rightarrow 0), \\
\sin x & (\Omega_{k0}<0),
\end{cases}
\]  and $E(z) = \frac{H(z)}{H_0}$ is the normalized Hubble parameter.

The dimensionless quantity $\Omega_{k0} = -\frac{k c^2}{a_0^2 H_0^2}$, known as the cosmic curvature density parameter is positive, negative or zero, corresponding to the spatial curvature $k = -1, +1, 0$, which signifies an open, closed, or flat Universe, respectively. From equation \eqref{ch7:D}, one can define the normalized transverse comoving distance as
\begin{equation}
D(z) = \frac{1}{\sqrt{\vert \Omega_{k0} \vert}}\sin\mbox{$n$} \left( \sqrt{\vert \Omega_{k0} \vert } \int_{0}^{z} \frac{\dif z'}{E(z')}\right),  \label{ch7:D_from_dc}
\end{equation} 
where \[ \sin nx= \begin{cases}
\sinh x &  (\Omega_{k0}>0), \\
~~x & (\Omega_{k0} \rightarrow 0), \\
\sin x & (\Omega_{k0}<0).
\end{cases}
\]

The deceleration parameter $q$, defined in Eq. \eqref{ch1:qdef}, can be represented as a function of redshift $z$, as 
\begin{equation}
q(z) = -1 + (1+z) \frac{H'}{H} = -1 + (1+z) \frac{E'}{E}  , \label{ch7:qz}
\end{equation} 
where a `prime' denotes derivative with respect to the redshift $z$.

The reduced Hubble parameter $E(z)$ and the normalized comoving distance $D(z)$ are related via equation \eqref{ch7:D_from_dc}, such that   
\begin{equation} \label{ch7:D'}
E(z)=\frac{\sqrt{1+\Omega_{k0}D^2}}{D'(z)}.
\end{equation} 

Finally, $q$ can be expressed as a function of the normalized comoving distance $D$ and its derivatives as
\begin{equation} 
q(z) = -1 + \frac{\Omega_{k0}DD'^2 - (1+\Omega_{k0}D^2)D''}{D'(1+\Omega_{k0}D^2)}(1+z) .	 \label{ch7:qnew}
\end{equation}

This will serve as the key equation for the non-parametric reconstruction of $q(z)$ using different combinations of the background datasets. 

The uncertainty associated with $q(z)$, $\sigma_q$, is obtained from Eq. \eqref{ch7:qnew} via the standard rule of error propagation. We have considered a zero mean and the Mat\'{e}rn ($\nu = \frac{9}{2}$, $p=4$) covariance function for the GP regression analysis. The reconstruction of $q$, in the present work, involves a two-step analysis. In the first step, we obtain the marginalized constraints on $M_B$ and $r_d$ in a cosmological model-independent framework. In the second step, these constraints are utilized in reconstructing $D(z)$, $D'(z)$, and $D''(z)$ for the corresponding combination of dataset. Finally, the deceleration parameter $q(z)$ is derived by using the reconstructed function $D(z)$, its derivatives $D'(z)$, and $D''(z)$ according to equation \eqref{ch7:qnew}.

\subsection{Constraints on $M_B$ and $r_d$} \label{sec7a}

\begin{figure}[t!]
	\begin{center}
		\includegraphics[angle=0, width=0.75\textwidth]{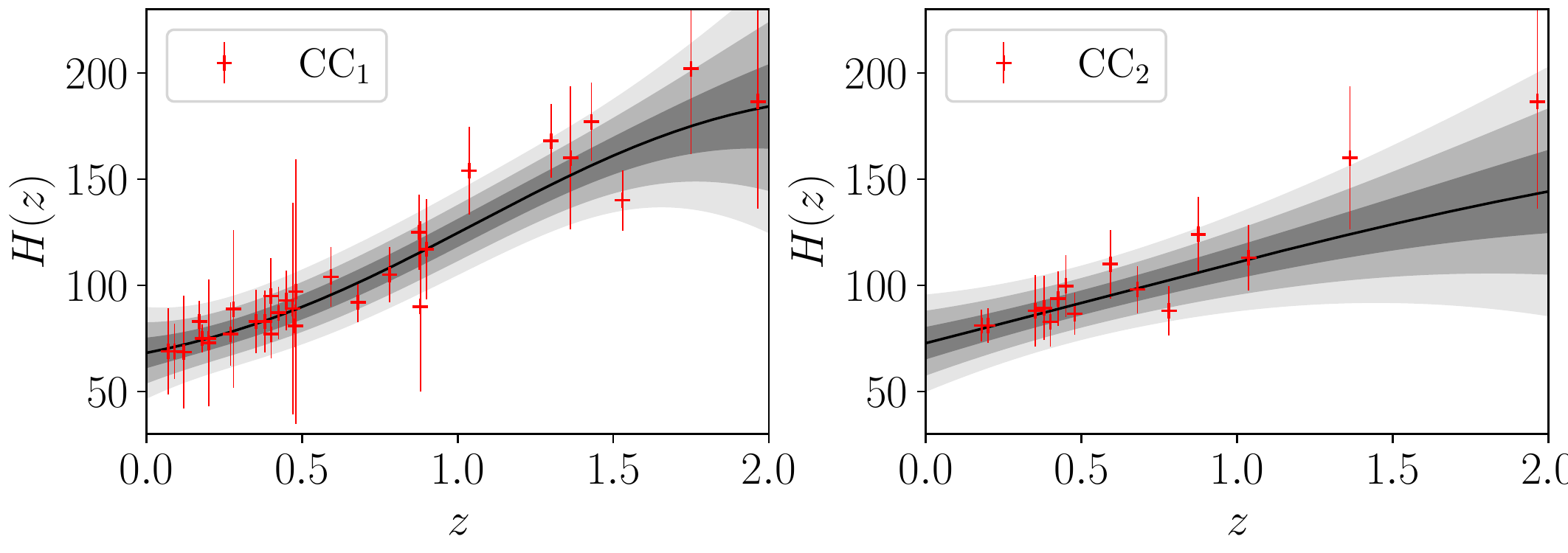}
	\end{center}
	\caption{{\small Plots for GP reconstructed $H(z)$ from the current updated CC$_1$ (left) and CC$_2$ (right) samples respectively. The solid black line represents the mean values of the reconstructed $H(z)$. }} \label{ch7:H_cc_plot}
\end{figure}

\begin{table}[t!]
	\caption{{\small Table showing the GP reconstructed $H_0$ (in units of km Mpc$^{-1}$ s$^{-1}$) from the latest updated CC data compilation.}}
	\begin{center}
		\resizebox{0.65\textwidth}{!}{\renewcommand{\arraystretch}{1.5} \setlength{\tabcolsep}{20pt}\centering 
			\begin{tabular}{l  c  c} 
				\hline \toprule  
				&   CC$_1$  & CC$_2$  \\
				\hline
				$H_0$ ~~~~&  $68.193 \pm 7.209$ ~~~~~& $72.776 \pm 7.636$ \\
\bottomrule				\hline 
			\end{tabular}
		}
	\end{center}
	\label{ch7:H0_cc_result}
\end{table}

We undertake a GP reconstruction of the Hubble parameter from the CC samples. All the systematic errors arising from the initial mass function (IMF) and stellar population synthesis (SPS) models, associated with the CC data were recently analyzed by Moresco \textit{et al}\cite{sps2}. The systematic errors linked with the CC dataset that are given in Table 3 of Moresco \textit{et al}\cite{sps2},  
have been added to the covariance matrices of the current CC data. We have interpolated this dataset to account for the error budget of the current measurements at each redshift due to these two extra sources. The covariance matrices, $\text{Cov}^{\text{IMF}}_{i,j}$ and $\text{Cov}_{i,j}^{\text{SPS}}$ are obtained as
\begin{equation}
{\text{Cov}}^{\text{X}}_{i,j}=\widehat{\eta^{\text{X}}}(z_i)H(z_i)\widehat{\eta^{\text{X}}}(z_j)H(z_j) , \label{ch7:Cov-CC-stimated}
\end{equation}
where $\widehat{\eta^{\text{X}}}(z)$'s are obtained by interpolation, and $H(z_i)$'s are CC measurements at different redshifts. The covariance matrices, $\text{Cov}^{\text{IMF}}_{i,j}$ and $\text{Cov}_{i,j}^{\text{SPS}}$ are then added to the statistical uncertainties, for obtaining the total covariance matrix of the current CC dataset. Plots for the reconstructed $H(z)$ from the updated CC$_1$ and CC$_2$ samples are shown in Fig. \ref{ch7:H_cc_plot}. The reconstructed $H_0$ values obtained from the CC$_1$ and CC$_2$ samples are given in Table \ref{ch7:H0_cc_result}.

With this smooth reconstructed function $H(z)$ from the CC data, we use a composite trapezoidal rule \cite{trapez} to obtain the integral 
\begin{eqnarray} 
\mathcal{I} &=& \int_{0}^{z} \frac{\dif z'}{H(z')} , \nonumber \\ 
& \approx & \frac{1}{2} \mathlarger{\mathlarger{\sum}}_{i=0}^{n-1} (z_{i+1} - z_i) \left[\frac{1}{H(z_{i+1})}+\frac{1}{H(z_{i})}\right]. \label{ch7:integral_H}
\end{eqnarray}

The statistical uncertainty in $\mathcal{I}$ is obtained by the error propagation formula
\begin{equation}  
\sigma^2_{\mathcal{I}} = \mathlarger{\mathlarger{\sum}}_{i=0}^{n} \frac{1}{4} (z_{i+1} - z_i)^2 \left[\frac{\sigma^2_{H_{i+1}}}{H^4_{i+1}}+\frac{\sigma^2_{H_{i}}}{H^4_{i}}\right]. \label{ch7:integral_sigH}
\end{equation}

Using equations \eqref{ch7:integral_H} and \eqref{ch7:integral_sigH}, we obtained a smooth function of the comoving distance $d_C$ and its associated uncertainty $\sigma_{d_C}$ from the CC Hubble data as
\begin{eqnarray} \label{ch7:D_recon_cc}
{d_C}_{\mbox{\tiny CC}}= \begin{cases}
\frac{c}{H_0 \sqrt{\Omega_{k0}}}\sinh \left[  H_0 \sqrt{\Omega_{k0}} ~ \mathcal{I}(z)\right] & \Omega_{k0}>0, \\
c ~\mathcal{I}(z) & \Omega_{k0}=0, \\
\frac{c}{ H_0 \sqrt{-\Omega_{k0}}}\sin \left[  H_0 \sqrt{-\Omega_{k0}} ~ \mathcal{I}(z)\right] &  \Omega_{k0}<0. 
\end{cases}
\end{eqnarray} 

The error $\sigma_{d_C}$ associated with the reconstructed $d_C$ from the CC Hubble data is given by
\begin{eqnarray} \label{ch7:sigD_recon_cc}
{\sigma_{d_C}}_{\mbox{\tiny CC}}= \begin{cases}
c \cosh \left[ H_0 \sqrt{\Omega_{k0}} ~ \mathcal{I}(z)\right] \sigma_{\mathcal{I}}(z) & \Omega_{k0}>0, \\
c ~\sigma_{\mathcal{I}}(z) & \Omega_{k0}=0, \\
c \cos \left[ H_0 \sqrt{-\Omega_{k0}} ~ \mathcal{I}(z)\right] \sigma_{\mathcal{I}}(z) &  \Omega_{k0}<0.
\end{cases}
\end{eqnarray}

This reconstructed ${d_C}_{\mbox{\tiny CC}}$ takes the role of a theoretical model which are further utilized to obtain the distance modulus from the CC Hubble data $\mu_{\mbox{\tiny CC}}$ using Eq. \eqref{ch1:mu} as 
\begin{equation}
\mu_{\mbox{\tiny CC}} = 5 \log_{10} \left[{d_C}_{\mbox{\tiny CC}}(1+z)\right] + 25 . \label{ch7:mu_CC}
\end{equation}

The associated 1$\sigma$ uncertainty $\sigma _{\mu_{\mbox{\tiny CC}}}$ is given by 
\begin{equation} 
\sigma _{\mu_{\mbox{\tiny CC}} }=\frac{5}{\ln 10}\frac{{\sigma _{d_C}}_{\mbox{\tiny CC}}}{{d_C}_{\mbox{\tiny CC}}} . \label{ch7:sigma_mu_CC}
\end{equation}

The distance modulus from the Pantheon SN compilation are combined with the CC $H(z)$ measurements to account for the degeneracy between the absolute magnitude $M_B$ of SNIa and the Hubble parameter at present epoch $H_0$. The corrected apparent magnitudes $m_B$ are reconstructed adopting another GP regression, and the constraints on $M_B$ are obtained by minimizing the $\chi^2$ function 
\begin{equation}
\chi^2 = \Delta \bm{\mu}^{\mbox{\small T}} \cdot \bm{\Sigma}^{-1} \cdot \Delta \bm{\mu}. \label{ch7:chi2}
\end{equation}
Here $\Delta \bm{\mu} = \bm{\mu}_{\mbox{\tiny SN}}- \bm{\mu}_{\mbox{\tiny CC}}$ and $\bm{\Sigma} = \bm{\Sigma}_{\mu_{\mbox{\tiny SN}}} + \sigma^2_{\mu_{\mbox{\tiny	CC}}}$ respectively. We get the best fit constraints on $M_B$ and the associated 1$\sigma$ uncertainties by a MCMC analysis with the assumption of a uniform prior distribution for $M_B \in [-25, -15]$.

\begin{table*}[t!]
	\caption{{\small Table showing the marginalized constraints on $M_B$ and $r_d$ (in units of Mpc) for different combinations of datasets.}}
	\begin{center}
		\resizebox{\textwidth}{!}{\renewcommand{\arraystretch}{1.5} \setlength{\tabcolsep}{18pt}\centering 
\begin{tabular}{l  c  c  c  c  c } 
				\hline
				\toprule 
				& & CC$_1$+SN & CC$_1$+SN+BAO & CC$_2$+SN & CC$_2$+SN+BAO \\
				\hline 
				& $M_B$  & $-19.409 \pm 0.010$  & $-19.412 \pm 0.007$ &  $-19.341 \pm 0.011$  & $-19.390 \pm 0.008$ \\[-0.75em]
				$\Omega_{k0}=0$ & \\[-0.75em]
				& $r_d$  & -	  &  $148.76 \pm 0.28$  &  	-   &   $149.61 \pm 0.39$  \\
				\hline 
				& $M_B$ &  $-19.412 \pm 0.014$  & $-19.413 \pm 0.009$ &  $-19.353 \pm 0.015$  & $-19.412 \pm 0.010$ \\[-0.75em]
				$\Omega_{k0}\neq 0$ & \\[-0.75em]
				& $r_d$ &  -	  &  $148.67 \pm 0.33$  &  	-   &   $150.22 \pm 0.47$  \\
				\bottomrule
				\hline
			\end{tabular}
		}
	\end{center}
	\label{ch7:MB_rd_tab}
\end{table*}

In order to introduce the BAO $H r_d$ measurements in combination with the CC and Pantheon data, we need to obtain the constraints on $r_d$, independent of any fiducial reference model. The volume-averaged BAO data are utilized for this purpose. We reconstruct $\frac{D_V}{r_d}$ via another GP and obtain the joint constraints on $M_B$ and $r_d$. One can evaluate the comoving distances from the reconstructed volume-averaged BAOs in combination with the reconstructed CC Hubble data, by means of Eq. \eqref{ch1:dilation_d} as, 
\begin{equation} \label{ch7:D_recon_bao}
{d_C}_{\mbox{\tiny BAO}}  = \left[\frac{D_V^3(z) H(z)}{c z}\right]^\frac{1}{2} .
\end{equation} 

This reconstructed ${d_C}_{\mbox{\tiny BAO}}$ along with its 1$\sigma$ uncertainty are combined with ${d_C}_{\mbox{\tiny CC}}$, following a similar manner as in Eq. \eqref{ch7:mu_CC}, \eqref{ch7:sigma_mu_CC} and \eqref{ch7:chi2} to simultaneously constrain $M_B$ and $r_d$ via a minimization of the combined $\chi^2$, employing another MCMC analysis assuming a uniform prior distribution with $r_d \in [135, 160]$. The best-fit results of $M_B$ and $r_d$ along with their respective 1$\sigma$ uncertainties are given in Table \ref{ch7:MB_rd_tab}. To examine the influence of spatial curvature, we consider $\Omega_{k0}=0$ as well as $\Omega_{k0} = -0.0106 \pm 0.0065$ from the Planck (TT,TE,EE+lowE+lensing) probe \cite{planck}.

\subsection{Reconstructing $D(z)$ and its derivatives} \label{sec7b}

The marginalized $M_B$ constraints, given in Table \ref{ch7:MB_rd_tab}, are used for computing the comoving distances $d_C$ of all supernovae in the Pantheon compilation via a transformation from the distance modulus $\mu$. The uncertainty matrix $\bm{\Sigma}_{d_C}$ associated with the SNIa comoving distance data is obtained from the total uncertainty matrix of distance modulus $\bm{\Sigma}_\mu$ given in Eq. \eqref{ch1:mu_error}. The comoving distances $d_C$ are identified as the training dataset that spans the function space. A GP regression of the SNIa comoving distances is then undertaken, and the target functions $d_C(z)$, ${d_C}^\prime(z)$ are reconstructed incorporating the condition $d_C(z = 0) = 0$ with zero uncertainty. This $d_C$, directly measured from SNIa, are utilized in Eq. \eqref{ch7:D'} to find $H_0$, 
\begin{equation} \label{ch7:H0_gp}
H_0 = c{{\left[{{d_C}^\prime}^2(0) - \Omega_{k0} ~{d_C}^2(0)\right]}}^{-\frac{1}{2}}.
\end{equation} 

The uncertainty associated with $H_0$ are propagated from the uncertainties associated with $d_C(0)$ and ${d_C}^\prime(0)$ and $\Omega_{k0}$ respectively. The estimated values of $H_0$, obtained from Eq. \eqref{ch7:H0_gp}, are shown in Table \ref{ch7:H0_gp_result}.

\begin{table*}[t!]
	\caption{{\small Table showing the estimated values of $H_0$ (in units of km Mpc$^{-1}$ s$^{-1}$) for different combinations of datasets computed from equation 
			\eqref{ch7:H0_gp}.}}
	\begin{center}
		\resizebox{\textwidth}{!}{\renewcommand{\arraystretch}{1.5} \setlength{\tabcolsep}{20pt}\centering 
\begin{tabular}{c  c  c  c  c  c} 
				\hline
				\toprule 
				&  & N1 & N3 & N2 & N4 \\
				\hline
				$\Omega_{k0} = 0$  & $H_0$ &  $ 68.711 \pm 0.414 $  & $ 68.395 \pm 0.412 $ &  $ 70.636 \pm 0.425 $  & $ 69.028 \pm 0.413 $ \\
				\hline
				$\Omega_{k0} \neq 0$ & $H_0$ & $ 68.397 \pm 0.415 $  & $ 68.396 \pm 0.413 $ &  $ 70.638 \pm \pm 0.428 $  & $ 68.395 \pm 0.416 $ \\
				\bottomrule
				\hline
			\end{tabular}
		}
	\end{center}
	\label{ch7:H0_gp_result}
\end{table*}

For computing the Hubble parameter from the BAO $H r_d$ measurements, we substitute the marginalized $r_d$ constraints (given in Table \ref{ch7:MB_rd_tab}) to the BAO $H r_d$ dataset. The resulting values of the Hubble parameter obtained are added with the CC $H(z)$ measurements to form the CC+BAO Hubble data. The total covariance matrix is obtained by appending the individual CC and BAO covariance matrices corresponding to the full $H(z)$ sample.

\begin{figure*}[t!]
	\begin{center}
		\includegraphics[angle=0, width=0.95\textwidth]{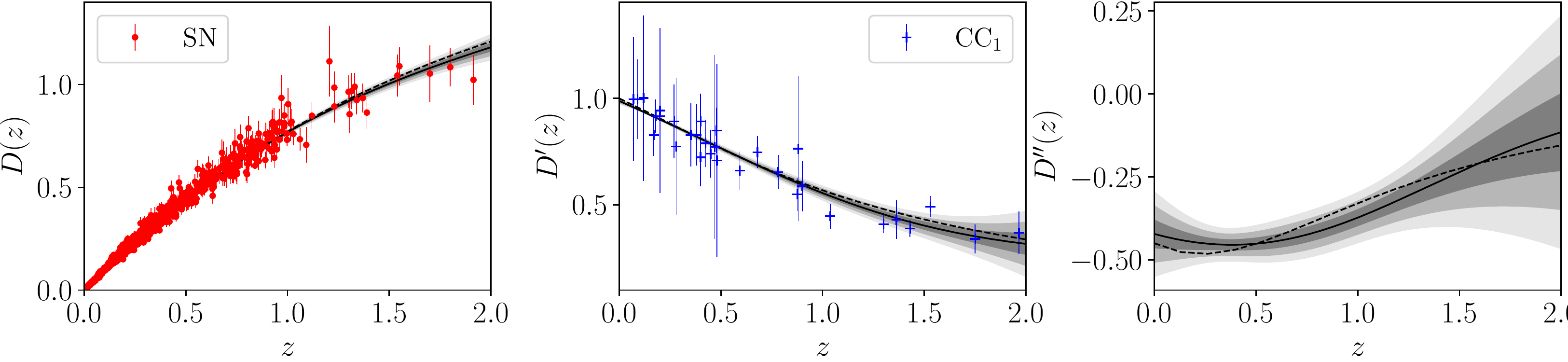}
	\end{center}
	\caption{{\small Plots for the reconstructed dimensionless comoving distance $D(z)$, its derivatives $D'(z)$ and $D''(z)$ using combined CCB+SN data (Set N1) for  
			a spatially flat Universe ($\Omega_{k0}=0$). The black solid line is the mean curve. The associated 1$\sigma$, 2$\sigma$ and 3$\sigma$ confidence regions 	
			are shown in lighter shades. The specific markers with error bars represent the observational data. The $\Lambda$CDM model with $\Omega_{m0} = 0.3$ is 
			represented by the dashed line.}} \label{ch7:D_N1}
\end{figure*}

\begin{figure*}[t!]
	\begin{center}
		\includegraphics[angle=0, width=0.95\textwidth]{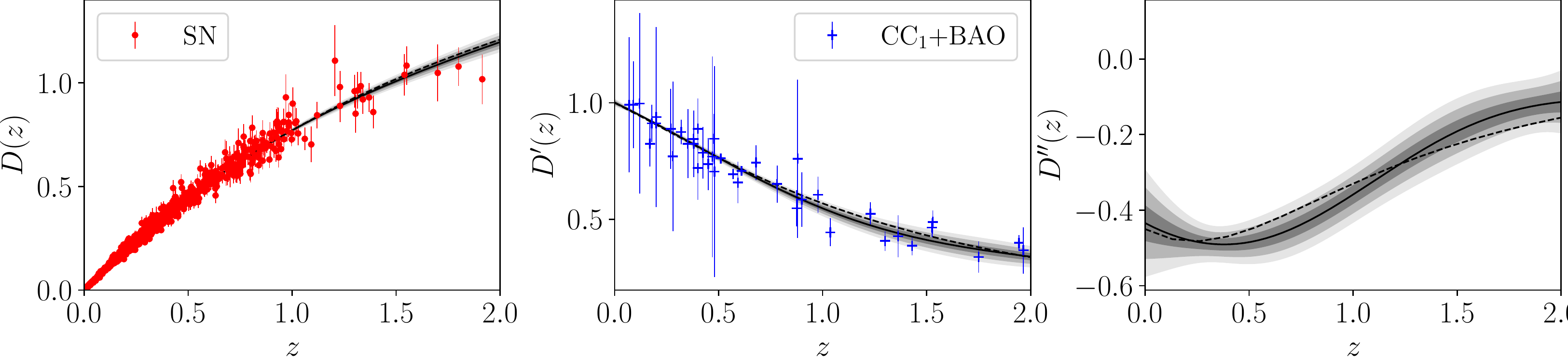}
	\end{center}
	\caption{{\small Plots for the reconstructed dimensionless comoving distance $D(z)$, its derivatives $D'(z)$ and $D''(z)$ using combined CCM+SN data (Set N2) for 
			a spatially flat Universe ($\Omega_{k0}=0$). The black solid line is the mean curve. The associated 1$\sigma$, 2$\sigma$ and 3$\sigma$ confidence regions 
			are shown in lighter shades. The specific markers with error bars represent the observational data. The $\Lambda$CDM model with $\Omega_{m0} = 0.3$ is 
			represented by the dashed line.}} \label{ch7:D_N2}
\end{figure*}

\begin{figure*}[t!]
	\begin{center}
		\includegraphics[angle=0, width=0.95\textwidth]{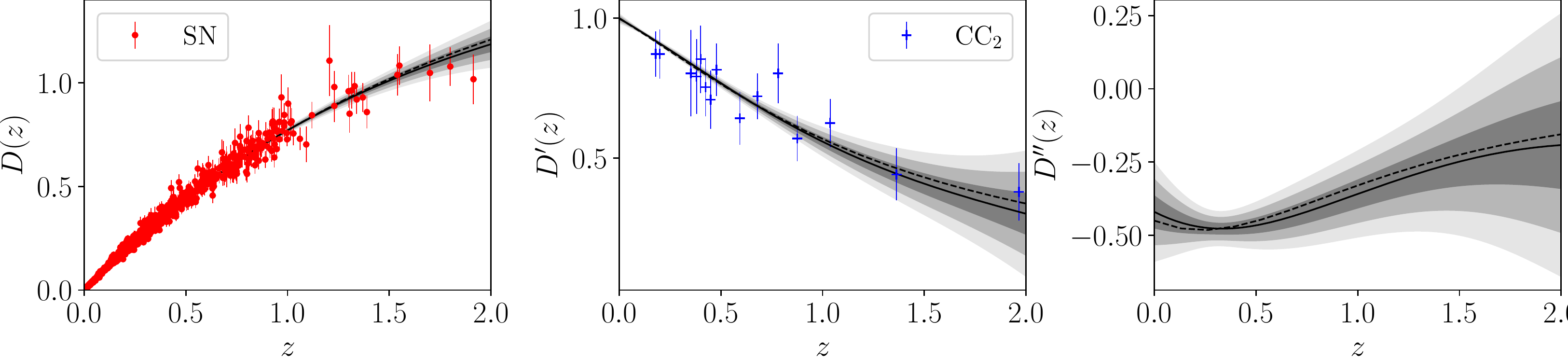}
	\end{center}
	\caption{{\small Plots for the reconstructed dimensionless comoving distance $D(z)$, its derivatives $D'(z)$ and $D''(z)$ using combined CCB+SN+BAO data (Set N3) for 
			a spatially flat Universe ($\Omega_{k0}=0$). The black solid line is the mean curve. The associated 1$\sigma$, 2$\sigma$ and 3$\sigma$ confidence regions are 
			shown in lighter shades. The specific markers with error bars represent the observational data. The $\Lambda$CDM model with $\Omega_{m0} = 0.3$ is represented 
			by the dashed line.}} \label{ch7:D_N3}
\end{figure*}

\begin{figure*}[t!]
	\begin{center}
		\includegraphics[angle=0, width=0.95\textwidth]{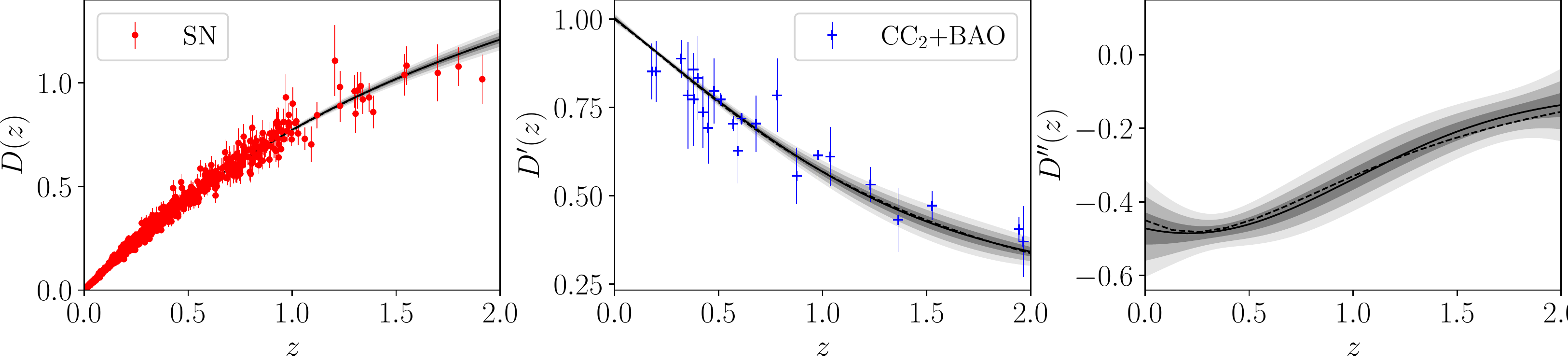}
	\end{center}
	\caption{{\small Plots for the reconstructed dimensionless comoving distance $D(z)$, its derivatives $D'(z)$ and $D''(z)$ using combined CCM+SN+BAO data (Set N4) for 
			a spatially flat Universe ($\Omega_{k0}=0$). The black solid line is the mean curve. The associated 1$\sigma$, 2$\sigma$ and 3$\sigma$ confidence regions are 
			shown in lighter shades. The specific markers with error bars represent the observational data. The $\Lambda$CDM model with $\Omega_{m0} = 0.3$ is represented 
			by the dashed line.}} \label{ch7:D_N4}
\end{figure*}

The CC and CC+BAO Hubble datasets are normalized with the $H_0$ values as given in Table \eqref{ch7:H0_gp_result} to obtain the reduced Hubble parameter $E$. Considering the error associated with $H_0$ to be $\sigma_{H_0}$, the uncertainty covariance matrix $\bm{\Sigma}_{E}$ associated with $E$, is evaluated as 
\begin{equation} \label{ch7:sigE_recon}
\bm{\Sigma}_{E} = \frac{\bm{\Sigma}_H}{ {H_0}^2} + \frac{H^2}{{H_0}^4}{\sigma_{H_0}}^2 , 
\end{equation} 
where $\bm{\Sigma}_{H}$ is the uncertainty covariance matrix of the Hubble data compilation. The comoving distances from the Pantheon compilation are normalized with the corresponding $H_0$ values from Table \eqref{ch7:H0_gp_result} to obtain the dimensionless comoving distances $D$ using Eq. \eqref{ch7:D_from_dc}. The uncertainty associated with training dataset $D$ are propagated from the uncertainties of $\mu$ ($\bm{\Sigma}_\mu$ in Eq. \eqref{ch1:mu_error}) and $H_0$ $(\sigma_{H_0})$ via the standard error propagation formula. These normalized comoving distances are later combined with the reduced CC or CC+BAO Hubble parameter measurements via equation \eqref{ch7:D'} as additional constraints on $D'(z)$ in the final GP regression analysis.

Thus, having acquired all the necessary training data, we proceed with a non-parametric GP reconstruction of the normalized comoving distance $D(z)$ and its derivatives $D'(z)$ and $D''(z)$ at different redshift $z$, for the following combination of datasets
\begin{itemize}
	\item Set N1 - CC$_1$+SN, \item Set N2 - CC$_2$+SN, \item Set N3 - CC$_1$+SN+BAO, \item Set N4 - CC$_2$+SN+BAO.
\end{itemize}

The hyperparameters in the Mat\'{e}rn 9/2 covariance function are obtained by maximizing the log-likelihood function, given in Eq. \eqref{ch1:lnlike}). With the trained hyperparameters, we reconstruct the mean values for the most probable continuous function $D(z)$ of the distance data and its derivatives, along with the associated confidence levels. Plots for the reconstructed $D(z)$, $D'(z)$ and $D''(z)$ versus $z$ are shown in Fig. \ref{ch7:D_N1}, \ref{ch7:D_N2}, \ref{ch7:D_N3} and \ref{ch7:D_N4} for N1, N2, N3 and N4 dataset combinations, respectively.

\subsection{Reconstruction of $q(z)$} \label{sec7c}

\begin{figure*}[t!]
	\begin{center}
		\includegraphics[angle=0, width=\textwidth]{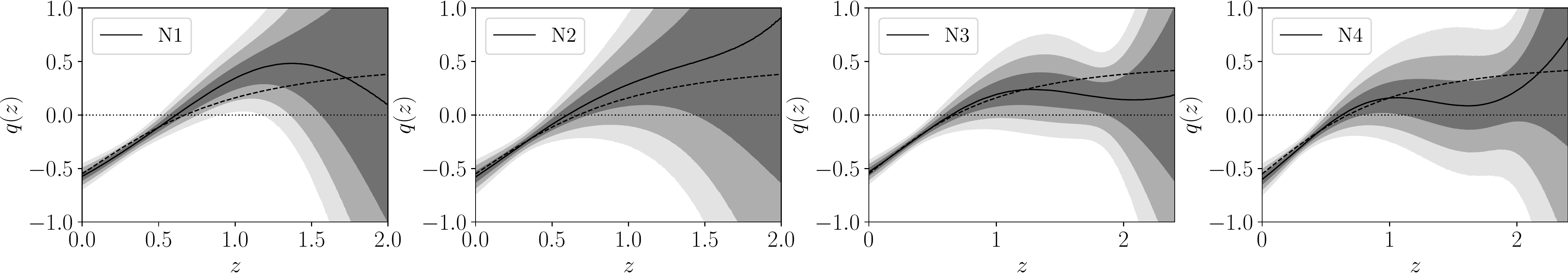}
	\end{center}
	\caption{{\small Plots for $q(z)$ reconstructed from the combined datasets N1, N2, N3, N4 for a spatially flat Universe ($\Omega_{k0}=0$). The solid black line 
			represents the mean values of the reconstructed $q(z)$. The black dashed line shows $q(z)$ corresponding to the $\Lambda$CDM model with $\Omega_{m0} = 
			0.3$.}} \label{ch7:qplot}
\end{figure*}

\begin{figure*}[t!]
	\begin{center}
		\includegraphics[angle=0, width=\textwidth]{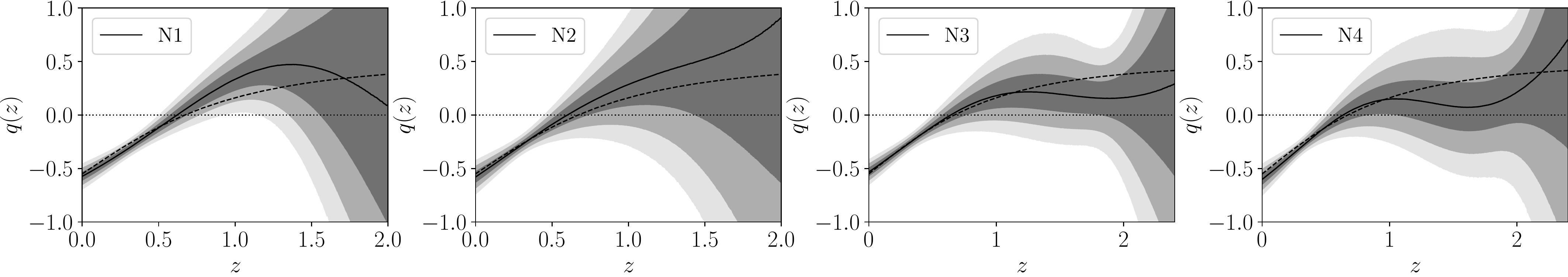}
	\end{center}
	\caption{{\small Plots for $q(z)$ reconstructed from the combined datasets N1, N2, N3, N4 for a Universe with a non-zero spatial curvature given by the Planck 
			2020 measurement $\Omega_{k0}= -0.0106 \pm 0.0065$ \cite{planck}. The solid black line represents the mean values of the reconstructed $q(z)$. The 
			black dashed line shows $q(z)$ corresponding to the $\Lambda$CDM model with $\Omega_{m0} = 0.3$.}} \label{ch7:qplot_ok}
\end{figure*}

\begin{table*}[t!]
	\caption{{\small Table showing the reconstructed mean values along with the 1$\sigma$ uncertainties for $q_0$ corresponding to the datasets N1, N2, N3 and N4. 
			An estimate for the late-time deceleration-acceleration transition redshift $z_t$ is also provided.}}
	\begin{center}
		\resizebox{\textwidth}{!}{\renewcommand{\arraystretch}{1.8} \setlength{\tabcolsep}{24 pt} \centering  
			\begin{tabular}{l c c c c c c }
				\hline \toprule
				&  &  N1 & N2 & N3 & N4  \\ 
				\hline 
				& $q_0$ & $-0.573^{ +0.041}_{-0.042}$ & $-0.580^{+0.055}_{-0.063}$ &  $-0.533^{+0.038}_{-0.038}$  & $-0.574^{+0.044}_{-0.045}$ \\ [-0.95em]
				$\Omega_{k0} = 0$ & \\[-0.95em]
				& $z_t$ & $0.611_{-0.045}^{+0.065}$ & $0.601_{-0.071}^{+0.140}$ &  $0.644_{ -0.064}^{ +0.092}$  & $0.602_{-0.050}^{+0.065}$ \\ 
				\hline 
				& $q_0$ &  $-0.571^{ +0.043 }_{ -0.044}$ & $-0.573^{+0.062}_{ -0.062}$ &  $-0.532^{ +0.041}_{-0.041}$  & $-0.573^{ +0.047}_{ -0.048}$ \\ [-0.95em]
				$\Omega_{k0} \neq 0$ & \\[-0.95em]
				& $z_t$ &  $0.621_{-0.046}^{+0.066}$ & $0.605_{-0.081}^{+0.182}$ &  $0.643_{-0.069}^{+0.094}$  & $0.610_{-0.055}^{+0.070}$  \\ 
				\bottomrule
				\hline 
			\end{tabular} 
		}
	\end{center}
	\label{ch7:q0_tab}
\end{table*}

Finally, we plot the cosmological deceleration parameter $q(z)$ using the reconstructed values of the comoving distance $D(z)$, its derivatives $D'(z)$ and $D''(z)$, at different $z$ according to equation \eqref{ch7:qnew}. In Fig. \ref{ch7:qplot} and \ref{ch7:qplot_ok}, we plot the reconstructed $q(z)$ within 3$\sigma$ uncertainty regions for the combined datasets N1, N2, N3 and N4 considering two prior choices on the spatial curvature, as $\Omega_{k0} = 0$, and $\Omega_{k0} = -0.0106 \pm 0.0065$ from Planck (TT, TE, EE+lowE+lensing) survey \cite{planck} respectively. The black solid lines represent the mean values of the reconstructed $q$ and the shaded regions correspond to the 68\% CL, 95\% CL and 99.7\% CL. The black dashed line shows the evolution of $q(z)$ assuming the $\Lambda$CDM model with $\Omega_{m0}=0.3$. The expected value of $q_{\Lambda\mbox{\tiny CDM}}$ at the present epoch is given by $q_{0_{\Lambda\mbox{\tiny CDM}}} = \frac{3}{2}\Omega_{m0} - 1 = -0.55$.

The mean values of the reconstructed $q_0$, along with the associated 1$\sigma$ uncertainties, corresponding to the datasets N1, N2, N3 and N4, are shown in Table \ref{ch7:q0_tab}. An estimate for the late-time transition redshift $z_t$ where the reconstructed $q(z)$ shows a signature flip is also provided. The plots show that the $\Lambda$CDM model is well allowed within a $2\sigma$ CL.

\subsection{Effect of $H_0$ priors} \label{sec-5.4}

\begin{figure*}[t!]
	\begin{center}
		\includegraphics[angle=0, width=\textwidth]{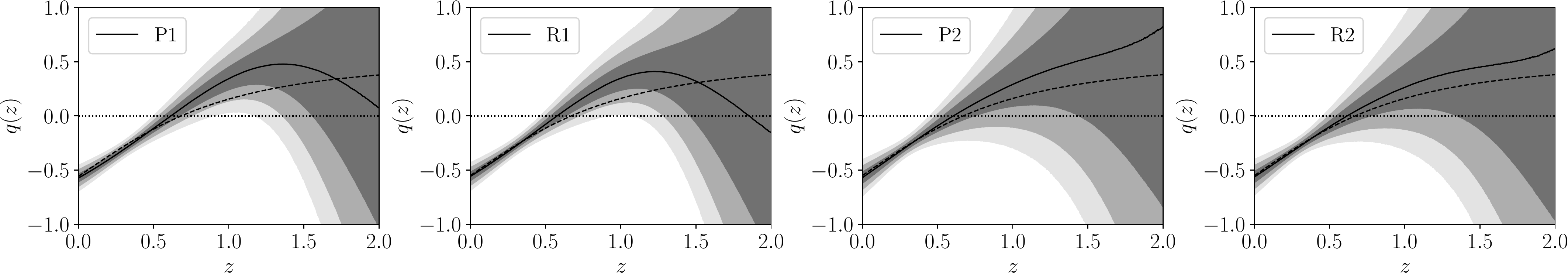}\\
		\includegraphics[angle=0, width=\textwidth]{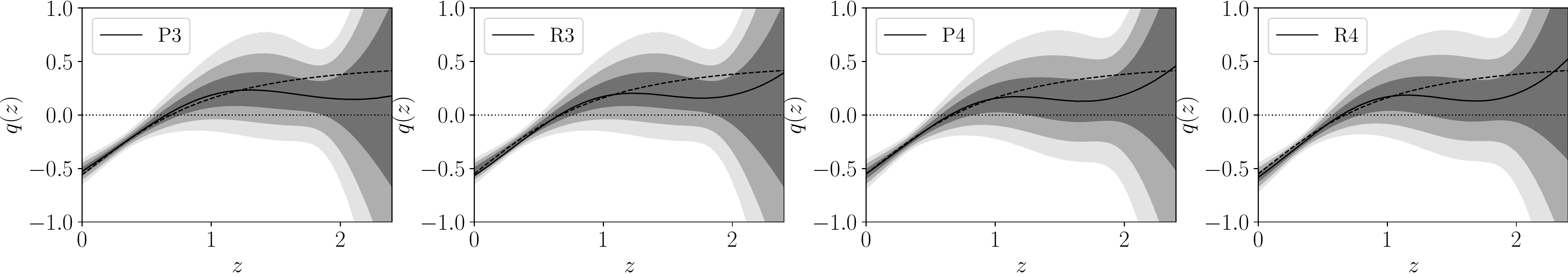}
	\end{center}
	\caption{{\small Plots for $q(z)$ reconstructed from the combined datasets P1, R1, P2, R2, P3, R3, P4 and R4 for a spatially flat Universe ($\Omega_{k0}
			=0$). The solid black line represents the mean values of the reconstructed $q(z)$. The black dashed line shows $q(z)$ corresponding to the 
			$\Lambda$CDM model with $\Omega_{m0} = 0.3$. A comparison among the four cases is shown in the extreme right column.}} \label{ch7:qplot_H0}
\end{figure*}

It is further examined if the two different strategies for determining the value of $H_0$, namely the P18 and R21 $H_0$ values, have any significant effect on the reconstruction. We proceed with the analysis following a similar methodology as discussed in Sec. \ref{sec7a}, \ref{sec7b} and finally Sec. \ref{sec7c}, the only exception being that we have added the P18 or R21 $H_0$ estimates to the CC $H(z)$ dataset in the beginning. Finally, we reconstruct $q(z)$ for the following combinations of datasets
\begin{multicols}{2}
\begin{itemize}
	\item Set P1 - P18+CC$_1$+SN, \item Set R1 - R21+CC$_1$+SN, \item Set P2 - P18+CC$_2$+SN, \item Set R2 - R21+CC$_2$+SN,
	\item Set P3 - P18+CC$_1$+SN+BAO, \item Set R3 - R21+CC$_1$+SN+BAO, \item Set P4 - P18+CC$_2$+SN+BAO, \item Set R4 - R21+CC$_2$+SN+BAO.
\end{itemize}
\end{multicols}

Plots for the reconstructed $q(z)$ using the combined datasets P1, R1, P2, R2, P3, R3, P4 and R4 along with their respective 1$\sigma$, 2$\sigma$ and 3$\sigma$ uncertainties are shown in Fig. \ref{ch7:qplot_H0}. It is seen that inclusion of the P20 or R21 $H 0$ measurements does not lead to any significant difference on the reconstruction of $q(z)$ in terms of allowing the $\Lambda$CDM model at the 2$\sigma$ CL. In case of the R1 combination, the mean reconstructed $q(z)$ shows the presence of a negative dip close to $z \approx 1.9$, indicating another stint of acceleration in the recent past. For the N1 and P1 combinations, the possibility of this negative dip in $q$ can be perceived at higher redshift values exceeding the domain of reconstruction. However, this behaviour may not be statistically too significant as a positive $q$ is comfortably included at the $1\sigma$ CL.

\section{Reconstruction from Perturbation data}

The evolution of matter density contrast $\delta$ is given by
\begin{equation} \label{ch7:delta_def}
\delta = \frac{\delta \rho_m}{\rho_m}.
\end{equation}

This $\delta$, in a linearized approximation, obeys the following second order differential equation 
\begin{equation} \label{ch7:perturb_eqn}
\ddot{\delta}+ 2H \dot{\delta} - 4\pi G \rho_m \delta = 0.
\end{equation} 
Here, $\rho_m$ is the background matter density and $\delta \rho_m$ gives the first-order matter perturbation. 

Rewriting Eq. \eqref{ch7:perturb_eqn} as a function of the redshift $z$, the reduced Hubble parameter $E(z)$ can be expressed as an integral over the perturbation $\delta$ and its derivative \cite{zhang_li} as
\begin{equation} \label{ch7:E2_rsd}
\small E^2(z) = \frac{(1+z)^2}{\delta ' (z)^2} \left[ \delta '(z=0) ^2  -  3 \Omega_{m0} \int_0^{z} \frac{\delta}{1+z} (-\delta ') \dif z \right].
\end{equation}

The RSD data measure the quantity $f \sigma_8$, known as the growth rate of structure. Here $f$ is the growth rate, defined as the derivative of the logarithm of perturbation $\delta$ with respect to logarithm of the scale factor  $a(t)$.
\begin{equation} \label{ch7:f_def}
f = \frac{\dif \, \textmd{ln} \delta}{\dif \, \textmd{ln} a}  = -(1+z) \frac{\dif \, \textmd{ln} \delta}{\dif \, z} = -(1+z) \frac{\delta '}{\delta} .
\end{equation}

The function $\sigma_{8}$ is known as the root-mean-square mass fluctuation within a sphere of radius $8h^{-1}$ Mpc, and is given by
\begin{equation} \label{ch7:s8_def}
\sigma_8 (z) = \sigma_8 (z=0) \frac{\delta (z)}{\delta (z=0)} .
\end{equation}

Therefore, the growth rate of structure can be derived from Eq. \eqref{ch7:f_def} and \eqref{ch7:s8_def} as
\begin{equation}
f \sigma_8 (z) = -\frac{\sigma_8 (z=0)}{\delta (z=0)} (1+z) \delta' . \label{ch7:fs8}
\end{equation}

Integrating Eq. \eqref{ch7:fs8} followed by some algebraic manipulation, one can obtain
\begin{equation} \label{ch7:delta_end}
\delta  = \delta (z=0) - \frac{\delta (z=0)}{\sigma_8 (z=0)} \int_0^{z} \frac{f \sigma_8}{1+z} \dif z .
\end{equation}

For the reconstruction of $q(z)$ using the RSD data requires calculation of the integral 
\begin{equation}
\mathcal{D} = \int_{0}^{z} \frac{f \sigma_8}{1+z} \dif z ,
\end{equation}
to obtain the perturbation $\delta$.

The statistical error associated with $E^2(z)$, defined in Eq. \eqref{ch7:E2_rsd}, can be expressed via the standard error propagation rule as 
\begin{equation}
\sigma_{E^2}(z) = \left[\left( \frac{\partial E^2}{\partial \delta'} \right)^2 \sigma_{\delta'}^2 + \left( \frac{\partial E^2}{\partial \mathcal{D}} 
\right)^2 \sigma_{\mathcal{D}}^2 \right] ^ \frac{1}{2}. 
\end{equation} 

Finally, the deceleration parameter $q(z)$ is reconstructed using the expression 
\begin{equation}
q(z) = -1 + \frac{1}{2}(1+z)\frac{\left[ E^2(z) \right]'}{E^2(z)}, \label{ch7:q_rsd_def} 
\end{equation}
where the uncertainty associated with $q(z)$ is propagated from the uncertainties in $E^2(z)$ and $\left[E^2(z)\right]'$ respectively.

\begin{figure}[t!]
	\begin{center}
		\includegraphics[angle=0, width=0.75\textwidth]{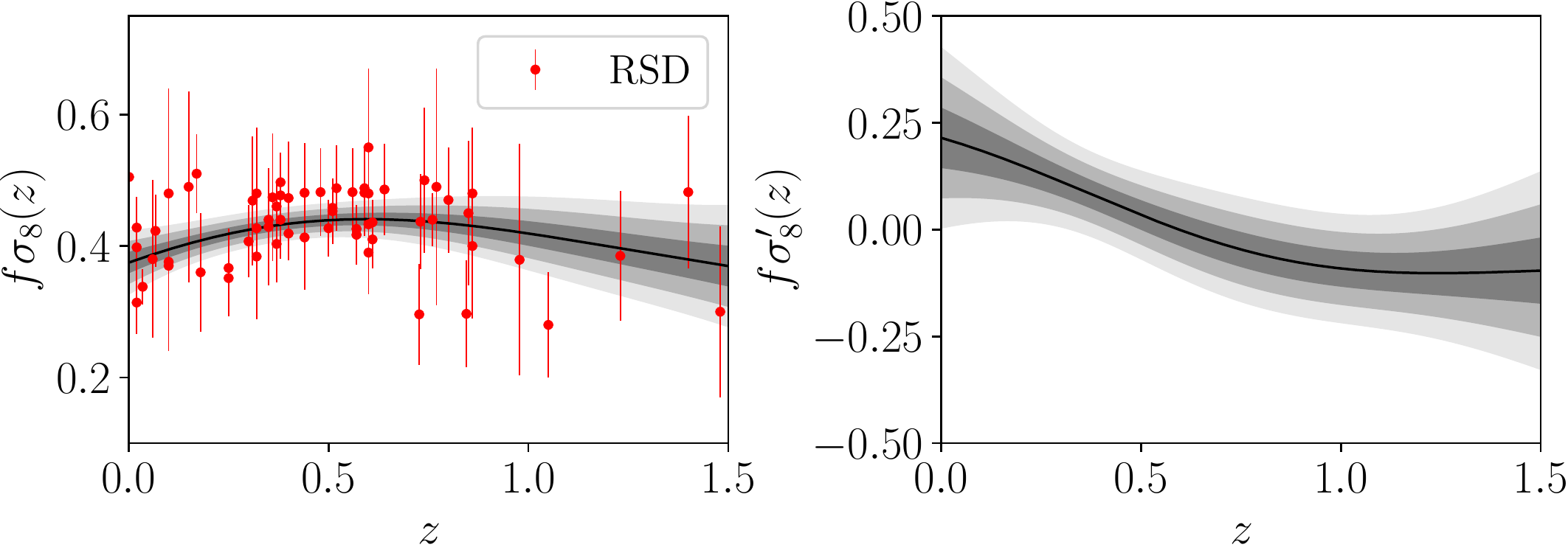}
	\end{center}
	\caption{{\small Plots for the GP reconstructed $f \sigma_8(z)$ and its derivative $\left[f \sigma_{8}\right]'$ from the RSD data. The solid black line represents the 
			mean values of the reconstructed functions. }} \label{ch7:fs8_plot}
\end{figure}

\begin{figure*}[t!]
	\begin{center}
		\includegraphics[angle=0, width=\textwidth]{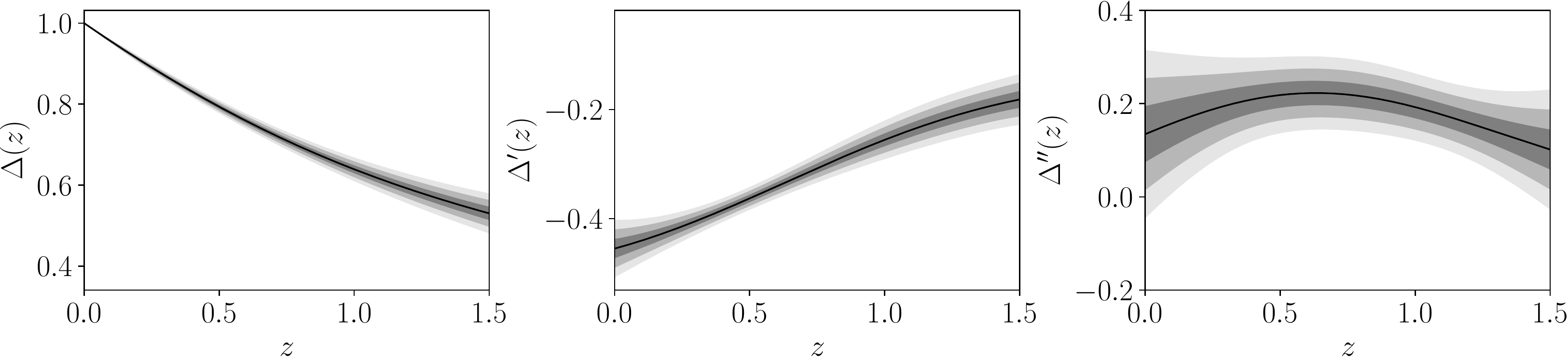}
	\end{center}
	\caption{{\small Plots for the reconstructed normalized perturbation $\Delta = \frac{\delta(z)}{\delta(z=0)}$ along with its higher derivatives $\Delta'(z)$ and 
			$\Delta''(z)$ from the RSD data. The solid black line represents the mean values of the reconstructed functions. }} \label{ch7:delta_plot}
\end{figure*}

A close observation on Eq. \eqref{ch7:q_rsd_def} suggests that the deceleration parameter is independent of the perturbation value $\delta$ at the present epoch $z = 0$. But $q(z)$ is directly dependent on the value of $\sigma_8$ and $\Omega_{m}$ at the present epoch, denoted as $\sigma_{8,0} = \sigma_{8}(z=0)$ and $\Omega_{m0}$ respectively. For a self-consistent reconstruction of $q(z)$ from the RSD data, we need to provide the accurate values for $\sigma_{8,0}$ and $\Omega_{m0}$. Instead of considering model-dependent estimates for $\sigma_{8,0}$ and $\Omega_{m0}$, attempts are made to constrain these quantities in a non-parametric way.

It is difficult to provide an analytical solution for Eq. \eqref{ch7:perturb_eqn}. Assuming the Universe to be spatially flat, an approximate solution is given in  \cite{peebles_fz,waga_fz,fry_fz,lightman_fz,wang_stein_fz,yggong_fz} as
\begin{equation} \label{ch7:f_soln}
f(z) = \Omega_{m}^\gamma ,
\end{equation} 
where $\Omega_{m}(z) = \frac{\Omega_{m0} (1+z)^3}{E^2(z)}$ and $\gamma$ is the growth index of perturbations corresponding to the background cosmological model. 

Therefore, $f \sigma_8$ in Eq. \eqref{ch7:fs8} can be written as 
\begin{equation} \label{ch7:fs8_final}
{f \sigma_8}^{\mbox{\tiny theo}} (z) = \sigma_{8,0} ~ \Omega_{m}^\gamma(z) \exp \left\lbrace \int_{0}^{z} - \frac{\Omega_{m}^\gamma(z')}{1+z'} \dif z'
\right\rbrace.
\end{equation}

We undertake a GP regression with the RSD data, and reconstruct the growth rate function $f\sigma_8(z)$, its derivative $\left[f\sigma_{8}\right]'(z)$, and plot the results in Fig. \ref{ch7:fs8_plot}. At the present epoch, the reconstructed values are $f \sigma_{8} (z = 0) = 0.3748 \pm 0.0164$ and $\left[f \sigma_{8}\right]'(z = 0) = 0.2148 \pm 0.0709$ respectively. The marginalized constraints on $\Omega_{m0}$ and $\gamma$ are obtained via a $\chi^2$ minimization between the theoretical $f \sigma_{8}^{\mbox{\tiny theo}}$ (incorporating the reconstructed $E(z)$ from the combined CC+SN datasets in equation \eqref{ch7:fs8_final}) and the GP reconstructed $f\sigma_{8}^{\mbox{\tiny obs}}$ measurements from the RSD data, as
\begin{align}
\chi^2 &= \Delta  \mathbf{V}^T ~\textbf{Cov}^{-1} ~ \Delta \mathbf{V}, \\
\Delta V_i &= f\sigma_{8}^{\mbox{\tiny obs}}(z_i) - f\sigma_{8}^{\mbox{\tiny theo}}(z_i) \\
\mathbf{Cov} &= \mathbf{Cov}^{\mbox{\tiny obs}} +\mathbf{Cov}^{\mbox{\tiny theo}},
\label{ch7:chi2_rsd}
\end{align}
where $\mathbf{Cov}^{\mbox{\tiny obs}}$ is the covariance matrix of $f\sigma_{8}^{\mbox{\tiny obs}}$ and $\mathbf{Cov}^{\mbox{\tiny theo}}$ is the covariance matrix of  $f\sigma_{8}^{\mbox{\tiny theo}}(z)$. 

The parameter $\sigma_{8,0}$ serves as an additional constraint, which can be estimated by substituting $z = 0$ in Eq. \eqref{ch7:fs8_final}, given by 
\begin{equation}
\sigma_{8,0} = \frac{f \sigma_{8} (0)}{\Omega_{m0}^\gamma} .  \label{ch7:s80-constraint}
\end{equation}

Adopting a MCMC analysis with the assumption of uniform priors for $\Omega_{m0} \in [0,1]$ and $\gamma \in [0.4,1.6]$, we obtain the marginalized constraints as $\Omega_{m0} = 0.265 \pm 0.027$ and $\gamma = 0.573 \pm 0.024$ respectively. The best-fit value of $\sigma_{8,0}$ is estimated from Eq. \eqref{ch7:s80-constraint} as $\sigma_{8,0} = 0.802 \pm 0.064$. With these parameter values we plot $\Delta(z)$, $\Delta'(z)$ and $\Delta''(z)$ in Fig. \ref{ch7:delta_plot} where $\Delta = \frac{\delta(z)}{\delta(z=0)}$ is the normalized matter perturbation.

\begin{figure*}[t!]
	\begin{center}
		\includegraphics[angle=0, width=\textwidth]{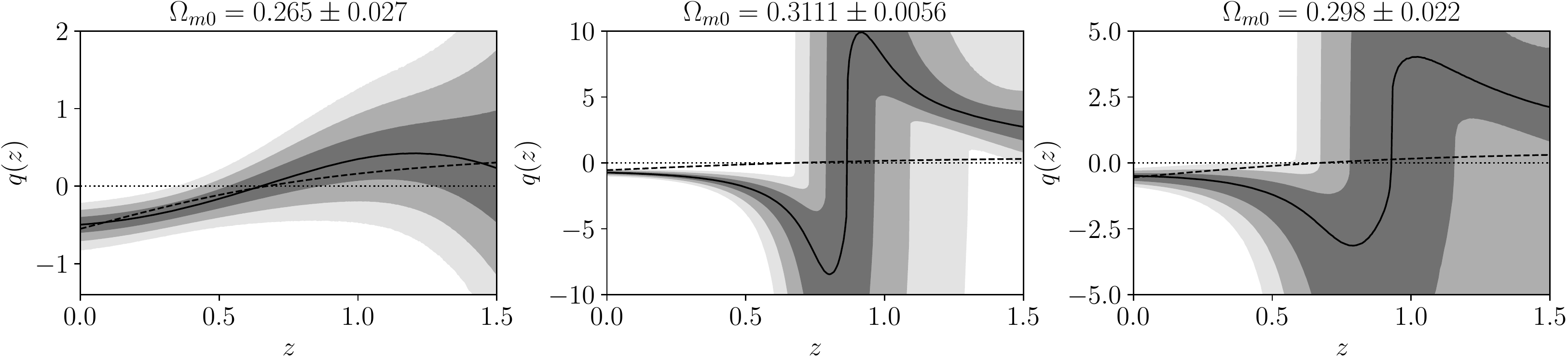}
	\end{center}
	\caption{{\small Plots for the deceleration parameter $q(z)$ reconstructed from the RSD dataset with different $\Omega_{m0}$ priors. The solid black lines 
			represent the mean values of the reconstructed $q(z)$. The black dashed line shows $q(z)$ corresponding to the $\Lambda$CDM model with $\Omega_{m0} 
			= 0.3$.}} \label{ch7:q_rsd_plot}
\end{figure*}

The plot for deceleration parameter $q(z)$ reconstructed from the RSD data using Eq. \eqref{ch7:q_rsd_def} is shown in the left panel of Fig. \ref{ch7:q_rsd_plot}. We observe that the deceleration parameter corresponding to the $\Lambda$CDM model is well contained at the 2$\sigma$ CL in the domain of reconstruction $0<z<1.5$. The reconstructed values of the deceleration parameter at the present epoch $q_0$ and the transition redshift $z_t$ are $q_0 = -0.496^{+0.098}_{-0.102} $ and $z_t = 0.651^{+0.213}_{-0.121}$ respectively.

To test the influence of $\Omega_{m0}$ on the reconstruction, two cases, $\Omega_{m0} = 0.3111 \pm 0.0056$ from the Planck\cite{planck} probe and $\Omega_{m0} = 0.298 \pm 0.022 0$ from the Pantheon SNIa \cite{pan} sample, have been considered as priors. For these two cases, the parameter $\sigma_{8,0}$ is considered to be $\sigma_{8,0} = 0.8102 \pm 0.0060$ from Planck\cite{planck} survey. We proceed with the GP reconstruction of $\Delta(z)$, $\Delta'(z)$ and $\Delta''(z)$ to obtain the cosmic deceleration parameter $q(z)$, arising from the above two cases. The results are shown in the middle and right panels of Fig. \ref{ch7:q_rsd_plot}. From the comparison shown in Fig. \ref{ch7:q_rsd_plot}, we find that the value of $\Omega_{m0}$ leads to contrasting evolutionary scenarios when reconstructing $q(z)$ with the RSD data.

The plots shown in the middle and right panels of Fig. \ref{ch7:q_rsd_plot} indicate a drastic change from a decelerated to an accelerated expansion of the Universe close to $z \approx 0.8$, which is
quite far from the transition redshift $z_t$ estimated from the combined background data. On the other hand, the left panel shows a more sedate transition at $z \approx 0.65$, much closer to the value of $z_t$ obtained on combining the background datasets. Thus, if the value of $z_t$ is more trusted, we find that the value of $\Omega_{m0}$ is relatively less than 0.3. This opens up a new possibility, the RSD data can help in constraining $\Omega_{m0}$ and the value of $z_t$ can itself be observationally used as a new discriminator for cosmological models \cite{zt_ref_lima}.

\section{Discussion}

This work aims to reconstruct the deceleration parameter $q$ from recent observational data without any parametrization ansatz. As mentioned in the introduction, there are already quite a few efforts in this direction. However, as new data are pouring in and new techniques are evolving, revisiting the nature of $q$ with newer datasets is quite imperative. The present work is an endeavour towards that. We focus on a better model-independent treatment of the SN and BAO data by including all the recently updated systematic uncertainties in the CC data. Reconstruction with the RSD data is an entirely new feature that has been included in the present work.

In all cases studied, the common feature is that the mean curve for the reconstructed $q$ shows that the present acceleration has set in quite recently, for $z>0.5$ but well below $z=1$. It should be emphasized that from $z=0$ to roughly $z=0.5$, no deceleration is allowed even in 3$\sigma$. The reconstructed $q(z)$ shows an approximately linear behaviour in $z$ for the redshift range $0<z<z_t$, closely resembling the $\Lambda$CDM behaviour. At higher redshift, beyond $z>1$, the reconstructed $q$ shows a non-monotonic behaviour for the combined CC+Pantheon datasets. The inclusion of BAO data gives rise to an oscillatory behaviour in the reconstructed $q$ at higher redshifts. The reconstructed $q(z)$ from the RSD data using $\Omega_{m0} = 0.265 \pm 0.027$, is similar to the results obtained from the combined CC+Pantheon datasets. We find that the $\Lambda$CDM model is always allowed within a $2\sigma$ CL.

We see that the perturbation value at the present epoch, $\delta_0$, does not effect the reconstruction of $q(z)$. The matter density parameter $\Omega_{m0}$ is found to have a noticeable influence on the reconstruction of $q(z)$ as shown in Fig. \ref{ch7:q_rsd_plot}. The mean values for the reconstructed $q_0$ and the late-time transition redshift $z_t$ are provided.

We repeated the same analysis with the squared exponential covariance function and got similar results, for example, allowing the $\Lambda$CDM model at the 2$\sigma$ CL. We find that this agreement with $\Lambda$CDM is much better at the low redshift regime. At higher $z$, the mean reconstructed curve deviates from the $\Lambda$CDM behaviour with large error bounds.

The two competing values of $H_0$, namely the P18 and R21, can hardly make any qualitative difference in the results as shown in Fig. \ref{ch7:qplot_H0}, except for the R1 combination where the mean values of the reconstructed $q(z)$ shows a negative dip. The N1 and P1 combinations show the possibility of this negative dip in $q$ at higher redshift values. However, this negative dip at high $z$ does not seem to have any high statistical significance, as the reconstructed $q$ in the recent past allows a decelerated expansion as well at the 1$\sigma$ CL for $z>z_t$.

The existing literature on the non-parametric reconstruction of $q$ indicates the presence of a dip in $q$ in the recent past. Bilicki and Seikel\cite{bilicki} worked with Union 2.1 SN, CC and $r$BAO data. Zhang and Xia\cite{xia} found that with the SN Union 2.1 or Union 2 data, a negative $q$ beyond a short-lived deceleration is not allowed in 2$\sigma$, but all the other data sets like CC, $r$BAO  and Gamma Ray Bursts (GRB) indicate a dip in $q$ towards a negative value. Jesus, Valentim, Escobal and Pereira\cite{jesus_nonpara} found constraints on the transition redshift $z_t$, along with a reconstruction of $q$ in a similar non-parametric GP framework with CC and Pantheon SN data. A combination of all the data sets was commonly avoided in \cite{bilicki,xia,jesus_nonpara}. Lin, Li and Tang\cite{lin} worked with the squared exponential covariance using a combination of the Pantheon SN and CC Hubble data. Lin, Li and Tang\cite{lin} found a negative dip in the best fit of reconstructed $q$, indicating an accelerated expansion in the recent past before a short-lived decelerated phase. Recently, G\'{o}mez-Valent\cite{adria} and Haridasu \textit{et al}\cite{haridasu} carried out two extensive analysis for the reconstruction of $q(z)$ using different combination of datasets. The Pantheon+MCT \cite{mct}, recent CC and BAO measurements, and the local R19 $H_0$ measurement \cite{riess} was considered for the reconstruction of $q$ in \cite{haridasu, adria}. Haridasu \textit{et al}\cite{haridasu} found no dip in the best fit values of the reconstructed $q$ although such a dip is allowed at the 1$\sigma$ CL. With the R19 data included, the presence of this dip in $q$ is quite clear in their work.

The present work also tests the possible effects of spatial curvature that has mostly been overlooked in literature mentioned, except in the work of Zhang and Xia\cite{xia}, which however, ignores the combination of datasets. Results show that there is hardly any significant difference between the reconstructed values of $q$ when a non-zero value of the curvature density parameter from the Planck probe\cite{planck} is taken into account.

We have opted for a better model-independent treatment of the Pantheon data, like estimating the marginalized $M_B$ constraints instead of fixing it to the best-fitting $\Lambda$CDM value, as done by Lin, Li and Tang\cite{lin}. Our analysis accounts for all systematic uncertainties within the CC data. We have also obtained the marginalized constraints on $r_d$ to eliminate the effect of any fiducial model dependence linked with BAO measurements. Constraints on $\Omega_{m0}$ have been obtained in a non-parametric way using the RSD data. We find that fine-tuning of these cosmological parameters, like $M_B$, $\Omega_{m0}$ and $r_d$, is desirable for a self-consistent combined analysis.

In conclusion, we can say that not only the nature of dark energy but also the evolutionary history of the Universe is yet to be correctly ascertained. As a general note, we can comment that we need more data, and perhaps a better model-independent treatment of the data as well. Thus, the reconstruction of kinematic parameters, like $q$, will have to be renewed time and again with newer datasets in search of a better understanding of the evolution.

\clearpage{}
\clearpage{}\chapter{Conclusion} \label{ch6}
\chaptermark{Conclusion}

The present thesis contains some investigations regarding the reconstruction of various quantities in cosmology from recent observational data. So, this 
exercise is relevant in the context of modelling the late-time cosmic acceleration. We start with an introduction to cosmology in the first chapter, where 
we describe various models for dark energy. Despite having different theoretical approaches for explaining the accelerated expansion of the Universe, till now,  
none of them has been universally accepted. A {\it reconstruction} is a reverse way of finding the viable cosmological model right from observations.

As discussed in section \ref{ch1:recon_methods}, a reconstruction in cosmology can be based on the parametric or non-parametric approach. For a 
parametric reconstruction, the relevant quantities are represented as simple functions of redshift, along with some model parameters that are 
estimated using observational data. A more unbiased approach is the non-parametric one, where the reconstruction is carried out without assuming 
any functional form.

In this thesis, the reconstructions are based on the non-parametric approach where the prime endeavour is to directly ascertain the evolution of 
different cosmological quantities like the deceleration parameter, jerk parameter, equation of state parameter, etc., from observational data. We 
have focused on the reconstruction of both kinematical as well as dynamical quantities in cosmology. The basic assumption made for reconstructing 
the kinematical quantities is that the Universe is homogeneous and isotropic, thus described by the FLRW metric. For reconstructing the dynamical 
quantities, we make use of the Einstein equations.

The method adopted is the Gaussian Process regression. For a given set of Gaussian-distributed observational data, we use Gaussian processes to reconstruct 
the most probable underlying continuous function describing that data along with its higher derivatives and also obtain the associated confidence levels, 
without limiting to any particular parametrization ansatz. The functions reconstructed via GP are characterized by a zero mean function and a covariance 
function. The latter depends on a set of \textit{hyperparameters} which are obtained by marginalizing the likelihood, given in equation \eqref{ch1:likelihood}. 
Different choices for the covariance function may have different effects on the reconstruction. We have used the squared exponential and the Mat\'{e}rn 
$\nu$ covariance functions in this thesis.

Various combinations of background datasets like the Cosmic Chronometer (CC) measurements of the Hubble parameter, recent compilations of the Type Ia 
Supernova distance modulus data (SN), the Pantheon Supernova compilation of CANDELS and CLASH Multi-Cycle Treasury (MCT) programs obtained by the HST, 
the radial and volume-averaged Baryon Acoustic Oscillation (BAO) data, the Cosmic Microwave Background (CMB) Shift parameter data, and the growth rate 
measurements from redshift space distortions (RSD) have been utilized. On account of the known tussle between the value of $H_0$ as given by the Planck 
2018 data \cite{planck} and that from the HST observations of 70 long-period Cepheids in the Large Magellanic Clouds by the SH0ES team \cite{riess,riess21}, 
reconstruction have been carried out separately, in addition to the primary analysis with the reconstructed value of $H_0$ from the datasets. A summary of 
all the datasets is given in section \ref{ch1:obsdata}.

To establish a standard relation between observational data and theoretical models, distance measurement in cosmology serves as an essential 
tool. Cosmography is strongly dependent on the validity of the cosmic distance duality relation (CDDR) given by Etherington\cite{etherin1993}, 
which connects the angular diameter distance $d_A$ with the luminosity distance $d_L$. The luminosity distance $d_L$ curve is obtained from the 
Pantheon SN-Ia data, and the angular distance $d_A$ curve is derived considering the volume-averaged BAO compilation in combination with the CC 
$H(z)$ measurements, in the same domain of redshift. The reconstruction is worked out avoiding any fiducial bias on the cosmological parameters 
(like the absolute magnitude $M_B$, Hubble parameter $H_0$ at the present epoch, the comoving sound horizon at the photon drag epoch $r_d$ and 
the matter density parameter $\Omega_{m0}$) included in the datasets. It is observed that the theoretical CDDR is in good agreement with the present 
analysis mostly within 1$\sigma$ and always in 2$\sigma$ of the reconstruction.

An important aspect emphasized in this thesis is the kinematic approach to the reconstruction. All cosmological quantities defined from the 
scale factor and its time derivatives are the kinematical quantities, for example, the Hubble parameter, the deceleration parameter, the 
jerk parameter, etc. Chapter \ref{ch4:chap4} focus on the non-parametric reconstruction of the cosmological deceleration $q$ and jerk $j$ parameters. 
The reconstructed deceleration parameter $q$ is seen to have a non-monotonic behaviour that becomes oscillatory at higher redshift. The reconstructed 
jerk parameter reveals the possibility of a non-monotonic evolution. The results are compared with those of the $\Lambda$CDM model. For various combinations 
of datasets, the deceleration and jerk parameters corresponding to the $\Lambda$CDM model are included in the $2\sigma$ CL. The two competing values of 
$H_0$, namely the P18 and R19, hardly make any qualitative difference in this regard. 

The effective equation of state $w_{\mbox{\small eff}}$ has been reconstructed in chapter \ref{ch4:chap4} by inserting the reconstructed values of $q$ 
in the Einstein's equations. Thus, this work can act as a bridge between the kinematic and dynamic approaches for reconstruction. 

In chapter \ref{ch5:chap5}, the possibility of an interaction between dark energy and dark matter has been investigated. The CC Hubble data, Pantheon 
SN-Ia compilation of CANDELS and CLASH MCT programs and $r$BAO Hubble data have been utilized for reconstructing the interaction in the dark sector as 
a function of redshift. We have investigated $\tilde{Q}$, which is the rate of transfer of energy between DE and DM, expressed in a dimensionless way, 
for three different versions of dark energy (i) an interacting vacuum with $w=-1$, (ii) a $w$CDM model where $w$ is close to $-1$ but not exactly equal 
to that and (iii) the CPL parametrization where $w(z) = w_0 + w_a \frac{z}{1+z}$ \cite{cpl_main}. The possibility of {\it no interaction} at all is quite 
likely. Also, the interaction, if any, is not really significant at the present epoch. However, if there is an exchange of energy between dark energy and 
dark matter, it appears that, this energy flows from the former to the latter, consistent with the thermodynamic requirement.

The evolution of the density parameters, ${\Omega}_m$ and ${\Omega}_d$, have been checked in the presence of this interacting scenario. It is observed that 
the dominance of dark energy, as indicated by the reconstruction, is a bit delayed than that in case of the $\Lambda$CDM model. In this chapter, attempts 
have also been made for testing the reconstructed interacting model against the laws of thermodynamics. The results hint towards a possibility that the 
Universe is undergoing a change from a thermodynamic non-equilibrium in the past towards an equilibrium state in the present epoch.

A non-parametric reconstruction of the deceleration parameter is revisited in chapter \ref{ch7:chap7}. As new data are pouring in and new techniques are 
evolving, revisiting the nature of $q$ with newer datasets is quite imperative. Chapter \ref{ch7:chap7} is an endeavour towards that. We have focussed on 
a better model-independent treatment of the Pantheon and BAO datasets, as well as the inclusion of all the recently updated systematic uncertainties in 
the CC dataset. Reconstruction with the RSD data is an entirely new feature that has been included in the present work. In the absence of a universally 
accepted form of dark energy, this kind of revisit is an essential tool for refining the present understanding of the accelerated expansion of the Universe. 
We have also estimated the late-time transition redshift $z_t$ where the reconstructed $q(z)$ shows a signature flip. The reconstructed $q(z)$ shows an 
approximately linear behaviour in $z$ for the redshift range $0<z<z_t$, closely mimicking the $\Lambda$CDM behaviour. Beyond $z>1$, the reconstructed $q$ 
shows a non-monotonic nature that becomes oscillatory at higher redshift. However, the $\Lambda$CDM model is well consistent at the 2$\sigma$ CL. The use of 
any prior $H_0$ measurement, namely the P18 and R21, does not make any qualitative difference in this regard. 

On examining the effect of a non-zero $\Omega_{k0}$ from the Planck survey, there are hardly any significant differences to note for the reconstructed values, 
in comparison to the $\Omega_{k0} = 0$ case. The matter density parameter $\Omega_{m0}$ is found to have a noticeable influence on the reconstruction with 
the RSD data. 
 
It deserves mention that the reconstructed functions in chapters \ref{ch2:chap2}, \ref{ch4:chap4}, \ref{ch5:chap5} and \ref{ch7:chap7} reveal a common 
interesting feature. The reconstructed quantities are better constrained in the low redshift range $0<z<0.5$ in comparison to the higher redshift. As the 
availability of data in the higher redshift range is much lower, the uncertainties associated with the mean values of the reconstructed functions are quite 
large. One possible way to overcome this problem is to work out the reconstruction at different redshift regimes separately. But there is no proper way to 
correlate between the different redshift bins. Future high-$z$ observations of CC, BAO, SN, RSD and other observables should be able to provide tighter 
constraints at higher redshift values.

\clearpage{}
\thispagestyle{empty}
\clearpage
\clearpage

\cleardoublepage

\bibliographystyle{mybst}
\addcontentsline{toc}{chapter}{Bibliography}
\bibliography{thesis}

\end{document}